\documentstyle[psfig]{elsart}
\input{amssym}

\newcommand{\da}{d_A}
\newcommand{\vir}{{\rm vir}}

\newcommand{\lin}{{\rm lin}}
\newcommand{\lens}{{\rm lens}}

\newcommand{\shell}{{\rm s}}
\def\sun{\hbox{$\odot$}}

\newcommand{\dirac}{{\rm D}}

\def\C{{\bf C}}

\newcommand{\bp}{{\cal C}} 

\newcommand{\bfx}{{\mathbf{x}}}
\newcommand{\bfl}{{\mathbf{l}}}

\newcommand{\veck}{{\bf k}}
\newcommand{\vecq}{{\bf q}}
\newcommand{\vecl}{{\bf l}}
\newcommand{\vecr}{{\bf r}}
\newcommand{\vecx}{{\bf x}}
\newcommand{\vecu}{{\bf u}}
\newcommand{\vecv}{{\bf v}}

\newcommand{\rms}{{\it rms}}
\newcommand{\cmb}{{\rm CMB}}

\newlength{\tskip}
\newlength{\colwidth}

\newcommand{\beq}{\begin{equation}}
\newcommand{\eeq}{\end{equation}}
\newcommand{\beqa}{\begin{eqnarray}}
\newcommand{\eeqa}{\end{eqnarray}}

\def\simgt{\gtrsim}
  
 \newcommand{\wj}{\left(
                          \begin{array}{ccc}
                          l_1  &  l_2  & l_3 \\
                            0  &  0    &  0
                          \end{array}
                          \right)}
\newcommand{\wjmp}[3]{\left(
                       \begin{array}{ccc}
       l_#1 & l_#2  & l_#3  \\
         m_#1 & m_#2  & m_#3
                         \end{array}
                   \right)}
\newcommand{\wjm}{\left(
                          \begin{array}{ccc}
                          l_1  &  l_2  & l_3 \\
                           m_1  &  m_2   &  m_3
                          \end{array}
                          \right)}
\newcommand{\wjma}[6]{\left(
                           \begin{array}{ccc}
         #1 & #2  & #3  \\
         #4 & #5  & #6
                           \end{array}
                   \right)}

\newcommand{\bi}{B_{l_1 l_2 l_3}}

\newcommand{\deld}{\delta^{\rm D}}
\newcommand{\bn}{\hat{\bf n}}
\newcommand{\bm}{\hat{\bf m}}
\newcommand{\bl}{\hat{\bf l}}
\newcommand{\bk}{\hat{\bf k}}
\newcommand{\rad}{r}    

\newcommand{\sky}{{\rm sky}}
\newcommand{\isw}{{\rm ISW}}

\newcommand{\Ylm}[1]{Y_{l_#1}^{m_#1}}
\newcommand{\Ylmn}{Y_{l}^{m}}
\newcommand{\alm}[1]{a_{l_#1 m_#1}}

\newcommand{\dsz}{{\rm kSZ}}
\newcommand{\sz}{{\rm SZ}}

\newcommand{\gal}{{\rm gal}}

\begin{document}
\begin{frontmatter}
\title{Halo models of large scale structure}
\author{Asantha Cooray}
\address{Theoretical Astrophysics, California Institute of Technology,
Pasadena CA 91125. E-mail: asante@caltech.edu}
\author{Ravi Sheth}
\address{Department of Physics \& Astronomy, University of Pittsburgh, 
PA 15260 \\ 
and\\
Theoretical Astrophysics Group, Fermi National
Accelerator Laborotary, Batavia IL 60637 USA. E-mail: sheth@fnal.gov}

\begin{abstract}

We review the formalism and applications of the halo-based description
of nonlinear gravitational clustering. In this approach, all mass is
associated with virialized dark matter halos; models of the number and 
spatial distribution of the halos, and the distribution of dark matter 
within each halo, are used to provide estimates of how the statistical 
properties of large scale density and velocity fields evolve as a result 
of nonlinear gravitational clustering.
We first describe the model, and demonstrate its accuracy
by comparing its predictions with exact results from numerical
simulations of nonlinear gravitational clustering.
We then present several astrophysical  applications of the halo model:
these include models of the spatial distribution of galaxies,
the nonlinear velocity, momentum and pressure fields, descriptions
of weak gravitational lensing,  and estimates of secondary contributions
to temperature fluctuations in the cosmic microwave background.
\end{abstract}
\end{frontmatter}



\newpage
\tableofcontents
\newpage

\section{Introduction}
 
This review presents astrophysical applications of an approach which
has its origins in papers by Jerzy Neyman \& Elizabeth Scott and
their collaborators nearly fifty years ago.  Neyman \& Scott 
\cite{NeySco52} were interested
in describing the spatial distribution of galaxies.  They argued that it
was useful to think of the galaxy distribution as being made up of
distinct clusters with a range of sizes.
Since galaxies are discrete objects, they described how to study
statistical properties of a distribution of discrete points;
the description required knowledge of the distribution
of cluster sizes, the distribution of points around the cluster center,
and a description of the clustering of the clusters \cite{NeySco52}.
At that time, none of these ingredients were known, and so in subsequent
work \cite{NeyScoSha53,NeyScoSha54}, they focussed on inferring these
parameters from data which was just becoming useful for statistical
studies.

\begin{figure*}[t]
\centerline{\psfig{file=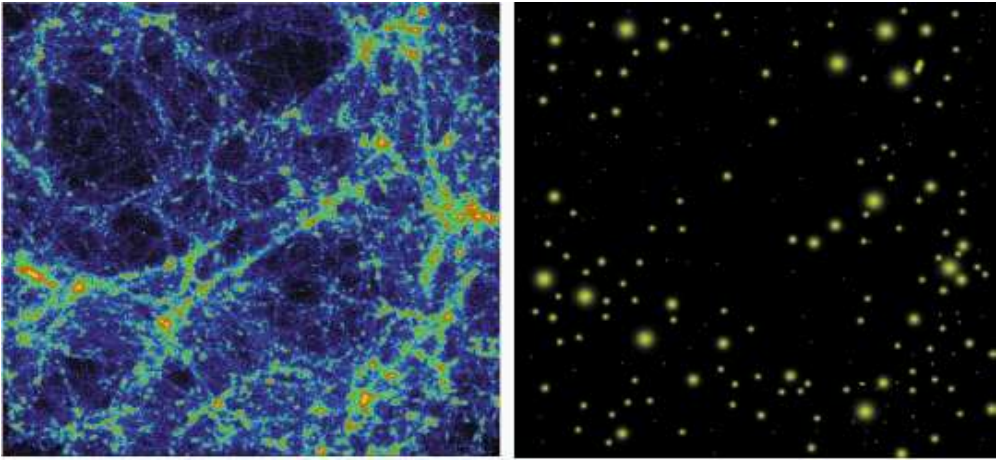,width=5.2in,angle=0}}
\caption{The complex distribution of dark matter (a) found in numerical
simulations can be easily replaced with a distribution of dark
matter halos (b) with the mass function following that found in
simulations and with a profile for dark matter within halos.}
\label{fig:halos}
\end{figure*}

Since that time, it has become clear that much of the mass in the Universe
is dark, and that this mass was initially rather smoothly distributed.
Therefore, the luminous galaxies we see today may be biased tracers of
the dark matter distribution.  That is to say, the relation between the
number of galaxies in a randomly placed cell and the amount of dark matter
the same cell contains, may be rather complicated.  In addition, there is 
evidence that the
initial fluctuation field was very close to a Gaussian random field.
Linear and higher order perturbation theory descriptions of gravitational
clustering from Gaussian initial fluctuations have been developed
(see Bernardeau et al. \cite{Beretal01} for a comprehensive review); these 
describe the
evolution and mildly non-linear clustering of the dark matter, but they
break down when the clustering is highly non-linear (typically, this
happens on scales smaller than a few Megaparsecs).
Also, perturbation theory provides no rigorous framework for describing 
how the clustering of galaxies differs from that of the dark matter.
 
The non-linear evolution of the dark matter distribution has also been
studied extensively using numerical simulations of the large scale 
structure clustering process.
These simulations show that an initially smooth matter distribution evolves
into a complex network of sheets, filaments and knots
(e.g., figure~\ref{fig:halos}).  The dense knots are often called dark
matter halos.  High resolution, but relatively small volume, simulations
have been used to provide detailed information about the distribution of
mass in and around such halos (i.e., the halo density profile of 
\cite{Navetal96,Mooetal99}), whereas larger volume, but lower resolution
simulations (e.g., the Hubble Volume simulations \cite{Evretal01} of the 
Virgo consortium\cite{Thoetal98}), have provided information about the 
abundance and spatial distribution of halos \cite{Jenetal01,Coletal99}.  
Simulations such as these show that the halo abundance, spatial distribution, 
and internal density profiles are closely related to the properties of the 
initial fluctuation field. 
When these halos are treated as the analogs of Neyman \& Scott's clusters,
their formalism provides a way to describe the spatial statistics of the 
dark matter density field from the linear to highly non-linear regimes.  
 
\begin{figure*}[t]
\centerline{\psfig{file=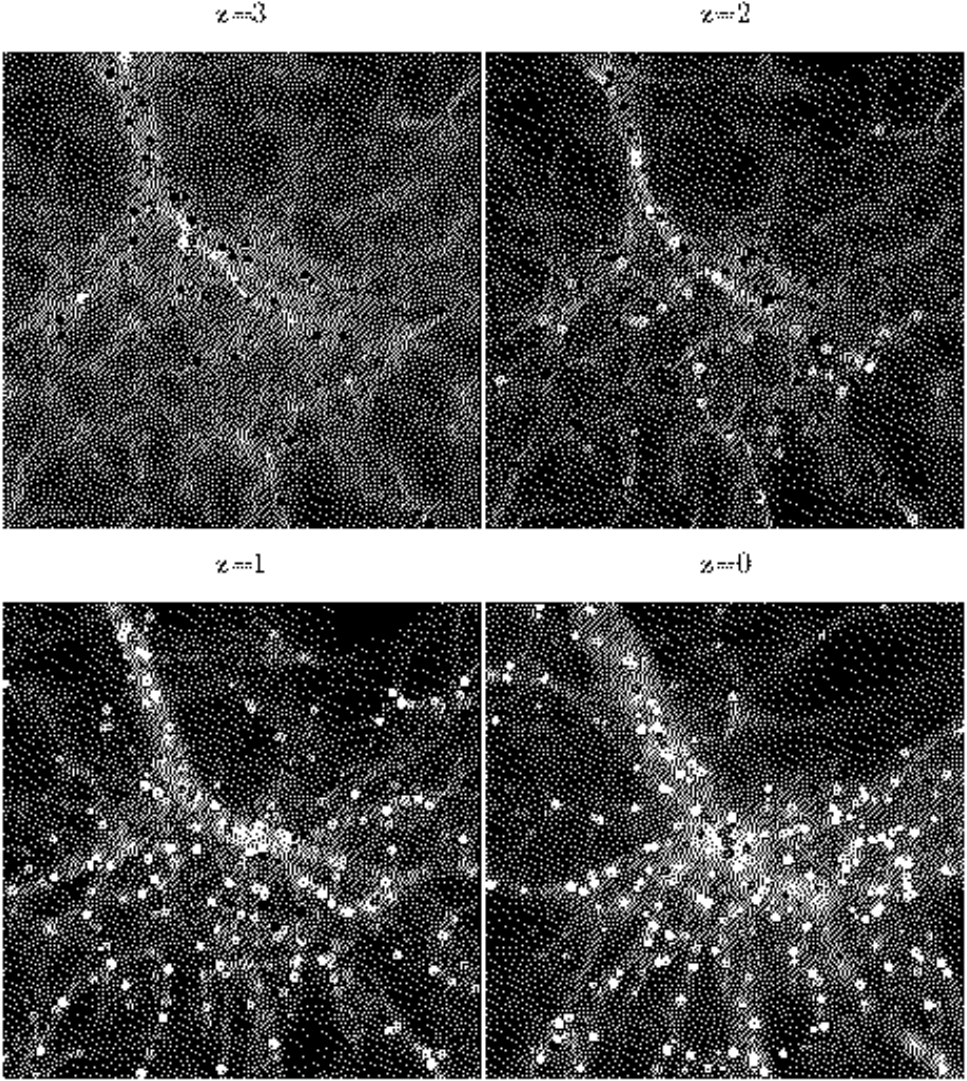,width=5.2in}}
\caption{Distribution of galaxies (in color)  superposed on the dark matter distribution (grey scale) in simulations run  by the GIF collaboration
\cite{Kauetal99}. Galaxy colors blue, yellow, green and red represent successively
smaller star formation rates.  Different panels show how the spatial
distributions of dark matter and galaxies evolve; the relation between
the two distributions changes with time, as do the typical star formation
rates.}
\label{fig:gif}
\end{figure*}

Such a halo based description of the dark matter distribution of large
scale structure is extremely useful because, following
White \& Rees \cite{WhiRee78}, the idea that galaxies form within such
dark matter halos has gained increasing credence.  In this picture,
the physical properties of galaxies are determined by the halos in which
they form.  Therefore, the statistical properties of a given galaxy
population are determined by the properties of the parent halo
population.  There are now a number of detailed semianalytic models
which implement this approach \cite{Kauetal99,SomPri99,Coletal00,Benetal01};
they combine simple physically motivated galaxy formation recipes with
the halo population output from a numerical simulation of the clustering
of the dark matter distribution to make predictions about how the galaxy
and dark matter distributions differ (see, e.g., Figure~\ref{fig:gif}).

In the White \& Rees based models, different galaxy types populate
different halos.  Therefore, the halo based approach provides a simple
and natural way of modelling the dependence of galaxy clustering on
galaxy type in these models.  It is also the natural way of modelling
the difference between the clustering of galaxies relative to dark
matter. 

Just as the number of galaxies in a randomly placed cell may be a
biased tracer of the amount of dark matter in it, other physical
properties such as the pressure, the velocity or the momentum, of a cell
are also biased tracers of the amount of dark matter a cell contains.
The assumption that dark matter halos are in virial equilibrium
allows one to estimate these physical properties for any given halo.
If the distribution of halos in a randomly chosen cell is known,
then the halo--based approach allows one to  estimate statistical
properties of, say, the pressure and the momentum, analogously to how
it transforms the statistics of the dark matter field into that for
galaxies.  

Data from large area imaging and redshift surveys of galaxies
(e.g., the 2dFGRS and the Sloan Digital Sky Survey) are now becoming 
available; these will provide constraints on the dark matter distribution 
on large scales, and on galaxy formation models on smaller scales.  
Weak gravitational lensing \cite{Kai92} provides a more direct probe of 
the dark matter density field.  The first generation of wide-field weak 
lensing surveys, which cover a few square degrees are now complete (see 
recent reviews by Bartlemann \& Schneider~\cite{BarSch01} and 
Mellier~\cite{Mel99}), and the next generation of lensing surveys will 
cover several hundreds of square degrees.  
The Sunyaev-Zel'dovich (SZ) effect~\cite{SunZel80}, due to the inverse-Compton 
scattering of cosmic microwave background (CMB) photons off hot electrons 
in clusters, is a probe of the distribution of the pressure on large scales.  
Several wide-field surveys of the SZ are currently planned 
(see review by Birkinshaw \cite{Bir99}). 
In addition, the next generation CMB experiments will measure temperature 
fluctuations on small angular scales.  On these small scales the density, 
velocity, momentum and pressure fields of the dark and/or baryonic matter 
leave their imprints on the CMB in a wide variety of ways.  For example, 
in addition to the thermal and kinematic SZ effects, the small scale 
temperature fluctuations are expected to be weakly lensed.   
The halo model provides a single self-consistent framework for 
modelling and interpretting all these observations.

The purpose of this review is twofold.  The first is to outline the
principles which underly the halo approach.  The second is to compare
the predictions of this approach with results from simulations and
observations.
Section~1 introduces background materials which are relevant for this
review.  Sections~2 to~5 present the halo approach to clustering.
What we know about dark matter halos is summarized in \S3, 
how this information is incorporated into the halo model is discussed
in \S4, and the first result, 
the halo model description of the dark matter density field is presented
in \S5.
The galaxy distribution is discussed in \S6, the velocity and momentum 
fields are studied in \S7,
weak gravitational lensing in \S8, and secondary effects on the
cosmic microwave background, including the thermal and kinetic
Sunyaev-Zel'dovich effects \cite{SunZel80,OstVis86}, and the non-linear
integrated Sachs-Wolfe effect \cite{SacWol67,ReeSci68}, are the
subject of \S9.  

We have chosen to discuss those aspects of the halo model which are
relevant for the statistical studies of clustering, such as the
two-point correlation functions and higher order statistics.
We do not discuss what cosmological and astrophysical information
can be deduced from the redshift distribution and evolution of halo
number counts.  The abundance of halos at high redshift is an important
ingredient in models of reionization and the early universe.  We do not
discuss any of these models here, since they are described in the recent
review by Barkana \& Loeb \cite{BarLoe01}.  
Finally, our description of the clustering of halos relies on some results
from perturbation theory which we do not derive in detail here.
For these, we refer the reader to the recent comprehensive review on the
pertubation theory description of gravitational clustering by
Bernardeau et al. \cite{Beretal01}.

\section{Background Materials}

This section describes the properties of adiabatic cold dark matter (CDM)
models which are relevant to the present review.

The expansion rate for adiabatic CDM cosmological models with a
cosmological constant is
\begin{equation}
H^2(z) = H_0^2 \left[ \Omega_m(1+z)^3 + \Omega_K (1+z)^2
              +\Omega_\Lambda \right]\,,
\label{eqn:hubble}
\end{equation}
where $H_0$ can be written as the inverse
Hubble distance today $cH_0^{-1} = 2997.9h^{-1} $Mpc.
The critical density is $\rho_{crit} = 3H^2/8\pi G$.
The total density is a sum over different components $i$, where
$i=c$ for the cold dark matter, $\Lambda$ for the cosmological 
constant, and
$b$ for baryons.  Our convention will be to denote the contribution
to the total density from component $i$, as $\Omega_i=\rho_i/\rho_{crit}$.
The contribution of spatial curvature to the expansion rate is
$\Omega_K=1-\sum_i \Omega_i$, and the matter density is
$\Omega_m=\Omega_c+\Omega_b$.

Convenient measures of distance and time include the conformal
distance (or lookback time) from the observer
at redshift $z=0$
\begin{equation}
\rad(z) = \int_0^z {dz' \over H(z')} \,,
\label{eqn:rad}
\end{equation}
(we have set $c=1$) and the angular diameter distance
\begin{equation}
\da = H_0^{-1} \Omega_K^{-1/2} \sinh (H_0 \Omega_K^{1/2} \rad)\,.
\label{eqn:da}
\end{equation}
Note that as $\Omega_K \rightarrow 0$, $\da \rightarrow \rad$
and we define $\rad(z=\infty)=\rad_0$.

\subsection{Statistical description of random fields} 

The dark matter density field in the adiabatic CDM model possesses
$n-$point correlation functions, defined in the usual way.
That is, in real coordinate space,  the $n-$point correlation function 
$\xi_n$ of density fluctuations $\delta(\vecx)$ is defined by
\begin{equation}
 \langle\delta(\vecx_1)\cdots \delta(\vecx_n) \rangle_c \equiv
 \xi_n(\vecx_1,\cdots ,\vecx_n).
\end{equation}
Here, we have expressed the density perturbations in the universe as fluctuations
relative to the background mean density, $\bar{\rho}$:
\begin{equation}
\delta(\vecr) = \frac{\rho(\vecr)}{\bar{\rho}} -1 \, .
\end{equation}

If all the $\vecx_i$ are the same, then
\begin{eqnarray}
 \langle\delta\rangle_c &=& \langle\delta\rangle \\
 \langle\delta^2\rangle_c &=&
   \langle\delta^2\rangle - \langle\delta\rangle_c^2 \equiv \sigma^2\\
 \langle\delta^3\rangle_c &=&
   \langle\delta^3\rangle - 
3\,\langle\delta^2\rangle_c\,\langle\delta\rangle_c
   - \langle\delta\rangle_c^3 \\
 \langle\delta^4\rangle_c &=&
   \langle\delta^4\rangle - 
4\,\langle\delta^3\rangle_c\,\langle\delta\rangle_c
   - 3\,\langle\delta^2\rangle_c^2
   - 6\langle\delta^2\rangle_c\,\langle\delta\rangle_c^2
   - \langle\delta\rangle_c^4.
\end{eqnarray}
We will almost always consider the case in which $\langle\delta\rangle=0$.

Many of the calculations to follow simplify considerably in Fourier 
space.  Throughout, we will use the following Fourier space conventions:
\begin{eqnarray}
 A({\bfx}) &=&
 \int \frac{d^3 \veck}{(2\pi)^3}\, A({\veck})\,\exp({\rm i}\veck\cdot\vecx)
 \qquad {\rm and}\nonumber \\
 \delta_\dirac(\veck_{1\ldots i}) &=& \int \frac{d^3 \vecx}{ (2\pi)^3} \,
              \exp[-{\rm i}\vecx\cdot (\veck_1 + \cdots + \veck_i)] \,
 \label{ftconvention}
\end{eqnarray}
for the Dirac delta function, which is not to be confused with the density
perturbation which does not have the subscript $D$.

Thus, the real space fluctuations in the density field is a sum over 
Fourier modes:
\begin{equation}
 \delta(\vecx) = \int {d^3\veck \over (2\pi)^3}\,\delta(\veck)\,
                      \exp({\rm i}\veck\cdot\vecx)
\end{equation}
and the two, three and four-point Fourier-space correlations are
\begin{eqnarray}
 \langle \delta(\veck_1)\,\delta(\veck_2) \rangle &=&
  (2\pi)^3\,\delta_\dirac (\veck_{12})\, P(k_1) \, , \\
 \langle \delta(\veck_1)\, \delta(\veck_2)\, \delta(\veck_3)\rangle &=&
  (2\pi)^3\, \delta_\dirac (\veck_{123})\,B(\veck_1,\veck_2,\veck_3) \, ,\\
 \langle \delta(\veck_1) \ldots \delta(\veck_4)\rangle_c &=&
 (2\pi)^3\,\delta_\dirac 
(\veck_{1234})\,T(\veck_1,\veck_2,\veck_3,\veck_4)\, ,
\label{eqn:pobitri}
\end{eqnarray}
where $\veck_{i\ldots j} = \veck_i + \ldots + \veck_j$. 
The quantities $P$, $B$ and $T$ are known as the power spectrum,
bispectrum and trispectrum, respectively.  Notice that
\begin{equation}
\xi_2(r) = \int \frac{d^3\veck}{(2\pi)^3} \, P(k)\,\exp({\rm 
i}\veck\cdot\vecr);
\end{equation}
the two-point correlation function and the power spectrum are
Fourier transform pairs.  Similarly, we can relate 
higher order correlations and their Fourier space analogies.

Rather than working with $P(k)$ itself, 
it is often more convenient to use the dimensionless quantity
\begin{equation}
\Delta(k) \equiv {k^3\,P(k)\over 2\pi^2} \, ,
\end{equation}
which is the power per logarithmic interval in wavenumber. Similarly, we can define a 
scaled dimensionless quantity for the Nth Fourier space correlation such that  it scales roughly as the logarithmic power spectrum defined above:
\begin{equation}
\Delta_N(\veck_1,...,\veck_N) = 	\frac{k^3}{2\pi^2} \left[P_N(\veck_1,...,\veck_N)\right]^{\frac{1}{N}} \, .
\end{equation}

We will also often use the quantity
\begin{equation}
 \sigma^2(R) = \int {dk\over k}\,{k^3\,P(k)\over 2\pi^2}\, |W(kR)|^2 ;
 \label{sigmaR}
\end{equation}
this is the variance in the smoothed density field when the smoothing
window has scale $R$.  If the window is a tophat in real space, then
$W(kR) = [3/(kR)^3] (\sin kR - (kR) \cos kR)$; it is $\exp(-k^2R^2/2)$ if
the real space smoothing window is a Gaussian: 
$\exp[-(r/R)^2/2]/\sqrt{2\pi R^2}$.

The initial perturbations due to inflation are expected to be Gaussian 
\cite{Sta82,Haw82,GutPi82,Baretal83,KolTur90}, so they can be characterized 
by a power spectrum  or a two point correlation function (Wick's theorem
states that, for a Gaussian field, correlations involving an odd number of 
density fluctuations are exactly zero).  
Thus, the bispectrum and trispectrum are defined so that, for a Gaussian 
field, they are identically zero; in the jargon, this means that only the
connected piece is used to define them, hence the subscript $c$ in the
expression equation~(\ref{eqn:pobitri}).  All these quantities evolve.  
Here and throughout, we do not explicitly write the redshift dependence 
when we believe no confusion will arise.

\subsection{Results from perturbation theory}\label{pthy}   

The large scale structure we see today is thought to be due to the 
gravitational evolution of initially Gaussian fluctuations
\cite{Pee82,Bluetal84,Davetal92}.
In an expanding universe filled with CDM particles, the action of
gravity results in the generation of higher order correlations:  the
initially Gaussian distribution becomes non-Gaussian.
The perturbation theory description of the gravitational evolution
of density perturbations is well developed \cite{Beretal01}.
Here we briefly summarize some of the results which are most relevant
for what is to follow.

The evolution of large scale structure density perturbations
 are governed by the continuity  equation,
\begin{equation}
 \frac{\partial \delta}{\partial t} +\frac{1}{a}\nabla \cdot (1+\delta) 
 \vecu =0 \, ,
 \label{eqn:continuity}
\end{equation}
and the Euler equation,
\begin{equation}
 \frac{\partial \delta}{\partial t}+H \vecu +\frac{1}{a}\left[(\vecu \cdot 
 \nabla)\vecu +\nabla \phi\right] =0 \, ,
 \label{eqn:euler}
\end{equation}
where the potential fluctuations due to density perturbations are related 
by the Poisson equation:
\begin{equation}
 \nabla^2 \phi = 4 \pi G \bar{\rho} a^2 \delta,
 \label{poisson}
\end{equation}
while the peculiar velocity  is 
related to the Hubble flow via
\begin{equation}
\vecu = \vecv-H\vecx \, .
\end{equation}

The linear regime is the one in which $\delta \ll 1$.  In the linear 
regime, the continuity and Euler equations may be combined to yield 
\cite{Pee80}
\begin{equation}
 \frac{\partial^2 \delta}{\partial t^2}+2H\frac{\partial \delta}{\partial 
 t} - 4\pi G \bar{\rho}\delta =0 \, .
 \label{eqn:combined}
\end{equation}
This is a second-order differential equation with two independent 
solutions; these correspond to modes which grow and decay with time.  
For our purposes, only the growing mode solution of 
equation~(\ref{eqn:combined}) is relevant.  This has the form \cite{Pee80}
\begin{equation}
 \delta(k,r)=G(r)\,\delta(k,0), 
\end{equation}
where 
\begin{eqnarray}
 G(r) &\propto& {H(r) \over H_0} \int_{z(r)}^\infty dz' (1+z') 
              \left[{H_0 \over H(z')} \right]^3 \nonumber \\
     &\approx& {5\over 2} {\Omega_m(z)/(1+z)\over 
     \Omega_m(z)^{4/7} - \Omega_\Lambda(z) + (1-\Omega_m(z)/2)(1 + \Omega_\Lambda(z)/70)}.
 \label{eqn:growth}
\end{eqnarray}
This shows that the linear theory density field may be scaled in time, 
or redshift, with the use of the growth solution $G(z)$.  
Note that $G \propto a=(1+z)^{-1}$ as $\Omega_m\to 1$. 
The approximation in the second line of equation~(\ref{eqn:growth}) is 
good to a few percent \cite{Lahetal91,LigSch90,CarPreTur92}.

In linear perturbation theory\footnote{It should be understood that
``$\lin$'' denotes here the lowest non-vanishing order of perturbation 
theory for the object in question. For the power spectrum, this is linear 
perturbation theory; for the bispectrum, this is second order 
perturbation theory, etc.}, the power spectrum of the
initial density fluctuation field is
\begin{equation}
 \frac{k^3P^\lin(k)}{2\pi^2} = \delta_H^2 \,
                               \left({k \over H_0} \right)^{n+3}T^2(k) \, .
 \label{eqn:linpk}
\end{equation}
Here, $n$ is said to be the slope of the initial spectrum.
A scale free form for $P(k)\sim k^n$ is rather generic; models of
inflation generally produce $n \sim 1$ (the so-called Harrison-Zel'dovich
spectrum \cite{Har70,Zel72,PeeYu70}).
The quantity $T(k)$, defined such that $T(0)=1$, describes departures
from the initially scale free form.  Departures are expected because the 
energy density of the Universe is dominated by radiation at early times 
but by matter at late times, and the growth rate of perturbations in the 
radiation dominated era differs from that in the matter dominated era.  
The transition from one to the other produces a turnover in the shape of 
the power spectrum \cite{BonEfs84,Bluetal84}.  Baryons and other species, 
such as massive neutrinos, leave other important features in the transfer 
function which can potentially be extracted from observational data.   
Accurate fitting functions for $T(k)$ which include these effects have 
been available for some time \cite{Baretal86,Hol89,EisHu98}.  
When illustrating calculations presented in this review, we use fits to 
the transfer functions given by \cite{EisHu99}.

When written in comoving coordinates, the continuity equation 
(\ref{eqn:continuity}) shows that the Fourier transforms of the linear 
theory (i.e., when $\delta\ll 1$) density and velocity fields are related 
\cite{Pee80}:
\begin{equation}
 \vecu(\veck) =  -i\, \dot{G}\, \delta(k)\, \frac{\veck}{k^2} \, ,
 \label{eqn:linvel}
\end{equation}
where the derivative is with respect to the radial distance $r(z)$ 
defined in equation~(\ref{eqn:rad}).  This shows that the power spectrum 
of linear theory velocities is
\begin{equation}
P^\lin_{\rm vel}(k) = \frac{\dot{G}^2}{k^2}P^\lin(k) \, .
\label{eqn:velpk}
\end{equation}
The fluctuations in the linear density field are also simply related to 
to those in the potential \cite{Bar80}.  In particular, the Fourier 
transform of the Poisson equation (\ref{poisson}) shows that
\begin{equation}
 \Phi(k) =\frac{3}{2} \Omega_m \left( \frac{H_0}{k}\right)^2 
          \left[1+3 \left(\frac{H_0}{k}\right)^2\Omega_K\right]^{-2}
          \left(\frac{G}{a}\right)\, \delta(k) \, .
 \label{eqn:poisson}
\end{equation}

Since gravity induces higher order correlations in the density field,
perturbation theory can be used to calculate them also.
The bispectrum, i.e., the Fourier transform of the three point correlation 
function of density perturbations, can be calculated using second 
order perturbation theory \cite{Goretal86,Maketal92,JaiBer96}:
\begin{eqnarray}
 B^\lin(\veck_p,\veck_q,\veck_r) &=& 
  2 F_2^{\rm s}(\veck_p,\veck_q)P(k_p)P(k_q) + 2\; {\rm Perm.}, \nonumber \\
 \label{eqn:bpt}
\end{eqnarray}
where 
\begin{equation}
 F_2^{\rm s}(\vecq_1,\vecq_2) = \frac{1}{2}
 \left[(1+\mu)+\frac{\vecq_1 \cdot \vecq_2}{q_1q_2}
 \left(\frac{q_1}{q_2}+\frac{q_2}{q_1}\right) + 
 (1-\mu)\frac{(\vecq_1 \cdot \vecq_2)^2}{q_1^2 q_2^2} \right]\, .
\label{eqn:f2}
\end{equation}
The bispectrum depends only weakly on $\Omega_m$, the only dependence 
coming from the fact that
$\mu\approx (3/7)\Omega_m^{-2/63}$ for $0.05<\Omega_m<3$ \cite{KamBuc99}.  

The expressions above show that, in perturbation theory, the bispectrum 
generally scales as the square of the power spectrum.  Therefore, it is 
conventional to define a reduced bispectrum:
\begin{equation}
 Q_{123} \equiv \frac{B_{123}}{P_1P_2+P_2P_3+P_3P_1}.
 \label{Q3}
\end{equation}
To lowest order in perturbation theory, $Q$ is independent of time and 
scale \cite{FrySel82,Fry84}.  When the $k$ vectors make an 
equililateral triangle configuration, then 
\begin{equation}
 Q_{\rm eq}(k) \equiv {1 \over 3}
 \left[ {\Delta_{\rm eq}^2(k) \over \Delta^2(k)} \right]^2,
 \quad {\rm where}\quad 
 \Delta_{\rm eq}^2(k) \equiv \frac{k^3}{2\pi^2} \sqrt{B(k,k,k)} 
\end{equation}
represents the bispectrum for equilateral triangle configurations.
In second order perturbation theory, 
\begin{equation}
 Q_{\rm eq}^{\rm PT} = 1 - \frac{3}{7}\Omega_m^{-2/63};
 \label{Q3PT}
\end{equation}
this should be a good approximation on large scales.  

Similarly, the perturbation theory trispectrum is 
\begin{eqnarray}
 && T^\lin = 4 \left[F_2^{\rm s}(\veck_{12},-\veck_1) 
                     F_2^{\rm s}(\veck_{12},\veck_3)\,P(k_1)P(k_{12})P(k_3)
                   + {\rm Perm.}\right] \nonumber \\
 &&\qquad \qquad 
           + 6 \left[F_3^{\rm s}(\veck_1,\veck_2,\veck_3)P(k_1)P(k_2)P(k_3) 
                     + {\rm Perm.}\right]\, ;
 \label{eqn:tript}
\end{eqnarray}
there are 12 permutations in the first set and 4 in the second \cite{Fry84}.
The function $F_3^{\rm s}$ can be derived through a recursion relation 
\cite{Goretal86,JaiBer96,Maketal92}
\begin{eqnarray}
 F_n^{\rm s}(\vecq_1,...,\vecq_n) &=& 
 \sum_{m=1}^{n-1} \frac{G_m^{\rm s}(\vecq_1,...,\vecq_m)}
 {(n-1)(2n+3)} \Big[(2n+1)\frac{\vecq_{1,n} \cdot \vecq_{1,m}}{\vecq_{1,m} 
 \cdot \vecq_{1,m}}F_{n-m}^{\rm s}(\vecq_{m+1},...,\vecq_n) \nonumber \\
 && \ + \ \frac{(\vecq_{1,n} \cdot \vecq_{1,n})(\vecq_{1,m}\cdot 
 \vecq_{m+1,n})}{(\vecq_{1,m} \cdot \vecq_{1,m})(\vecq_{m+1,n} \cdot 
 \vecq_{m+1,n})}G_{n-m}^{\rm s}(\vecq_{m+1},...,\vecq_n) \Big]
\end{eqnarray}
with $\vecq_{a,b} = \vecq_a+...+\vecq_b$, $F_1^{\rm s}=G_1^{\rm s}=1$ and
\begin{equation}
 G_2^{\rm s}(\vecq_1,\vecq_2) = \mu+\frac{1}{2}\frac{\vecq_1 \cdot 
 \vecq_2}{q_1q_2}\left(\frac{q_1}{q_2}+\frac{q_2}{q_1}\right)+(1-\mu)\frac{(\vecq_1 
 \cdot \vecq_2)^2}{q_1^2 q_2^2} \, ,
\end{equation}
where $\mu$ has the dependence on $\Omega_m$ as in $F_2^{\rm s}$ 
(equation~\ref{eqn:f2}).
The factor of 2 in equation~(\ref{eqn:bpt}) and the factors of
4 and 6 in equation~(\ref{eqn:tript}) are due to the use of symmetric
forms of the $F_n^{\rm s}$.  Once again, it is useful to define 
\begin{equation}
 Q_{1234} \equiv 
 \frac{T_{1234}}{[P_1P_2P_{13} + {\rm cyc.}] + [P_1P_2P_3 + {\rm cyc.}]},
\label{Q4}
\end{equation}
where the permutations include 8 and 4 terms respectively in the ordering of
$(k_1,k_2,k_3,k_4)$.  For a square configuration, 
\begin{equation}
Q_{\rm sq}(k) \equiv
\frac{T(\veck,-\veck,\veck_\perp,-\veck_\perp)}{[8P^2(k)P(\sqrt{2}k)][4P^3(k)]}\,.
\end{equation}
In perturbation theory, $Q_{\rm sq} \approx 0.085$. 

The perturbation theory description of clustering also makes predictions
for correlations in real space.  For clustering from Gaussian initial
conditions, the higher order moments of the dark matter distribution
in real space satisfy
\begin{equation}
 \langle\delta^n\rangle = S_n\,\langle\delta^2\rangle^{n-1},
          \qquad{\rm if}\qquad \langle\delta^2\rangle \ll 1,
\label{dnsn}
\end{equation}
where the $S_n$ are numerical coefficients which are approximately
independent of scale over a range of large scales on which
$\langle\delta^2\rangle\ll 1$.  These coefficients are \cite{Ber94b}
\begin{eqnarray}
 S_3^\lin &=& \frac{34}{7}+\gamma_1, \qquad
 S_4^\lin = \frac{60712}{1323} +\frac{62}{3} \gamma_1 +
  \frac{7}{3}\gamma_1^2 +\frac{2}{3}\gamma_2 , \quad{\rm and}\nonumber \\
 S_5^\lin &=& \frac{200575880}{305613}+\frac{1847200}{3969} \gamma_1 +
              \frac{1490}{63}\gamma_2 + \frac{50}{9} \gamma_1 \gamma_2 +
              \frac{10}{27} \gamma_3 \, ,
\label{PTsn}
\end{eqnarray}
where
\begin{equation}
\gamma_j \equiv \frac{d^j \ln \sigma^2(R)}{d (\ln R)^j} \,
\end{equation}
and $\sigma(R)$ is defined by inserting the linear theory value of
$P^\lin(k)$ in equation~(\ref{sigmaR}). 
Note that $\gamma_1 = -(n+3)$ and $\gamma_i=0$ for $i>1$
if the initial spectrum is a power-law with slope $n$.
For the CDM family of spectra, one can neglect derivatives of 
$\sigma^2(R)$ with respect to scale for $R \le 20$ $h^{-1}$ Mpc.
Also note that the $S_n^\lin$ depend only slightly on cosmology: 
e.g., the skewness is $S_3^\lin = 4+\frac{6}{7}\Omega_m^{-2/63}+\gamma_1$ 
\cite{Bouetal92,Hivetal95,FosGaz98}. 

Although all derivations to follow will be general, we will often 
illustrate our results with the currently favored $\Lambda$CDM cosmological 
model.
Following the definition of the expansion rate (equation~\ref{eqn:hubble})
and the power spectrum of linear density field (equation~\ref{eqn:linpk}),
the relevant parameters for this model are $\Omega_c=0.30$,
$\Omega_b=0.05$, $\Omega_\Lambda=0.65$, $h=0.65$, and $n=1$.

The associated power spectrum of linear fluctuations is normalized to
match the observed anisotropy in the cosmic microwave background at the
largest scales, i.e., those measured by the COBE mission.  This means 
\cite{BunWhi97} that we set $\delta_H=4.2 \times 10^{-5}$.  
A constraint on the shape of the spectrum also follows from specifying
the amplitude of the power spectrum on a smaller scale.  This additional
constraint is usually phrased as requiring that $\sigma_8$, the rms value
of the linear density fluctuation field when it is smoothed with a tophat
filter of scale $R=8 h^{-1}$ Mpc (i.e., $\sigma_8$ is calculated using
equation~\ref{sigmaR} with $R=8 h^{-1}$ Mpc), has a specified value.
This value is set by requiring that the resulting fluctuations are
able to produce the observed abundance of galaxy clusters.
Uncertainties in the conversion of X-ray flux, or temperature,
to cluster mass, yield values for $\sigma_8$ which are in the range
(0.5 -- 0.6)$\Omega_m^{-(0.5 - 0.6)}$ \cite{Ekeetal96,ViaLid99}.
Another constraint on the value of $\sigma_8$ is that when properly
evolved to past, the same density power spectrum should also match
associated fluctuations in the CMB; the two constraints are generally in
better agreement in a cosmology with a cosmological constant than in an
open universe.  The use of a realistic value for $\sigma_8$ is important
because higher order correlations typically depend non-linearly on the
amplitude of the initial linear density field.

\subsection{Beyond perturbation theory}
\label{sec:pdfit}

The following ansatz, due to \cite{Hametal91}, provides a good description
of the two-point correlations in real space, $\xi_2(r)$, and in Fourier 
space, $P(k)$, 
even in the regime
where perturbation theory becomes inaccurate.  The argument is that
non-linear gravitational evolution rescales all lengths, so pairs initially
separated by scale $r_{\rm L}$ will later be separated by a different 
scale, say $r_{\rm NL}$.  The initial and final scales are related by
\begin{equation}
 r_{\rm NL} = {r_{\rm L}\over [1+\bar\xi_{\rm NL}(r_{\rm NL})]^{1/3}}
 \qquad {\rm where} \qquad
 \bar\xi_{\rm NL}(x) = {3\over x^3} \int_0^x dr\ r^2\,\xi_2(r).
\end{equation}
The ansatz is that there exists some universal function
\begin{equation}
 \bar\xi_{\rm NL}(r_{\rm NL}) = 
 X_{\rm NL}\Bigl[\bar\xi_{\rm L}(r_{\rm L})\Bigr].
\end{equation}
This is motivated by noting that
$\bar\xi_{\rm NL}(r_{\rm NL}) \propto \bar\xi_{\rm L}(r_{\rm L})$
in the linear regime where $\bar \xi_{\rm NL}\ll 1$, and
$\bar\xi_{\rm NL}(r_{\rm NL}) \propto [\bar\xi_{\rm L}(r_{\rm L})]^{3/2}$
in the highly non-linear regime where $\bar\xi\gg 1$.  The 3/2 scaling
comes if $\bar\xi_{\rm NL}\propto a^3$ in the highly non-linear regime
(this is expected if, on the smallest scales, the expansion of the
background is irrelevant), whereas $\bar\xi_{\rm L}\propto a^2$.
In the intermediate regime, it has been argued that 
$\bar\xi_{\rm NL}(r_{\rm NL}) \propto [\bar\xi_{\rm L}(r_{\rm L})]^3$
\cite{NitPad94}. The exact transitions between the different regimes, 
however, must be calibrated using numerical simulations.
 
For similar reasons, one expects a scaling for the non-linear power 
spectrum
\begin{equation}
 k_{\rm L} = {k_{\rm NL}\over [1 + \Delta_{\rm NL}(k_{\rm NL})]^{1/3}}
 \qquad {\rm with}\qquad
 \Delta_{\rm NL}(k_{\rm NL}) = f_{\rm NL}\Bigl[\Delta_{\rm L}(k_{\rm L})\Bigr].
\end{equation}
Fitting formulae for $X_{\rm NL}$ and $f_{\rm NL}$, obtained by calibrating
to numerical simulations, are given in \cite{PeaDod94,JaiMoWhi95}; in what 
follows we will use the fits given by \cite{PeaDod96} for the power spectrum 
with
\begin{equation}
 f_{\rm NL}(x) = x\left[\frac{1+B\beta x+\left(Ax\right)^{\alpha \beta}}
  {1+\left(\left[Ax\right]^\alpha g^3/\left[Vx^{1/2}\right]\right)^\beta}
  \right]^{1/\beta} \, ,
\label{eqn:pdfit}
\end{equation}
where,
\begin{eqnarray}
 A &=& 0.482 \left(1+\frac{n}{3}\right)^{-0.947} \; \; \;
 B = 0.226 \left(1+\frac{n}{3}\right)^{-1.778} \nonumber \\
 \alpha &=& 3.310 \left(1+\frac{n}{3}\right)^{-0.244} \; \; \;
 \beta = 0.862 \left(1+\frac{n}{3}\right)^{-0.287} \nonumber \\
 V &=& 11.55 \left(1+\frac{n}{3}\right)^{-0.423} \, ,
\end{eqnarray} 
and
\begin{equation}
 n(k_L) = \frac{d\, {\rm ln}\,P}{d\, {\rm ln}\,k}\Big|_{k=k_L/2} \, .
\end{equation}
Note that in equation~(\ref{eqn:pdfit}), the redshift dependence comes
only from the factor of $g^3$, where $g$ is the growth suppression factor 
relative to an $\Omega_m=1$ universe: $g=(1+z)\,G(z)$, with $G(z)$ the 
linear growth factor given in equation~(\ref{eqn:growth}).  
The parameters of the above fit come from a handful of simulations and 
are valid for a limited number of cosmological models. 
Fits to the non-linear power spectrum in some cosmological models 
containing dark energy are provided by \cite{Maetal99}.  

Although this ansatz, and the associated fitting function represents a 
significant step beyond perturbation theory, there have been no successful 
extensions of it to higher order clustering statistics.  In addition, it 
is not obvious how to extend it to describe fields other than the density 
of dark matter.

Hyper-extended perturbation theory (HEPT; \cite{ScoFri99}) represents 
a reasonably successful attempt to extract what is known from
perturbation theory and apply it in the highly nonlinear regime.  
This model makes specific predictions about higher order clustering.  
For example, 
\begin{equation}
Q_{\rm eq}^{\rm HEPT}(k) = \frac{4 - 2^{n}}{1+2^{n+1}} 
\label{Q3hept}
\quad {\rm and}\quad  
Q_{\rm sq}^{\rm sat} = \frac{1}{2}\left[\frac{54 - 27\cdot 2^n +
2\cdot 3^n + 6^
n}{1+6\cdot2^n + 3\cdot 3^n + 6\cdot 6^n}\right],
\end{equation}
where $n(k)$ is the {\it linear} power spectral index at $k$.  
Fitting functions for $Q_{\rm eq}(k)$ for 
$0.1\lesssim k\lesssim 3h$~Mpc$^{-1}$, calibrated from numerical 
simulations, can be found in \cite{ScoCou00}.

\section{Dark Matter Halo Properties}
\label{sec:halos}

In this section, and the next, we will describe an approach which
allows one to describe all $n-$point correlations of large scale
structure.  This description can be used to study clustering
of a variety of physical quantities, including the dark matter density
field, the galaxy distribution, the pressure, the momentum, and others.
 
The approach assumes that all the mass in the Universe is partitioned up
into distinct units, which we will often call halos.  If distinct halos
can be identified, then it is likely that they are small compared to the
typical distances between them.  This then suggests that the statistics
of the mass density field on small scales are determined by the spatial
distribution within the halos; the precise way in which the halos
themselves may be organized into large scale structures is not important.
On the other hand, the details of the internal structure of the halos
cannot be important on scales larger than a typical halo; on large scales,
the important ingredient is the spatial distribution of the halos.
This realization, that the distribution of the mass can be studied in
two steps: the distribution of mass within each halo, and the spatial
distribution of the halos themselves, is the key to what has come to be
called the halo model.

The halo model assumes that, in addition to thinking of the spatial
statistics in two steps, it is useful and accurate to think of the physics
in two steps also.  In particular, the model assumes that the regime in
which the physics is not described by perturbation theory is confined to
regions within halos, and that halos can be adequately approximated by
assuming that they are in virial equilibrium.
 
Clearly, then, the first and the most important step is to find a suitable 
definition
of the underlying units, i.e. the halos. This section describes
what we know about the abundance, spatial distribution, and internal density
profiles of halos.  All these quantities depend primarily on halo mass.
In the next section, we combine these ingredients together to build the
halo model of large scale structure.

\subsection{The spherical collapse model}
\label{scoll}

The assumption that non-linear objects formed from a spherical collapse
is a simple and useful approximation.  The spherical collapse of an
initially tophat density perturbation was first studied in 1972 by
Gunn \& Gott~\cite{GunGot72}; see~\cite{FilGol84,Ber85} for a
discussion of spherical collapse from other initial density profiles.

In the tophat model, one starts with a region of initial, comoving
Lagrangian size $R_0$.
Let $\delta_i$ denote the initial density within this region.
We will suppose that the initial fluctuations were Gaussian with
an rms value on scale $R_0$ which was much less than unity.
Therefore, $|\delta_i|\ll 1$ almost surely.
This means that the mass $M_0$ within $R_0$ is
$M_0 = (4\pi R_0^3/3) \bar\rho(1+\delta_i)\approx (4\pi R_0^3/3)\bar\rho$
where $\bar\rho$ denotes the comoving background density.
 
As the Universe evolves, the size of this region changes.
Let $R$ denote the comoving size of the region at some later time.
The density within the region is $(R_0/R)^3\equiv (1+\delta)$.
In the spherical collapse model there is a deterministic relation
between the initial comoving Lagrangian size $R_0$ and density of
an object, and its Eulerian size $R$ at any subsequent time.
For an Einstein--de Sitter universe, one can obtain a parametric solution
to $R(z)$ in terms of $\theta$:
\begin{equation}
{R(z)\over R_0} = {(1+z)\over (5/3)\,|\delta_0|}\,{(1-\cos\theta)\over 2}
\quad {\rm and}\quad
{1\over 1+z} = \left({3\over 4}\right)^{2/3}
{(\theta-\sin\theta)^{2/3}\over (5/3)\,|\delta_0|} ,
\label{scmodel}
\end{equation}
where $\delta_0$ denotes the initial density $\delta_i$ extrapolated
using linear theory to the present time (e.g. \cite{Pee80}).
If $\delta_i<0$, then $(1-\cos\theta)$ should be replaced with
$(\cosh\theta-1)$ and $(\theta-\sin\theta)$ with
$(\sinh\theta-\theta)$.
 
In the spherical collapse model, initially overdense regions collapse: with
$\theta=0$ at start, they `turnaround' at $\theta=\pi$, and have
collapsed completely when $\theta = 2\pi$.  Equation~(\ref{scmodel})
shows that the size of an overdense region evolves as
\begin{equation}
{R_0\over R(z)} =
{6^{2/3}\over 2}{(\theta-\sin\theta)^{2/3}\over (1-\cos\theta)}.
\end{equation}
At turnaround, $\theta=\pi$, so $[R_0/R(z_{\rm ta})]^3 = (3\pi/4)^2$;
when an overdense region turns around, the average density within it
is about 5.55 times that of the background universe.
 
At collapse, the average density within the region is even higher:
formally, $R(z_{\rm col})=0$, so the density at collapse is infinite.
In practice the region does not collapse to vanishingly small size:
it virializes at some non-zero size.  The average density within the
virialized object is usually estimated as follows.  Assume that after
turning around the object virializes at half the value of the turnaround
radius in physical, rather than comoving units.  In the time between
turnaround and collapse, the background universe expands by a factor of
$(1+z_{\rm ta})/(1+z_{\rm col}) = 2^{2/3}$ (from equation~\ref{scmodel}),
so the virialized object is eight times denser than it was at turnaround
(because $R_{\rm vir}=R_{\rm ta}/2$).  The background density at
turnaround is $(2^{2/3})^3 = 4$ times the background density at
$z_{\rm vir}$.  Therefore, the virialized object is
\begin{equation}
\Delta_{\rm vir} \equiv (9\pi^2/16) \times 8 \times 4 = 18\pi^2 \, ,
\label{Dvir}
\end{equation}
times the density of the background at virialization.
 
What was the initial overdensity of such an object?
The first of equations~(\ref{scmodel}) shows that if the region
is to collapse at $z$, the average density within it must have had
a critical value, $\delta_{\rm sc}$, given by
\begin{equation}
{\delta_{\rm sc}(z)\over 1+z} = {3\over 5}\,\left({3\pi\over 
2}\right)^{2/3}.
\label{dsc}
\end{equation}
Thus, a collapsed object is one in which the initial overdensity,
extrapolated using linear theory to the time of collapse, was
$\delta_{\rm sc}(z)$.  At this time, the actual overdensity is
significantly larger than the linear theory prediction.   Although
the formal overdensity is infinite, a more practical estimate
(equation~\ref{Dvir}) says that the object is about 178 times denser
than the background.
 
There is an important feature of the spherical collapse model which
is extremely useful.  Since $(1+\delta)=(R/R_0)^3$, the equations
above provide a relation between the actual overdensity $\delta$,
and that predicted by linear theory, $\delta_0$, and this relation
is the same for all $R_0$.  That is to say, it is the ratio $R/R_o$
which is determined by $\delta_i$, rather than the value of $R$ itself.
Because the mass of the object is proportional to $R_0^3$, this means
that the critical density for collapse $\delta_{\rm sc}$ is the same
for all objects, whatever their mass.  In addition, the evolution of
the average density within a region which is collapsing is also
independent of the mass within it (of course, it does depend on the
initial overdensity).
 
To see what this relation is, note that the parametric solution of
equation~(\ref{scmodel}) can be written as a formal series expansion,
the first few terms of which are \cite{Ber94a}
\begin{equation}
{\delta_0\over 1+z} = \sum_{k=0}^\infty a_k\,\delta^k =
       \delta - {17\over 21}\,\delta^2 + {341\over 567}\,\delta^3
         - {55805\over 130977}\,\delta^4 + \ldots
\label{bndo}
\end{equation}
To lowest order this is just the linear theory relation:
$\delta$ is the initial $\delta_0$ times the growth factor.
A good approximation to the spherical collapse relation
$\delta_0(\delta)$, valid even when $\delta\gg 1$, is \cite{MoWhi96}
\begin{equation}
{\delta_0\over 1+z} = {3\,(12\pi)^{2/3}\over 20}
     - {1.35\over (1+\delta)^{2/3}}
     - {1.12431\over(1+\delta)^{1/2}} + {0.78785\over 
(1+\delta)^{0.58661}}.
\label{mow}
\end{equation}

While these are all convenient estimates of the parameters of collapsed
objects, it is important to bear in mind that the collapse is seldom
spherical, and that the estimate for the virial density is rather adhoc.
Descriptions of ellipsoidal collapse have been considered 
\cite{Ick73,BonMye96a,Mon95,Sheetal01a}, as have
alternative descriptions of the $\delta_0(\delta)$ relation \cite{Engetal00}.
In most of what follows, we will ignore these subtleties.

Though we have used an Einstein-de Sitter model to outline
several properties related to spherical collapse,
our discussion remain qualitatively similar in cosmologies
for which $\Omega_m \le 1$ and/or $\Omega_\Lambda \ge 0$.
The actual values of $\delta_{\rm sc}$ and $\Delta_{\rm vir}$ 
depend on cosmology:  fitting functions for these are available in 
the literature \cite{Ekeetal96,Navetal96,NakSut97,BryNor98,Hen00}.

\subsection{The average number density of halos}
\label{nofm}

Let $n(m,z)$ denote the comoving number density of bound objects,
halos, of mass $m$ at redshift $z$.  (Some authors use $dn/dm$ 
to denote this same quantity, and we will use the two notations 
interchangeably.)
Since halos formed from regions in
the initial density field which were sufficiently dense that
they later collapsed, to estimate $n(m,z)$, we must first estimate
the number density of regions in the initial fluctuation field which
were dense enough to collapse.  A simple model for this was provided by 
Press \& Schechter  in Ref. \cite{PreSch74}:
\begin{equation}
 {m^2 n(m,z) \over\bar\rho}\,{{\rm d}m\over m} =
          \nu f(\nu)\,{{\rm d}\nu\over\nu},
\label{nmfv}
\end{equation}
where $\bar\rho$ is the comoving density of the background with
\begin{equation}
 \nu f(\nu) = \sqrt{\nu\over 2\pi}\ \exp(-\nu/2),\qquad {\rm and}\qquad
        \nu\equiv {\delta^2_{\rm sc}(z)\over \sigma^2(m)}.
\label{fps}
\end{equation}
Here $\delta_{\rm sc}(z)$ is the critical density required for
spherical collapse at $z$, extrapolated to the present time using
linear theory. In an Einstein-de Sitter cosmology, $\delta_{\rm sc}(z=0)=1.686$ while in other cosmologies, $\delta_{\rm sc}$ depends weakly on
$\Omega_m$ and $\Omega_\Lambda$ \cite{Ekeetal96}.
In equation~(\ref{fps}), $\sigma^2(m)$ is the variance in the
initial density fluctuation field when smoothed with a tophat
filter of scale $R = (3m/4\pi\bar\rho)^{1/3}$, extrapolated
to the present time using linear theory:
\begin{equation}
 \sigma^2_\lin(m) \equiv 
     \int \frac{{\rm d}k}{k}\, {k^3P^\lin(k)\over 2\pi^2}\, |W(kR)|^2,
\label{eqn:sigmalin}
\end{equation}
where $W(x)=(3/x^3)\,[\sin(x)-x\cos(x)]$.
 
A better fit to the number density of halos in simulations of
gravitational clustering in the CDM family of models is given
by Sheth \& Tormen \cite{SheTor99}:
\begin{equation}
 \nu f(\nu) = A(p)\left(1 + (q\nu)^{-p}\right)
 \,\left({q\nu\over 2\pi}\right)^{1/2}\,\exp(-q\nu/2),
 \label{fgif}
\end{equation}
where $p\approx 0.3$,
$A(p) = [1 + 2^{-p}\Gamma(1/2-p)/\sqrt{\pi}]^{-1} \approx 0.3222$,
and $q\approx 0.75$.
If $p=1/2$ and $q=1$, then this expression is the same as that in
equation~(\ref{fps}).   At small $\nu\ll 1$, the mass
function scales as $\nu f(\nu) \propto \nu^{0.5-p}$.  Whereas the
small mass behavior depends on the value of $p$, the exponential
cutoff at $\nu\gg 1$ does not.  The value $\nu=1$ defines a
characteristic mass scale, which is usually denoted $m_*$:
$\sigma(m_*) \equiv \delta_{\rm sc}(z)$ and is 
$\sim 2 \times 10^{13}$ M$_{\sun}$ at $z=0$; 
note that halos more massive than $m_*$ are rare.

Elegant derivations of equation~(\ref{fps}) in 
\cite{Eps83,PeaHea90,Bonetal91} show that it can be related to a model 
in which halos form from spherical collapse.  When extended to the 
ellipsoidal collapse model described by \cite{BonMye96a}, the same 
arguments give equation~(\ref{fgif}) \cite{Sheetal01a,SheTor02}.  
Alternative models for the the shape of $n(m,z)$ are available in the 
literature \cite{AppJon90,ManSal95,LeeSha98,Han01};
we will not consider them further, however, as 
equation~(\ref{fgif}) has been found to provide a good description of the
mass function in numerical simulations.  

\begin{figure*}[t]
\centerline{\psfig{file=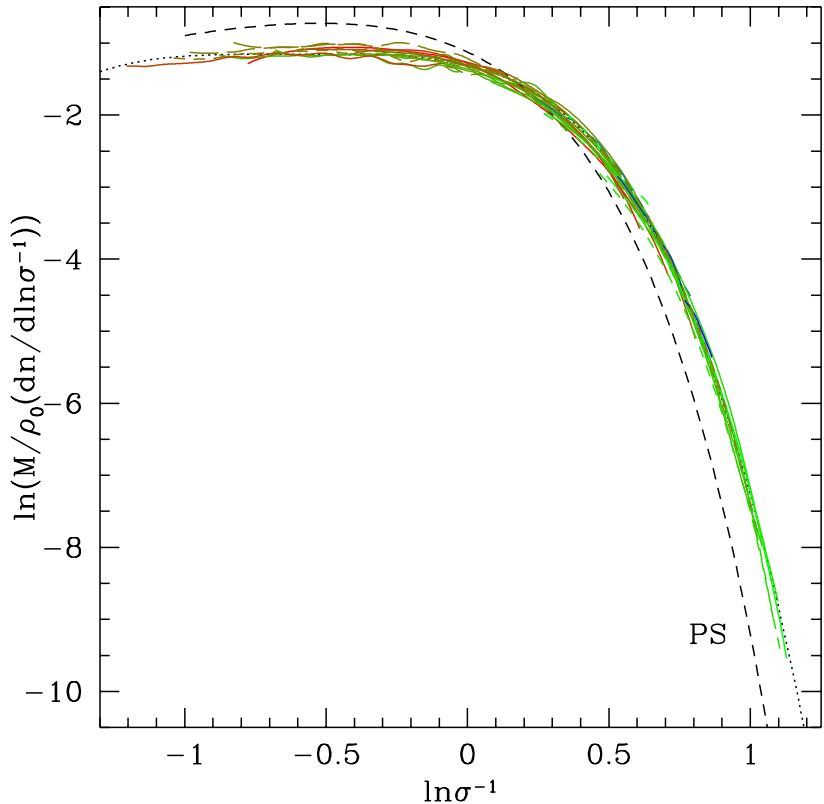,width=4.5in}}
\caption{The halo mass function in numerical simulations of the 
Virgo collaboration. The measured mass distribution is show in color;  
dashed line shows the Press-Schechter mass function; dotted line is 
a fitting formula which is similar to the Sheth-Tormen mass function. 
The figure is from \cite{Jenetal01}.}
\label{fig:massfunction}
\end{figure*} 

This is shown in Figure~\ref{fig:massfunction}, which is taken
from numerical simulations run by the Virgo collaboration \cite{Jenetal01}.
The jagged lines show the mass function at various output times
in the simulation rescaled from mass $m$ to $\sigma(m)$.
The figure shows that, when rescaled in this way,
\begin{equation}
f(\sigma,z) \equiv \frac{m}{\bar{\rho}} \frac{d n(m,z)}{d \ln \sigma^{-1}} \, ,
\end{equation}
is a universal curve (results from all output times in the simulations
trace out approximately the same curve).  The dashed line shows that
this distribution of halo masses is not so well described by
equation~(\ref{fps}).  The dotted line shows
\begin{equation}
f(\sigma) = 0.315 \exp \left( - | \ln \sigma^{-1} + 0.61 |^{3.8}\right) \,;
\end{equation}
this fitting formula is accurate to 20\% in the range
$-1.2\leq\ln\sigma^{-1}\leq 1.05$ \cite{Jenetal01}.
It is very well described by equation~(\ref{fgif}), which is physically
motivated, and so it is equation~(\ref{fgif}) which we will use in what
follows.
 
\subsection{The number density of halos in dense regions}
\label{bofm}

Suppose we divide space up into cells of comoving volume $V$.
The different cells may contain different amounts of mass $M$, which
means they have different densities:  $M/V\equiv \bar\rho(1+\delta)$.
Let $N(m,z_1|M,V,z_0)$ denote the average number of $m$ halos which
collapsed at $z_1$, and are in cells of size $V$ which contain mass
$M$ at $z_0$.  The overdensity of halos in such cells is
\begin{equation}
\delta_{\rm h}(m,z_1|M,V,z_0) = {N(m,z_1|M,V,z_0)\over n(m,z_1)V} - 1.
\label{dh}
\end{equation}
Since we already have a model for the denominator, to proceed, we
need a good estimate of $N(m,z_1|M,V,z_0)$.
 
A halo is a region which was sufficiently overdense that it collapsed.
So the number of halos within $V$ equals the initial size of $V$ times
the number density of regions within it which were sufficiently dense
that they collapsed to form halos.  If $V$ is overdense today, it's
comoving size is smaller than it was initially;
the initial comoving size was $M/\bar\rho = V(1+\delta)$.
If we write $N(m,z_1|M,V,z_0)=n(m,z_1|M,V,z_0)V(1+\delta)$, then we
need an estimate of the number density $n(m,z_1|M,V,z_0)$.
 
The average number density of halos $n(m,z_1)$ is a function of the
critical density required for collapse at that time:
$\delta_{\rm sc}(z_1)$.  In the present context, $n(m,z)$ should be
thought of as describing the number density of halos in extremely
large cells which are exactly as dense as the background (i.e., cells
which have $M\to\infty$ and $\delta = 0$).
Denser cells may be thought of as regions in which the critical
density for collapse is easier to reach, so a good approximation to
$n(m,z_1|M,V,z_0)$ is obtained by replacing $\delta_{\rm sc}(z)$ in
the expression for $n(m,z)$ with
 $\delta_{\rm sc}(z_1) - \delta_0(\delta,z_0)$ \cite{MoWhi96}.
Note that we cannot use $\delta$ itself, because $\delta_{\rm sc}(z_1)$
has been extrapolated from the initial conditions using linear theory,
whereas $\delta$, being the actual value of the density, has been
transformed from its value in the initial conditions using non-linear
theory.  Equations~(\ref{bndo}) and~(\ref{mow}) show the spherical
collapse model for this non-linear $\delta_0(\delta,z)$ relation.
Here, $\delta_0(\delta,z_0)$ denotes the initial density, extrapolated
using linear theory, which a region must have had so as to have density
$\delta$ at $z_0$.
 
Thus, a reasonable estimate of the density of $m$-halos which
virialized at $z_1$ and are in cells of size $V$ with mass $M$
at $z_0$ is
\begin{equation}
{m^2 n(m,z_1|M,V,z_0)\over\bar\rho} {{\rm d}m\over m} =
 \nu_{10}\,f(\nu_{10})\,{{\rm d}\nu_{10}\over\nu_{10}}
\ {\rm where}\ \nu_{10} =
{[\delta_{\rm sc}(z_1)-\delta_0(\delta,z_0)]^2\over 
\sigma^2(m)-\sigma^2(M)}, \nonumber \\
\label{nmM}
\end{equation}
and $f(\nu)$ is the same functional form which described the
unconditional mass function (equation~\ref{fps} or~\ref{fgif}).

Two limits of this expression are interesting.  As $V\to\infty$,
$\delta\to\infty$ and $\delta_0\to\delta_{\rm sc}(z_0)$ independent of the
value of $M$.  A region of small size which contains mass $M$, however, is what
we call a halo, with mass $M$.  Thus, if we are given a halo of mass
$M$ at $z_0$, then $N(m,z_1|M,V=0,z_0)$ is the average number of subclumps
of mass $m$ it contained at the earlier time when $z_1\ge z_0$.  This limit
of equation~(\ref{nmM}) gives what is often called the conditional or
progenitor mass function \cite{Bow91,Bonetal91,LacCol93,SheTor01}.
The opposite limit is also very interesting.
As $V\to\infty$, $M\to\infty$ as well: in this limit,
$\sigma^2(M)\to 0$ and $|\delta|\to 0$, and so equation~(\ref{nmM})
reduces to $n(m,z_1)$, as expected.
 
Suppose that we are in the large cell limit.  By large, we mean that
the rms density fluctuation in these cells is much smaller than unity.
Thus, $|\delta|\ll 1$ in most cells and we can use
equation~(\ref{bndo}) for $\delta_0(\delta)$.
Large cells contain large masses, so $M$ in these cells is much larger
than the mass $m_*$ of a typical halo.  In this limit,
$\sigma(M)\ll \sigma(m)$ for most values of $m$ allowing one to
set $\sigma(M)\to 0$.
This leads to
\begin{equation}
n(m,z_1|M,V,z_0) \approx  n(m,z_1) - \delta_0(\delta,z_0)
 \left({\partial n(m,z_1)\over \partial\delta_{\rm 
sc}}\right)_{\delta_{\rm sc}(z_1)} + \ldots \, ,
\end{equation}
such that
\begin{equation}
\delta_{\rm  h}(m,z_1|M,V,z_0)\approx\delta-(1+\delta)\,\delta_0(\delta,z_0)\
 \left({\partial\ {\rm ln}\,n(m,z_1)\over\partial\delta_{\rm 
sc}}\right)_{\rm
 \delta_{\rm sc}(z_1)}.
\end{equation}
Inserting equation~(\ref{fgif}) for $n(m,z)$ and keeping terms to
lowest order in $\delta$ give (e.g.,  
\cite{ColKai89,MoWhi96,SheLem99,SheTor99})
\begin{equation}
\delta_{\rm h}(m,z_1|M,V,z_0) \approx \delta\,
\left(1 + {q\nu-1\over\delta_{\rm sc}(z_1)} + {2p/\delta_{\rm sc}(z_1)\over
1+(q\nu)^p}\right) = b_1(m,z_1)\,\delta,
\label{b1m}
\end{equation}
where $\nu\equiv \delta_{\rm sc}^2(z_1)/\sigma^2(m)$.
This expression states that the overdensity of halos in very large cells
to be linearly proportional to the overdensity of the mass; the constant
of proportionality, $b_1(m,z_1)$, depends on the masses of the halos,
and the redshifts they virialized, but is independent of the size of the 
cells.
 
If $q=1$ and $p=0$, then massive halos (those which have $\nu > 1$ or masses greater than
the characteristic mass scale of $m_*$)
have $b_1(m,z_1)>1$ and are said to be biased relative to the dark matter,
while less massive halos ($\nu < 1$) are anti-biased.
Notice that $b_1$ can be very large for the most massive halos, but it
is never smaller than $1-1/\delta_{\rm sc}(z_1)$.
Equation~(\ref{dsc}) shows that halos which virialized at the present
time (i.e., $z_1=0$), have bias factors which are never less than
$\approx 0.41$.   Since $\delta_{\rm sc}(z_1)\gg 1$, in 
equation~(\ref{dsc}),
halos that virialized at early
times have bias factors close to unity  (See 
\cite{NusDek93,Fry96,TegPee98} 
for a derivation of this limiting case which uses
the continuity equation.)
 
Since $M/V\equiv \bar\rho(1+\delta)$, the results above show that,
in large cells, $n(m|\delta)\approx [1+b_1(m)\delta]\,n(m)$.
Since $b_1(m)\gg 1$ for the most massive halos, they
occupy the densest cells.  It is well known
that the densest regions of a Gaussian random field are more strongly
clustered  than cells of average density
\cite{Ric54,Kai84a,Baretal86}. 
Therefore, the most massive halos must also be more strongly clustered
than low mass halos.  This is an important point to which we will return
shortly.
 
\subsection{The distribution of halos on large scales: Deterministic 
biasing}
 
The linear bias formula is only accurate on large scales.  If we write
\begin{equation}
\delta_{\rm h}(m,z_1|M,V,z_0) = \sum_{k>0} b_k(m,z_1)\,\delta^k
\label{dhbkdk}
\end{equation}
then inserting equation~(\ref{nmM}) in equation~(\ref{dh}),
setting $\sigma(M)\to 0$, and expanding gives \cite{Scoetal01,Moetal97b}
\begin{eqnarray}
b_1(m,z_1) &=& 1 + \epsilon_1 + E_1, \nonumber\\
b_2(m,z_1) &=& 2(1+a_2)(\epsilon_1 + E_1) + \epsilon_2 + E_2, \nonumber\\
b_3(m,z_1) &=& 6(a_2+a_3)(\epsilon_1+E_1) + 3(1+2a_2)(\epsilon_2 + E_2)
                                      + \epsilon_3 + E_3, \nonumber\\
b_4(m,z_1) &=& 24(a_3+a_4)(\epsilon_1+E_1) +
           12[a_2^2 + 2(a_2+a_3)](\epsilon_2+E_2) \nonumber\\
       & & \qquad\qquad\qquad\qquad\qquad
           + 4(1+3a_2)(\epsilon_3 + E_3) + \epsilon_4 + E_4
\label{bk}
\end{eqnarray}
for the first few coefficients.  Here
\begin{eqnarray}
\epsilon_1 &=& {q\nu - 1\over\delta_{\rm sc}(z_1)}, \qquad
\epsilon_2 = {q\nu\over\delta_{\rm sc}(z_1)}\,
             \left({q\nu - 3\over\delta_{\rm sc}(z_1)}\right), \qquad
\epsilon_3 = {q\nu\over\delta_{\rm sc}(z_1)}\,
             \left({q\nu - 3\over\delta_{\rm sc}(z_1)}\right)^2, 
\nonumber\\
\epsilon_4 &=& \left({q\nu\over\delta_{\rm sc}(z_1)}\right)^2
               \left({q^2\nu^2 - 10q\nu + 15\over\delta_{\rm 
sc}(z_1)}\right),
\end{eqnarray}
and
\begin{eqnarray}
E_1 &=& {2p/\delta_{\rm sc}(z_1)\over 1 + (q\nu)^p}, \qquad
{E_2\over E_1} = {1 + 2p\over\delta_{\rm sc}(z_1)} + 2\epsilon_1,\qquad
{E_3\over E_1} = {4(p^2 - 1) + 6pq\nu\over\delta^2_{\rm sc}(z_1)}
                 + 3\epsilon_1^2, \nonumber \\
{E_4\over E_1} &=& {2q\nu\over\delta_{\rm sc}^2(z_1)}
\left({2q^2\nu^2\over\delta_{\rm sc}(z_1)} - 15\epsilon_1\right) 
\nonumber\\
  & & \qquad\qquad + 2{(1+p)\over\delta^2_{\rm sc}(z_1)}
\left({4(p^2-1) + 8(p-1)q\nu + 3\over\delta_{\rm sc}(z_1)} + 
6q\nu\epsilon_1\right).
\end{eqnarray}
If $p=0$, all the $E_k$s are also zero, and these expressions reduce
to well known results from \cite{Moetal97b}. 
By construction, note that the bias parameters obey
consistency relations:
\begin{equation}
 \int dm \, {m\,n(m,z)\over\bar\rho}\, b_k(m,z) =
   \left\{ \begin{array}{ll}
                        1 & {\rm if}\ k=1\\
                        0 & {\rm if}\ k>1
           \end{array}\right. .
 \label{biascon}
\end{equation}

\begin{figure*}[t]
\vspace{-5cm}
\centerline{\psfig{file=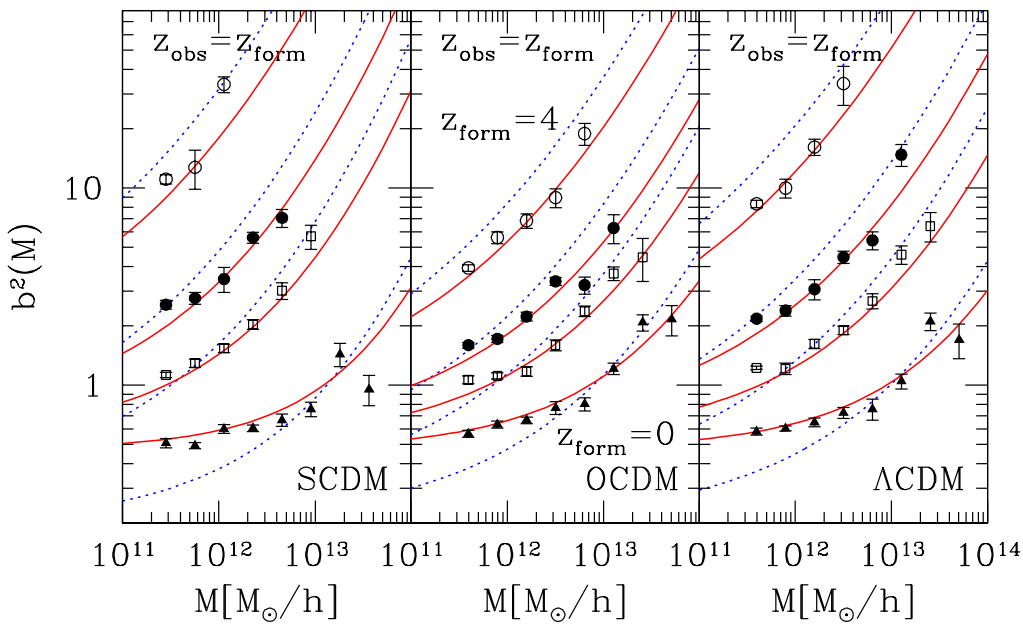,width=5.5in}}
\caption{Large scale bias relation between halos and mass
(from \cite{SheTor99}).  Symbols show the bias factors at $z_{\rm obs}$
for objects which were identified as virialized halos at
$z_{\rm form}=4$,2,1 and 0 (top to bottom in each panel).
Dotted and solid lines show predictions based on the Press-Schechter
and Sheth-Tormen mass functions.  }
\label{fig:bias}
\end{figure*}

Figure~\ref{fig:bias} compares these predictions for the halo bias
factors with measurements in simulations (from \cite{SheTor99}).
Note that more massive halos tend to be more biased, and that
halos of the same mass were more strongly biased at high redshift
than they are today.  The solid and dotted lines show the predictions
from equation~(\ref{bk}) with equations~(\ref{fps}) and~(\ref{fgif})
for the halo mass function, respectively.  Figure~\ref{fig:massfunction}
shows that equation~(\ref{fps}) predicts too few massive halos;
as a result, it predicts a larger bias factor for these massive
halos than is seen in the simulations.  Equation~(\ref{fgif}) provides
an excellent fit to the halo mass function; the associated bias factors
are also significantly more accurate.

The expressions above for the bias coefficients are obtained from our
expression for the mean number of halos in cells $V$ which contain
mass $M$.  If the relation between $\delta_{\rm h}$ and $\delta$
is deterministic, that is, if the scatter around the mean number of
halos at fixed $M$ and $V$ is small, then the distribution of halo
overdensities is related to that of the dark matter overdensities
by a non-linear transformation; the coefficients $b_k$ describe
this relation.  Thus, if the dark matter distribution
at late times is obtained by a transformation of the initial
distribution, then the halos are 
also related to the initial distribution through a non-linear
transformation.
 
While a deterministic relation between $\delta_{\rm h}$ and $\delta$
is a reasonable approximation on large scales, on smaller scales the
scatter is significant \cite{MoWhi96}.  On small scales, the bias is 
both non-linear and stochastic.  Accurate analytic models for this 
stochasticity are presented in \cite{SheLem99,Casetal01}, but we will 
not need them for what follows.  Also, ignored in what follows are:
i) the deterministic bias coefficients $b_k(m,z_1)$ which follow from
the assumption that halos are associated with peaks in the initial
density field \cite{Moetal97b};
ii) the deterministic bias coefficients which are motivated
by perturbation theory rather than the spherical collapse model
\cite{Catetal98}. (Also,  see \cite{GazFos98} for a discussion of 
the relation between perturbation theory and the coefficients $a_k$ 
in the expressions above).
 
On large scales where deterministic biasing is a good approximation,
the variance of halo counts in cells is
\begin{equation}
\Bigl\langle \delta_{\rm h}(m,z_1|M,V,z_0)^2\Bigr\rangle =
\Bigl\langle \left(\sum_{k>0} b_k(m,z_1)\,\delta^k\right)^2\Bigr\rangle
            \approx b^2_1(m,z_1)\, \Bigl\langle \delta^2\Bigr\rangle_V \, .
\end{equation}
Thus, to describe the variance of the halos counts
we must know the variance in the dark matter on the same scale:
$\langle\delta^2\rangle_V$.
On very large scales, it should be a good approximation to replace
$\langle\delta^2\rangle$ by the linear theory estimate.
On slightly smaller scales, it is better to use the perturbation theory
estimates of \cite{Scoetal01}.  Given these, the variance of halo counts
in cells on large scales is straightforward to compute.
 
The higher order moments can also be estimated if the biasing is
deterministic.  This is because equation~(\ref{dhbkdk}) allows one to
write the higher order moments of the halo distribution,
$\langle\delta_{\rm h}^n\rangle$, in terms of those of the dark matter,
$\langle\delta^n\rangle$.
Quasi-linear perturbation theory shows that
$\langle\delta^n\rangle = S_n\,\langle\delta^2\rangle^{n-1}$ if
$\langle\delta^2\rangle \ll 1$ (see equation~\ref{dnsn}).
The $S_n$ are numerical coefficients which are approximately independent
of scale over a range of scales on which $\langle\delta^2\rangle\ll 1$;
for clustering from Gaussian initial conditions, the $S_n$ are given by
equation~(\ref{PTsn}).
By keeping terms to consistent order, one can show that \cite{FryGaz93}
\begin{equation}
 \Bigl\langle\delta_{\rm h}^n(m,z_1|M,V,z_0)\Bigr\rangle =
 H_n \Bigl\langle\delta_{\rm h}^2(m,z_1|M,V,z_0)\Bigr\rangle^{n-1},
\end{equation}
where
\begin{eqnarray*}
 H_3 &=& b_1^{-1}(S_3+3c_2), \qquad
 H_4 = b_1^{-2}(S_4 +12c_2S_3 +4c_3+12c_2^2), \nonumber\\
 H_5 &=& b_1^{-3}\Bigl[ S_5 + 20c_2S_4 + 15c_2S_3^2 +
         (30c_3+120c_2^2)S_3 + 5c_4 + 60c_3c_2 + 60c_2^3 \Bigr] ,
\end{eqnarray*}
$c_k=b_k/b_1$ and we have not bothered to write explicitly
that $H_n$ depends on halo mass and on the cell size $V$.
 
\begin{figure*}[t]
\centerline{\psfig{file=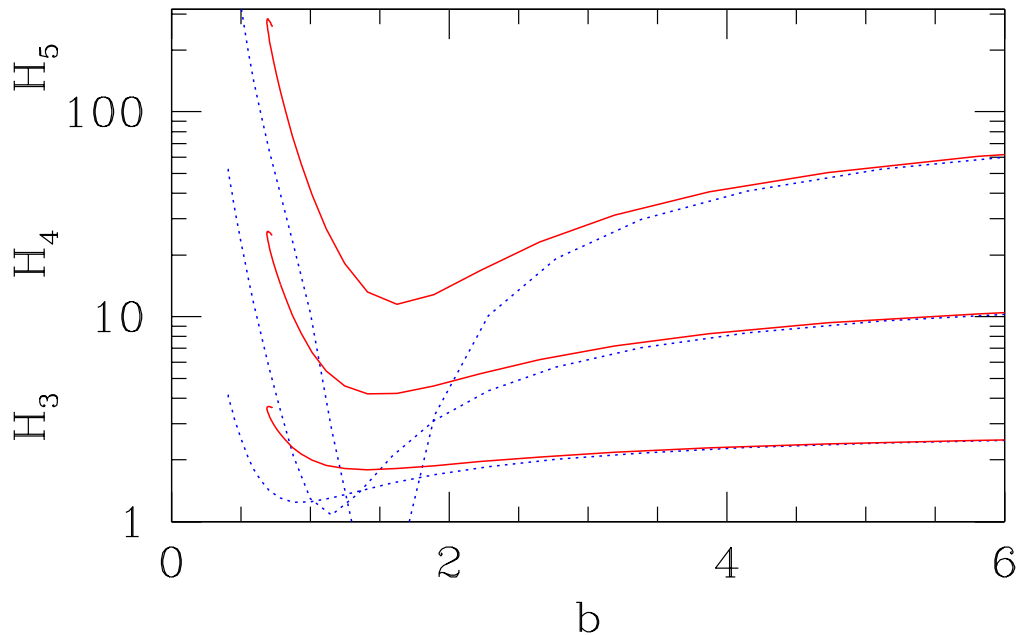,width=5.5in}}
\vspace{-5cm}
\caption{Higher order moments of the halo distribution if the
initial fluctuation spectrum is scale free and has slope $n=-1.5$.
Dotted and solid curves show the result of assuming the halo mass
function has the Press--Schechter and Sheth--Tormen forms. }
\label{fig:h3h4}
\end{figure*}

The expressions above show that the distribution of halos
depends explicitly on the distribution of mass.  On large scales
where the relation between halos and mass is deterministic, one might
have thought that the linear theory description of the mass distribution
can be used.  For clustering from Gaussian initial conditions, however,
linear theory itself predicts that $S_n=0$ for all $n>2$.  Therefore, to
describe the halo distribution, it is essential to go beyond linear theory to
the quasi-linear perturbation theory.
 
Figure~\ref{fig:h3h4} shows an example of how the first few $H_n$
depend on halo mass, parameterized by $b(m)$.  Note that, in general,
the less massive halos (those for which $b<1$) have larger values
of $H_n$.  This is a generic feature of halo models.  At high masses
($b\gg 1$) both sets of curves asymptote to $H_n=n^{n-2}$.

Before moving on to the next subsection, consider some asymptotic
properties of the $b_k$ in equation~(\ref{bk}), and of the high order
moments $H_n$ derived from them.
For small halos ($\nu\ll 1$) identified at early times ($z_1\gg 1$),
$b_1\approx 1$ and $b_k\approx 0$ for $k>1$.  Therefore $H_n=S_n$ and such
halos are not biased relative to mass.
In contrast, when $\nu\gg 1$ and $z_1$ is not large, i.e.
for massive halos identified at low redshift, $b_k=b_1^k$ for $k>1$.
In this limit, the $H_n$ are independent of both $S_n$ and $a_k$.
Therefore, the spatial distribution of these halos is determined
completely by the statistical properties of the {\it initial} density
field and are not modified by the dynamics of gravitational clustering.
In the limit of $\nu\gg 1$, for an initially Gaussian random field,
$H_n = n^{n-2}$; these are the coefficients of a Lognormal distribution 
which has small variance.  This shows that the most massive halos, or 
the highest peaks in a Gaussian field, are not Gaussian distributed.  
 
For small halos identified at low redshift ($\nu\ll 1$ and $z_1\ll 1$),
$b_1\approx 1-1/\delta_{\rm sc}(z_1)$ and
$b_k\approx -k!(a_{k-1}+a_k)/\delta_{\rm sc}(z_1)$ for $k\ge 2$.
In this case $H_n$ may depend significantly on the dynamical evolution
of the underlying mass density field.  The skewness of such halos, $H_3$,
can be larger than $S_3$.  On the other hand, for halos with $\nu=1$,
the skewness is $H_3=S_3-6/\delta_{\rm sc}^2(z_1)$, which is
substantially smaller than $S_3$ unless $z_1$ is high.
 
The most important result of this subsection is that, in the limit in
which biasing is deterministic, the bias parameters which relate the
halo distribution to that of the mass are completely specified if the
halo abundance, i.e., the halo mass function, is known.  If
perturbation theory is used to describe the distribution of the mass,
then these bias parameters allow one to describe the distribution of
the halos.  The perturbation theory predictions and the halo mass function
both depend on the shape of the initial power spectrum.  Thus, in this
model, the initial fluctuation spectrum is used to provide a complete
description of halo biasing.

\subsection{Halo density profiles}

Secondary infall models of spherical collapse \cite{FilGol84,Ber85}
suggest that the density profile around the center of a collapsed halo
depends on the initial density distribution of the region which collapsed.
If halos are identified as peaks in the initial density field
\cite{Kai84a,HofSha85}, then massive halos correspond to higher peaks
in the initial fluctuation field.  The density run around a high peak is
shallower than the run around a smaller peak \cite{Baretal86}:
high peaks are less centrally concentrated.
Therefore one might reasonably expect massive virialized halos to also
be less centrally concentrated than low mass halos.  Such a trend is
indeed found \cite{Navetal96}.

\begin{figure*}[t]
\centerline{\psfig{file=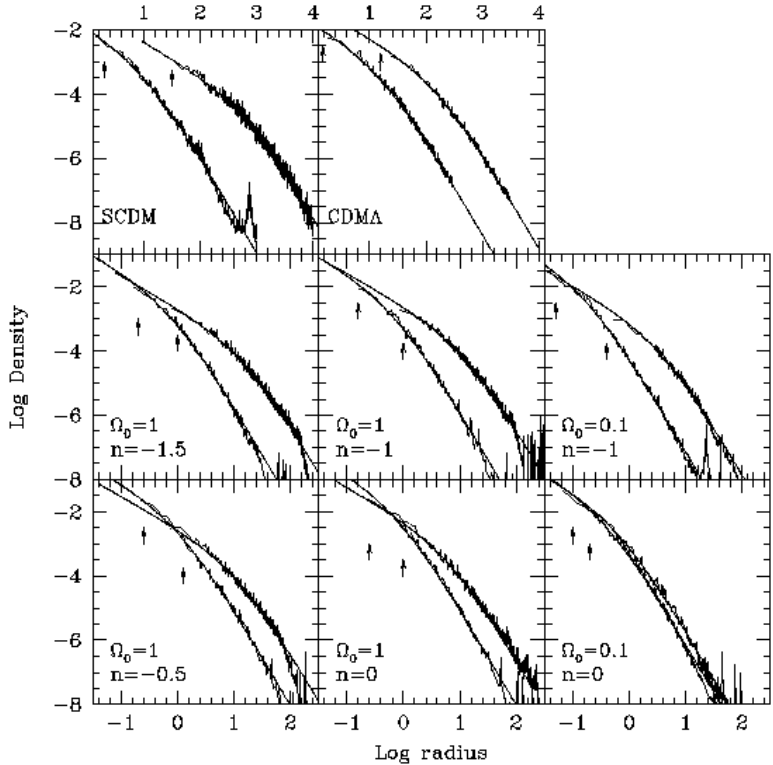,width=4.5in}}
\caption{Distribution of dark matter around halo centers
(from \cite{Navetal96}).  The density is in units of
$10^{10}$ M$_{\sun}$/kpc$^3$ and radii are in kpc.
Different panels show the density profile around the least and most
massive halos in simulations of a wide variety of cosmological models
and initial power spectra (labelled by the density parameter and
spectral index).  Arrows show the softening length; measurements on
scales smaller than this are not reliable.  Solid lines show the NFW
fit to the density distribution is extremely accurate.  }
\label{fig:haloprofile}
\end{figure*}

\begin{figure*}[t]
\centerline{\psfig{file=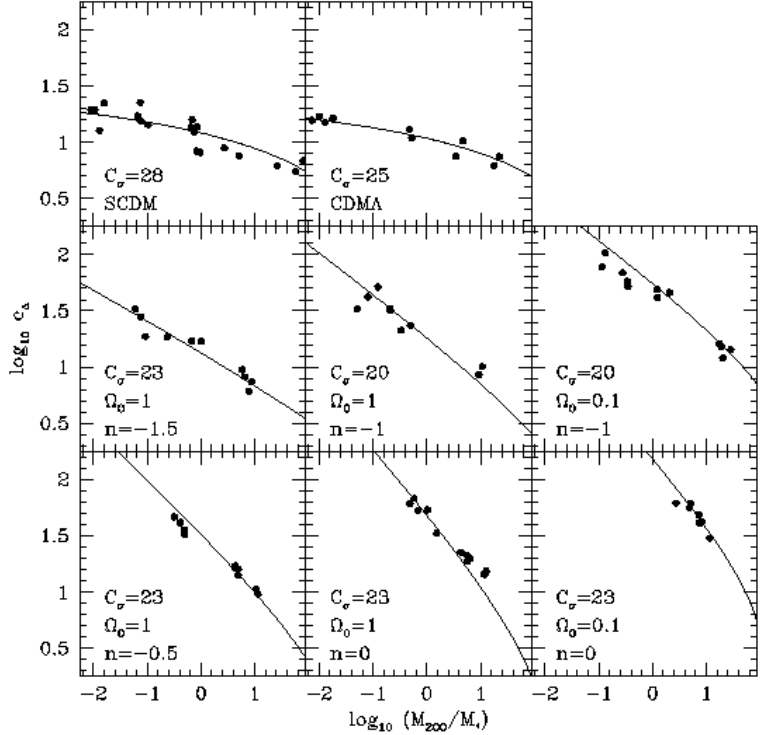,width=4.5in}}
\caption{Mean concentration at fixed mass, $c_\Delta = r_{vir}/r_s$,
for dark matter halos as a function of halo mass (from results 
presented in \cite{Ekeetal01}).  Different panels show the same 
cosmological models and power spectra as Figure~\ref{fig:haloprofile}.}
\label{fig:haloconc}
\end{figure*}

Functions of the form
\begin{equation}
 \rho(r|m) = {\rho_s\over (r/r_s)^\alpha(1 + r/r_s)^\beta} \quad{\rm 
or}\quad
 \rho(r|m) = {\rho_s\over (r/r_s)^\alpha[1 + (r/r_s)^\beta]}\, ,
\label{eqn:profile}
\end{equation}
have been extensively studied as models of elliptical galaxies
\cite{Her90,Zha96}.
Setting $(\alpha,\beta)=(1,3)$ and $(1,2)$ in the
expression on the left gives the Hernquist~\cite{Her90} and
NFW~\cite{Navetal96} profiles, whereas $(\alpha,\beta)=(3/2,3/2)$ in
the expression on the right is the M99 profile~\cite{Mooetal99}.
 
\begin{figure*}[t]
\centerline{\psfig{file=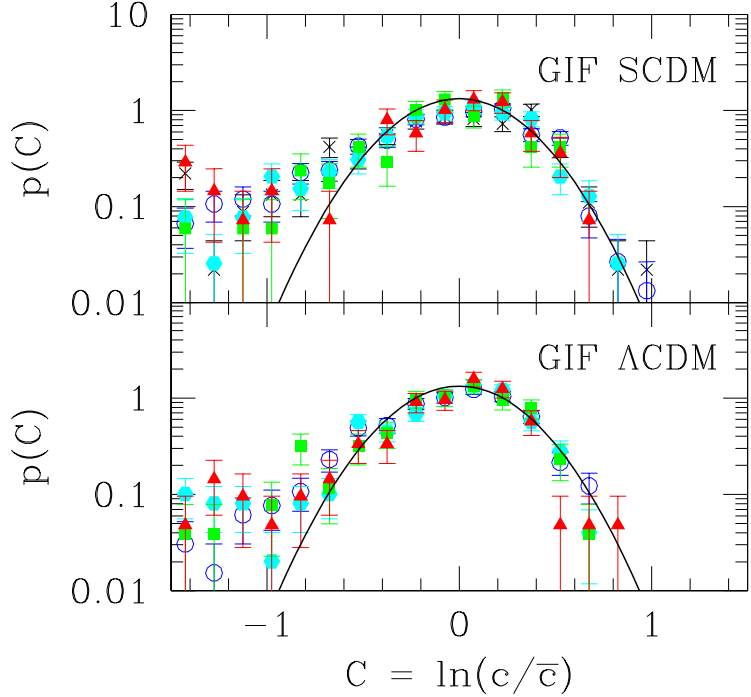,width=4.5in}}
\caption{Distribution of concentrations at fixed mass for dark matter
halos fit to NFW profiles.  Different symbols show results for halos
in different mass bins.  When normalized by the mean concentration in
the bin, the distribution is well described by a log-normal function
(equation~\ref{eqn:pconc}).
The pile up of halos at small values of the concentration is due to
numerical resolution of the GIF simulations.}
\label{fig:pconc}
\end{figure*}

The NFW and M99 profiles provide very good descriptions of the density
run around virialized halos in numerical simulations 
(figure~\ref{fig:haloprofile}). The two profiles differ on small 
scales, $r\ll r_s$, and whether one provides a better description of 
the simulations than the other is still being hotly debated.  Both 
profiles are parameterized by $r_s$ and $\rho_s$, which define a scale
radius and the density at that radius, respectively.  
Although they appear to provide a
two-parameter fit, in practice, one finds an object of given mass $m$
and radius $r_{vir}$ in the simulations, and then finds that $r_s$ which
provides the best fit to the density run.  This is because the edge of
the object is its virial radius $r_{vir}$, while the combination of $r_s$
and the mass determines the characteristic density, $\rho_s$, following
\begin{equation}
 m\equiv \int_0^{r_{vir}} dr\, 4\pi r^2\rho(r|m) .
\end{equation}
For the NFW and M99 profiles,
\begin{equation}
 m = 4 \pi \rho_s r_s^3 \,\left[ \ln(1+c) - \frac{c}{1+c}\right]
  \quad {\rm and}\quad
 m = 4 \pi \rho_s r_s^3\, {2\ln(1 + c^{3/2})\over 3},
 \label{eqn:deltamass}
\end{equation}
where $c \equiv r_{vir}/r_s$ is known as the concentration parameter.
Note that we have explicitly assumed that the halo profile is truncated
at $r_{vir}$, even though formally, the NFW and M99 profiles extend to
infinity.  Because these profiles fall as $r^{-3}$ at large radii, the
mass within them diverges logarithmically.  Our decision to truncate
the profile at the virial radius insures that the mass within the profile
is the same as that which is described by the halo mass function
discussed previously.
 
Since most of the mass is at radii much smaller than $r_{vir}$, the fitted
value of $r_s$ is not very sensitive to the exact choice of the boundary
$r_{vir}$.  The simulations show that for halos of the same mass, there
is a distribution of concentrations $c=r_{vir}/r_s$ which is well-fit
by a log-normal distribution \cite{Jin00,Buletal01}:
\begin{equation}
 p(c|m,z)\, dc = \frac{d\ln c}{\sqrt{2 \pi \sigma_c^2}} \,
 \exp\left[-\frac{\ln^2[c/\bar{c}(m,z)]}{2\,\sigma_{\ln c}^2}\right]  .
\label{eqn:pconc}
\end{equation}
Although the mean concentration $\bar{c}(m,z)$ depends on halo mass, the 
width of the distribution does not.  This is shown in 
Figure~\ref{fig:pconc}, which is taken from \cite{SheTor02}.  
The Figure shows that the distribution of $c/\bar{c}$ is indeed well 
approximated by a log-normal function.  

For the NFW profile,
\begin{equation}
 \bar{c}(m,z) = {9\over 1+z} \left[\frac{m}{m_*(z)}\right]^{-0.13}
 \qquad {\rm and} \qquad \sigma_{\ln c}\approx 0.25\,,
 \label{eqn:concentration}
\end{equation}
where $m_*(z)$ is characteristic mass scale at which $\nu(m,z)=1$.
A useful approximation, due to \cite{PeaSmi00}, is that
$\bar{c}[{\rm M99}]\approx (\bar{c}[{\rm NFW}]/1.7)^{0.9}$.
Equation~(\ref{eqn:concentration}) quantifies the tendency
for low mass halos to be more centrally concentrated, on average, 
than massive halos.  

In what follows, it will be useful to have expressions for the
normalized Fourier transform of the dark matter distribution within 
a halo of mass $m$:
\begin{equation}
 u(\veck|m) = {\int d^3\vecx\, \rho(\vecx|m)\, e^{-i\veck \cdot \vecx} 
                 \over \int d^3\vecx\, \rho(\vecx|m)}.
\end{equation}
For spherically symmetric profiles truncated at the virial radius, 
this becomes 
\begin{equation}
 u(k|m) = \int_0^{r_{vir}} dr\ 4\pi r^2\,{\sin kr\over kr}\ 
{\rho(r|m)\over m}.
\label{eqn:yint}
\end{equation}
Table~\ref{tab:rhoruk} contains some $\rho(r|m)$ and $u(k|m)$ pairs 
which will be useful in what follows.  

\begin{figure*}[t]
\centerline{\psfig{file=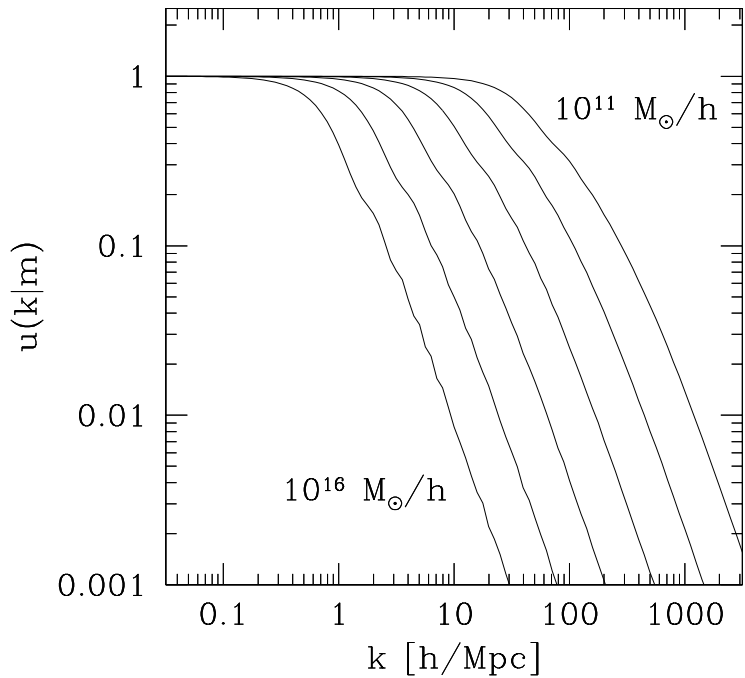,width=4.5in}}
\caption{Fourier transforms of normalized NFW profiles $u(k|m)$, for a 
variety of choices of halo mass, at the present time (redshift $z=0$). 
Equation~(\ref{eqn:concentration}) for the halo mass--concentration 
relation has been used.  The curves show that the most massive halos 
contribute to the total power only at the largest scales, whereas 
smaller halos contribute power even at small scales. }
\label{fig:ykm}
\end{figure*}

For the NFW profile, 
\begin{eqnarray}
 u(k|m) &=&  \frac{4\pi\rho_s r_s^3}{m}\, \Biggl\{ \sin(k r_s)\,
 \Bigl[ {\rm Si}([1+c]kr_s) - {\rm Si}(kr_s)\Bigr]
  -\frac{\sin(c kr_s)}{(1+c)kr_s}  \nonumber \\
 && \qquad\qquad\qquad + \cos(kr_s)\,
               \Bigl[{\rm Ci}([1+c]kr_s)-{\rm Ci}(kr_s)\Bigr]  \Biggr\} \, , 
 \label{uknfw}
\end{eqnarray}
where the sine and cosine integrals are 
\begin{equation}
 {\rm Ci}(x) = -\int_x^\infty \frac{\cos t}{t}\,dt
 \quad{\rm and}\quad {\rm Si}(x) = \int_0^x \frac{\sin t}{t}\, dt \, .
\end{equation}
Figure~\ref{fig:ykm} shows $u(k|m)$ as a function of $m$ for NFW halos. 
In general, the shape of the Fourier transform depends both on the halo 
concentration parameter, $c$, and the mass $m$.  The figure shows a 
trend which is to all the profiles in Table~\ref{tab:rhoruk}:  the 
small scale power is dominated by low mass halos.  

There are no complete explanations for why the NFW or M99 profiles
fit the dark matter density distribution of dark matter in numerical 
simulations, although there are reasonably successful models of why 
the concentrations depend on mass \cite{Navetal96,NusShe99,Wecetal02}. 
In the present context, the reason why they fit is of secondary importance;
what is important is that these fits provide simple descriptions of the 
density run around a halo.  In particular, what is important is that 
the density run around a dark matter halo depends mainly on its mass; 
though the density profile also depends on the concentration, the 
distribution of concentrations is determined by the mass.  

\begin{table}
\begin{center}
\caption{Density profiles and associated normalized Fourier transforms.  
Distances are in units of the scale radius:  $x = r/r_s$, $c=r_{vir}/r_s$ 
and $\kappa=kr_s$, and, when truncated, the boundary of the halo is 
$r_{vir}$.  The sine and cosine integrals are defined in the main 
text.}
\label{tab:rhoruk}
\begin{tabular}{ccc}
\\
\hline
 $\rho(x)$ & range & $u(\kappa)$ \\
\hline
 $ (2\pi)^{-3/2}\exp(-x^2/2)$ & & $\exp(-\kappa^2/2)$ \\
 $\exp(-x)/8\pi$ & & $(1 + \kappa^2)^{-2}$ \\
 $\exp(-x)/(4\pi x^2)$ & & ${\rm atan}(\kappa)/\kappa$ \\
 $x^{-2} (1 + x^2)^{-1}/(2\pi^2)$ & & $[1 - \exp(-\kappa)]/\kappa$ \\
 $3/(4\pi c^3)$ & $x\le c$ & 
   $3\,[\sin(c\kappa) - c\kappa\,\cos(c\kappa)]/(c\kappa)^3$ \\
 $(4\pi c\, x^2)^{-1}$ & $x\le c$ & ${\rm Si}(c\kappa)/c\kappa$ \\
 $x^{-1} (1 + x)^{-2}$ & $x\le c$ & Equation~(\ref{uknfw}) \\
\hline
 & & \\
\end{tabular}
\end{center}
\end{table}

\section{Halos and large scale structure}
\label{sec:halomodel}
 
At this point, we have formulae for the abundance and spatial
distribution of halos, as well as for the typical density run around
a halo.  This means that we are now in a position to construct the
halo model.  The treatment below will be completely general.
To make the model quantitative, one simply inserts their favorite
formulae for the halo profile, abundance and clustering (such as
those presented in the previous section) into the expressions below.

The formalism written down by Neyman \& Scott \cite{NeySco52},
which we are now in a position to consider in detail, had three drawbacks.
First, it was phrased entirely in terms of discrete statistics; some
work is required to translate it into the language of continuous density
fields.  Second, it was phrased entirely in terms of real coordinate
space quantities.  As we will see shortly, many of the formulae in
the model involve convolutions which are considerably easier to
perform in Fourier space.  And, finally, the particular model they
assumed for the clustering of halos was not very realistic.
 
Scherrer \& Bertschinger \cite{SchBer91} appear to have been the first
to write the model for a continuous density field, using Fourier space
quantities, in a formulation which allows one to incorporate more
general and realistic halo--halo correlations into the model.
It is this formulation which we describe below.
 
\subsection{The two-point correlation function}
In the model, all mass is bound up into halos which have a range of
masses and density profiles.  Therefore, the density at position ${\bfx}$ is
given by summing up the contribution from each halo:
\begin{eqnarray}
 \rho({\bfx}) &=& \sum_{i} f_i({\bfx}-{\bfx}_i)
 = \sum_{i} \rho({\bfx}-{\bfx}_i|m_i)
 \equiv \sum_{i} m_i\ u({\bfx}-{\bfx}_i|m_i) \nonumber\\
 &=& \sum_i \int dm\, d^3 x'\, \delta(m-m_i)\,\delta^3({\bfx}'-{\bfx}_i) \
     m\,u({\bfx}-{\bfx}'|m),
\end{eqnarray}
where $f_i$ denotes the density profile of the $i$th halo which is
assumed to be centered at ${\bfx}_i$.  The second equality follows
from assuming that the density run around a halo depends only on its
mass; this profile shape is parameterized by $\rho$, which depends on
the distance from the halo center and the mass of the halo.  The third
equality defines the normalized profile $u$, which is $\rho$ divided
by the total mass contained in the profile:
$\int d^3{\bfx}'\,u({\bfx}-{\bfx}'|m)=1$.
 
The number density of halos of mass $m$ is
\begin{equation}
 \Bigl\langle \sum_i \delta(m-m_i)\,\delta^3({\bfx}'-{\bfx}_i)\Bigr\rangle
 \equiv n(m),
\end{equation}
where $\langle ... \rangle$ denotes an ensemble average.
The mean density is
\begin{eqnarray}
 \bar\rho &=& \Bigl\langle \rho({\bfx}) \Bigr\rangle
     = \Bigl\langle \sum_{i} m_i\ u({\bfx}-{\bfx}_i|m_i)\Bigr\rangle
     = \int dm\,n(m) m \int d^3{\bfx}'\, u({\bfx}-{\bfx}'|m) \nonumber \\
    &=& \int dm\,n(m) m ,
\end{eqnarray}
where the ensemble average has been replaced by an average over the
halo mass function $n(m)$ and an average over space.
 
The two-point correlation function is
\begin{equation}
 \xi({\bfx}-{\bfx}') = \xi^{1h}({\bfx}-{\bfx}') + \xi^{2h}({\bfx}-{\bfx}')
\end{equation}
where
\begin{eqnarray*}
 \xi^{1h}({\bfx}-{\bfx}') &=& \int dm\,{m^2\,n(m)\over\bar\rho^2}
           \int d^3{\bf y}\,u({\bf y}|m)\,u({\bf y}+{\bfx}-{\bfx}'|m) 
\nonumber\\
 \xi^{2h}({\bfx}-{\bfx}') &=& \int dm_1\, {m_1\,n(m_1)\over\bar\rho}
       \int dm_2\, {m_2\,n(m_2)\over\bar\rho}
       \int d^3{\bfx}_1\, u({\bfx}-{\bfx}_1|m_1)\nonumber \\
 && \qquad\qquad \times \ \int d^3{\bfx}_2\, u({\bfx}'-{\bfx}_2|m_2)\
                                \xi_{hh}({\bfx}_1-{\bfx}_2|m_1,m_2) \,;
\label{xi12}
\end{eqnarray*}
the first term describes the case in which the two contributions to
the density are from the same halo, and the second term represents the
case in which the two contributions are from different halos.
Both terms require knowledge of how the halo abundance and density
profile depend on mass.  The second term also requires knowledge
of $\xi_{hh}({\bfx}-{\bfx}'|m_1,m_2)$, the two-point correlation function
of halos of mass $m_1$ and $m_2$.
 
The first term is relatively straightforward to compute:  it is just
the convolution of two similar profiles of shape $u(r|m)$, weighted by
the total number density of pairs contributed by halos of mass $m$.
This term was studied in the 1970's, before
numerical simulations had provided accurate models of the halo abundances
and density profiles \cite{Pee74,McCSil78}.  
The more realistic values of these inputs were
first used to model this term some twenty years later \cite{SheJai97}.
 
The second term is more complicated.  If $u_1$ and $u_2$ were extremely
sharply peaked, then we could replace them with delta functions; the
integrals over ${\vecx_1}$ and ${\vecx_2}$ would yield
 $\xi_{hh}({\bfx}-{\bfx}'|m_1,m_2)$.  Writing
 ${\bfx}_1-{\bfx}_2 = 
({\bfx}-{\bfx}')+({\bfx}'-{\bfx}_2)-({\bfx}-{\bfx}_1)$,
shows that this should also be a reasonable approximation
if $\xi_{hh}(r|m_1,m_2)$ varies slowly on scales which are
larger than the typical extent of a halo.  Following the discussion 
in the previous section with respect to halo bias, on large scales
where biasing is deterministic,
\begin{equation}
 \xi_{hh}(r|m_1,m_2) \approx b(m_1)\,b(m_2)\ \xi(r)
 \label{xihhapprox}
\end{equation}
Now $\xi(r)$ can be taken outside of the integrals over $m_1$ and
$m_2$, making the two integrals separable.  The consistency relations
(equation~\ref{biascon}) show that each integral equals unity.
Thus, on scales which are much larger than the typical halo
$\xi^{2h}(r)\approx \xi(r)$.  However, on large scales,
$\xi(r)\approx\xi^\lin(r)$, and so the two-halo term is really very
simple:  $\xi^{2h}(r)\approx\xi^\lin(r)$.
 
Setting
 $\xi_{hh}(r|m_1,m_2) \approx b(m_1)\,b(m_2)\,\xi(r)$ will overestimate
the correct value on intermediate scales.  Furthermore, on small scales
the halo-halo correlation function must eventually turn over (halos
are spatially exclusive---so each halo is like a small hard sphere);
assuming that it scales like $\xi(r)$ is a gross overestimate.
Using  $\xi_{hh}(r|m_1,m_2)\approx b(m_1)\,b(m_2)\ \xi^\lin(r)$,
i.e., using the linear, rather than the non-linear correlation function,
even on the smallest scales, is a crude but convenient way of accounting
for this overestimate.  Although the results of \cite{SheLem99} allow
one to account for this more precisely, it turns out that great accuracy
is not really needed since, on small scales, the correlation function is
determined almost entirely by the one-halo term anyway.
Although almost all work to date uses this approximation, it is
important to bear in mind that it's form is motivated primarily by
convenience.  For example, if volume exclusion effects are only important
on very small scales, then setting $\xi(r)\approx\xi^{\rm 1-loop}(r)$
rather than $\xi^\lin(r)$, i.e., using the one--loop perturbation theory
approximation rather than the simpler linear theory estimate, may provide
a better approximation.
 
Because the model correlation function involves convolutions, it is
much easier to work in Fourier space: the convolutions of the real-space
density profiles become simple multiplications of the Fourier transforms
of the halo profiles.  Thus, we can write the dark matter power spectrum
as
\begin{eqnarray}
 P(k) &=&  P^{1h}(k) + P^{2h}(k), \qquad {\rm where} \nonumber \\
 P^{1h}(k) &=& \int dm\,n(m) \left(\frac{m}{\bar{\rho}}\right)^2\,
                             |u(k|m)|^2 \nonumber\\
 P^{2h}(k) &=& \int dm_1\, n(m_1) 
\left(\frac{m_1}{\bar{\rho}}\right)\,u(k|m_1) \nonumber \\
	&& \quad \quad 
            \int dm_2\,n(m_2) \left(\frac{m_2}{ \bar{\rho}}\right)\, 
u(k|m_2)\,
            P_{hh}(k|m_1,m_2) \, . \nonumber \\
 \label{pkm}
\end{eqnarray}
Here, $u(k|m)$ is the Fourier transform of the dark matter distribution
within a halo of mass $m$ (equation~\ref{eqn:yint}) and
$P_{hh}(k|m_1,m_2)$ represents the power spectrum of halos
of mass $m_1$ and $m_2$.  Following the discussion of the halo--halo
correlation function (equation~\ref{xihhapprox}), we approximate this by
\begin{equation}
 P_{hh}(k|m_1,m_2) \approx \prod_{i=1}^{2} b_i(m_i)P^\lin(k) \nonumber \\
\end{equation}
bearing in mind that the one-loop perturbation theory estimate may be
more accurate than $P^\lin(k)$.

\subsection{Higher-order correlations}
Expressions for the higher order correlations may be derived similarly.
However, they involve multiple convolutions of halo profiles.
This is why it is much easier to work in Fourier space: the convolutions
of the real-space density profiles become simple multiplications of the
Fourier transforms of the halo profiles.
Similarly, the three-point and four-point correlations include
terms which describe the three and four point halo power spectra.
The bi- and tri- spectra of the halos are
\begin{eqnarray}
 B_{hhh}(\veck_1,\veck_2,\veck_3;m_1,m_2,m_3) &=& \prod_{i=1}^{3}b_i(m_i)
  \Biggl[B^\lin(\veck_1,\veck_2,\veck_3)\nonumber\\
 && \qquad + \frac{b_2(m_3)}{b_1(m_3)}P^\lin(k_1)P^\lin(k_2)\Biggr]\, ,
 \nonumber \\
 T_{hhhh}(\veck_1,\veck_2,\veck_3,\veck_4;m_1,m_2,m_3,m_4) &=&
 \prod_{i=1}^{4}b_i(m_i) \Biggl[T^\lin(\veck_1,\veck_2,\veck_3,\veck_4)
 \nonumber \\
 && \quad + 
\frac{b_2(m_4)}{b_1(m_4)}P^\lin(k_1)P^\lin(k_2)P^\lin(k_3)\Biggr]\, . 
\nonumber \\
\end{eqnarray}
Notice that these require the power, bi- and trispectra of the mass,
as well as mass-dependent $i$th-order bias coefficients $b_i(m)$.
Whereas $P$, $B$ and $T$ come from perturbation theory (\S~\ref{pthy}),
the bias coefficients are from the non-linear spherical or ellipsoidal
collapse models and are given in \S~\ref{bofm}.
 
Using this information, we can write the dark matter bispectrum as
\begin{equation}
B_{123}= B^{1h}+B^{2h}+B^{3h}\, ,
\end{equation}
where,
\begin{eqnarray}
 &&B^{1h} = \int dm\, n(m) \left(\frac{m}{\bar{\rho}}\right)^3\
 \prod_{i=1}^3 u({\vec k}_i|m)   \nonumber\\
&&B^{2h} = \Big[\int dm_1\,n(m_1) 
\left(\frac{m_1}{\bar{\rho}}\right)\,u(k_1|m_1)
             \int dm_2\,n(m_2)\left(\frac{m_2}{\bar{\rho}}\right)^2\, 
u(k_2|m_2) u(k_3|m_2)  \nonumber \\
&& \quad \quad \times P_{hh}(k_1|m_1,m_2) +  {\rm cyc.} \Big]\nonumber\\
 && B^{3h} = \left[ \prod_{i=1}^3 \int dm_i\,u(k_i|m_i) n(m_i) 
\left(\frac{m_i}{\bar{\rho}}\right)\right]
  B_{hhh}^{123}(m_1,m_2,m_3) \, , \nonumber \\
\label{eqn:bi}
\end{eqnarray}
where  $B_{hhh}^{123}(m_1,m_2,m_3)\equiv 
B_{hhh}(k_1,k_2,k_3|m_1,m_2,m_3)$  and  denotes the
bispectrum of halos of mass $m_1,m_2$ and $m_3$.
 
Finally, the connected part of the trispectrum can be written as the 
sum of four terms
\begin{equation}
T_{1234}=T^{1h}+T^{2h}+T^{3h}+T^{4h} \, ,
\end{equation}
where 
\begin{eqnarray}
&&T^{1h} =\int dm\,n(m) \left(\frac{m}{\bar{\rho}}\right)^4\
 \prod_{i=1}^4 u({\vec k}_i|m) \nonumber\\
 && T^{2h} = \Big[ \int dm_1\,n(m_1)\left( 
\frac{m_1}{\bar{\rho}}\right)\,u(k_1|m_1)
             \int dm_2\,n(m_2)\left(\frac{m_2}{\bar{\rho}}\right)^3\, 
\nonumber \\
&& \quad \quad \times u(k_2|m_2) u(k_3|m_2) u(k_4|m_2)  
P_{hh}(k_1|m_1,m_2) + {\rm cyc.} \Big] \nonumber \\
&+& \Big[ \int dm_1\,n(m_1)\left( 
\frac{m_1}{\bar{\rho}}\right)^2\,u(k_1|m_1)u(k_2|m_2)
             \int dm_2\,n(m_2)\left(\frac{m_2}{\bar{\rho}}\right)^2\, 
u(k_3|m_2) u(k_4|m_2)  \nonumber \\
&& \quad \quad \times P_{hh}(|\veck_1+\veck_2||m_1,m_2) + {\rm cyc.} \Big] 
\nonumber \\
&&T^{3h} =  \int dm_1\,n(m_1) 
\left(\frac{m_1}{\bar{\rho}}\right)\,u(k_1|m_1)
\int dm_2\,n(m_2) \left(\frac{m_2}{\bar{\rho}}\right)\,u(k_2|m_2) 
\nonumber \\
&& \quad \quad \times
             \int dm_3\,n(m_3)\left(\frac{m_3}{\bar{\rho}}\right)^2\, 
u(k_3|m_3) u(k_4|m_3)  
B_{hhh}(\veck_1,\veck_2,\veck_3+\veck_4|m_1,m_2,m_3)\nonumber \\
&& T^{4h} = \left[ \prod_{i=1}^4 \int dm_i\,u(k_i|m_i) n(m_i) 
\left(\frac{m_i}{\bar{\rho}}\right)\right]
  T_{hhhh}^{1234}(m_1,m_2,m_3,m_4) .
\end{eqnarray}
 
For simplicity, we reduce the notation related to integrals over 
the Fourier transform of halo
profiles and write the power spectrum,
\begin{eqnarray}
 &&P(k) =  P^{1h}(k)+P^{2h}(k) \nonumber \\
 &&P^{1h}(k) =M_{02}(k,k)  \nonumber \\
&&P^{2h} =   P^\lin(k) \left[ M_{11}(k)\right]^2  \, ,
 \label{power}
\end{eqnarray}
bispectrum,
\begin{eqnarray}
&&B_{123}=  B^{1h}+B^{2h}+B^{3h}\nonumber \\
&&B^{1h} = M_{03}( k_1, k_2, k_3) \nonumber \\
&&B^{2h} = M_{11}( k_1)M_{12}( k_2, k_3)
P^\lin(k_1) + {\rm cyc.} \nonumber \\
&&B^{3h} =  \left[\prod_{i=1}^3 M_{11}( k_i) \right]
B^\lin_{123} + M_{11}( k_1)M_{11}( k_2)M_{21}( k_3)
P^\lin(k_1) P^\lin(k_2) +  {\rm cyc.} \, ,  \nonumber \\
\label{bisp}
\end{eqnarray}
and trispectrum,
\begin{eqnarray}
&&T_{1234}=  T^{1h}+T^{2h}+T^{3h}+T^{4h}\nonumber \\
&&T^{1h} = M_{04}( k_1, k_2, k_3, k_4) \nonumber \\
&&T^{2h} = \Big[M_{11}( k_1)M_{13}( k_2, k_3, k_4)
P^\lin(k_1) + {\rm cyc.} \Big] \nonumber \\
&& \quad \quad + \Big[ M_{12}( k_1, k_2)M_{12}( k_3, k_4)
P^\lin(|\veck_1+\veck_2|) + {\rm cyc.} \Big]\nonumber \\
&&T^{3h} =  M_{11}( k_1)M_{11}( k_2)M_{12}( k_3, k_4)
B^\lin(\veck_1,\veck_2,\veck_3+\veck_4) \nonumber \\
&& \quad \quad + \Big[  M_{11}( k_1)M_{11}( k_2)M_{22}( k_3, k_4)
P^\lin(k_1) P^\lin(k_2) +  {\rm cyc.}\Big]  \nonumber \\
&&T^{4h} =  \left[\prod_{i=1}^4 M_{11}( k_i) \right]
T^\lin_{1234} \nonumber \\
&& \quad +M_{11}( k_1)M_{11}( k_2)M_{11}( k_3)M_{21}( k_4)
P^\lin(k_1) P^\lin(k_2) P^\lin(k_3) +  {\rm cyc.} \, .
\label{eqn:tri}
\end{eqnarray}
Here, $b_0\equiv 1$ and
\begin{equation}
 M_{ij}(k_1,\ldots,k_j) \equiv
 \int dm n(m)\left(\frac{m}{\bar{\rho}}\right)^j b_i(m) [u(k_1|m)\ldots 
u(k_j|m)] \, , \nonumber \\
 \label{Iij}
\end{equation}
with the three-dimensional Fourier transform of the
halo density distribution, $u(k|m)$, following equation~(\ref{eqn:yint}).
 
The one-point moments, smoothed on scale $R$, can also be obtained by an
integral with the appropriate window function $W(kR)$. 
In the case of variance,
\begin{eqnarray}
\sigma^2(R) &=& \int \frac{k^2 dk}{2\pi^2}P(k) |W(kR)|^2 \nonumber \\
&=&  \int \frac{k^2 dk}{2\pi^2} P^\lin(k) \left[ M_{11}(k)\right]^2  
|W(kR)|^2 +
 \int \frac{k^2 dk}{2\pi^2} M_{02}(k,k)   |W(kR)|^2 \nonumber \\
&\approx& \sigma^2_\lin(R) + \int dm 
n(m)\left(\frac{m}{\bar{\rho}}\right)^2 \overline{u^2}(R|m)\,
\end{eqnarray}
where
\begin{equation}
\overline{u^n}(R|m)= \int \frac{k^2 dk}{2\pi^2} u^n(k|m) |W(kR)|^2 \, .
\end{equation}

In simplifying, we have written the fully non-linear power spectrum of 
the density field in terms of the halo model (equation~\ref{power}) 
and taken the large-scale limit $M_{11} \approx 1$.  This is a 
reasonable approximation because of the consistency conditions 
(equation~\ref{biascon}). Here, $\sigma^2_\lin(R)$ follows from 
equation~(\ref{eqn:sigmalin}). With similar approximations, we can derive
higher-order connect moments (see \cite{Scoetal01} for details).

With the same general integral defined in \cite{Scoetal01},
\begin{equation}
A_{ij}(R) \equiv
 \int dm n(m)\left(\frac{m}{\bar{\rho}}\right)^2 b_i(m)  
\overline{u^2}(R|m) \overline{u^j}(R|m) \, ,
\label{Aij}
\end{equation}
we can write the one point moments as
\begin{eqnarray}
 &&\langle \delta^2 \rangle  \equiv \sigma^2 = \sigma_\lin^2 +A_{00} 
\nonumber \\
&& \langle \delta^3\rangle = S_3^\lin \sigma_\lin^4 + 3 \sigma_\lin^2 
A_{10}
+A_{01} \nonumber \\
&& \langle \delta^4 \rangle_c  =
S_4^\lin \sigma_\lin^6 + 6 \frac{S_3^\lin}{3} \sigma_\lin^4 A_{10} + 7
\frac{4\sigma_\lin^2}{7} A_{11} +A_{02} \nonumber \\
&&\langle \delta^5 \rangle_c =
S_5^\lin \sigma_\lin^8 + 10 \frac{S_4^\lin}{16} \sigma_\lin^6 A_{10} +
25 \frac{3S_3^\lin}{5} \sigma_\lin^4 A_{11} + 15
\frac{\sigma_\lin^2}{3}  A_{12} +A_{03} \, , \nonumber \\
\label{S4}
\end{eqnarray}
where the terms in $\langle \delta^n \rangle_c$ are ordered
from $n$-halo to 1-halo contributions. The coefficient of an $m$-halo
contribution to $\langle\delta^n\rangle_c$ is given by $s(n,m)$
(e.g. $6$ and $7$ in the second and third terms of equation~\ref{S4}), the
Stirling number of second kind, which is the number of ways of putting
$n$ distinguishable objects ($\delta$) into $m$ cells (halos), with no
cells empty \cite{SchBer91}.
 
In general, we can write the $n$th moment as
\begin{equation}
\langle\delta^n \rangle_c = S_n^{\rm PT} \sigma^{2(n-1)}_L +
\sum_{m=2}^{n-1} s(n,m)\ \alpha_{nm} S_{m}^{\rm PT}\ \sigma^{2(m-1)}_L
A_{1n-m-1} + A_{0n-2},
\label{Sn}
\end{equation}
where the first term in equation~(\ref{Sn}) represents the
$n-$halo term, the second term is the contribution from $m$-halo terms, and
the last term is the 1-halo term. The coefficients $\alpha_{nm}$ measure
how many of the terms contribute as $A_{1n-m-1}$, with the other
contributions being subdominant. For example, in equation~(\ref{S4}), the
2-halo term has a total contribution of 7 terms, 4 of them contain 3
particles in one halo and 1 in the other, and 3 of them contain 2
particles in each. The factor $4/7$ is included to take into account
that the $3-1$ amplitude dominates over the $2-2$ amplitude. Note that
in these results, we neglected all  contributions from the
non-linear biasing parameters in view of the consistency conditions 
given in equation~(\ref{biascon}).  The $S_n^{\rm PT}$ were defined in
equation~(\ref{PTsn}).  

\subsection{An illustrative analytic example}\label{simple}
To introduce the general behavior of the halo based predictions, we 
first consider a simple illustrative example. We assume that the initial 
spectrum of the density fluctuation field is $P_0(k) = A/k^{3/2}$.  
This is not a bad approximation to the shape of the power spectrum on 
cluster-like scales in CDM models.  If we set 
$\Delta^2_0(k) \equiv k^3\,P_0(k)/(2\pi^2)$ then the variance on scale 
$R$ is
\begin{equation}
 \sigma^2(R) = \int {{\rm d}k\over k} \Delta^2_0(k)\, W^2_{\rm TH}(k R) 
             = {16\sqrt{\pi}\over 15} {A\over 2\pi^2}\,{R^{-3/2}}.
\end{equation}
Setting $\sigma(R_*)\equiv\delta_{\rm sc}$ means 
\begin{equation}
 \Delta^2_0(k) = {15\,\delta_{\rm sc}^2\over 16\sqrt{\pi}}\,(kR_*)^{3/2}.
\end{equation}
We will approximate the fraction of mass in virialized halos of mass $m$
using equation~(\ref{fps}):
\begin{displaymath}
 f(m)\,{\rm d}m = {mn(m)\,{\rm d}m\over\bar\rho} = {{\rm d}\nu\over\nu}\,
                  \sqrt{\nu\over 2\pi}\,\exp\left(-{\nu\over 2}\right),
 \label{fmps}
\end{displaymath}
where $\nu\equiv \delta^2_{\rm sc}/\sigma^2(m) = (m/m_*)^{1/2}$
and $m_* = 4\pi R_*^3\bar\rho/3$.
We will assume that the density run around the center of a virialized 
halo scales as
\begin{equation}
 {\rho(r|m)\over\bar\rho} = {2\Delta_{\rm nl}\over 3\pi}\, c^3(m)\,
                            {y^{-2}\over 1+y^2},
\end{equation}
when $\bar\rho$ is the background density,
$y = r/r_s$, $c=r_{vir}/r_s$,
$\Delta_{\rm nl}=(R/r_{vir})^3$, and $m/\bar\rho = 4\pi R^3/3$.
Here $R$ is the initial size of the halo, $r_{vir}$ is the
virial size, and $r_s$ is the core radius.
Since the profile falls more steeply than $r^{-3}$ at large $r$,
the total mass is finite: $4\pi\int {\rm d}r\ r^2 \rho(r|m) = m$.
We will assume that the core radius depends on halo mass:  
$c(m) = c_*\,(m_*/m)^\gamma$.
 
The normalized Fourier transform of this profile is
\begin{eqnarray}
 u(k|m) &=& {\int {\rm d}r\ r^2 \rho(r|m)\,\sin(kr)/(kr)\over
             \int {\rm d}r\ r^2 \rho(r|m)} 
         = {1 - {\rm e}^{-kr_s}\over kr_s}, \nonumber\\
 {\rm where} && kr_s = {kR_*\over\Delta^{1/3}_{\rm nl}c_*}
               \left({m\over m_*}\right)^{\gamma+1/3}
               \equiv \kappa\,\left({m\over m_*}\right)^{\gamma+1/3}.
\end{eqnarray}
Note that at large $k$, $u(k|m)$ decreases as $1/k$. 
 
If we set $\gamma=1/6$ (so more massive halos are less concentrated), 
then $kr_s = \kappa\nu$, and the integrals over the mass function which 
define the power spectrum can be done analytically.  For example, the 
contribution to the power from particles which are in the same halo is
\begin{displaymath}
 \Delta^2_{1h}(k) = \int {\rm d}m\,{m^2n(m)\over \bar\rho^2}\,|u(k|m)^2| 
                  = {2\,\Delta_{\rm nl}\over 3\pi}\,c_*^3\,
                      \kappa\left(1 + {1\over\sqrt{1+4\kappa}} - 
                                  {2\over\sqrt{1+2\kappa}}\right),
\end{displaymath}
and the contribution from pairs in separate halos is
\begin{displaymath}
 \Delta^2_{2h}(k) = B^2(k)\,\Delta^2_0(k) 
                  = B^2(k)\,{15\,\delta_{\rm sc}^2\over 16\sqrt{\pi}}\,
                            (\Delta_{\rm nl}^{1/3}c_*)^{3/2}\,\kappa^{3/2},
\end{displaymath}
where
\begin{displaymath}
 B(k) = {1\over\kappa}\Biggl[{2\over\delta_{\rm sc}} - 1 +
           \sqrt{1 + 2\kappa}\left(1 - {1\over\delta_{\rm sc}}\right) -
           {1/\delta_{\rm sc}\over\sqrt{1 + 2\kappa}}\Biggr].
\end{displaymath}
Here $B(k)\equiv \int {\rm d}m\,[mn(m)/\bar\rho]\,b(m)\,u(k|m)$ 
and we have used the fact that, if the mass function is given by 
equation~(\ref{fps}), then $b(m) = 1 + (\nu-1)/\delta_{\rm sc}$ 
from equation~(\ref{b1m}).

At small $\kappa$, the one-halo term is $2\Delta_{\rm nl}\,c^3_*/\pi$
times $\kappa^3$, whereas the two-halo term is $\Delta^2_0(k)$ times
$B^2(k)\to 1 - (2/\delta_{\rm sc}+1)\kappa$;
the effect is to multiply the linear spectrum by a $k$ dependent factor 
which is less than unity.  Thus, at small $k$ most of the power comes 
from the two-halo term.  At large $k$, the two-halo term is 
$2(1-1/\delta_{\rm sc})^2/\kappa$ times the linear spectrum, 
so it grows as $\kappa^{1/2}$.  On the other hand, the one-halo term 
is $2\Delta_{\rm nl}c^3_*/3\pi$ times $\kappa$.  
Thus, the power on small scales is dominated by the one-halo term.  
 
If, on the other hand, $\gamma=-1/3$ (so more massive halos are 
more concentrated, unlike numerically simulated halos), 
then $\kappa = kr_s = kR_*/(c_*\Delta_{\rm nl}^{1/3})$
is independent of $m$, and so $u(k|m)$ is also independent of $m$.  
Since $\int {\rm d}m\ (m/\bar\rho)^2\,n(m) = 3\,(m_*/\bar\rho)$, 
the two power spectrum terms are 
\begin{displaymath}
 \Delta^2_{1h}(k) = {2\Delta_{\rm nl}\over\pi}\,
          c_*^3\,\kappa^3\left({1 - {\rm e}^{-\kappa}\over\kappa}\right)^2
                \ {\rm and}\ \
 \Delta^2_{2h}(k) = \left({1 - {\rm e}^{-\kappa}\over\kappa}\right)^2
                             \Delta^2_0(k).
\end{displaymath}
This shows how the variation of the central concentration with halo 
mass changes the contribution to the total power from the two terms.  
In addition, changing the halo mass function would obviously change the 
final answer.  And, for $\gamma=-1/3$, changing the initial power spectrum 
only changes the prefactor in front of $\Delta^2_{1h}(k)$.  The prefactor
in front of $\Delta^2_{2h}(k)$ is unaffected, though, of course,
$\Delta^2_0(k)$ has been changed.

\subsection{Compensated profiles}\label{cmpnst}
This section considers density profiles which are `compensated'; 
these are combinations of over- and under-dense perturbations, 
normalized so that the mass in each of the two components is the same.  
The reason for considering such profiles is to illustrate a curious 
feature of the halo model:  when only positive perturbations are present, 
then, as $k\to 0$, the single halo contribution to the power tends to a 
constant:  
\begin{equation}
 P^{1h}(k\to 0) \to \int dm\, n(m)\,\left({m^2\over \bar\rho^2}\right) 
\end{equation}
In CDM--like spectra, the linear power-spectrum is $\propto k$ at small $k$, 
so that the single halo term eventually dominates the power.  
This problem is also present in the higher order statistics such as 
the bi- and trispectra.  For the power spectrum, this constant is like 
a mean square halo mass, so that this excess large scale power resembles 
a shot-noise like contribution.  This suggests that it must be 
subtracted--off by hand.  However, subtracting the same constant at all 
$k$ is not a completely satisfactory solution, because the power at 
sufficiently large $k$ can be very small, in which case subtracting 
off $P^{1h}(k=0)$ might lead to negative power at large $k$.  
The compensated profile model is designed so that the one-halo term is 
well behaved at small $k$.  On the other hand, as we show, compensated 
models suffer from another problem:  they have no power on large scales!  

Consider the correlation function which arises from a random, Poisson,
distribution of density perturbations, in which all perturbations are
assumed to be identical.  We will consider what happens when we allow
perturbations to have a range of sizes later, and correlations
between perturbations will be included last.
The density at a distance $r$ from a compensated perturbation can
be written as the sum of two terms:
\begin{equation}
 \rho(r) = \rho_+(r) + \rho_-(r).
\end{equation}
So that we have a concrete model to work with, we will assume that 
\begin{equation}
 {\rho_+(x)\over \bar\rho} = a\,\exp \left({-x^2\over 2\sigma_+^2}\right)
 \qquad{\rm and}\qquad
 {\rho_-(x)\over\bar\rho} = 1 - \exp \left({-x^2\over 2\sigma_-^2}\right),
\end{equation}
where $\bar\rho$ denotes the mean density of the background in which
these perturbations are embedded.
(The Gaussian is a convenient choice because it is a monotonic function 
for which the necessary integrals are simple.)
We will require $a>1$, so that the positive perturbation
is denser than the background.  We will discuss the scales $\sigma_+$
and $\sigma_-$ of the two perturbations shortly.  For now, note that
the negative perturbation is bounded between zero and one:
$\rho_-(x)$ is always less than the mean density.
The reason for this is that we are imagining that the perturbation
can be thought of as an initially uniform density region of size
$\sigma_-$ from which mass has been scooped out according to $\rho_-(x)$,
and replaced by mass which is distributed as $\rho_+(x)$.
The total density fluctuation is
\begin{equation}
 \delta(x) \equiv {\rho(x)\over\bar\rho} - 1
           = a\,\exp\left(-x^2\over 2\sigma_+^2\right) -
               \exp\left(-x^2\over 2\sigma_-^2\right).
\end{equation}
The integral of $\delta$ over all space is
\begin{equation}
 g \equiv 4\pi \int_0^\infty {\rm d}x\ x^2\,\delta(x) =
        (2\pi)^{3/2}\, \Bigl(a\,\sigma_+^3 - \sigma_-^3 \Bigr) \, ,
\end{equation}
and depends on the amplitudes and scales of the positive and
negative perturbations.
 
If we set
\begin{equation}
 \sigma_- = a^{1/3}\,\sigma_+
\label{sigmas}
\end{equation}
then $g=0$.
This corresponds to the statement that the positive perturbation contributes
exactly the same amount of mass which the negative perturbation removed.
The only difference is that the mass has been redistributed into the
form $\rho_+(x)$.  It is in this sense that the profiles are compensated.
 
We can build a toy model of evolution from this by defining
$\sigma_+(t)/\sigma_- \equiv R(t)$.  We will imagine that, at some
initial time, $R(t)\approx 1$, and that it decreases thereafter.
This is supposed to represent the fact that gravity is an attractive
force, so the mass which was initially contained with $\sigma_-$ is
later contained within the smaller region $\sigma_+$.  If mass is
conserved as $\sigma_+$ shrinks (equation~\ref{sigmas}), then
it must be that $a(t) = R(t)^{-1}$.  Thus, the amplitude $a$ is
related to the ratio of the initial and final sizes of $\sigma_+$.
The fact that $R(t)\approx 1$ initially reflects the assumption
that the initial density field was uniform.  Today, the mass is in
dense clumps---the positive perturbations.  Each positive perturbation
assembled its mass from a larger region in the initial conditions.
Our particular choice of setting $g$ to zero comes from requiring
that all the mass in the positive perturbation came from the negative
one.
 
Note that we haven't yet specified the exact form for the evolution of the
profile, $R(t)$.  Independent of this evolution, we can use the
formalism in \cite{McCSil77b} to compute the correlation functions
as a function of given $t$. 
The correlation function is the number density of profiles, $\eta$,
times the convolution of such a profile with itself, $\lambda(r)$.
Since all the mass is in the positive perturbations, and each positive
perturbation contains mass $m=(2\pi)^{3/2}\,\bar\rho\,\sigma_-^3$, we
can set $\eta=\bar\rho/m$.  In our compensated halo model, $\lambda$
is the sum of three terms:
\begin{equation}
\lambda(r) = \lambda_{++}(r) + \lambda_{--}(r) - 2\lambda_{+-}(r),
\label{lambda}
\end{equation}
where $\lambda_{++}$,$\lambda_{--}$, and $\lambda_{+-}$,
denote the various types of convolutions.
For the Gaussian profiles we are considering here,
\begin{eqnarray}
 \lambda_{++}(r) &=& a^2\,\pi^{3/2}\,\sigma_+^3\,
                 \exp\left({-r^2\over 4\sigma_+^2}\right), \qquad
 \lambda_{--}(r)  =  \pi^{3/2}\,\sigma_-^3\,
                 \exp\left({-r^2\over 4\sigma_-^2}\right), \nonumber\\
 \lambda_{+-}(r) &=& a\,\pi^{3/2}\,
    \left(2\,\sigma_+^2\,\sigma_-^2\over \sigma_+^2+\sigma_-^2\right)^{3/2}
    \,\exp\left({-r^2/2\over \sigma_+^2 + \sigma_-^2}\right).
\label{eqn:3terms}
\end{eqnarray}
Inserting equation~(\ref{sigmas}) makes the factors in front of the
exponentials resemble each other more closely.
 
It is a simple matter to verify that these compensated profiles
satisfy the integral constraint:
\begin{equation}
 \eta\,4\pi \int_0^\infty {\rm d}r\ r^2\,\lambda(r) = 0.
 \label{intconst}
\end{equation}
If the correlation function were always positive, this integral constraint 
would not be satisfied.  Because of the minus sign in 
equation~(\ref{lambda}) above, the correlation function in compensated 
halo models can be negative on large scales.
 
The power spectrum is obtained by Fourier-transforming the correlation 
function.  Since the Fourier transform of a Gaussian is a Gaussian, 
equation~(\ref{eqn:3terms}) shows that, in these compensated models, 
the power spectrum is the sum of three Gaussians:
\begin{eqnarray}
 P(k) &=& {1\over 2\pi^2}
         \int {\rm d}r\ r^2\,\eta\lambda(r)\,{\sin kr\over kr}\nonumber \\
      &=& \eta\,(a\,\sigma_+^3)^2 \,
\left({\rm e}^{-k^2\sigma_+^2} + 
{\rm e}^{-k^2\sigma_-^2} - 2\,{\rm e}^{-k^2\sigma_+^2/2}{\rm 
e}^{-k^2\sigma_-^2/2}\right).
\label{eqn:samemass}
\end{eqnarray}
The form of this expression is easily understood, since convolutions
in real space are multiplications in Fourier space such that the power
spectrum is simply a sum of products of Gaussians.  Indeed, if we let 
$U(k)$ and $W(k)$ denote the Fourier transforms of the positive and 
negative perturbations, then
\begin{equation}
 P(k) = \Bigl[U(k) - W(k)\Bigr]^2.
\label{pkcomp}
\end{equation}
 
It is interesting to compare this expression with the case of a
positive perturbation only.  In this case, $\lambda = \lambda_{++}$,
and $P(k) = U(k)^2$.  Since $\lambda_{++}\ge 0$ always, such a
model does not satisfy the integral constraint.  Analogously,
uncompensated profiles have $P(k)\to$ constant at small $k$.
If one thinks of the compensated profile as providing a correction
factor to the power spectrum of positive perturbations, then the
expression above shows that this correction is $k$ dependent:
simply subtracting-off a constant term from $U(k)$ is incorrect.
In the compensated Gaussian model above, $W(k)\to 0$ at large $k$,
so $P(k)\to U(k)$ on small scales.  However, $P(k)\to 0$ at small
$k$---there is no power on large scales.
 
\begin{figure}
\centering
\mbox{\psfig{figure=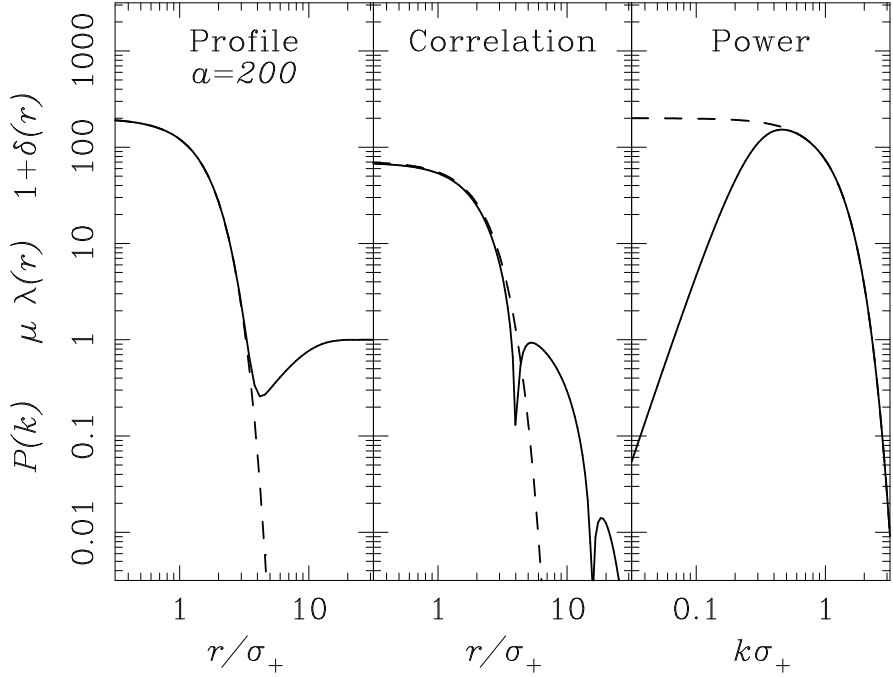,height=9cm,bbllx=87pt,bblly=56pt,bburx=608pt,bbury=452pt}}
\caption{Density profiles and correlation functions associated with
uncompensated (dashed) and compensated (solid) Gaussian perturbations.
The correlation function $\lambda$ is negative for $r/\sigma_+$ in the
range $3-15$ or so, so we have plotted $|\lambda|$ instead.}  
\label{cmpgaus}
\end{figure}
Fig.~\ref{cmpgaus} shows all this explicitly.
The panels show density profiles, correlation functions and power
spectra for uncompensated (dashed curves) and compensated (solid curves)
Gaussian perturbations which have $a=200$ and $\sigma_+=1$.
Notice how, for compensated profiles, the correlation function
oscillates about zero.  Notice also how $P(k)$ for the two cases
tends to very different limits at small $k$.
 
So far we have assumed that all profiles had the same shape,
parameterized by $\sigma_-$.  Because $m\propto\sigma_-^3$,
allowing for a range of masses is the same as allowing for a
range of profile shapes.  Thus,
\begin{equation}
 P(k) = \int {\rm d}m\ n(m)\,P(k|m),
\end{equation}
where $n(m)$ is the number density of perturbations which have mass
$m$, and $P(k|m)$ is the power spectrum for perturbations which contain
this mass. Since each of the $P(k|m)$s tends to zero at small $k$, this 
will also happen for $P(k)$.  The shape of $n(m)$ depends on the initial 
spectrum of fluctuations.  If we insert the shape of $n(m)$ associated 
with an initial $P(k)\propto 1/k$ spectrum, and use the Gaussian profiles
above, then the integral over $m$ can be done analytically.
Using, $\delta^2_{\rm sc}/\sigma^2(m) = (m/m_*)^{2/3} = \mu$,
and $\sigma_*$ to denote $\sigma_-$ for an $m_*$ halo,
$m_* = (2\pi)^{3/2}\bar\rho \sigma^3_*$, we get
\begin{eqnarray}
 k^3P(k) &=& k^3\int {\rm d}m\ n(m)\,\left(m\over\bar\rho\right)^2
        \left( {\rm e}^{-k^2\sigma_+^2} + {\rm e}^{-k^2\sigma_-^2} - 
2\,{\rm
        e}^{-k^2\sigma_+^2/2}{\rm e}^{-k^2\sigma_-^2/2}\right) \nonumber\\
      &=& \left(k^3m_*\over\bar\rho\right)
  \int {{\rm d}\mu\over\mu} \mu^{3/2+1/2}{\exp(-\mu/2)\over\sqrt{2\pi}}
  \nonumber \\
&& \qquad \times
\left( {\rm e}^{-(k\sigma_*/a^{1/3})^2 \mu} + {\rm e}^{-k^2\sigma_*^2 \mu}
-2\,{\rm e}^{-(k\sigma_*/a^{1/3})^2\mu/2}\,{\rm 
e}^{-k^2\sigma_*^2\mu/2}\right)
\nonumber \\
&=& 8\pi\kappa^3 \left({1\over [1 + 2\kappa^2/a^{2/3}]^2}
                     + {1\over [1 + 2\kappa^2]^2}
                       - {2\over [1 + \kappa^2+ \kappa^2/a^{2/3}]^2}\right)
\end{eqnarray}
where we have set $\kappa\equiv k\sigma_*$.

This spectrum is different from the one in which all halos had the same 
mass (equation~\ref{eqn:samemass}).  The power associated with any given
halo mass falls exponentially at large $k$; the result of adding up the 
contributions from all halos means that $P(k)$ only decreases as $k^{-4}$ 
at large $k$.  This is a consequence of the fact that the less massive 
halos are smaller and much more numerous than the massive halos.
 
We can also work out these relations for tophat perturbations.  Here,
\begin{eqnarray}
\delta(r) &=& A-1 \qquad{\rm if}\ \  0\le r\le R_+ \nonumber \\
          &=& -1  \qquad{\rm if}\ \  R_+< r\le R_- ,
\end{eqnarray}
and it equals zero for all $r > R_-$.  If we require the mass in
the positive perturbation cancel the mass in the negative one, then
$A = (R_-/R_+)^3$.  The various convolution integrals that should
be substituted in equation~(\ref{lambda}) are
\begin{eqnarray}
 \lambda_{++}(r) &=& A^2\,{4\pi R_+^3\over 3}
 \left[1-{3\over4}{r\over R_+} + {1\over 16}\left({r\over R_+}\right)^3\right]
 \qquad {\rm if}\ \  0\le r\le 2 R_+ \nonumber \\
 \lambda_{--}(r) &=& {4\pi R_-^3\over 3}
 \left[1-{3\over4}{r\over R_-} + {1\over 16}\left({r\over R_-}\right)^3\right]
 \qquad {\rm if}\ \  0\le r\le 2 R_- \nonumber \\
 \lambda_{+-}(r) &=& A\,{4\pi R_+^3\over 3}
                    \qquad{\rm if}\ \  0\le r\le (R_- - R_+) \nonumber \\
 &=& {A\pi\over 12 r} (R_-+R_+-r)^2
 \Bigl(r^2 + 2 r (R_- + R_+) - 3 (R_- - R_+)^2 \Bigr)
\end{eqnarray}
when $(R_--R_+)\le r\le (R_+ + R_-)$.
It is now straightforward to verify that the resulting expression for
$\lambda(r)$ satisfies the integral constraint given in 
equation~(\ref{intconst}).
 
Allowing for a range of profile shapes means that
 $\xi(r) = \int {\rm d}m\ p(m)\,\eta\lambda(r|m)$, 
where $m$ parameterizes the profile shape, and $p(m)$ is the
probability that a perturbation had shape $m$.  Note that in most models 
of current interest, the profile shape is a function of the mass
contained in the halo.
Since each of the $\lambda(r|m)$s satisfies the integral constraint,
$\xi(r)$ will also.  Similarly, the contribution to $P(k)$ at
small $k$ will be zero.

Thus, in contrast to positive perturbations, compensated profiles
satisfy the integral constraint on the correlation function, and
have vanishing power at small $k$.  Both these are physically
desirable improvements on the positive perturbation alone model.

The model with only positive perturbations is the only one which has 
been studied in the literature to date.  One consequence of this
is that, in these models,
$P^{\rm 1h}(k)\to \int dm\,{m^2 n(m)/\bar\rho^2}\ne 0$ as $k\to 0$.
Since $P^{\rm 2h}(k)$ tends to the linear perturbation theory value
in this limit, the sum of the two terms is actually inconsistent with
linear theory on the largest scales.  The discrepancy is small in models
of the dark matter distribution, but, for rare objects, the shot-noise-like
contribution from the $P^{\rm 1h}(k)$ term can be large \cite{Sel00}.
How to treat this discrepancy is an open question \cite{Scoetal01}.
One might have thought that compensating the profiles provides a
natural way to correct for this.  Unfortunately, compensated profiles
are constructed so that $u(k|m)\to 0$ at small $k$.  Since $P^{\rm 2h}(k)$
also depends on $u(k|m)$, in compensated models it too tends to zero at
small $k$.  Therefore, whereas uncompensated profiles lead to a little
too much power on large scales, compensating the profiles leads to no
large scale power at all!  (The physical reason for this is clear:  
because they are compensated, the total mass in the profiles integrates 
to zero.  Therefore, the profiles represent only local rearrangements 
of the mass; on scales larger than the typical perturbation, this 
rearrangement can be ignored---hence, the models have no large scale 
power.)  This drawback of the model which should be borne in mind 
when making predictions about the power on large scales.

\section{Dark Matter Power Spectrum, Bispectrum and Trispectrum}
We will now discuss results related to the dark matter distribution.
We show how the power spectrum is constructed under the halo model, 
discuss some aspects of higher order clustering, and include a calculation 
of correlations in estimates of the power spectrum.  We conclude this 
section with a discussion of the extent to which the halo model can be 
used as a astrophysical and cosmological tool, and suggest some ways 
in which the model can be extended.  

\subsection{Power Spectrum}

\begin{figure*}[t]
\centerline{\psfig{file=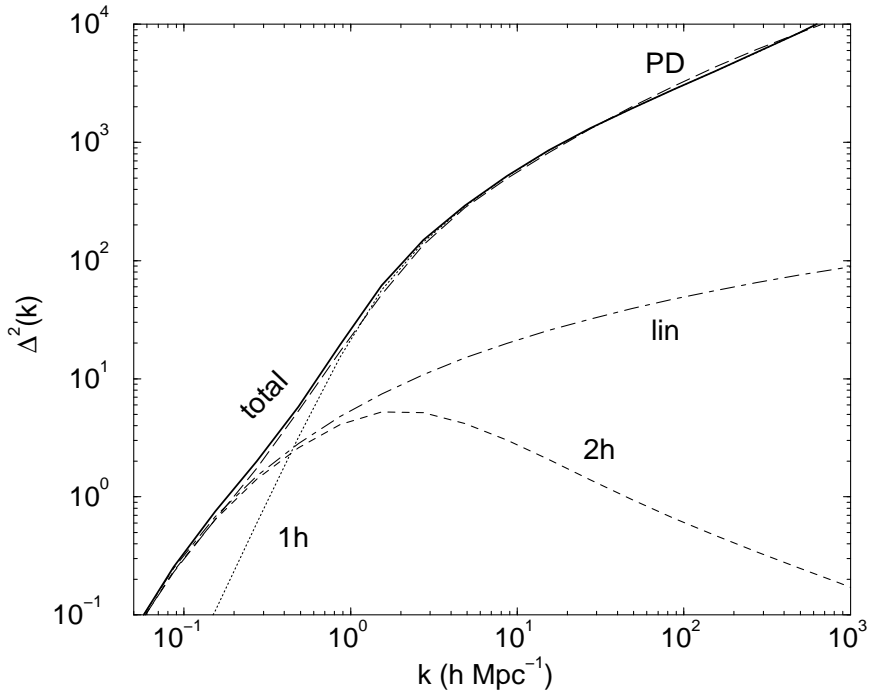,width=4.5in,angle=-90}}
\caption{Power spectrum of the dark matter density field at the
present time.  Curve labeled `PD' shows the fitting formula of
\cite{PeaDod96}.  Dot dashed curve labeled `lin' shows the linear
$P^\lin(k)$.  Dotted and short dashed curves show the two terms which
sum to give the total power (solid line) in the halo model.}
\label{fig:dmpower}
\end{figure*}

Figure~\ref{fig:dmpower} shows the power spectrum of the dark matter
density field at the present time ($z=0$).
Dotted and short dashed lines show the contributions to the power from
the single and two halo terms.  Their sum (solid) should be compared to the
power spectrum measured in numerical simulations, represented here by
the dashed curve labeled `PD' which shows the fitting function of 
equation~(\ref{eqn:pdfit}).  (At the largest $k$ shown, this fitting
function represents an extrapolation well beyond what has actually 
been measured in simulations to date, so it may not be reliable.)  
In computing the halo model curves we have included the effect of the 
scatter in the halo concentrations (equation~\ref{eqn:concentration}).  
Although ignoring the scatter is actually a rather good approximation, 
for precise calculations, the scatter is important, especially for 
statistics which are dominated by massive halos.

In general, the linear portion of the dark matter power spectrum, 
$k \le 0.1h$ Mpc$^{-1}$, results from the correlations between dark 
matter halos and reflects the halo--mass dependent bias prescription.  
The spherical or ellipsoidal collapse based models described 
previously describe this regime reasonably well at all redshifts.  
At $k \sim 0.1-1h$ Mpc$^{-1}$, the one- and two--halo terms are 
comparable; on these scales, the power comes primarily from halos more 
massive than $M_\star$. At higher $k$'s, the power comes mainly from 
individual halos with masses below $M_\star$.

The small scale behavior of the power spectrum is sensitive to 
assumptions we make with regarding the halo profile. If we change 
the shape of the density profile, e.g., from NFW to M99, then 
$P(k)$ will change.  However, if we also modify the mean 
mass--concentration relation, then the difference between the two 
$P(k)$s can be reduced substantially.  
We discuss the effect of allowing a distribution $p(c|m)$ of 
concentrations at fixed mass (i.e., allowing some scatter around the 
mean mass--concentration relation) at the end of this section.  

\subsection{Bispectrum and trispectrum}

\begin{figure*}[t]
\centerline{\psfig{file=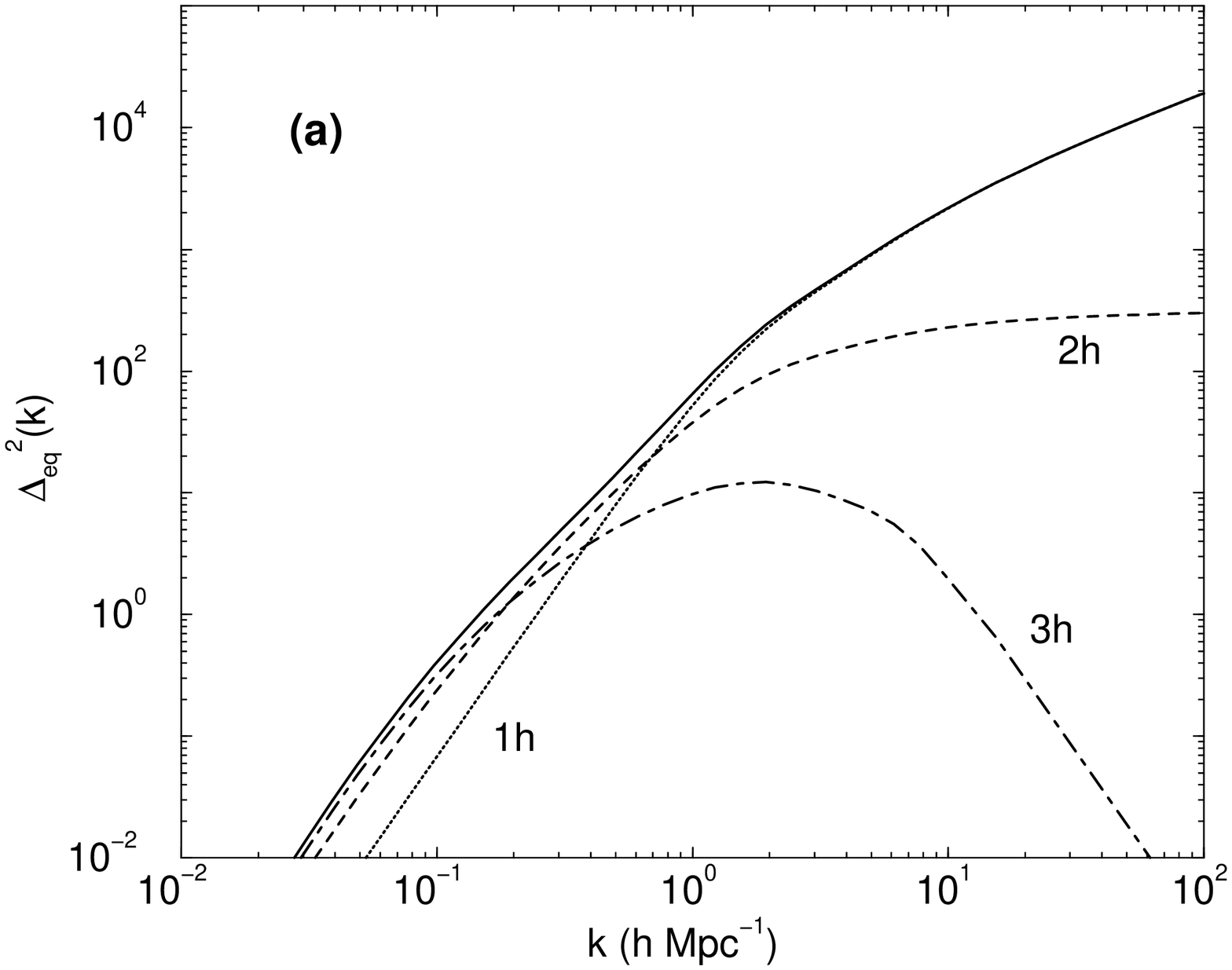,width=2.5in,angle=-90}
            \psfig{file=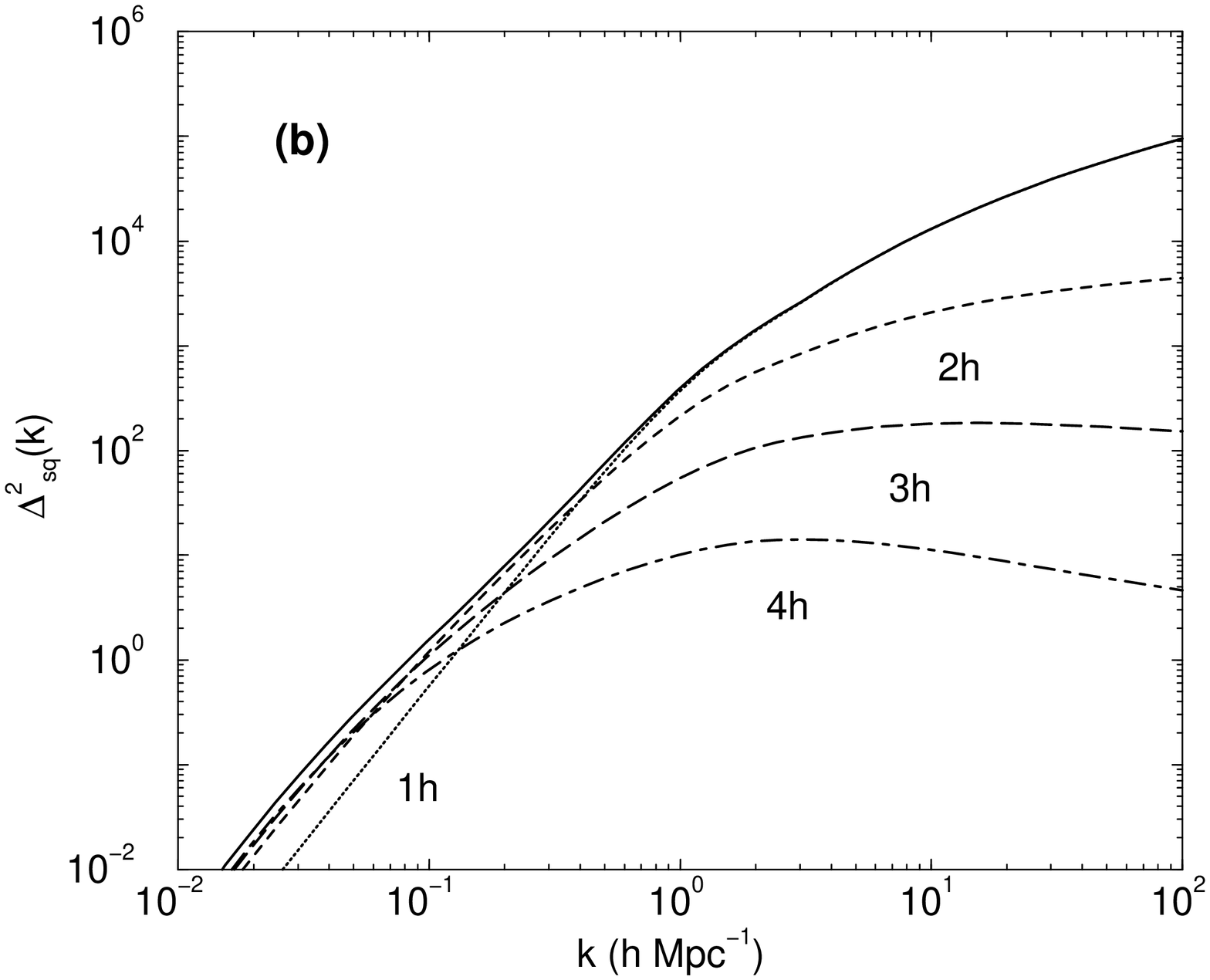,width=2.5in,angle=-90}}
\caption{The (a) equilateral bispectrum and (b) square
trispectrum of the dark matter in the halo model.  Solid 
lines show the total bispectrum and trispectrum, and the 
different line styles show the different contributions to the 
total.}
\label{fig:dmbi}
\end{figure*}

\begin{figure*}[t]
\centerline{\psfig{file=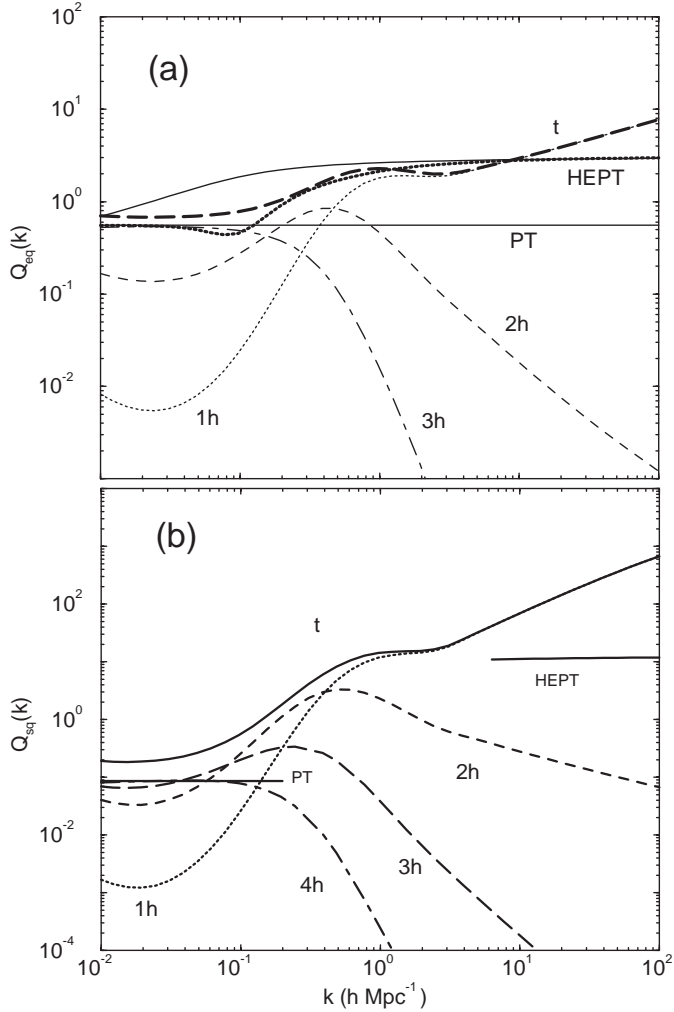,width=3.5in}}
\caption{(a) $Q_{\rm eq}(k)$ and (b) $Q_{\rm sq}$ at $z=0$.  
Different lines styles show the different contributions to the total 
(bold dashed) in the halo model description.  Thin solid lines show 
the second order perturbation theory (PT) and HEPT values. In (a), the 
thick solid line shows the fitting formula for $Q_{\rm eq}$ from 
\cite{ScoCou00}.  Notice that on linear scales, the halo model 
prediction is about twenty percent larger than the PT value in (a), 
and about a factor of two larger than the PT value in (b).}
\label{fig:dmq}
\end{figure*}

Figures~\ref{fig:dmbi}(a) and (b) show the bispectrum and trispectrum 
of the density fluctuation field at $z=0$. Since the bispectrum and 
trispectrum depend on the shape of the triangle and quadrilateral, 
respectively, the figure is for configurations which are equilateral 
triangles and squares.  Since the power spectra and equilateral 
bispectra share similar features, it is more instructive to study 
$Q_{\rm eq}(k)$, defined by equation~(\ref{Q3}).  
Figure~\ref{fig:dmq}(a) compares the halo model estimate of $Q_{\rm eq}$ 
with the second order perturbation theory (PT) and HEPT predictions 
(equations~\ref{Q3PT} and~\ref{Q3hept}). 
In the halo prescription, $Q_{\rm eq}$ at 
$k \simgt 10 k_{\rm nonlin} \sim 10 h$~Mpc$^{-1}$ 
arises mainly from the single halo term. 
Figure~\ref{fig:dmq}(a) also shows the fitting function for 
$Q_{\rm eq}(k)$ from \cite{ScoCou00}, which is based on simulations 
in the range $0.1h \lesssim k \lesssim 3h$ Mpc$^{-1}$. This function is
designed to converge to the HEPT value at small scales and the PT
value at large scales. Notice that the HEPT prediction is considerably  
smaller than the halo model prediction on small scales.

\begin{figure*}[t]
\centerline{\psfig{file=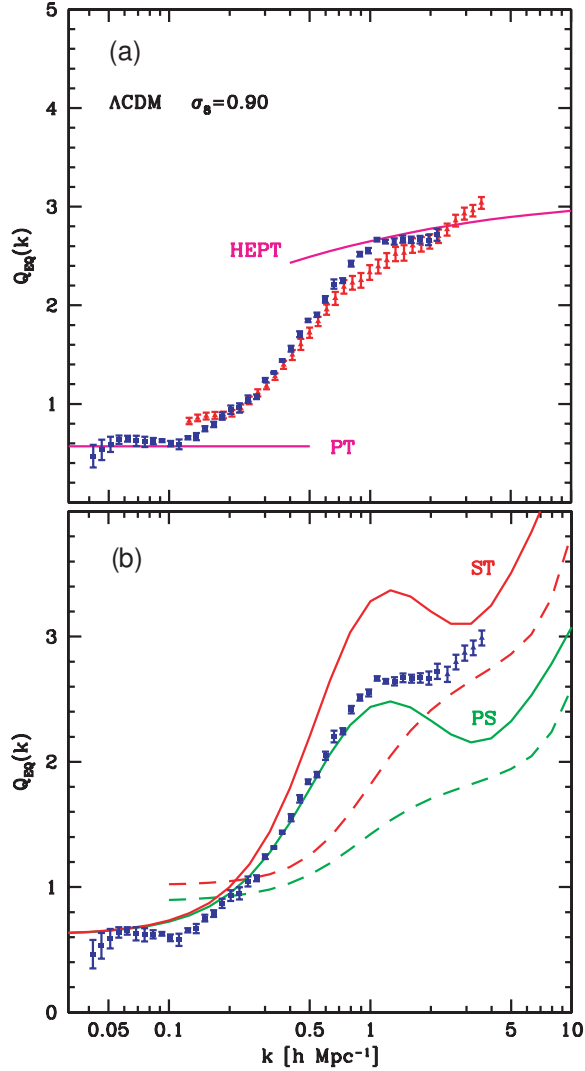,width=3.in}}
\caption{$Q_{\rm eq}(k)$  as a function of scale, 
(a) measured in numerical simulations and 
(b) compared to halo model predictions. In (a), the triangles are
measurements in a box of size 100$h^{-1}$ Mpc while squares denote
measurements in box sizes of 300$h^{-1}$ Mpc.  The linear 
perturbation theory (PT) and hyperextended perturbation 
theory (HEPT) values are show as solid lines.  
In (b), the halo model predictions associated with Press-Schechter 
and Sheth-Tormen mass functions generally bracket the measurements. 
The dashed lines show the result of only including contributions 
from halos less massive than $10^{14}\,h^{-1}M_\odot$.  They lie 
significantly below the solid curves, illustrating that massive 
halos provide the dominant contributions to these statistics. 
The figure is taken from \cite{Scoetal01}.}
\label{fig:dmbisimul}
\end{figure*}

Figure~\ref{fig:dmbisimul} (from \cite{Scoetal01}) compares the 
predicted $Q_{\rm eq}$s with measurements in numerical simulations. 
To the resolution of the simulations, the data are consistent with 
perturbation theory at the largest scales, and with HEPT in the 
non-linear regime.  The halo model predictions based on the two 
mass function choices (Press-Schechter and Sheth-Tormen) generally 
bracket the numerical simulation results, assuming the same halo 
profile and concentration--mass relations are the same in both cases.  
The most massive halos are responsible for a significant fraction of 
the total non-Gaussianity in the non-linear density field.  This is 
shown in the bottom panel of Figure~\ref{fig:dmbisimul}; when halos 
more massive than 10$^{14}\,M_\odot/h$ are absent, $Q_{\rm eq}$ is 
reduced substantially (compare dashed and solid curves). 

The halo based calculation suggests $Q_{\rm eq}$ increases, whereas 
HEPT suggests that $Q_{\rm eq}$ should remain approximately constant, 
on the smallest scales.  These small scales are just beyond the reach of 
numerical simulations to date.  
As we discuss later, the scales where the two predictions differ 
significantly are not easily probed with observations either, at 
least at the present time.  

For the trispectrum, and especially the contribution of trispectrum to
the power spectrum covariance as we will soon discuss, we are mainly 
interested in terms of the form $T(\veck_1,-\veck_1,\veck_2,-\veck_2)$, 
i.e. parallelograms which are defined by either the length $k_{12}$ 
or by the angle between $\veck_1$ and $\veck_2$.  To illustrate, our 
results, we will take $k_1=k_2$ and the angle to be $90^\circ$ 
($\veck_2=\veck_\perp$) so that the parallelogram is a square.
It is then convenient to define
\begin{equation}
\Delta^2_{\rm sq}(k) \equiv \frac{k^3}{2\pi^2}
T^{1/3}(\veck,-\veck,\veck_\perp,-\veck_\perp) \, .
\end{equation}
This quantity scales roughly as $\Delta^2(k)$.  This spectrum is 
shown in figure~\ref{fig:dmbi}(b) with the individual contributions 
from the 1h, 2h, 3h, 4h terms shown explicitly.  At 
$k\simgt 10k_{\rm nonlin}\sim 10h$Mpc$^{-1}$, $Q_{\rm sq}$ is 
due mainly from the single halo term. 

As for $Q_{\rm eq}$, the halo model predicts that $Q_{\rm sq}$ will 
increase at high $k$.  Numerical simulations do not quite have enough 
resolution  to test this \cite{Scoetal01}.  

Figures~\ref{fig:dmq}(a) and~(b) show that as one considers higher 
order statistics, the halo model predicts a substantial excess in power 
at linear scales compared to the perturbation theory value.  
This is another manifestation of the problem, noted in \S~\ref{cmpnst}, 
that, in positive perturbation models, the single-halo contribution 
to the power does not vanish as $k\to 0$.  
While this discrepancy appears large in the $Q_{\rm sq}$ statistic, 
it does not affect the calculations related to the covariance of 
large scale structure power spectrum measurements since, on linear 
scales, the Gaussian contribution usually dominates the non-Gaussian 
contribution. However, we caution that dividing the halo model 
calculation on linear scales by the linear power spectrum to obtain, 
say halo bias or galaxy bias, may lead to errors.

\subsection{Power Spectrum Covariance}
 
\begin{table*}
\begin{center}
\caption{\label{tab:dmcorr}}
{\sc Dark Matter Power Spectrum Correlations\\}
\begin{tabular}{lrrrrrrrrrr}
\hline
$k$   & 0.031 &  0.058 & 0.093 & 0.110 & 0.138 & 0.169
& 0.206 & 0.254 & 0.313 & 0.385 \\
\hline
0.031 &  1.000 & 0.041 & 0.086 & 0.113 & 0.149 & 0.172
& 0.186&  0.186 & 0.172 & 0.155 \\
0.058 &  (0.023) & 1.000 & 0.118 & 0.183 & 0.255 &
0.302 & 0.334 & 0.341 & 0.328 & 0.305 \\
0.093 &  (0.042) & (0.027) & 1.000 & 0.160 & 0.295
& 0.404 & 0.466 & 0.485 & 0.475 & 0.453 \\
0.110 &  (0.154) & (0.086) & (0.028) & 1.000 &
0.277 & 0.433 & 0.541 & 0.576 & 0.570 & 0.549 \\
0.138 &  (0.176) & (0.149) & (0.085) & (0.205) &
1.000 & 0.434 & 0.580 & 0.693 & 0.698 & 0.680 \\
0.169 &  (0.188) & (0.138) & (0.177) & (0.251) &
(0.281) & 1.000 & 0.592 & 0.737 & 0.778 & 0.766 \\
0.206 &  (0.224) & (0.177) & (0.193) & (0.314) &
(0.396) & (0.484) & 1.000 & 0.748 & 0.839 & 0.848 \\
0.254 &  (0.264)  & (0.206) & (0.261) & (0.355) &
(0.488) & (0.606) & (0.654) & 1.000 & 0.858 & 0.896 \\
0.313 &  (0.265)  & (0.202) & (0.259) & (0.397) &
(0.506) & (0.618) & (0.720) & (0.816) & 1.000 & 0.914 \\
0.385 &  (0.270)  & (0.205) & (0.262) & (0.374) &
(0.508) & (0.633) & (0.733) & (0.835) & (0.902) & 1.000 \
\\
\hline
$\sqrt{\frac{C_{ii}}{C_{ii}^{G}}}$ & 1.00 & 1.02& 1.04& 1.07& 1.14& 1.23& 1.38& 1.61& 1.90& 2.26 \\
\hline
 & & & & & & & \\
\\
\end{tabular}
\end{center}
\footnotesize
NOTES.---%
Diagonal normalized covariance matrix of the binned dark matter
density field power spectrum with $k$ in units of $h$ Mpc$^{-1}$.
Upper triangle displays the covariance found under the halo model.
Lower triangle (parenthetical numbers) displays the covariance found
in numerical simulations by \cite{MeiWhi99}.  Final line shows
the fractional increase in the errors (root diagonal covariance) due to
non-Gaussianity as calculated using the halo model.\\
\end{table*}

The trispectrum is related to the variance of the estimator of the 
binned power spectrum \cite{Scoetal99,MeiWhi99,EisZal01,CooHu01b}:
\begin{equation}
 \hat P_i = {1 \over V } \int_{\shell i} {d^3 k \over V_{\shell i}}
             \delta^*(-\veck) \delta(\veck)  \, ,
\end{equation}
where the integral is over a shell in $k$-space centered around $k_i$,
$V_{\shell i} \approx 4\pi k_i^2 \delta k$ is the volume of the shell
and $V$ is the volume of the survey.  Recalling that $\delta({\bf
0})\rightarrow V/(2\pi)^3$
for a finite volume,
\begin{equation}
C_{ij} \equiv \left< \hat P_i \hat P_j \right> - 
      \left< \hat P_i \right> \left< \hat P_j \right>  
       = {1\over V} \left[ {(2\pi)^3 \over V_{\shell i} } 
          2 P_i^2 \delta_{ij} + T_{ij} \right]  \, ,
\end{equation}
where
\begin{equation}
T_{ij} \equiv \int_{\shell i} {d^3 k_i \over V_{\shell i}}
      \int_{\shell j} {d^3 k_j \over V_{\shell j}}
      T(\veck_i,-\veck_i,\veck_j,-\veck_j) \,.
\label{eqn:covarianceij}
\end{equation}
Although both terms scale in the same way with the volume of the 
survey, only the (first) Gaussian piece necessarily decreases with 
the volume of the shell.  For the Gaussian piece, the sampling error 
reduces to a simple root-N mode counting of independent modes in a 
shell.  The trispectrum quantifies the non-independence of the
modes both within a shell and between shells.  Therefore, calculating 
the covariance matrix of the power spectrum estimates reduces to 
averaging the elements of the trispectrum across configurations in 
the shell.  For this reason, we now turn to the halo model 
description of the trispectrum.  

To test the accuracy of the halo trispectrum, we compare dark matter 
correlations predicted by our method to those from numerical 
simulations by \cite{MeiWhi99} (see also, \cite{Scoetal99}). 
Specifically, we calculate the covariance matrix $C_{ij}$ from
equation~(\ref{eqn:covarianceij}) with the bins centered at $k_i$ 
and volume $V_{\shell i} = 4\pi k_i^2 \delta {k_i}$ corresponding to 
their scheme.  We also employ the parameters of their $\Lambda$CDM 
cosmology and assume that the parameters that defined the halo
concentration properties from our fiducial $\Lambda$CDM model holds
for this cosmological model also. The physical differences between the
two cosmological model are minor, though normalization differences can
lead to large changes in the correlation coefficients.

\begin{figure*}[t]
\centerline{\psfig{file=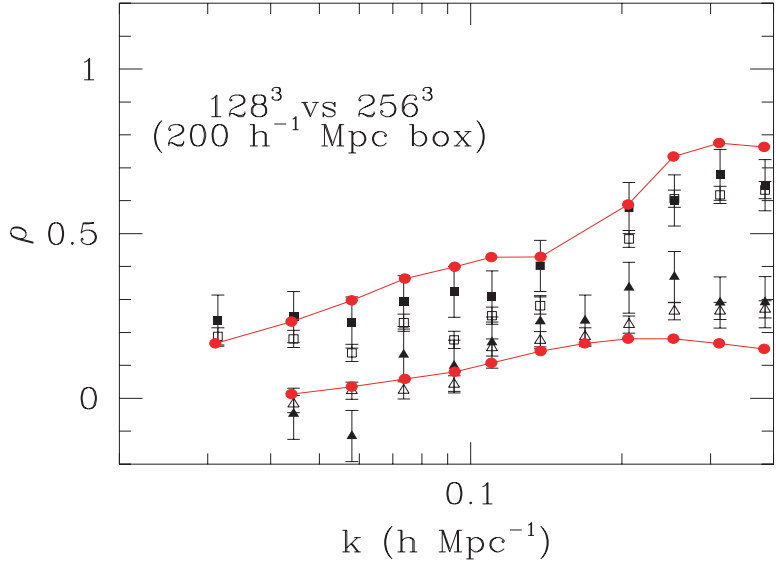,width=4.0in}}
\caption{The correlations in dark matter power spectrum between bands 
centered at $k$ (see Table~1) and those centered at $k=0.031$h Mpc$^{-1}$ 
a (lower triangles) and $k=0.169$h Mpc$^{-1}$ ((upper squares). The open 
and filled symbols in these cases are for 200 h$^{-1}$ Mpc  box 
simulations with 128$^3$ and 256$^3$ particles, respectively. The solid 
lines with filled circles represent the halo model predictions for same 
bands and are consistent with numerical
simulations at the level of 10\% or better. The figure is reproduced from 
\cite{MeiWhi99}.}
\label{fig:dmcorr}
\end{figure*}

Table \ref{tab:dmcorr} compares the halo model predictions for the 
correlation coefficients
\begin{equation}
\hat C_{ij} = {C_{ij} \over \sqrt{C_{ii} C_{jj}}}
\end{equation}
with those measured in the simulations.  Agreement in the off--diagonal 
elements is typically better than $\pm 0.1$, even in the region where
non-Gaussian effects dominate, and the qualitative features such as 
the increase in correlations across the non-linear scale are reproduced. 
The correlation coefficients for two bands in the linear 
($0.031h$ Mpc$^{-1}$) and non-linear ($0.169h$ Mpc$^{-1}$) regimes 
are shown in Figure~\ref{fig:dmcorr}.  Triangles and squares 
show the values measured in the simulations, and filled circles and 
solid lines show the halo model predictions.  The halo model is in 
agreement with numerical measurements over a wide range of scales, 
suggesting that it provides a reasonable way of estimating the 
covariance matrix associated with the dark matter power spectrum.
In contrast, perturbation theory can only be used to describe the
covariance and correlations in the linear regime while in the 
non-linear regime, and although the HEPT provides a reasonable 
description when $k_i \sim k_j$, it results in large discrepancies 
when $k_i \gg k_j$ \cite{MeiWhi99,Scoetal99,Ham00}.

\begin{figure*}[t]
\centerline{\psfig{file=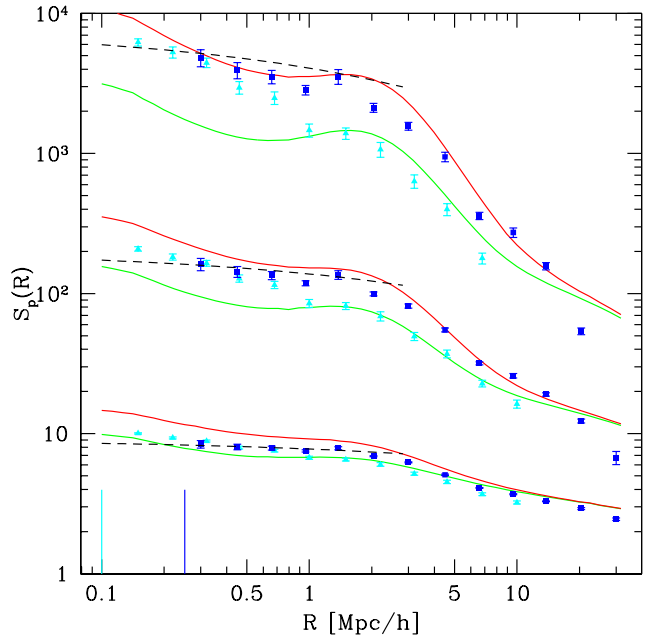,width=3.5in}}
\caption{Real-space moments with $p=3$ (skewness), 4 (kurtosis) 
and 5 as a function of smoothing scale. Squares and triangles show 
measurements in high and low resolution simulations, illustrating 
how difficult it is to make the measurement.  Solid lines show the 
predictions based on the NFW profile but with Press-Schechter 
(lower) and Sheth-Tormen (upper) mass functions. 
Dashed line shows the HEPT prediction. 
The figure is from \cite{Scoetal01}.}
\label{fig:sp}
\end{figure*}

A further test of the accuracy of the halo approach is to consider 
higher order real-space moments such as the skewness and kurtosis.  
Figure~\ref{fig:sp} compares measurements of  higher order moments 
in numerical simulations with halo model predictions:  the halo 
model is in good agreement with the simulations.  

\subsection{Can we trust the halo model?}
\begin{figure}[t]
\centerline{\psfig{file=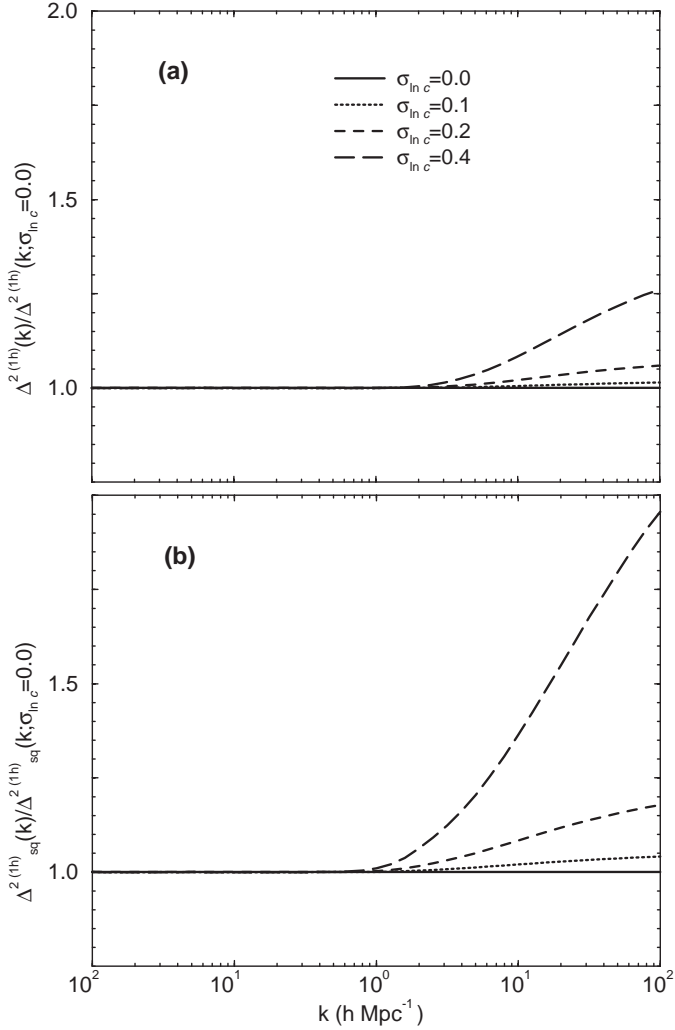,width=3.5in}}
\caption{Ratio of the single halo term contribution to the 
total power when the distribution of concentrations at fixed 
mass is lognormal with width $\sigma_{{\rm ln} c}$, to that
when $\sigma_{\ln c}\rightarrow 0$ for the power spectrum (a) 
and trispectrum (b). The small scale behavior, particularly of 
the higher order statistics, is sensitive to the high concentration 
tails of the $p(c|m)$ distribution.}
\label{fig:conc}
\end{figure}

The halo model provides a physically motivated means of estimating
the two-point and higher order statistics of the dark matter density
field.  However, it has several limitations which should not be forgotten
when interpreting results. As currently formulated, the approach assumes all
halos share a parameterized smooth spherically-symmetric profile which 
depends only on halo mass.  However, we know that halos of the same mass 
have a distribution of concentration parameters, so that there is some 
variation in halo profile shape, even at fixed mass.  In addition, halos 
in simulations are rarely smooth, and they are often not spherically 
symmetric.  

It is straightforward to incorporate the distribution of halo 
concentrations into the formalism \cite{CooHu01b,ScoShe01}.  
In essence, a distribution $p(c|m)$ leads to changes in power at 
non-linear scales $k\gtrsim 1h$~Mpc$^{-1}$.   
This is shown in Figure ~\ref{fig:conc}(a):  the power on large linear 
scales is unaffected by a distribution $p(c|m)$, but the large $k$ 
power increases as the width of $p(c|m)$ increases.  
Increasing $\sigma_{\ln\,c}$ increases the power at small scales, 
because of the increased probabilty of occurence of high 
concentrations from the tail of the distribution.
(Recall that simulations suggest $\sigma_{\ln\,c}\approx 0.25$.)  
Higher order statistics depend even more strongly on $\sigma_{\ln\,c}$, 
because they weight the large $c$ tails heavily.  To illustrate, 
Figure~\ref{fig:conc}(b) shows how the trispectrum, with 
$\veck_1=\veck_2=\veck_3=\veck_4$, changes as $\sigma_{\ln\,c}$ increases.  

\begin{figure*}[t]
\centerline{\psfig{file=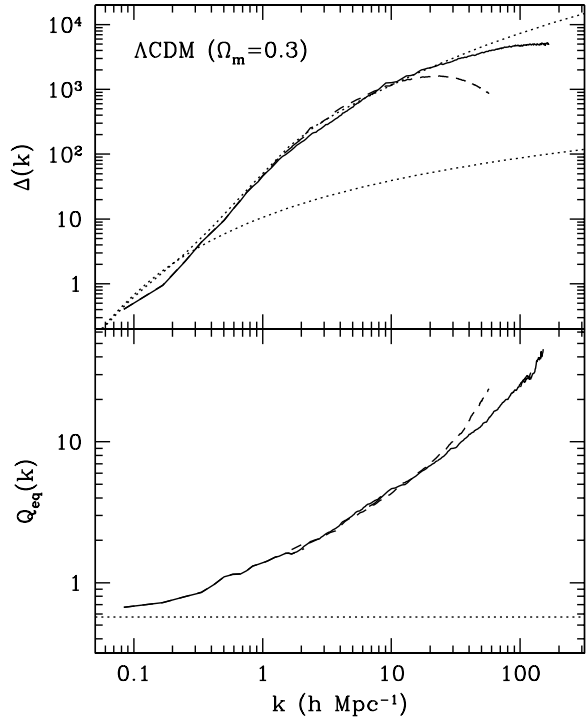,width=4.2in}}
\caption{Dark matter power spectrum (top) and reduced bispectrum 
for equilateral configurations (bottom) in numerical simulations 
(solid lines). Dashed lines show result of replacing halos with 
smooth M99 profiles and remaking the measurements.  
The replacement agrees with the original 
measurements up to the resolution limit of the simulation: 
$k \sim 10h$ Mpc$^{-1}$. Dotted curves are the linear and nonlinear
expectation, based on fitting functions, in the case of the power 
spectrum and the linear perturbation theory result of $4/7$ in the 
case of $Q_{\rm eq}$.  The figure is from \cite{MaFry00b}.}
\label{fig:mafry}
\end{figure*}

Substructure is expected to contribute about 15\% of the total dark matter 
mass of a halo (e.g., \cite{Toretal98,Ghietal00}), and it will affect the 
power spectrum and higher order correlations on small scales.  
Measurements of $P(k)$ and $B(k)$ for equilateral triangles in which the
actual clumpy nonspherical halo profiles in numerical simulations were
replaced by smooth NFW or M99 halo profiles suggest that for
$k\le 10k_{\rm nonlin}$ or so, substructure and asphericities are not
important (see Figure~\ref{fig:mafry}).  A detailed discussion of how to 
account for this substructure is in \cite{SheJai02}.  

No models to date account for departures from spherical symmetry, but 
this is mainly because until recently \cite{JinSut02}, there was no 
convenient parametrization of profile shapes which were not spherically 
symmetric.  There is no conceptual reason which prevents one from including 
ellipsoidal halos in the model.  Until this is done, note that spherically
averaged profiles are adequate for modelling the power spectrum and other
statistics which average over configurations, such as the $S_n$ parameters.
The bispectrum is the lowest order statistic which is sensitive to
the detailed shape of the halos.  
The dependence of bispectrum configuration on the spherical assumption
was shown in some detail by \cite{Scoetal01}; they found that the spherical
assumption may be the cause of discrepancies at the $\sim 20-30\%$ level
between the halo model predictions and configuration dependence
of the bispectrum in the mildly non-linear regime measured in simulations.
Uncertainties in the theoretical mass function also produce variations at
the 20\% to 30\% level (see, \cite{Jenetal01}).

Improvements to the halo model that one should consider include:\\
(1) Introduction of the asphericity of dark matter halos through a
randomly inclined distribution of prolate and oblated ellipsoids.
Recent work has shown that simply modifying the spherically symmetric 
profile shape to have different scale lengths along the three principal 
axes provides a reasonable parametrization of the ellipsoidal profiles 
of halos in numerical simulations, with the distribution of axis ratios 
depending on halo mass \cite{JinSut02}.  This makes it relatively 
straightforward to include asphericity in the model.  Since the same 
ellipsoidal collapse model \cite{Sheetal01a} which predicts the correct 
shape for the halo mass function (equation~\ref{fgif}), can also be used 
to predict the distribution of halo axis ratios, it would be interesting 
to see if this distribution matches that in simulations.  
At the present time, shape information from X-ray observations of galaxy 
clusters is  limited \cite{Coo00a}, although \cite{Jeteatl02} argue
that departures from spherical symmetry are necessary to correctly
interpret their data.\\
(2) Incorporation of the effects of halo substructure. 
See \cite{SheJai02} for a first step in this direction, which incorporates 
simple models of what is seen in numerical simulations 
\cite{Ghietal00,BlaShe00}.\\
(3) Solution of the integral constraint problem at large scales
discussed in \S~\ref{cmpnst}. \\

\begin{figure}[t]
\centerline{\psfig{file=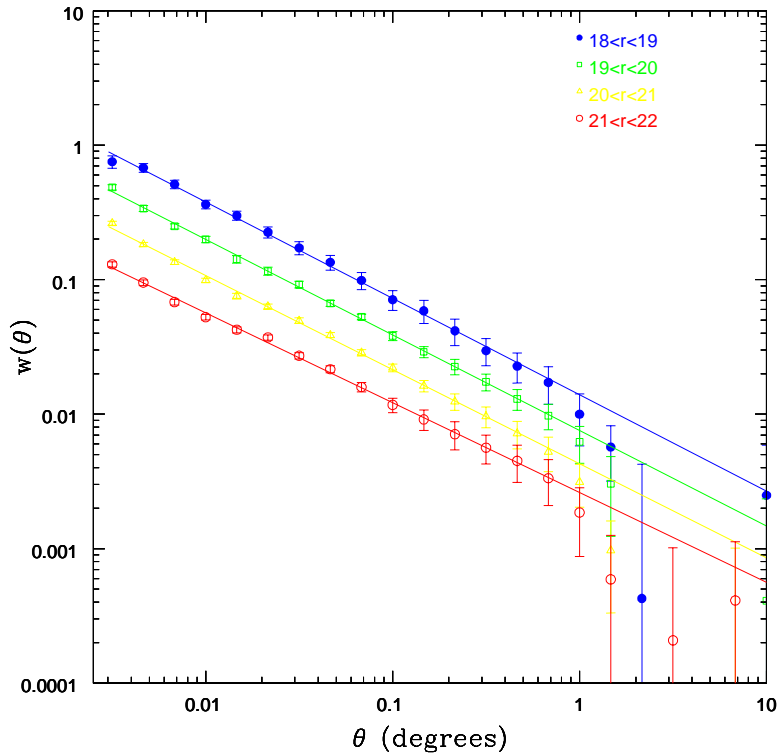,width=4.2in,angle=0}}
\caption{The angular two-point correlation function of galaxies in the
SDSS early release data, for a number of bins in apparent $r^*$ band 
magnitude. In all cases, the correlation function is quite well 
described by a power law:  $w(\theta)\propto \theta^{-0.7}$. 
The figure is from \cite{Conetal01}.}
\label{fig:sdsswtheta}
\end{figure}

\section{From Dark Matter to Galaxies}
\label{sec:galaxy}

\begin{figure}[t]
\centerline{\psfig{file=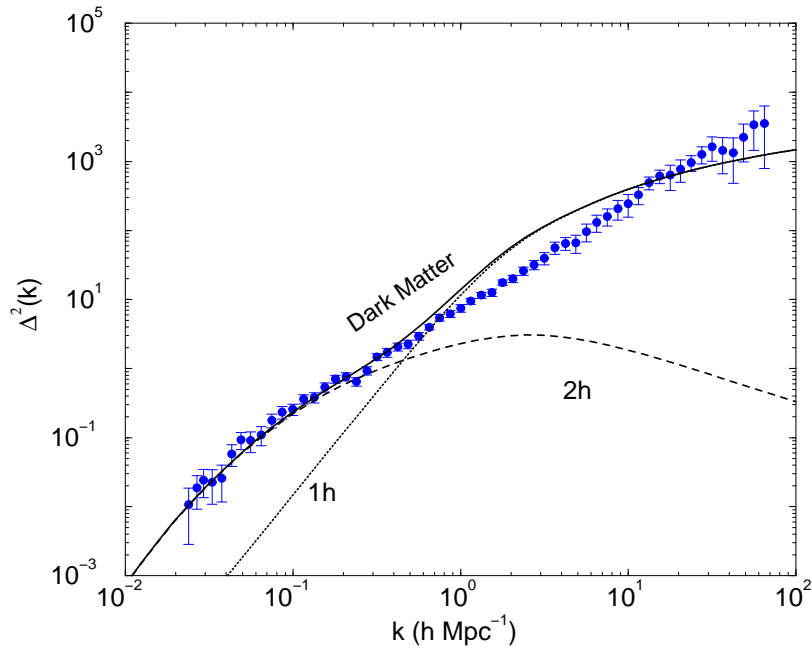,width=4.2in,angle=-90}}
\caption{The PSC$z$ galaxy power spectrum (symbols, from \cite{HamTeg00})
compared to the dark matter power spectrum in a $\Lambda$CDM model
(solid curve).  We have fixed the amplitude of the dark matter power
spectrum so that it matches the data on large scales.  The discrepancy
on smaller nonlinear scales suggests that the bias between the galaxies
and the dark matter must be scale dependent.  }
\label{fig:galaxypower}
\end{figure}

\begin{figure}[t]
\centerline{\psfig{file=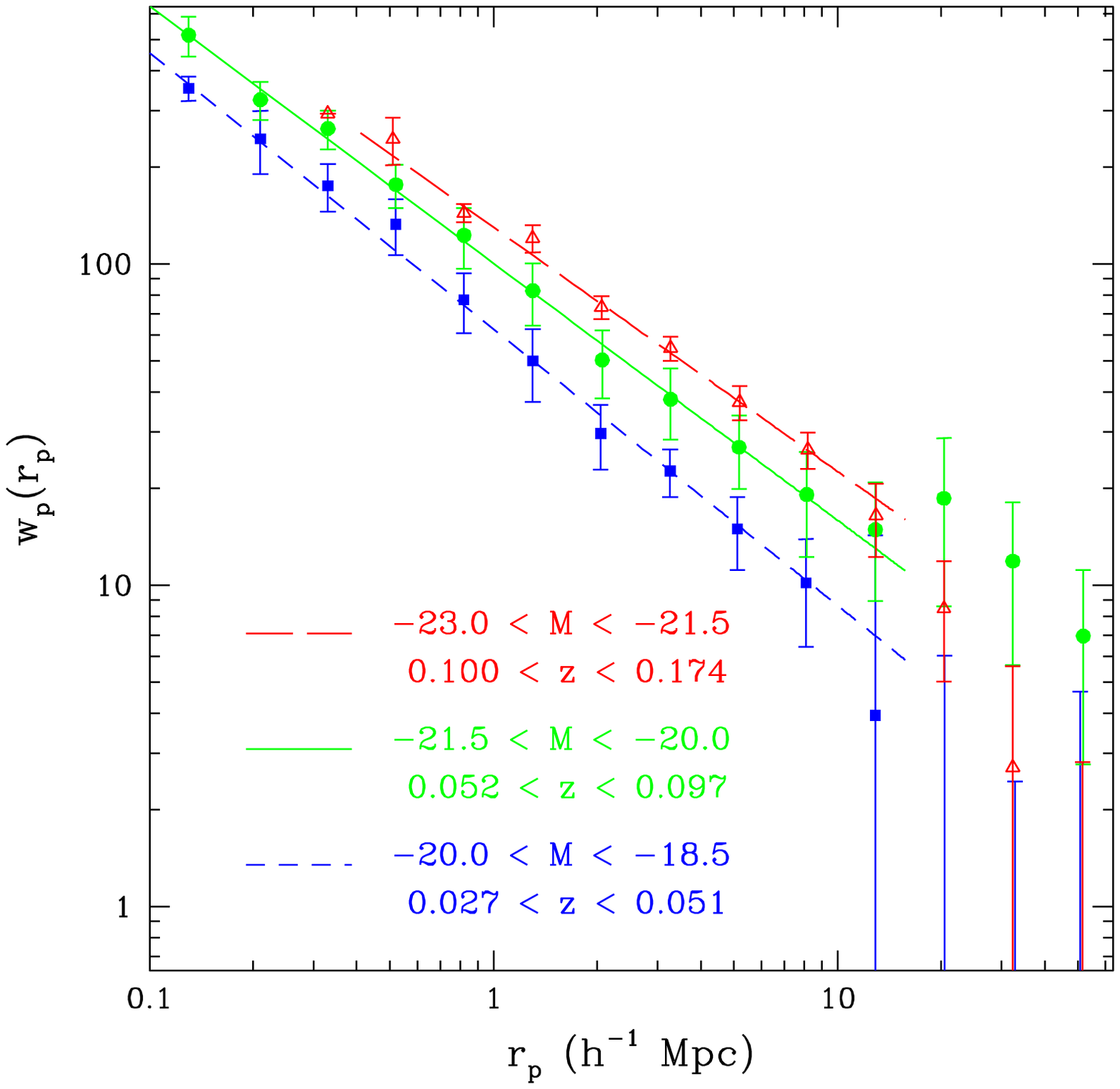,width=2.75in,angle=0}
            \psfig{file=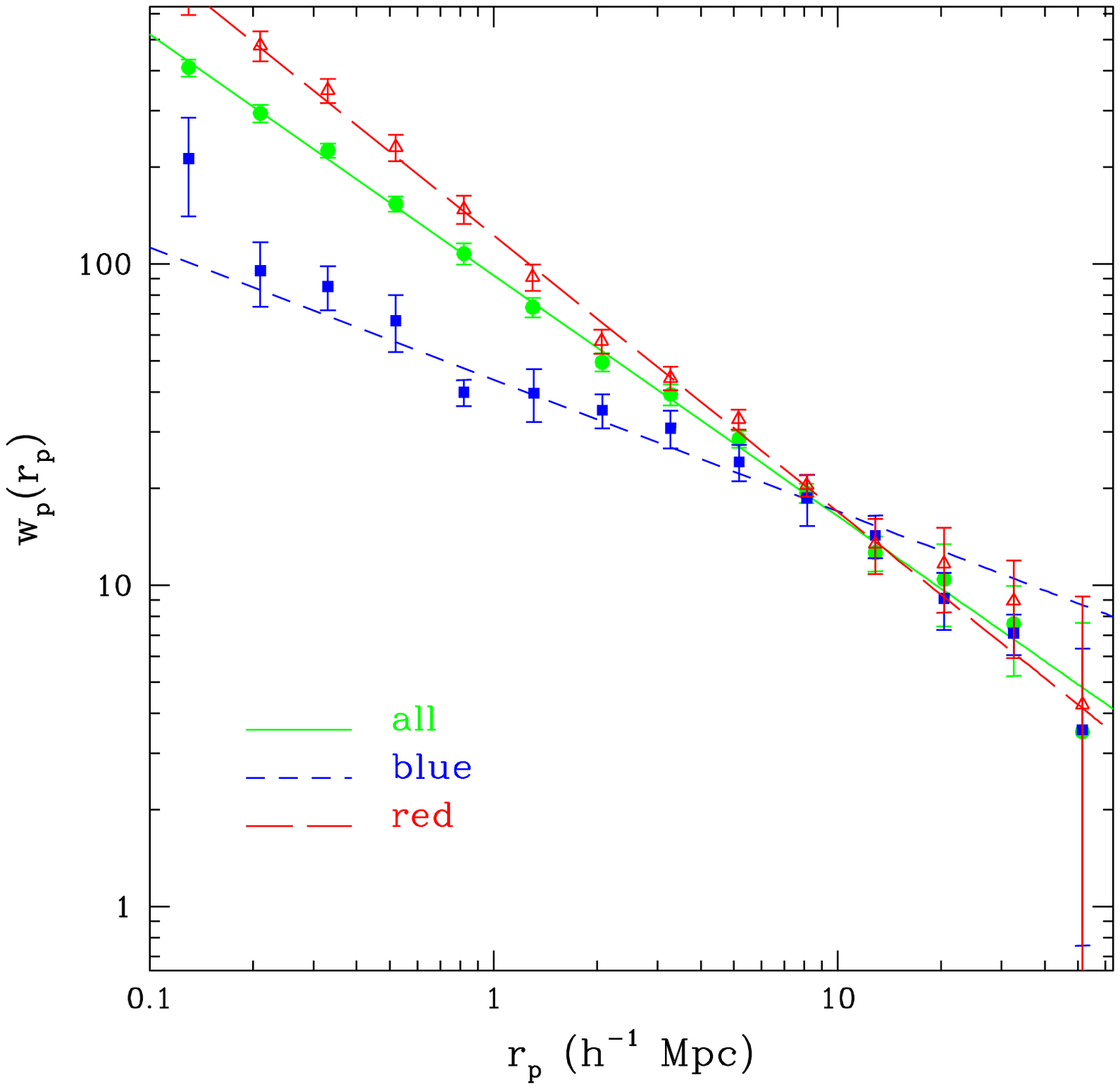,width=2.75in,angle=0}}
\caption{Projected two-point correlation function of galaxies 
with absolute magnitude and redshift ranges indicated (left) and 
for different bins in color (right).  In panel on left, squares, 
circles and triangles show results for faint, intermediate and 
luminous galaxies respectively.  Although the more luminous galaxies 
are more strongly clustered, the same power-law slope provides a 
reasonable fit at all luminosities.  In constrast, the slope of 
the power-law is a strong function of color.  Both panels are from 
\cite{Zehetal02}.}
\label{fig:sdsswrp}
\end{figure}

We have known since the late 1960's that the angular correlation function 
of optically selected galaxies is a power law:
 $w(\theta)\propto \theta^{-(\gamma-1)}$, with $\gamma\approx 1.8$
\cite{TotKih69}.  Figure~\ref{fig:sdsswtheta} shows a recent measurement 
of $w(\theta)$ from the SDSS collaboration \cite{Conetal01}:
it is also well described by this power law.  This suggests that the
three-dimensional correlation functions and power-spectra should also
be power laws.  The symbols in Figure~\ref{fig:psczpower}
show that the power-spectrum of galaxies in the PSC$z$ survey as measured
by \cite{HamTeg00} is accurately described by a power-law over a
range of scales which spans about three orders of magnitude.  More 
recently, the 2dFGRS \cite{Noretal01} and SDSS \cite{Zehetal02} data 
show that, although more luminous galaxies cluster more strongly,
for a wide range of luminosities, the three-dimensional correlation
function is indeed close to a power law.  Figure~\ref{fig:sdsswrp}, 
from \cite{Zehetal02}, shows that although the slope of the power-law 
is approximately independent of luminosity (left), it is a strong 
function of galaxy color; on small scales, redder galaxies have 
steeper correlation functions.  

In contrast, a generic prediction of CDM models is that, at the present
time, the two-point correlation function of the dark matter, and its
Fourier transform, the dark matter power spectrum, are not power laws
(see, e.g., the solid curve in Figure~\ref{fig:psczpower}).
Why is the clustering of galaxies so different from that of the
dark matter?
 
\subsection{The clustering of galaxies}
In the approach outlined by White \& Rees \cite{WhiRee78}, baryonic gas
can only cool and form stars if it is in potential wells such as those
formed by virialized dark matter halos.  As a result, all galaxies are
expected to be embedded in dark halos (see figure~\ref{fig:gif}).
More massive halos may contain many galaxies, in which case it is
natural to associate the positions of galaxies with subclumps within
the massive halo; some, typically low mass, halos may contain no
galaxies; but there are no galaxies which are without halos.
Within this framework, the properties of the galaxy population are
determined by how the gas cooling rate, the star formation rate, and the
effects of stellar evolution on the reservoir of cooled gas, depend on
the mass and angular momentum of the parent halo.
There are now a number of different prescriptions for modeling these
`gastrophysical' effects \cite{WhiFre91,Kauetal99,Coletal00}.
 
Within the context of the halo model, the gastrophysics determines
how many galaxies form within a halo, and how these galaxies are
distributed around the halo center.  Thus, the halo model provides a
simple framework for thinking about and modeling why galaxies cluster
differently than dark matter 
\cite{Jinetal98,Sel00,PeaSmi00,Scoetal01,SheDia01,Scr02,BerWei01}.
 
Suppose we assume that the number of dark matter particles in a halo
follows a Poisson distribution, with mean proportional to the halo mass
such that $\langle N_{dm}|m\rangle \propto m$, and
$\langle N_{dm}(N_{dm}-1)|m\rangle \propto m^2$. Note that these 
proportionalities are the origin of
the weighting by $m$ and $m^2$ in equation~(\ref{pkm}) for
$P^{2h}_{dm}(k)$ and $P^{1h}_{dm}(k)$.  To model the power
spectrum of galaxies, we, therefore, simply modify 
equation~(\ref{pkm}) to read
\begin{eqnarray}
 P_\gal(k) &=& P^{1h}_{\gal}(k) + P^{2h}_{\gal}(k)\, ,
               \qquad\qquad{\rm where}\nonumber\\
 P^{1h}_{\gal}(k) &=& \int dm \, n(m) \,
                    \frac{\left< N_\gal(N_\gal-1)|m\right>}{\bar{n}_\gal^2}\,
                    |u_\gal(k|m)|^p \,,\nonumber\\
 P^{2h}_{\gal}(k) &\approx& P^\lin(k) \left[ \int dm\, n(m)\, b_1(m)\,
 \frac{\left< N_\gal|m\right>}{\bar{n}_\gal}\, u_\gal(k|m)\right]^2 .
 \label{eqn:gal1h}
 \label{eqn:gal2h}
\end{eqnarray}
Here,
\begin{equation}
 \bar{n}_\gal = \int dm \, n(m)\, \left< N_\gal|m \right>
 \label{eqn:barngal}
\end{equation}
denotes the mean number density of galaxies.
On large scales where the two-halo term dominates and
$u_\gal(k|m)\to 1$, the galaxy power spectrum simplifies to
\begin{equation}
 P_{\gal}(k) \approx b_\gal^2\, P^\lin(k),
\end{equation}
where
\begin{equation}
 b_\gal = \int dm\, n(m)\, b_1(m)\,
          \frac{\left< N_\gal|m\right>}{\bar{n}_\gal}
\end{equation}
denotes the mean bias factor of the galaxy population.
 
In addition to replacing the weighting by mass (i.e., the number
of dark matter particles) with a weighting by number of galaxies,
there are two changes with respect to equation~(\ref{pkm}).
First, $u_\gal(k|m)$ denotes the Fourier transform of the density
run of galaxies rather than dark matter around the halo center.
Although a natural choice is to approximate this integral by using the 
subclump distribution within a halo, we will show shortly that setting 
it to be the same as that of the dark matter (equation~\ref{eqn:yint}) 
is a reasonable approximation \cite{Sheetal01b}.  
Second, in the single-halo term, the simplest model is to set $p=2$
for $P^{1h}_{dm}(k)$.  However, in halos which contain only a single
galaxy, it is natural to assume that the galaxy sits at the center of
its halo.  To model this, one would set $p=2$ when
$\left< N_\gal(N_\gal-1)\right>$ is greater than unity and $p=1$
otherwise.  

It is worth considering a little more carefully where these scalings 
in the one-halo term come from.  Suppose that in a halo which contains 
$N_\gal$ galaxies, one galaxy sits at the halo centre.  Each of the 
galaxies contributes a factor of $u_\gal$ to the power, except for 
the central galaxy which contributes a factor of unity.  Pairs which come 
from the same halo are of two types: those which include the central galaxy, 
and those which do not.  Since only the galaxies which are not at the 
centre get factors of $u_\gal$, the weighting must be proportional to 
\begin{displaymath}
 \sum_{N_\gal>1} p(N_\gal|m)\,\left[
  (N_\gal-1)\,u_\gal(k|m) 
  + {(N_\gal-1)((N_\gal-2)\over 2}\, u_\gal(k|m)^2\right];
\end{displaymath}
where $p(N_\gal|m)$ is the probability an $m-$halo contains 
$N_\gal$ galaxies, and the sum is from $N_\gal>1$ because, 
to contribute pairs, there must be at least two galaxies in the halo.  
The first term is the contribution from pairs which include the central 
galaxy, and the second term is the contribution from the other pairs.  
The sums over $N_\gal$ yield 
\begin{displaymath}
 \Bigl[\langle N_\gal-1|m\rangle + p(0|m)\Bigr]\,
 \Bigl[u_\gal(k|m) - u_\gal(k|m)^2\Bigr] 
\end{displaymath}
\begin{displaymath}
 + \langle N_\gal(N_\gal-1)/2|m\rangle\, u_\gal(k|m)^2 .
\end{displaymath}
Evidently, to compute this term requires knowledge of $p(0|m)$.  
However, if we are in the limit where most halos contain no galaxies, 
then the leading order contribution to the sum above is 
 $p(2|m)\,u_\gal(k|m)$.  But, in this limit, 
 $\langle N_\gal(N_\gal-1)|m\rangle\equiv 
  \sum N_\gal(N_\gal-1)\,p(N_\gal|m) \approx 2\,p(2|m)$, 
so this leading order term should be well approximated by 
$\langle N_\gal(N_\gal-1)/2|m\rangle\,u_\gal(k|m)$.  
In the opposite limit of a large number of galaxies per halo, it should 
be accurate to set $p(0|m)\ll 1$.  Then the expression above reduces to  
$\langle N_\gal-1|m\rangle\,[u_\gal(k|m) - u_\gal(k|m)^2] 
 + \langle N_\gal(N_\gal-1)/2\rangle\, u_\gal(k|m)^2$.  
For Poisson counts, 
 $\langle n(n-1)\rangle = \langle n\rangle^2$.  If this is indicative 
of other count models also, then this shows that the dominant term is 
the one which comes from the second factorial moment.  Therefore, it 
should be reasonable to approximate the exact expression above by 
$\langle N_\gal(N_\gal-1)|m\rangle\,u_\gal(k|m)$ when 
$\langle N_\gal(N_\gal-1)|m\rangle\le 1$, and by 
$\langle N_\gal(N_\gal-1)|m\rangle\,u_\gal(k|m)^2$ otherwise.  
Notice that the two limits differ only by one factor of $u_\gal$.  
 
The expressions above show explicitly that if $\langle N_\gal|m\rangle$
and $\langle N_\gal (N_\gal -1)|m\rangle$ are not proportional to $m$
and $m^2$ respectively, then the clustering of galaxies will be different 
from that of the dark matter, even if $u_\gal(k|m)=u_{dm}(k|m)$.  
Because the one- and two-halo terms are modified (with respect to the 
dark matter case) by two different functions, it may be possible to adjust 
them separately in such a way that they sum to give the observed power 
law.  

Thus, the halo model shows that the distribution $p(N_\gal|m)$ determines 
whether or not $P_\gal(k)$ is a power law.  Although the analysis above 
assumed that $p(N_\gal|m)$ depends only on $m$, it is very likely that 
other properties of a halo, than simply its mass, determine the number of 
galaxies in it.  For example, $N_\gal$ almost certainly depends on the 
halo's formation history.  Since the concentration $c$ of the halo density 
profile also depends on the formation history 
\cite{Navetal96,Ekeetal01,Wecetal02}, 
a convenient way to incorporate the effects of the formation history is 
to set $p(N_\gal|m,c)$, and then integrate over the lognormal scatter in 
halo concentrations when computing the halo model predictions.  In what 
follows, we will ignore this subtlety.  

\begin{figure}
\vspace{-5cm}
\centerline{\psfig{file=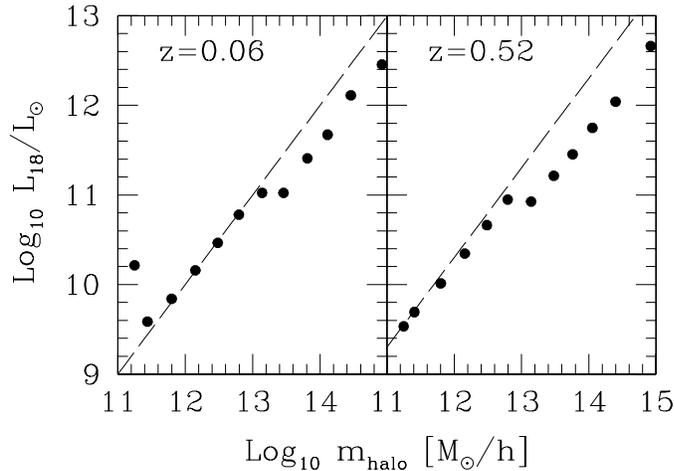,width=4.2in}}
\caption{The total luminosity in galaxies brighter than $M_{r^*}<-18$ 
which are in a halo, as a function of the total mass of the halo, 
from the semi-analytic galaxy formation models of \cite{Kauetal99}.  
Dashed lines show lines of constant mass-to-light ratio:  the value 
of $M/L_{18}$ at $z=0.5$ shown is a factor of two smaller than at 
$z=0$.}
\label{fig:M2L}
\end{figure}

Although the exact shape of $p(N_\gal|m)$ is determined by gastrophysics, 
there are some generic properties of it which are worth describing.  
Since galaxies form from baryons, a simple first approximation would be 
to assume that the first moment, $\langle N_\gal|m\rangle$, should be 
proportional to the mass in baryons, which, in turn, is likely to be a 
fixed fraction of the mass in dark matter of the parent halo.  
If we assume that $\langle N_\gal|m\rangle\propto m^\alpha$
(so $\mu = 1$ is the scaling of the dark matter), then there are two
reasons why we might expect $\alpha\le 1$.
Firstly, for the very massive halos, it is natural to associate galaxies
with subclumps within the halo.  The total number of subclumps within
a massive parent halo which are more massive than a typical galaxy scales
as $\alpha\approx 0.9$ \cite{Ghietal00}. 
Halo substructure as a plausible model for the galaxy distribution is 
discussed by \cite{Coletal99,Klyetal99}. If one identifies all subclumps 
in CDM haloes which had velocity dispersions larger than about 100 km/s 
(which is typical for a small galaxy sized halo), then the correlation 
function of these objects is a power law of about the same slope and 
amplitude as that of optically selected galaxies.  Remarkably, the slope 
and amplitude of this power law are approximately the same whether one 
identifies the subclumps at redshifts as high as $3$ or as low as $0$ 
(see \cite{Bag98} for a clear discussion of why this happens, and 
Figure~\ref{fig:mwhite} below).    

Secondly, galaxy formation depends on the ability of baryons to cool.
Since the velocity dispersion within a halo increases with halo mass,
the efficiency of cooling decreases. This might lead to a reduction 
in the efficiency of galaxy formation at the high mass end relative to the 
low mass end.  Such a mass dependent efficiency for galaxy formation has 
been used to explain the observed excess of entropy in galaxy clusters 
relative to smaller groups \cite{Bry00}.
At the low mass end, one might imagine that there is a minimum dark
halo mass within which galaxies can be found.  This is because the energy
feedback from supernovae which explode following an initial burst of
star formation may be sufficient to expel the baryons from the shallower 
potential wells of low mass halos. Also, during the epoch of reionization 
at $z<6$, photoionization may increase the gas temperature.  
The temperature of the reheated gas may exceed the virial temperature 
of low mass halos, thus suppressing star formation in them 
\cite{KauWhiGui93,Buletal00,Benetal01}.  

\begin{figure}
\vspace{-5cm}
\centerline{\psfig{file=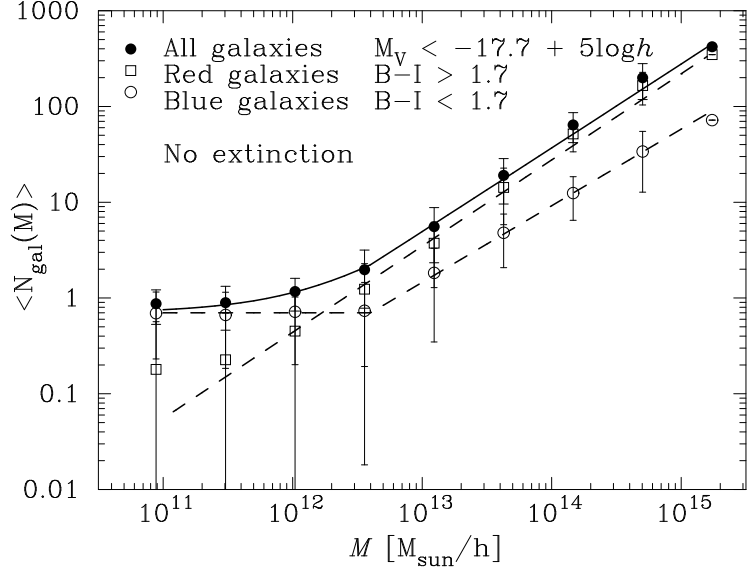,width=4.2in}}
\caption{The average number of galaxies as a function of dark matter
halo mass in the semi-analytic galaxy formation models of \cite{Kauetal99}.
Curves show the fits in equation~(\ref{eqn:galcounts}).}
\label{fig:munich}
\end{figure}
 
Detailed semi-analytic galaxy formation models allow one to quantify
these effects \cite{Kauetal99,Benetal00}.
The symbols in Figure~\ref{fig:munich} show how $\langle N_\gal|m\rangle$
depends on galaxy type and luminosity in the models of \cite{Kauetal99}.
The lines show simple fits (from \cite{SheDia01}):
\begin{eqnarray}
\langle N_{\rm Blue}|m \rangle &=& 0.7 \qquad\qquad {\rm if}\
10^{11}\,M_{\sun} h^{-1} \leq m\leq M_{\rm Blue}\nonumber \\
                &=& 0.7\,(m/M_{\rm Blue})^{\alpha_B}\qquad
                   {\rm if}\ m>M_{\rm Blue} \nonumber \\
\langle N_{\rm Red}|m \rangle &=& (m/M_{\rm Red})^{\alpha_R} \qquad m
\geq 10^{11}\,M_{\sun} h^{-1}
\nonumber \\
\langle N_\gal|m \rangle &=& \langle N_{\rm Blue}|m \rangle +
\langle N_{\rm Red}|m \rangle,
\label{eqn:galcounts}
\end{eqnarray}
where  $M_{\rm Blue} = 4\times 10^{12}\,M_\odot/h$, $\alpha_{\rm B} = 0.8$,
$M_{\rm Red} = 2.5\times 10^{12}\,M_\odot/h$, and $\alpha_{\rm R} = 0.9$.

\begin{figure}[t]
\centerline{\psfig{file=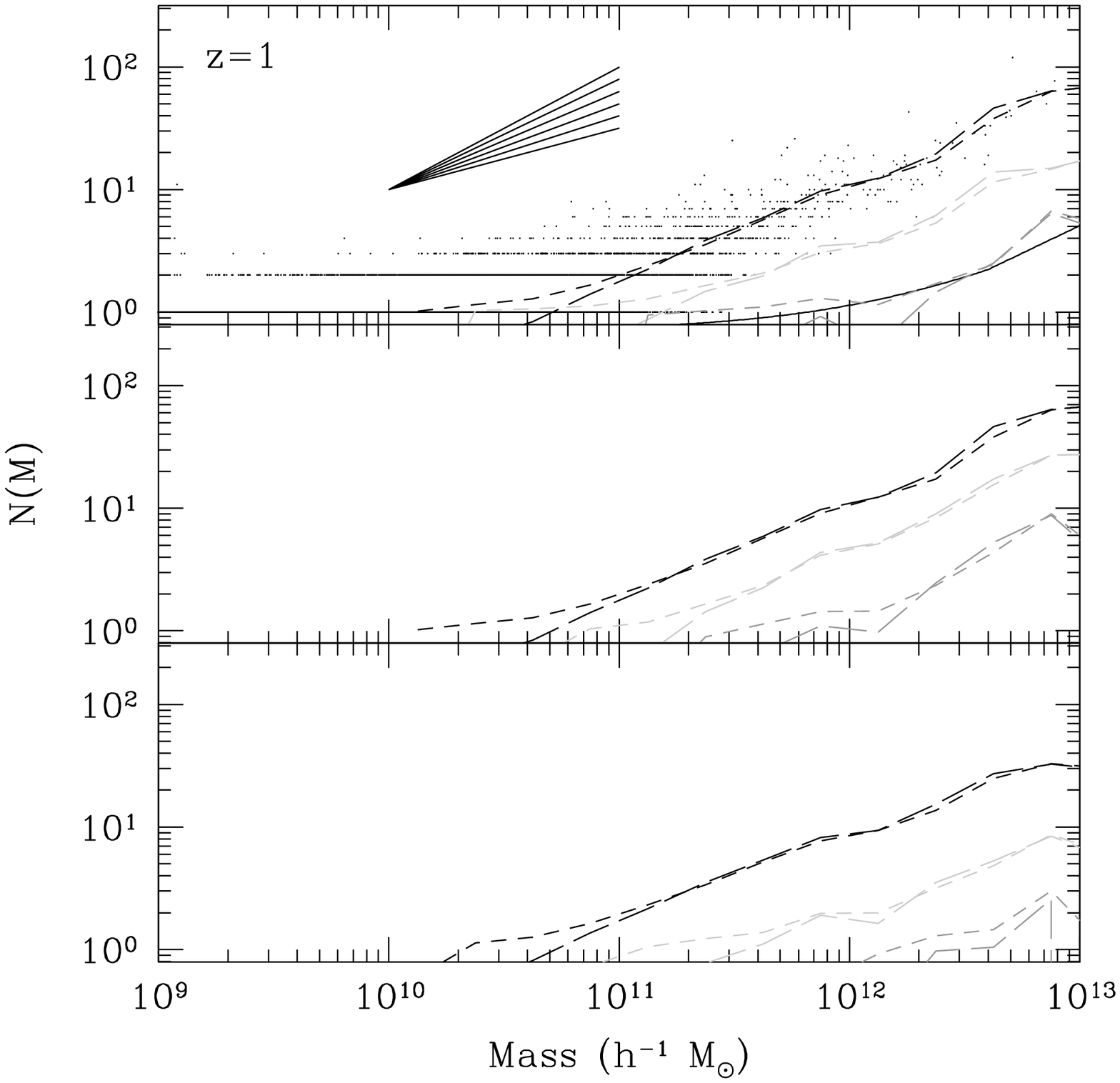,width=2.9in,angle=0}
            \psfig{file=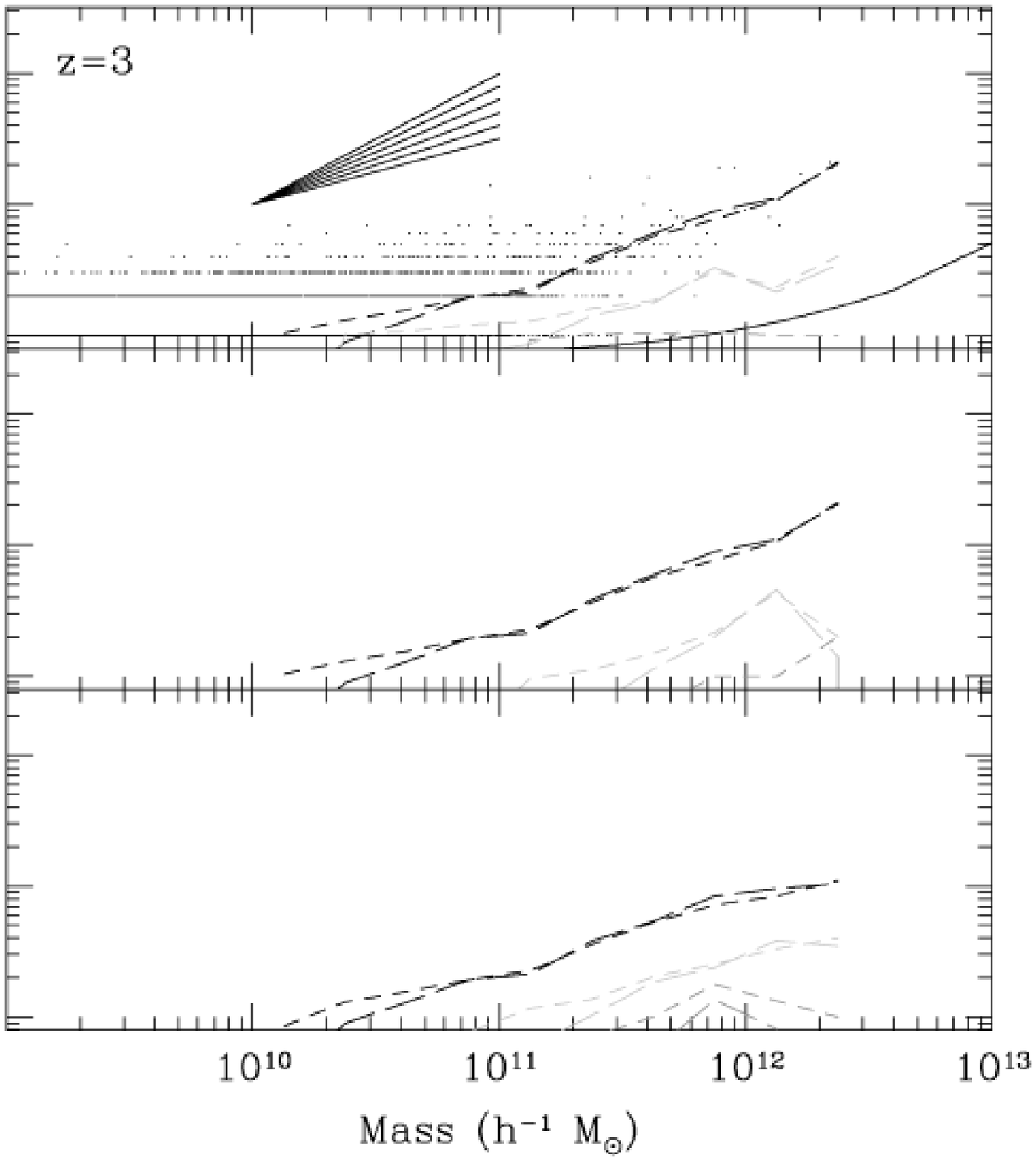,width=2.45in,angle=0}}
\caption{The number of subclumps in a halo as a function of parent 
halo mass in a simulation at $z=1$ (left) and 3 (right).
Top panel shows $\langle N\rangle$ (long dashed) and 
$\sqrt{\langle N(N-1)\rangle}$ (short dashed) as a function
of mass for: all subclumps (upper lines) and for subclumps with mass 
greater than $10^{10}$ (middle) and $10^{11}h^{-1}M_\odot$ (lower), 
respectively.  The lower solid line shows equation~(\ref{eqn:galcounts}).
Middle panel is similar, but with cuts on stellar mass:  all subclumps 
(upper lines) and subclumps with stellar mass greater than 
$10^{9}$ (middle) and $10^{10}h^{-1}M_\odot$ (lower).
Bottom panel shows cuts on star-formation rate:  all subclumps 
(upper lines), and for  subclumps with star formation rates greater 
than 1 (middle) and 10 (lower) $M_\odot$/yr.}
\label{fig:mwhite}
\end{figure}

Figure~\ref{fig:mwhite} compares the distribution of subclumps in the 
numerical simulations of \cite{Whietal01} with the expected number 
counts of galaxies within halos (equation~(\ref{eqn:galcounts}).  
The number of semianalytic galaxies per halo scales similarly to the 
dark matter halo subclumps when the mass limit of subclumps are above 
$10^{11}h^{-1}$ M$_\odot$, suggesting that identifying halo subclumps 
with galaxies is a reasonable model.
 
\begin{figure}[t]
\centerline{\psfig{file=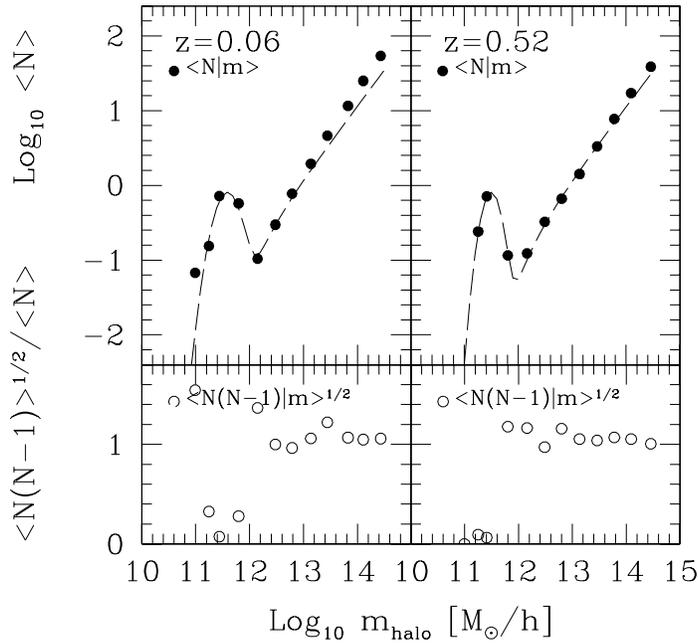,width=4.2in,angle=0}}
\caption{The mean, $\langle N\rangle$, and second factorial moment, 
$\langle N(N-1)\rangle$ of the distribution of the number of galaxies 
per halo as a function of the halo mass $m$.  Symbols show measurements 
in the semianalytic model of \cite{Kauetal99} and we have selected 
objects which are predicted to have absolute magnitudes between 
$-19$ and $-20$ in the SDSS $r^*-$band.  
Results for absolute magnitudes in the range $-17$ to $-18$, 
and $-18$ to $-19$ are qualitatively similar, although the peak 
for the lower luminosity bins shifts to lower masses.}
\label{fig:galn}
\end{figure}

Another interesting feature of these models in shown in 
Figure~\ref{fig:galn}.  The top and bottom panels show 
$\langle N_\gal|m \rangle$ and $\langle N_\gal(N_\gal -1)|m \rangle$ 
from the GIF models, but now we only show counts for galaxies which 
have absolute magnitudes in the range $-19\le M_{r^*}\le -20$.  
The top panels show that there is a pronounced peak in the number of 
galaxies per halo when $\langle N_\gal|m\rangle\le 1$; in this regime, 
there is a relatively tight correlation between the luminosity of a 
galaxy and the mass of its parent halo.  In the more massive halos 
which contain many galaxies, there is no correlation between luminosity 
and halo mass, and the number of galaxies scales approximately linearly 
with halo mass.  Figure~\ref{fig:munich} is built up from a number of 
curves like those shown here.  

The bottom panels in Figure~\ref{fig:galn} are also interesting.  
If $p(N_\gal|m)$ were Poisson, then 
 $\langle N_\gal(N_\gal -1)|m \rangle = \langle N_\gal|m \rangle^2$.
While the Poisson model is reasonably accurate at large
$\langle N_\gal|m \rangle$, the scatter in $N_\gal$ at fixed $m$ can be
substantially less than Poisson at the low mass end.
This is largely a consequence of mass conservation \cite{SheLem99}:
the Poisson model allows an arbitrarily large number of galaxies to be
formed from a limited amount of dark matter.  For this reason,
\cite{Scoetal01} argued that a binomial distribution should provide a
convenient approximation to $p(N_\gal|m)$.
A binomial is specified by its mean and its second moment.
To match the semianalytic models, the mean must be given by
equation~(\ref{eqn:galcounts}), and the second moment by
\begin{equation}
 \langle N_\gal(N_\gal -1)\rangle^{1/2} =
 \alpha(m)\,\langle N_\gal|m\rangle \, ,
\end{equation}
where $\alpha(m) = \log \sqrt{m/10^{11} h^{-1} M_{\sun}}$ for
$m < 10^{13} h^{-1} M_{\sun}$ and $\alpha(m)=1$ thereafter.
The Binomial assumption allows one to model higher order correlations,
since, by analogy with the two point correlation function, the halo
model for $\xi_n$ depends on the $n-$th moment of $p(N_\gal|m)$.
For example, the bi- and trispectra require knowledge of the third and 
fourth moments of $p(N_\gal|m)$.
 
\begin{figure}[t]
\centerline{\psfig{file=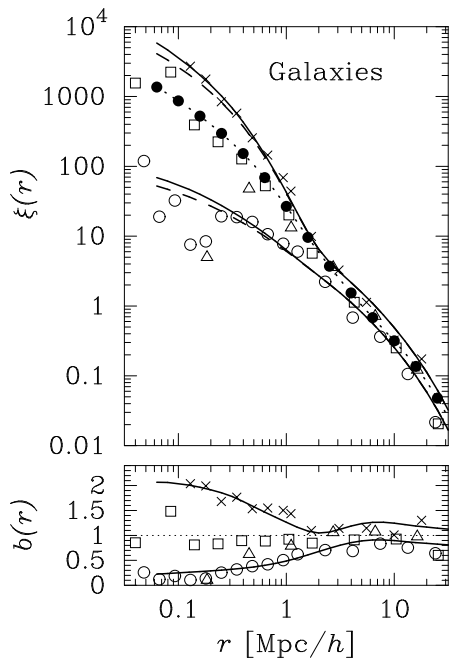,width=4.2in,angle=0}}
\caption{Correlation functions of different tracers of the dark matter 
density field in the GIF $\Lambda$CDM semianalytic galaxy formation 
model.  Filled circles are for the dark matter, crosses are for red 
galaxies,squares for galaxies which have low star formation rates, 
triangles for galaxies with high star formation rates, and open 
circles for blue galaxies.  The two solid curves show the halo model 
predictions for the red and blue galaxies, and the dashed curves 
show what happens if we use the second factorial moment of the galaxy 
counts, rather than the second moment when making the model predicition.  
For comparison, the dotted curve shows the predicted dark matter 
correlation function. Bottom panel shows how the bias factor:  
$\sqrt{\xi(r)/\xi_{dm}(r)}$ depends on scale.  
The figure is from \cite{Sheetal01d}.}
\label{fig:redblue}
\end{figure}

Figure~\ref{fig:redblue} shows the result of inserting the 
$N_\gal-m$ relations shown in Figure~\ref{fig:munich} 
(equation~(\ref{eqn:galcounts}) in the halo model, and changing 
nothing else (i.e., the red and blue galaxies are both assumed to 
follow the same NFW profile as the dark matter).  The symbols show 
measurements in the GIF semianalytic models which 
equation~(\ref{eqn:galcounts}) describes, and the curves, which provide 
a good fit, show the halo model prediction.  On small scales, the 
redder galaxies have a steeper correlation function than the blue 
galaxies, in qualitative agreement with the SDSS measurements shown 
in Figure~\ref{fig:sdsswrp}.  The agreement between the simulations 
and the halo model calculation suggests that almost the entire 
difference between the clustering of red and blue galaxies is a 
consequence of the $N_\gal-m$ relation.  The smaller additional 
effect which comes from allowing the red and blue galaxies to be 
distributed differently around the parent halo centre (e.g., if the 
reds are more centrally concentrated), is studied in some detail 
in \cite{Scr02}.

The dependence of clustering on luminosity (Figure~\ref{fig:sdsswrp}) 
is also straightforward to understand.  If luminous galaxies reside 
in the more massive halos (this is a natural prediction of most 
semianalytic models), then, because the more massive halos are 
more strongly clustered (Figure~\ref{fig:bias}), the more luminous 
galaxies should be more strongly clustered.  The halo model also 
shows clearly that, in magnitude limited surveys such as the SDSS, 
this sort of luminosity dependent clustering must be taken into 
account when interpreting how the angular clustering strength depends 
on the magnitude limit, and when inverting $w(\theta)$ to estimate $P(k)$.  
If not, then the fact that the more strongly clustered luminous 
galaxies which are seen out to larger distances, and hence contribute 
to the largest scale power, will lead to erroneous conclusions about 
the true amount of large scale power.  

Whereas most implementations of the halo model have concentrated on 
the $p(N_\gal|m)$ relation derived from semianalytic galaxy formation 
models (e.g., Figure~\ref{fig:munich}), information about $p(N_\gal|m)$ 
is encoded in the luminosity functions of galaxies and clusters.  
For example, \cite{PeaSmi00} argue that observations of the 
mass-to-light ratio in groups (e.g., plots like Figure~\ref{fig:M2L}, 
but made using data rather than semianalytic models), 
the combined  luminosity function of galaxies in groups and clusters, 
and the galaxy luminosity function itself, can together be used to 
determine the mean number of galaxies per halo mass.  The idea is 
to use the galaxy luminosity function to estimate a characteristic 
luminosity; use it to estimate the number of galaxies in a group 
by matching to the luminosity function of galaxies in groups and 
clusters; assign a mass to galaxy groups and clusters by requiring
the observed number density of groups from the halo mass function 
agree with that obtained from the luminosity function. This leads to 
a measured mean number of galaxies of the form
 $\langle N_\gal|m \rangle \propto m^{0.92}$ which is close to 
that shown in, e.g., Figure~\ref{fig:munich}.  The SDSS galaxy cluster 
catalogs offer a promising opportunity to exploit this approach.  

It is remarkable that this simple $N_\gal-m$ parametrization of 
the semianalytic models is all that is required to understand how 
and why the clustering depends on galaxy type.  It is this 
fact which has revived interest in the halo model.

\subsection{Galaxy--dark matter cross power spectrum}
The halo model also suggests a simple parameterization of the
cross-correlation between the galaxy and dark matter distributions
\cite{Sel00,GuzSel00}:
\begin{eqnarray}
P_{\gal-\rm dm}(k) &=& P^{1h}_{\gal -\rm dm}(k)+P^{2h}_{\gal -\rm dm}(k)
\qquad\qquad {\rm where} \nonumber\\
P^{1h}_{\gal-dm}(k) &=& \int dm\, {mn(m)\over\bar\rho}\,
                       \frac{\left< N_\gal|m \right>}{\bar{n}_\gal}\,
                       |u_{dm}(k|m)\,|\,|u_\gal(k|m)|^{p-1} \,\nonumber\\
P^{2h}_{\gal -dm}(k) &=& P^\lin(k) \left[ \int dm\, {mn(m)\over \bar\rho}\,
                                            b_1(m) u(k|m)\right]\nonumber\\
&&\qquad\times \, \left[ \int dm\, n(m)\, b_1(m)
               \frac{\left< N_\gal\right>}{\bar{n}_\gal} u_\gal(k|m)\right] ,
\label{eqn:galdm1h}
\end{eqnarray}
and, as before, one sets $p=1$ if $\left< N_\gal\right> < 1$ and
one is interested in requiring that one galaxy always sits at the
halo center.  This expression is easily generalized to the
cross-correlation between two galaxy samples.

If one galaxy always sits at the halo centre, then these expressions 
must be modified.  To see the effect of this on the two-halo term, we must 
average both pieces of the two halo term over $p(n|m)$, with the requirement 
that $n>0$.  This requires evaluation of sums of the form 
\begin{displaymath}
\sum_{n>0} [1 + (n-1) u(k|m)]\, p(n|m) 
 = 1 - p(0|m) + \langle n-1|m\rangle\,u(k|m) + u(k|m)\,p(0|m)
\end{displaymath}
which we could also have written as 
\begin{displaymath}
 N_{\rm eff}(k|m) \equiv [1 - p(0|m)]\,[1 - u_{\rm gal}(k|m)] 
 + \langle n|m\rangle\, u_{\rm gal}(k|m) .
\end{displaymath}
Since both factors in the first term are positive, this shows clearly 
that there is an enhancement in power which comes from always placing one 
galaxy at the halo centre.  Since $u(k|m)$ decreases as $k$ increases, 
the enhancement in power is largest on small scales (large $k$).  
In sufficiently massive halos one might expect to have many galaxies, 
and so $p(0|m)\ll 1$.  In this limit, the expression above becomes 
 $1-u(k|m) + \langle n|m\rangle\,u(k|m) = 1+\langle n-1|m\rangle\,u(k|m)$.  
On the other hand, if most halos have no galaxies, then $p(1|m)$ is 
probably much larger than all other $p(n|m)$ with $n\ge 2$.  Then the 
leading order term in the sum above is $p(1|m)$.  Since 
$\langle n|m\rangle \equiv \sum np(n|m) \approx p(1|m)$, we have that 
$N_{\rm eff}(k|m)\approx\langle n|m\rangle$.  In this limit, only a 
fraction $\langle n|m\rangle\ll 1$ of the halos contain a galaxy, and 
the galaxy sits at the halo centre, so there is no factor of $u$.  

The contribution of the galaxy counts to the one halo term of the 
galaxy--mass correlation function is similar.  Using the expressions 
above yields 
\begin{eqnarray}
 P_{\rm gm}^{\rm 1h}(k) &=& \int {\rm d}m\,n(m)\,{m\over\bar\rho}\,|u(k|m)|
  \, {N_{\rm eff}(k|m)\over \bar n_{\rm gal}} \nonumber\\
 P_{\rm gm}^{\rm 2h}(k) &\approx& P_{\rm lin}(k)\ 
 \left[\int {\rm d}m\,n(m)\,{m\over\bar\rho}\,b(m)\,u(k|m)\right]\ \nonumber\\
 && \qquad\qquad \times \left[\int {\rm d}m\,n(m)\,b(m)\, {N_{\rm eff}(k|m)\over \bar n_{\rm gal}}
 \right] .
\label{pkmg}
\end{eqnarray}
If the run of galaxies around the halo centre is not the same as of the 
dark matter, then one simply uses $u_{\rm gal}$ instead of $u$ in 
$N_{\rm eff}$.  If the two-halo term usually does not dominate the 
power on small scales (this is almost always a good approximation), it 
is reasonable to ignore the enhancement in power associated with the central 
galaxy, and to simply set 
 $N_{\rm eff}(k|m)\approx \langle n|m\rangle\, u_{\rm gal}(k|m)\approx 
 \langle n|m\rangle$.  
The one-halo term requires knowledge of $p(0|m)$.  
Since $p(0|m)$ is usually unknown, the approximation above 
interpolates between the two limits discussed earlier by setting 
$N_{\rm eff}=\langle n|m\rangle\, u(k|m)$ if $\langle n|m\rangle \ge 1$, 
and $N_{\rm eff}=\langle n|m\rangle$ if $\langle n|m\rangle < 1$.

In what follows, it will be convenient to define the cross-correlation
coefficient:
\begin{equation}
r(k) \equiv \frac{P_{\gal -\rm dm}(k)}{\sqrt{P_{dm}(k)\,P_{\gal}(k)}} \, .
\end{equation}
Note that $r(k)$ may depend on scale $k$.

Because we cannot measure the clustering of dark matter directly,
the galaxy--dark matter cross-correlation is not observable.
However, if one cross-correlates the galaxy distribution with weak
lensing shear measurements, then the resulting signal is sensitive to
this cross-correlation \cite{Sel00,GuzSel00}.
We discuss this further in \S~\ref{sec:galaxy-mass}.

\subsection{Discussion}

\begin{figure}[t]
\centerline{\psfig{file=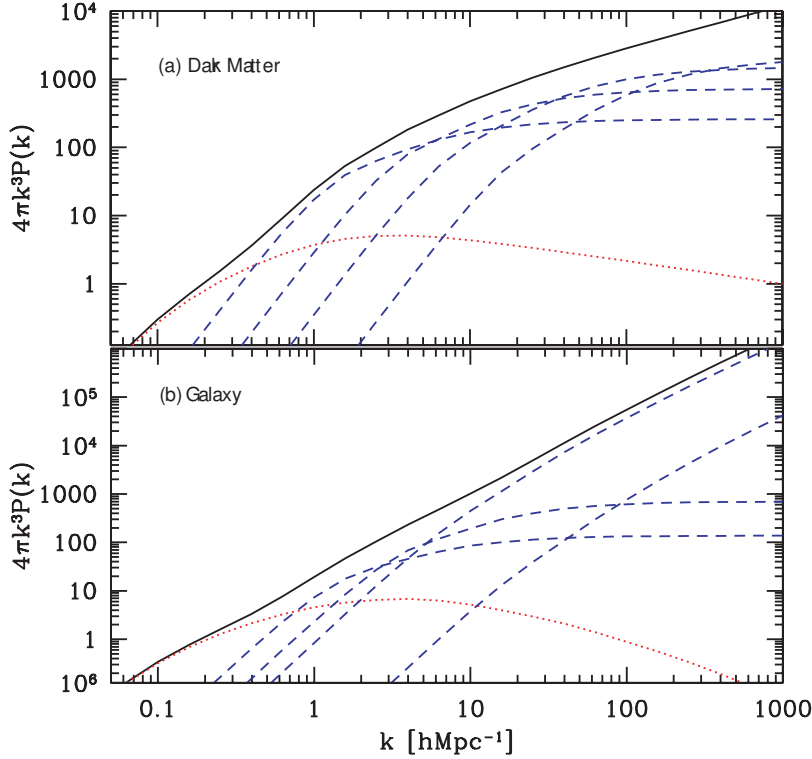,width=4.2in}}
\caption{(a) Contributions by halo mass to the one-halo term 
in the halo model description of the power spectrum.  
(b) The same, but for the dark matter-galaxy cross-correlation 
power spectrum.  At small scales, there is effectively no 
contribution from the smallest mass halos for the galaxy power 
spectrum. This figure is from \cite{Sel00}.}
\label{fig:galaxymass}
\end{figure}
 
\begin{figure}[t]
\centerline{\psfig{file=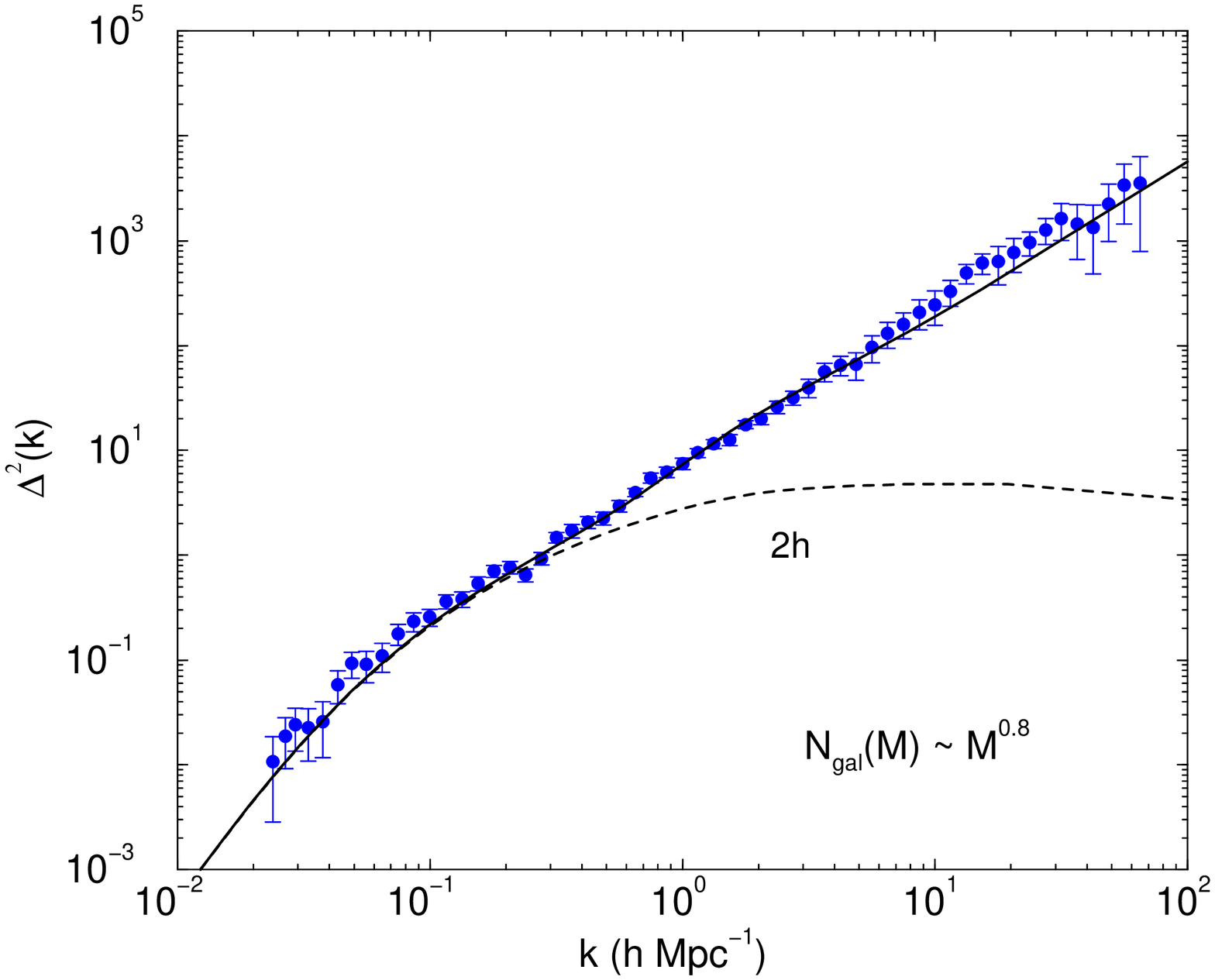,width=4.2in,angle=-90}}
\caption{The PSC$z$ galaxy power spectrum (symbols) and the result of 
tuning the first two moments of $p(N_\gal|m)$ so as to produce the 
power law like behavior (solid curve).}
\label{fig:psczpower}
\end{figure}
 
Figure~\ref{fig:galaxypower} showed the galaxy power spectrum from 
the PCSZ survey \cite{HamTeg00}.  The nonlinear dark matter power
spectrum, scaled with a constant ($k-$independent bias factor) to 
match the linear regime cannot also match the power on small scales:  
this shows that the bias between dark matter and galaxies must 
depend on scale.  

The top panel in Figure~\ref{fig:galaxymass} shows the contributions 
to the dark matter power spectrum as a function of halo mass.  
The halo model description of the galaxy power spectrum shows clearly 
that the $p(N_\gal|m)$ distribution changes the relative contributions 
of low and high--mass halos to the total power, and so modifies the 
shape of the power spectrum in a way which depends on $k$.  

The main change to the amplitude of the small-scale contribution to 
the galaxy power spectrum, the change which results in a power-law 
shape, comes from the halos which contain at least one galaxy, or, 
effectively halos containing what are called {\it field} or 
{\it isolated} galaxies. The assumption that these galaxies are at 
the center of the halo they occupy results in a power-law at small 
scales \cite{PeaSmi00}. As discussed in \cite{Scoetal01}, the 
contribution to the total power from such halos is very sensitive 
to the low-mass cutoff in the galaxy-mass relation.  
Thus, the small scale clustering of galaxies essentially allows one to 
constrain certain parameters related to the galaxy formation, such as the 
minimum mass in which a galaxy can exist. 

Also, as shown in figure~\ref{fig:galaxymass}, at intermediate scales, 
the massive halos contribute less to the galaxy-dark matter power 
spectrum than to the dark matter power spectrum. This is because the 
$N_\gal \sim M^{0.8-0.9}$ weighting suppresses the contribution 
from the high mass end of dark matter halos.  Figure~\ref{fig:psczpower} 
compares the associated galaxy power spectrum with that measured in 
the PCSZ survey.  Note the power law behavior of the galaxy power 
spectrum over three to four decades in wavenumber. 

In the halo model, galaxy power spectra and higher order correlations, 
when studied as a function of galaxy type or environment, allow one to 
extract certain galaxy properties such as the mean $N_\gal-m$ relation, 
and the mean mass of dark matter halos in which galaxies reside.  This 
information may be helpful for understanding the galaxy formation and 
evolution processes.
In \cite{Scoetal01} and \cite{ScoShe01}, constraints on $p(N_\gal|m)$ 
were obtained by comparing halo model predictions with the measured 
variance and higher order correlations of galaxies in the APM 
\cite{Madetal90} and PSCZ surveys. 
The halo based contraints of galaxy formation models are likely to 
increase with ongoing wide-field surveys such as the Sloan Digital Sky 
Survey (SDSS) and the 2dFGRS.  The halo approach to galaxy clustering 
has already become helpful for interpreting the SDSS two-point 
galaxy correlation function \cite{Dodetal01} and the lensing-mass 
correlation \cite{GuzSel00}.

\section{Velocities}
\label{sec:velocities}

One of the great strengths of the halo-based approach is that it provides 
a clear prescription for identifying the scale on which perturbative 
approaches will break down, and non-linear effects dominate.  
The separation of linear and non-linear scales 
is an important tool when describing large scale 
velocities and related statistics.
We now  present the halo model description of velocities by extending
\cite{She96,SheDia01}.  

\subsection{Velocities of and within halos}
In the model, all dark matter particles are assumed to be in approximately 
spherical, virialized halos.  The velocity of a dark matter particle is 
the sum of two terms, 
\begin{equation}
 v = v_{\rm vir} + v_{\rm halo}:
\label{vsum}
\end{equation}
the first is due to the velocity of the particle about the center of 
mass of its parent halo, and the second is due to the motion of the 
center of mass of the parent.  We will assume that each of these 
terms has a dispersion which depends on both halo mass and on the 
local environment, so that 
\begin{equation}
 \sigma^2(m,\delta) = 
 \sigma^2_{\rm vir}(m,\delta) + \sigma^2_{\rm halo}(m,\delta).  
\label{sigtot}
\end{equation}
The expression above assumes two things:  the rms velocities depend 
on halo mass and local density only, and that the rms virial velocity 
within a halo is independent of the motion of the halo 
itself.  Presumably both assumptions break down if the dark matter is 
collisional and/or dissipative.  
For collisionless matter, the assumption that the virial motions within 
a halo is independent of the halo's environment is probably reasonably 
accurate.  It is not clear that the same is true for  halo speeds.  
Indeed, it has been shown that halos in dense regions move 
faster than those in underdense regions
\cite{Coletal00,SheDia01}. It will turn out, however, that 
the fraction of regions in which $\sigma^2_{\rm halo}(m,\delta)$ is 
significantly different from $\sigma^2_{\rm halo}(m,0)$ is quite small.  
This means that neglecting the density dependence of halo velocities
should be a reasonable approximation.  

Consider the first term, $v_{\rm vir}$.  
We will assume that virialized halos are isothermal spheres, so 
that the distribution of velocities within them is Maxwellian.  
This is in reasonable agreement with measurements of virial velocities 
within halos in numerical simulations.  
If $\sigma_{vir}$ denotes the rms speeds of particles within a halo, 
then the virial theorem requires that 
\begin{equation}
 {Gm\over r}\propto \sigma_{vir}^2 \propto 
 {H(z)^2\over 2}\,\Delta^{1/3}_{vir}(z)
 \left(3m\over 4\pi\rho_{crit}(z)\right)^{2/3} \, 
 \label{vcirc}
\end{equation}
where the final proportionality comes from the fact that all halos have 
the same density whatever their mass:  
$m/r^3\propto \Delta_{vir}\,\rho_{crit}$.  This shows that 
$\sigma_{vir}\propto m^{1/3}$:  the more massive halos are expected 
to be `hotter'.  At fixed mass, the constant of proportionality depends 
on time and cosmology, and on the exact shape of the density profile of 
the halo.  A convenient fitting formula is provided by \cite{BryNor98}:  
\begin{equation}
 \sigma_{\rm vir}(m,z) = 102.5\,g_\sigma\, \Delta^{1/6}_{\rm vir}(z)\,
 \left(H(z)\over H_0\right)^{1/3}
 \left({m\over 10^{13} M_\odot/h}\right)^{1/3}\ {{\rm km}\over {\rm s}},
 \label{sigmavir}
\end{equation}
where $g_\sigma = 0.9$, and
\begin{equation}
\Delta_{\rm vir} = 18\pi^2 + 60 x - 32 x^2, \qquad {\rm with}\quad 
               x = \Omega(z)-1  \, 
\end{equation}
and $\Omega(z) = [\Omega_m\,(1+z)^3]\, [H_0/H(z)]^2$.  
This fitting formula for the average density within a 
virialized object, $\Delta_{vir}\,\rho_{crit}$, generalizes the value 
$18\pi^2$ given previously for an Einstein de-Sitter universe in
equation~(\ref{Dvir}).  

In \cite{SheDia01}, it was shown that this virial relation 
between mass and velocities
is  independent of the local environment.
In practice, however, $\sigma_{vir}$ may depend on position within the halo.  
Accounting for the fact that halos really have more complicated 
density and velocity profiles is a detail which complicates the 
analysis, but not the logic of the argument.  
If the virialized halo is an isothermal sphere, the density run around 
the halo center falls as the square of the distance, then $\sigma_{vir}$ 
is the same everywhere within the halo.  In practice, halos are not quite 
isothermal, but we will show later that the scaling above is still 
both accurate and useful.  

We now turn to the second term, $v_{\rm halo}$.  
It will prove more convenient to first study halo speeds after 
averaging over all environments, before considering the speeds as a 
function of local density.  This is similar to the order in which we 
discussed the halo mass function and its dependence on density.  
We first consider a halo of size $r$ at the present time.  
Because the initial density fluctuations were small, the particles 
in this halo must have been drained from a larger region $R$ in the 
initial conditions:  $R/r\approx \Delta_{\rm vir}^{1/3}$, 
where $\Delta_{\rm vir}\approx 200$ or so.  This means, for example, 
that massive halos were assembled from larger regions than less massive 
halos.  Suppose we compute the rms value of the initial velocities of all the 
particles which make up a given halo  and extend to include all
halos of mass $m$, then we have effectively computed the rms velocity 
in linear theory, smoothed on the scale $R(m)\propto m^{1/3}$.  

It is well known that the linear theory prediction for the evolution 
of velocities is more accurate than the linear theory prediction for 
the evolution of the density \cite{Pee80}.  
In what follows, we will assume that at the present time, the velocities 
of halos are reasonably well described by extrapolating the velocities 
of peaks and are smoothed on the relevant scale, $R\propto m^{1/3}$, using linear 
theory.  For Gaussian initial conditions, this means that any given value 
of $v_{\rm halo}$ is drawn from a Maxwellian with dispersion 
$\sigma^2_{\rm halo}(m)$ given by:
\begin{equation}
 \sigma_{\rm halo}(m) = H_0\Omega_m^{0.6}\ \sigma_{-1} \ 
 \sqrt{1 - \sigma_0^4/\sigma_1^2\sigma_{-1}^2} ,
\label{sigmapeak}
\end{equation}
where,
\begin{displaymath}
 \sigma_j^2(m) = {1\over 2\pi^2} \int {\rm d}k\ 
 k^{2 + 2j}\,P^\lin(k)\, W^2[kR(m)] ,
\end{displaymath}
and $W(x)$ is the Fourier transform of the smoothing window.  The factor $H_0 \Omega_m^{0.6}$ comes from a well-known approximation to the derivative of the growth function, with $d\log G/d \log a \sim \Omega_m^{0.6}$ when $a$ is the scale factor.
Notice that the predicted rms velocity depends both on cosmology 
and on the shape of the power spectrum.  The term under the 
square-root arises from the peak constraint \cite{Baretal86}---it tends to unity as 
$m$ decreases:  the peak constraint becomes irrelevant for the less 
massive, small $R$, objects.  

\begin{figure}
\centering
\mbox{\psfig{figure=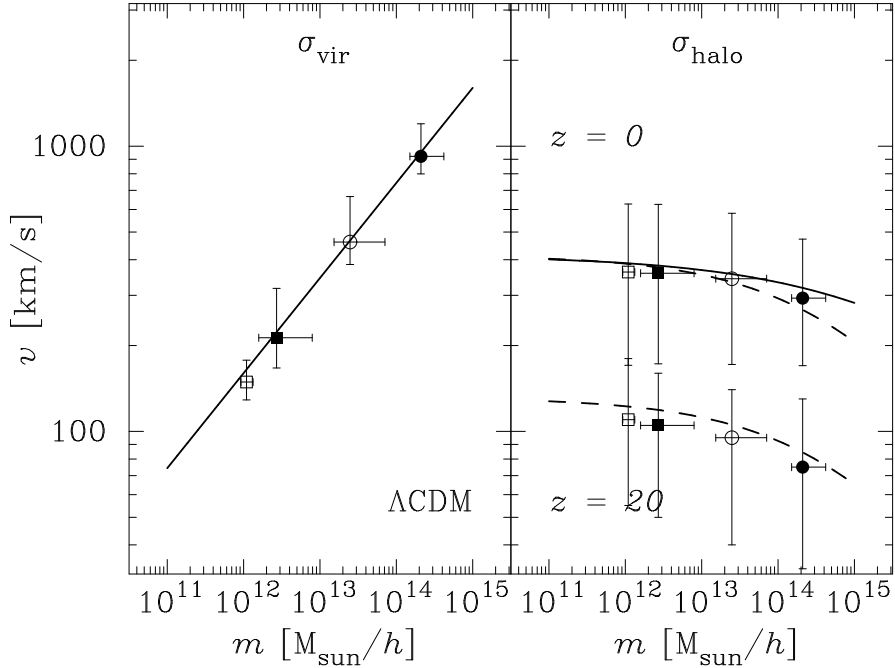,height=9cm,bbllx=72pt,bblly=58pt,bburx=629pt,bbury=459pt}}
\caption{Dependence on halo mass of the non-linear ($\sigma_{\rm vir}$) 
and linear theory ($\sigma_{\rm halo}$) terms in our model.  
Solid curves show the scaling we assume, and symbols show the 
corresponding quantities measured in the $z=0$ output time of the 
$\Lambda$CDM GIF simulation.  Error bars show the 90 percentile ranges 
in mass and velocity.  Dashed curve in panel on right shows the expected 
scaling after accounting for the finite size of the simulation box.  
Symbols and curves in the bottom of the panel on the right show
the predicted and actual velocities at $z=20$.}
\label{assumel}
\end{figure}

In figure~\ref{assumel}, we compare the dependence on mass for the
two velocity terms in numerical simulations by the GIF 
collaboration \cite{Kauetal99} and the dependences we have discussed above.
The symbols with error bars show the median and ninety percentile 
ranges in mass and velocity.  
Open squares, filled squares, open circles and filled 
circles show halos which have $60-100$, $100-10^3$, $10^3-10^4$ and 
$10^4-10^5$ particles, respectively.  There are two sets of symbols 
in the panel on the right.  For the time being, we are only interested 
in the symbols in the upper half which show halo velocities at $z=0$.  
The solid curves in the two panels show the scalings we assume.  

Although the scaling of the virial term with mass is quite accurate, 
it appears that the extrapolated linear theory velocities are 
slightly in excess of the measurements in the simulations.  
This is almost entirely due to the finite size of the simulation 
box.  The upper dashed curve shows the effect of using 
equation~(\ref{sigmapeak}) to estimate the rms speeds of halos, 
after setting $P(k)=0$ for $k<2\pi/L$, where $L$ is the box-size:  
$L=141$ Mpc/$h$.  
Thus, the two panels show that our simple estimates of the two 
contributions to the variance of the velocity distribution are 
reasonably accurate.  

Notice that the two terms scale differently with halo mass; indeed, 
to a first approximation, one might even argue that halo speeds 
are independent of halo mass.  Figure~\ref{assumel} 
shows that $\sigma_{\rm halo}(m) < \sigma_{\rm vir}(m)$ for massive 
halos.  Since massive halos have larger dispersions than less massive 
halos, the large velocity tail of $f(v)$ is determined primarily by the 
non-linear virial motions within massive halos, rather than by the 
peculiar motions of the halo centers of mass.  For this reason, 
the large velocity tail of $f(v)$, at least, is unlikely to be 
sensitive to inaccuracies in our treatment of halo velocities, or to 
our neglect of the possibility that halo speeds may depend on the 
environment.  
Before moving on, note that our finding that massive halos are hotter, 
whereas the speeds with which halos move is approximately independent of 
mass, suggests that if massive halos occupy denser 
regions, then, we expect a temperature--density relation such that
denser regions should be hotter.  We will return to this later.  

\subsection{The distribution of non-linear velocities}

In an ideal gas, the distribution of particle velocities $f(v)\,{\rm d}v$ 
has the Maxwell-Boltzmann form:  each cartesian component of the velocity 
is drawn from an independent Gaussian distribution.  Because of the action 
of gravity, the dark matter distribution at the present time is certainly 
not an ideal gas; numerical simulations show that $f(v)\,{\rm d}v$ is very 
different from a Maxwell--Boltzmann \cite{Sasetal90}; the distribution of 
each component of the velocity has an approximately Gaussian core with 
exponential wings.  The halo model decomposition of peculiar velocities 
into linear and non-linear contributions (equation~\ref{vsum}), provides 
a simple explanation for why this is so \cite{She96,SheDia01}.  

Let $p(v|m)\,{\rm d}v$ denote the probability that a particle in a halo 
of mass $m$ has velocity in the range d$v$ about $v$.  
Then the total distribution is given by summing up the various 
$p(v|m)$ distributions, weighting by the fraction of particles 
which are in halos of mass $m$:  
\begin{equation}
 f(v) = {\int {\rm d}m\,mn(m)\,p(v|m)\over \int {\rm d}m\,mn(m)}
      = \int {\rm d}m\,{mn(m)\over\bar\rho}\,p(v|m),
 \label{fv}
\end{equation}
where $n(m)\,{\rm d}m$ is the number density of halos that have mass in 
the range d$m$ about $m$.  The weighting by $m$ reflects the fact that the 
number of dark matter particles in a halo is supposed to be proportional 
to the halo mass.  This expression holds both for the size of the 
velocity vector itself, which we will often call the speed, as well as 
for the individual velocity components.  

To proceed, we need a model for the actual shape of $p(v|m)$.  
Since $v$ is the sum of two random variates (equation~\ref{vsum}), 
we study each in turn.  The virial motions are assumed to be Maxwellian.  
Also, for Gaussian initial density fluctuations, the linear peaks theory 
model of the halo motions means that they too are Maxwellian.  
Thus, in the model, each of the three cartesian components of the 
velocity of a dark matter particle in a clump of mass $m$ is given by 
the sum of two Gaussian distributed random variates, one with dispersion 
$\sigma^2_{\rm vir}(m)/3$ and the other with $\sigma^2_{\rm halo}(m)/3$.  
If we further assume that the motion around the clump center is 
independent 
of the motion of the clump as a whole, then these two Gaussian variates are 
independent and $p(v|m)$ is a Maxwellian with a dispersion which is the 
sum of the individual dispersions given by the sum in quadrature of 
equations~(\ref{sigmavir}) and~(\ref{sigmapeak}).  

In practice, we are only likely to observe velocities along the line of 
sight.  Thus, we will eventually be interested in the 
distribution of $f(v)$ projected along the line of sight.  
Projection changes the Maxwellian $p(v|m)$ distributions into Gaussians:
\begin{equation}
 p(v|m) = {{\rm e}^{-[v/\sigma(m)]^2/2}\over\sqrt{2\pi\sigma^2(m)}} \qquad
 {\rm where}\ \sigma^2(m) = {\sigma^2_{vir}(m)\over 3} + 
                            {\sigma_{halo}^2(m)\over 3};
 \label{pvm}
\end{equation}
i.e., $\sigma^2(m)$ is one third of the sum in quadrature of 
equations~(\ref{sigmavir}) and~(\ref{sigmapeak}).  

Now, $\sigma^2_{vir}(m)/\sigma^2_v(m*) = (m/m*)^{2/3}$, whereas 
$\sigma_{halo}^2$ is independent of halo mass (Figure~\ref{assumel}).  
Therefore, the characteristic function of $f(v)$ is 
\begin{eqnarray}
 \int dv\ {\rm e}^{ivt} f(v) &=& \int dm\ mn(m)\ \int dv\ {\rm e}^{ivt} 
p(v|m)
 \nonumber\\ 
&=& \int dm\ mn(m)\ {\rm e}^{-t^2\sigma_{vir}^2(m)/6}\ 
                    {\rm e}^{-t^2\sigma_{halo}^2/6} \nonumber\\ 
&=& {\rm e}^{-t^2\sigma_{halo}^2/6} \int {d\nu\over \nu}\sqrt{\nu\over 
2\pi}\,
    {\rm e}^{-\nu/2}\ {\rm e}^{-t^2\sigma_{vir}^2(\nu)/6} \nonumber\\
&=& {{\rm exp}(-t^2\sigma_{halo}^2/6)\over [1 + 
t^2\sigma_{vir}^2(m*)/3]^{1/2}}.
\label{pvanalytic}
\end{eqnarray}
The penultimate expression uses equation~(\ref{fps}) for the halo mass 
function and assumes that the initial spectrum of 
fluctuations was scale free with $P(k)\propto k^{-1}$, which should be 
a reasonable approximation to the CDM spectrum on cluster scales.  
The final expression is quite simple:  it is the product of the Fourier 
transforms of a Gaussian and a $K_0-$Bessel function.  Therefore, $f(v)$ 
is the convolution of a Gaussian with a $K_0-$Bessel function.  The 
Bessel function has exponential wings.  Because the dispersion of the 
Gaussian and the Bessel function are similar, equation~(\ref{pvanalytic}) 
shows that $f(v)$ should have a Gaussian core which comes from the linear 
theory halo motions, with exponential wings which come from the non-linear 
motions within halos.  

\begin{figure}
\centering
\mbox{\psfig{figure=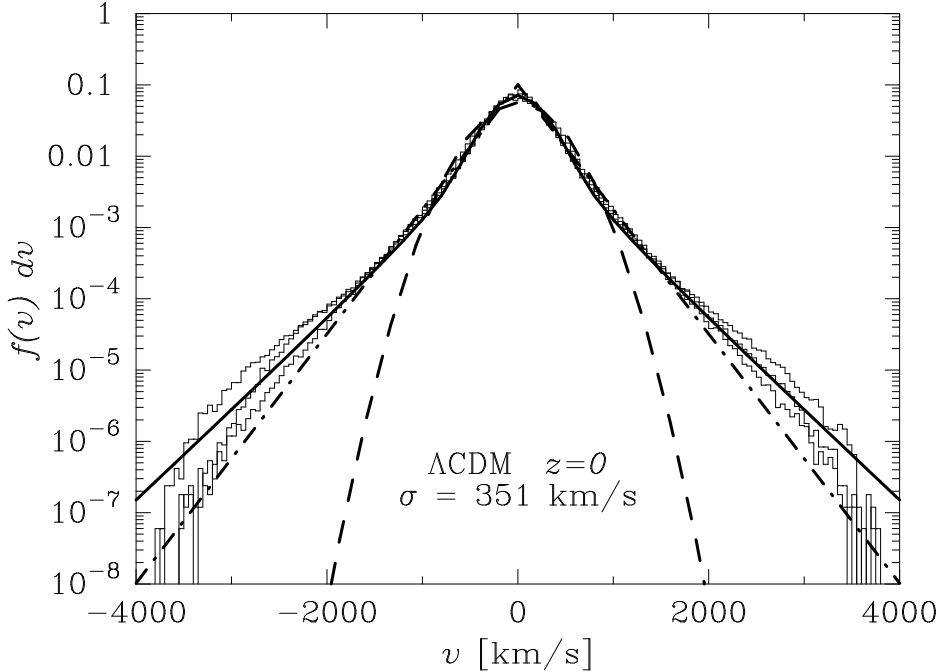,height=9cm,bbllx=72pt,bblly=58pt,bburx=629pt,bbury=459pt}}

\caption{The distribution of one-dimensional peculiar velocities for 
dark matter particles in a $\Lambda$CDM cosmology.  Histograms show the 
distribution of the three cartesian components measured in GIF 
simulations.  Dashed and dot-dashed curves show Gaussian and exponential 
distributions which have the same dispersion.  The solid curve shows the 
distribution predicted by our model, after accounting for the finite 
size of the simulation box.  The exponential wings are almost entirely 
due to virial motions within halos.}
\label{pvlcdm}
\end{figure}

Figure~\ref{pvlcdm} shows the one-dimensional $f(v)$ distribution given 
by inserting equations~(\ref{pvm}) and~(\ref{fgif}) in equation~(\ref{fv}) 
for the same cosmological model presented in Figure~\ref{assumel}.
The histograms show the distribution measured in GIF simulations.
For comparison, the dashed and dot-dashed curves in each panel show 
Gaussians and exponential distributions which have the same dispersion. 
The solid curves show the distribution predicted by the halo-based model:  
note the exponential wings, and the small $|v|$ core that is more Gaussian 
than exponential.  The exponential wings are almost entirely due to 
non-linear motions within massive halos, so they are fairly insensitive to 
our assumptions about how fast these halos move.   

It is worth emphasizing that $\sigma(m)$ in equation~(\ref{pvm}) is set 
by the cosmological model and the initial conditions. Thus, the 
second moment of the distribution in Figure~\ref{pvlcdm} is not a free 
parameter.  This halo model for $p(v|m)$ can be thought of as a simple 
way in which contributions to the velocity distribution statistic 
are split up into a part which is due to non-linear effects, given by the 
first term 
in equation~\ref{vsum}, and a part which follows from extrapolating 
linear theory to a later time, denoted by the second term of that equation.  
The agreement with simulations suggests that this simple treatment of 
non-linear and linear contributions to the statistic are quite accurate.  

Before moving on, note that the second moment of this distribution gives 
the mass-weighted velocity dispersion:  
\begin{equation}
\sigma_{vel}^2 =  \int dm \frac{m n(m)}{\bar{\rho}} \left[\sigma^2_{vir}(m) + 
                            \sigma_{halo}^2(m)\right]\, .
\label{eqn:disp}
\end{equation}
This quantity is a measure of the total kinetic energy in the Universe, 
and hence is directly related to the Layzer-Irvine Cosmic Energy equation 
\cite{Sheetal01b}.  Observational estimates of this quantity are discussed 
by \cite{DavMilWhi97}.  
Because the virial velocities within massive halos are substantially 
larger the motions of the halos themselves (Figure~\ref{assumel}), setting 
\begin{eqnarray}
\sigma_{halo}^2(m) &\approx& 
            H^2 f(\Omega_m)^2 \int \frac{dk}{2\pi^2} P(k\rad) |W(kR(m)|^2 \, 
\end{eqnarray}
(i.e., ignoring the peak constraint and simply assuming that halo 
velocities trace the linear velocity field smoothed at the scale from 
which halos collapsed) is a reasonable approximation.  
This expression for $\sigma_{vel}$ will be useful in the analyses of 
the CMB which follow.  

\subsection{Pairwise velocities}
It is reasonable to expect that, as a result of their gravitational 
interaction, pairs of particles will, on average, approach each other.  
The gravitational attraction depends on separation, and it must fight 
the Hubble expansion which also depends on separation, so one might 
expect the mean velocity of approach to depend on the separation scale 
$r$.  In fact, pair conservation provides a relation between the rate 
at which the correlation function on scale $r$ evolves, and the mean 
pairwise motion at that separation.  In particular, pair conservation 
requires the mean peculiar velocity between a pair of particles at 
separation $r$ to satisfy \cite{Pee80}
\begin{equation}
 -\frac{v_{12}(r)}{Hr} = 
    \frac{1}{3[1+\xi(r)]}\frac{\partial (1+\bar{\xi})}{\partial \ln a} \, ,
\label{eqn:v12}
\end{equation}
where $\bar{\xi}(x)$ is the volume averaged correlation function on proper
scale $x$, $\bar{\xi}(x) = 3x^{-3}\int_0^x dy y^2 \xi(y)$. 
Since we have an accurate model for $\xi(r)$, we can use it to estimate 
$v_{12}(r)$.  
Before we do so, it is useful to see what linear theory would predict.  

\begin{figure}
\centerline{\psfig{file=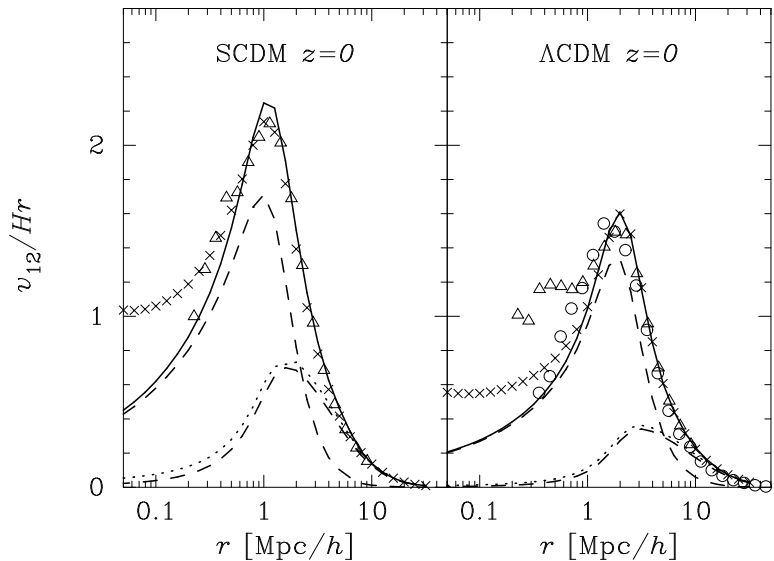,width=4.2in}}
\caption{The ratio of mean streaming velocity of dark matter particles
at scales separated by $r$, to the Hubble expansion at that scale.
Triangles show the Virgo simulation measurements,
circles show the GIF $\Lambda$CDM simulation,
and dot-dashed curves show the Hubble expansion velocity.
Crosses show the result of using the \cite{PeaDod96} formulae
for the correlation function in the fitting formula provided by 
\cite{Jusetal99}.
Solid curves show the halo model described here which accounts for
the fact that the nonlinear evolution is different from what linear
theory predicts, and then weights the linear and nonlinear scalings
by the relative fractions of linear and nonlinear pairs.
Dashed curves show the two contributions to the streaming motion
in the halo model; the curves which peak at large $r$ are for pairs in
two different halos.  Dotted curve shows the approximation of using
the linear theory correlation function to model this two-halo term. }
\label{fig:v12plot}
\end{figure}

In linear theory, 
$\partial \bar{\xi}/\partial \ln a \approx 2 f(\Omega_m) \bar{\xi}$, 
where $f(\Omega_m)$ comes from the usual approximation to the derivative 
of the growth function:  
$f(\Omega_m)\equiv \partial \ln G/\partial \ln a \approx \Omega_m^{0.6}$. 
Thus, in linear theory, 
\begin{equation}
 -\frac{v_{12}(r)}{Hr} =  \frac{2f(\Omega_m)\bar{\xi}(r,a)}{3[1+\xi(r,a)]} \, .
 \label{eqn:v12simple}
\end{equation}
On large scales where linear theory analyses can still be applied, 
$\xi(r,a) \ll 1$.  If we can drop this term from the denominator then 
this expression for the mean pairwise velocity is the same as that 
obtained directly from linear theory \cite{Hametal91,NitPad94,Jusetal99}. 
This linear theory expression underestimates velocities on small nonlinear 
scales by a factor of $\sim$ 3/2.  

If we use the halo model decomposition $\xi = \xi_{1h} + \xi_{2h}$, and 
then use the fact that $\xi_{2h}$ scales like the linear theory 
correlation function, then equation~(\ref{eqn:v12}) becomes 
\begin{equation}
 -\frac{v_{12}(r)}{Hr} =  \frac{1}{3[1+\xi(r,a)]}\left[
                2f(\Omega_m)\bar{\xi}_{2h}(r,a)
                +\frac{\partial \bar{\xi}_{1h}}{\partial \ln a}\right] \, .
\label{eqn:v12halo}
\end{equation}
The next step is to compute the derivative of the single halo term.  
Since $\xi_{1h}$ depends on the halo mass function and density profiles, 
the derivative can be computed directly \cite{MaFry00c,Sheetal01b}:  
\begin{eqnarray}
 {\partial \bar\xi_{1h} \over \partial {\rm ln} a} &=&
 {\partial {\rm ln} m_* \over \partial {\rm ln} a}
 \Bigl[\bar\xi_{1h} (r,a) - \xi_{1h} (r,a)\Bigr] \nonumber \\
 &&\ \ + \ {3\over r^3} \int_0^r {\rm d}r'\,{r'}^2
 \int_0^\infty {\rm d}m {n(m)\over\bar\rho}  {\lambda(r|m)\over\bar\rho}
 {\partial {\,\rm ln}\lambda \over \partial {\,\rm ln} a} \, ,
 \label{v12profile}
\end{eqnarray}
where $\lambda(r|m)$ denotes the convolution of the density profile with 
itself:
\begin{equation}
 \lambda(r|m) = 2\pi \int dy\, y^2\, \rho(y|m) 
              \int_{-1}^{1}d\beta\, \rho(z|m)|_{z^2=y^2+r^2-2yr\beta} \, .
\end{equation}
Now, $\partial\ln\lambda/\partial\ln a \approx 
 (\partial\ln\lambda/\partial\ln c)(\partial\ln c/\partial\ln a)_{m/m_*}$; 
since the time dependence of $c$ only comes from its dependence on $m_*$ 
and the derivative is taken by keeping $m_*$ constant, this term is zero. 
The piece which remains depends on $\partial\ln m_*/\partial \ln a$.  
If $P(k) \propto k^n$, then 
$\partial\ln m_*/\partial \ln a = f(\Omega_m) 6/(3+n)$ and \cite{Sheetal01b}
\begin{eqnarray}
-{v_{\rm 12}\over Hr} &=& {f(\Omega)\over 3[1+\xi(r,a)]}
\Biggl[ 2\bar\xi_{2h} (r,a) +  {6\over 3 + n_*}
\Bigl[\bar\xi_{1h} (r,a) - \xi_{1h} (r,a)\Bigr]\Biggr],
\end{eqnarray}
where $n_*=-1.53$ is the slope of the power spectrum on scale $m_*$ for 
the $\Lambda$CDM cosmology.  Figure~\ref{fig:v12plot} compares the 
mean pairwise velocities from this halo model calculation with 
measurements in numerical simulations.

Extending the approach to estimate how the mean pairwise velocity of 
halos or of galaxies depends on scale requires modeling the halo or 
galaxy correlation function.  This in turn requires estimating how 
the halo-mass dependent bias factors evolve.  The evolution of the bias 
factors is straightforward to compute \cite{Sheetal01d}.  The resulting   
2-halo contribution to $v_{12}$ is 
\begin{equation}
{v_{12}^{\rm 2i}(r)\over Hr} \approx
{v_{12}^{\rm dm}(r)\over Hr}\ b_i\
\left[{1+\xi^\lin_{\rm dm}(r)\over 1 + \xi_i(r)}\right] \, ,
\label{v12twog}
\end{equation}
where $i$ represents a tracer of halo with a large scale bias $b_i$ with 
respect to the linear density field. The 1-halo contribution to the 
pairwise peculiar velocities follow similar to the relation for dark matter 
in equation~(\ref{v12profile}), but with the galaxy or halo correlation 
function substituted for the dark matter.

\begin{figure}[t]
\centerline{\psfig{file=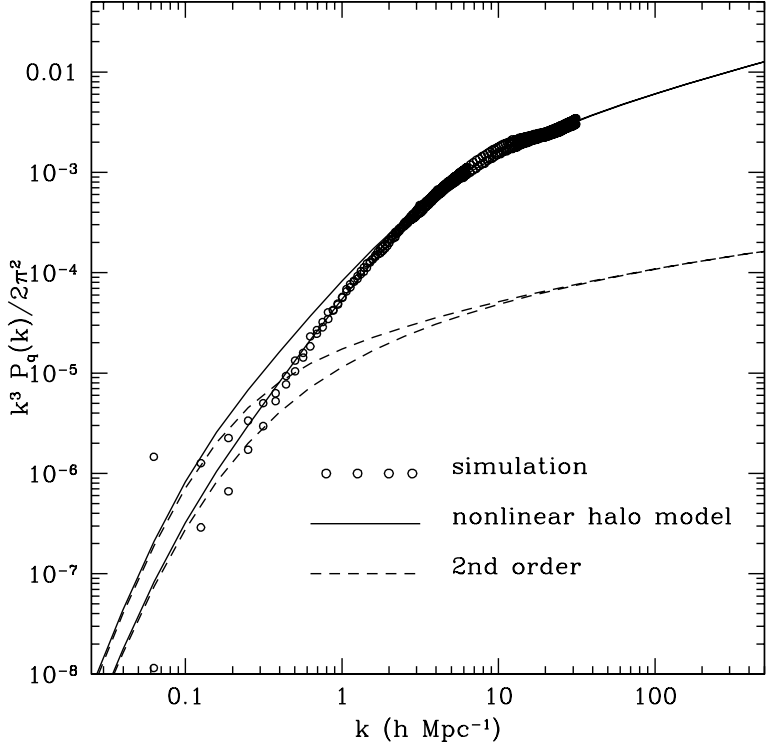,width=4.2in}}
\caption{Three dimensional power spectrum of the momentum density field 
parallel (upper) and perpendicular (lower) to the wavevector $\veck$.
The halo model estimate is compared to the numerical simulations and 
calculations based on 2nd order perturbation theory. Note that the 
parallel component contributes to the time-derivative of the density
field (\S~\ref{sec:isw}) through the continuity equation, whereas the 
perpendicular component (involving the momentum density field of baryons, 
rather dark matter) contributes to the kinetic Sunyaev-Zel'dovich effect 
(\S~\ref{sec:ksz}).  This figure is from \cite{MaFry01}.}
\label{fig:mafrymom}
\end{figure}

When combined with the BBGKY hierarchy, the halo model of the two- 
and three-point correlation functions allows one to estimate how the 
pairwise velocity dispersion depends on scale.  Although this 
calculation can, in principle, be done exactly, a considerably 
simpler but reasonably accurate approximation is sketched in 
\cite{Sheetal01c}.  A halo model calculation of the full distribution 
of pairwise velocities on small scales is in \cite{She96}; when 
combined with results from \cite{SheDia01} and \cite{Sheetal01c}, 
it can be extended to larger scales, although this has yet to be done.

\subsection{Momentum and Velocity Power Spectra}\label{sec:mom}

We have already provided an estimate of the mass weighted velocity 
dispersion (equation~\ref{eqn:disp}).  Since mass times velocity defines 
a momentum, we will now study the statistics of the momentum field.  
Specifically, define the momentum $p\equiv(1+\delta)v$.  
The divergence of the momentum is 
\begin{eqnarray}
i\veck \cdot p({\bf k}) 
        & =& i\veck \cdot {\bf v}(\veck) + \int \frac{d^3\veck'}{(2\pi)^3}
        \delta(\veck-\veck') i\veck \cdot {\bf v}(\veck')\, .
\end{eqnarray}
The first term involving the velocity field gives the contribution
from the velocity field in the linear scale limit, $\delta \ll 1$, 
while the non-linear aspects are captured in the term involving the 
convolution of the $\delta v$ term.  We can write the power spectrum 
of the divergence of the momentum density field \cite{MaFry01}, as
\begin{eqnarray}
 k^2P_{pp}(k) &=& k^2P_{vv}^\lin(k) + k^2\int
 \frac{d^3{\bf
 k'}}{(2\pi)^3} \mu'^{2} P_{\delta \delta}(|\veck-\veck'|) P_{vv}(k')
\nonumber \\
 &&+\ k^2\ \int \frac{d^3{\bf k'}}{(2\pi)^3} 
 \frac{(k-k'\mu')\mu'}{|\veck-\veck'|}
 P_{\delta v}(|\veck-\veck'|) P_{\delta v}(k')
\end{eqnarray}
In the non-linear regime integrating over angles yields, with
$\veck-\veck' \sim \veck$, 
\begin{equation}
 k^2P_{pp}(k)  = k^2P_{vv}^\lin(k) + {k^2 P(k)\over 3}
                 \int \frac{k'^2 dk'}{2\pi^2} P_{vv}(k') .
\label{eqn:momdiv}
\end{equation}
This latter result is similar to the one proposed 
by \cite{Hu00a} to calculate the momentum density field associated 
with the baryon field.  Here, one replaces the density field power 
spectrum with the non-linear power spectrum, either from the halo 
model or from perturbation theory.   In figure~\ref{fig:mafrymom}, 
we summarize results from \cite{MaFry01}, which shows that the halo model calculation is 
in good agreement with the numerical measurements.

The halo model provides a simple description why the above approach
works \cite{Sheetal01b}.  In general, we can separate the contributions to 1- and 2-halo 
terms, so that
\begin{equation}
 P_{pp}(k) = P^{1h}_{pp}(k) + P^{2h}_{pp}(k) \, .
\end{equation}
Using  equation~(\ref{eqn:disp}), we can write the two terms by noting that
\begin{eqnarray}
 k^2P_{pp}^{1h}(k) &=& k^2P_{vv}^{1h}(k) \ +
 \int {\rm d}m\,{m^2n(m)\over \bar\rho^2}\,
 \ {k^2\sigma^2_{\rm halo}(m)\over f^2(\Omega)H^2}\,
 {k^3\bigl|u(kr_{\rm vir}|m)\bigr|^2\over 2\pi^2},\nonumber\\
 k^2P_{pp}^{2h}(k)  &=&  k^2P_{vv}^{2h}(k) +
 {k^2\sigma^2_{\rm halo}(m_*)\over f^2(\Omega)H^2}
 P_{\delta\delta}^{2h}(k),
\label{dkpp}
\end{eqnarray}
where $u(k)$ is the same density profile factor when computing
the power in the density field. The second factor in the two-halo term
comes from using the fact that $\sigma_{\rm halo}$ depends only weakly
on $m$ (see, figure~\ref{assumel}),
so we approximate it by setting it equal to its value at $m_*$.
Similarly, the 1 and  2-halo terms of the velocity power spectra are
\begin{eqnarray}
 k^2P_{vv}^{1h}(k) &=&
 \int {\rm d}m\,{m^2n(m)\over \bar\rho^2\,(\Delta_{\rm nl}/\Omega)}\,
 {k^2\sigma^2_{\rm halo}(m)\over f^2(\Omega)H^2}\, 
 {k^3W^2(kr_{\rm vir}|m)\over 2\pi^2}\qquad {\rm and} \nonumber \\
 k^2P_{vv}^{2h}(k)\! &=&\! P^\lin(k)
 \left[\int\!{\rm d}m\,{mn(m)\over\bar\rho}\,W(kR|m)\right]^2,
\label{dkvv}
\end{eqnarray}
where $W(x)$ is the Fourier transform of a tophat window, 
$(r_{\rm vir}/R)^3 = \Omega/\Delta_{\rm nl}$,
with $R(m)=(3m/4\pi\bar\rho)^{1/3}$.

In the halo model, the 1-halo contribution to the momentum density
field is similar to the approximation introduced by \cite{Hu00a}, where
one sets $P_{pp}(k) \approx P(k) V^2_\lin$ where 
$V^2_\lin = \int dk P^\lin(k)/2\pi^2$. 
The single-halo contribution integrates over the linear-theory 
velocity power spectrum that is smoothed with a filter at the scale 
of the initial size of the halo. Since the halo velocity is 
independent of mass (panel on right of Figure~\ref{assumel}), 
one obtains a reasonably accurate result by simply setting 
$\sigma_{halo}$ to the value at $m_*$.  In this approximation, 
$P_{pp}(k)^{1h} \approx [\sigma_{halo}(m_*)/fH]^2 P^{1h}(k)$ and 
at non-linear scales, $P(k) \approx P^{1h}(k)$; thus, at non-linear 
scales, the ratio of power in momentum to velocity is a constant.

\begin{figure}[t]
\centerline{\psfig{file=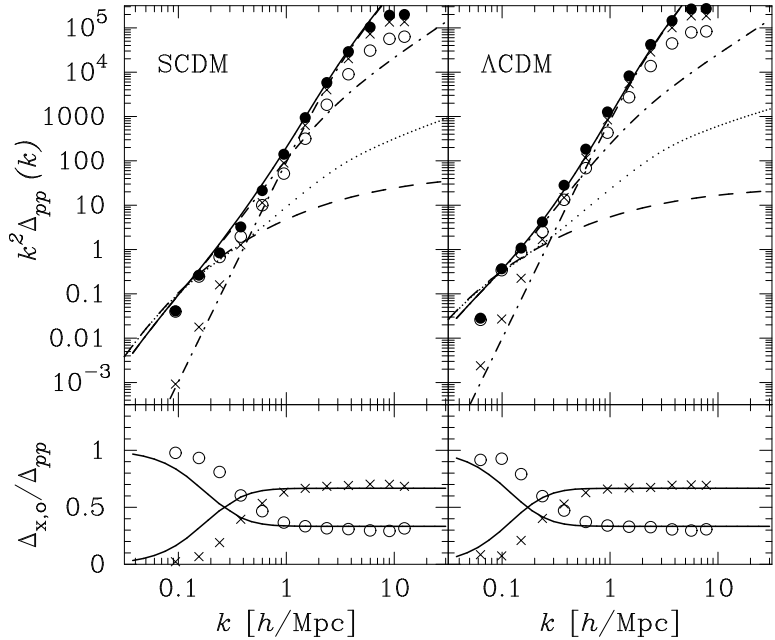,width=4.2in}}
\caption{Power spectrum of the momentum: $k^2\,\Delta_{pp}(k)$.
Filled circles show the sum of the power spectrum of the divergence
(open circles) and the curl (crosses) of the momentum fields in the
simulations.  Dashed curves show the linear theory prediction, and solid
curves show our nonlinear theory prediction for the total power.
The solid curves are obtained by summing the dot-dashed curves, which
represent the contributions to the power from the single-halo and
two-halo terms discussed in the text.  Bottom panel shows the fraction
of the total power contributed by the divergence (open circles) and
the curl (crosses) components, and solid lines show what our
model predicts. }
\label{fig:Ppp}
\end{figure}

Figure~\ref{fig:Ppp} compares this model with measurements 
in the GIF simulations.  The turnover in the measurements at 
$k\sim 5h/$Mpc in these figures is not real; it is due to the finite 
grid on which the power spectra have been evaluated.
On the larger scales (smaller $k$) where the grid is not important, our
model provides a good description of the power spectrum of the momentum.
The halo model predicts that the power spectrum of the curl should equal
$(2/3)\,k^2P^{1h}_{pp}(k)$.  In addition, one must account for
the curl which comes from the second term in the two-halo contribution
to the power.  We have done this by assuming that two thirds of this term
is in the curl component.  The bottom panels show that this provides a
good description of how the power is divided up between the divergence
and the curl on small scales, but the agreement is not good on large
scales.  Presumably this is because our assignment of 2/3 of the power
from the second term in $P_{pp}^{2h}$ is an overestimate.

We can extend the discussion to also consider cross power spectra between
velocities, momentum and the density fields. The 2-halo terms
associated with these correlations are simple given the fact that all 
these three field trace each other:  
\begin{equation}
P^{2h}_{pv}(k) = \sqrt{P^{2h}_{pp}(k)P^{2h}_{vv}(k)},
\end{equation}
and similarly for the other pairs.  The single-halo terms are only
slightly more complicated:
\begin{eqnarray}
P^{1h}_{\delta v}(k) &=&
\int {\rm d}m\,{m^2n(m)\over \bar\rho^2\sqrt{D_{\rm nl}/\Omega}}\,
{\sigma_{\rm halo}(m)\over f(\Omega)H}\,
{k^3\bigl|u(kr_{\rm vir}|m)\bigr|\bigl|W(kR|m)\bigr|\over 2\pi^2},\nonumber\\
P^{1h}_{pv}(k) &=& P_{vv}^{1h}(k) \ +
\int {\rm d}m\,{m^2n(m)\over \bar\rho^2\sqrt{D_{\rm nl}/\Omega}}\,
{\sigma^2_{\rm halo}(m)\over f^2(\Omega)H^2}\,
{k^3\bigl|u(kr_{\rm vir}|m)\bigr|\bigl|W(kR|m)\bigr|\over 2\pi^2},\nonumber\\
P^{1h}_{p\delta}(k) &=&
P_{\delta v}^{1h}(k) \ + \int {\rm d}m\,{m^2n(m)\over \bar\rho^2}\,
 {\sigma_{\rm halo}(m)\over f(\Omega)H}\,
 {k^3\bigl|u(kr_{\rm vir}|m)\bigr|^2\over 2\pi^2}.
\end{eqnarray}
If one again ignores the weak mass dependence of $\sigma_{\rm halo}$, 
$P_{p\delta}^{1h}(k)\approx
V_{\rm rms}P^{1h}(k)$ at large $k$, where
$V_{\rm rms} = \sigma_{\rm halo}(m_*)$.  This closely resembles the
corresponding approximation for the momentum spectrum:
$P_{pp}(k)\approx V^2_{\rm rms}P(k)$,
and so provides a simple way of using $P(k)$
to estimate $P_{pv}$.

\begin{figure}[t]
\centerline{\psfig{file=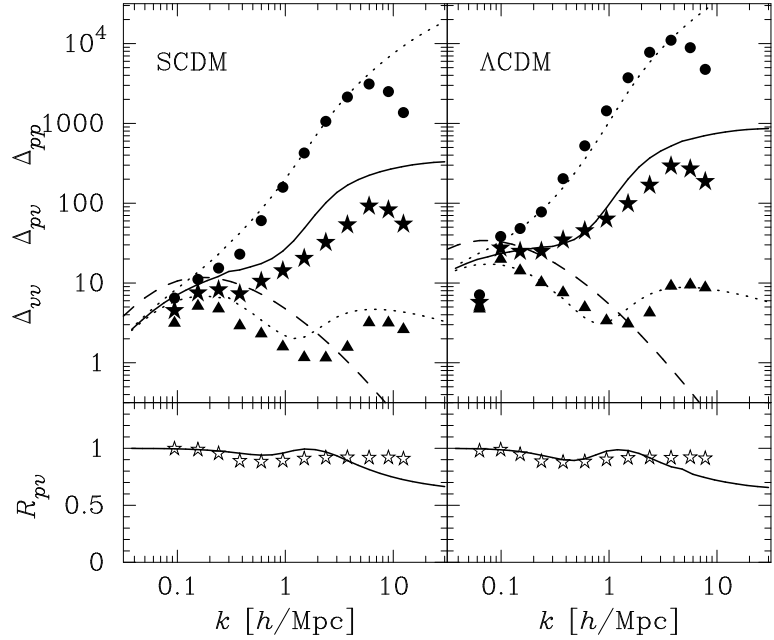,width=4.2in}}
\caption{Cross-spectrum of the momentum and the velocity.
Filled circles show $\Delta_{pp}$, triangles show $\Delta_{vv}$
and stars show $\Delta_{pv}$.  Dashed curves show the linear theory
prediction, and dotted curves show our nonlinear theory predictions
for the momentum and the velocity, and solid curves show our prediction
for the cross spectrum.  Bottom panel shows the ratio of the
cross spectrum to the square root of the product of the individual
spectra, and solid lines show what our model predicts.  }
\label{fig:Ppv}
\end{figure}

If we define $R_{p\delta}\equiv P_{p\delta}/\sqrt{P_{pp}P_{\delta\delta}}$
and similarly for the other pairs, then the expressions above show that
$R=1$ at small $k$.  If we ignore the mass dependence of $\sigma_{\rm halo}$,
then $R_{p\delta}\approx 1$ at both small and large $k$, so it depends
on scale only over a limited range of scales.  Fig.~\ref{fig:Ppv} shows our
predictions for $P_{pv}$ and $R_{pv}$ fit the simulations quite well;
note that $R_{pv}$ is always quite close to unity, even at large $k$.

\subsection{Redshift-Space Power Spectrum}

Our description of virial velocities provide a mechanism to calculate the redshift space distortions in the non-linear regime of clustering. Following \cite{Kai87}, we can write the redshift space fluctuation, $\delta_g^z$, of galaxy density field as
\begin{equation}
\delta_g^z(\veck) = \delta_g(\veck)+\delta_v \mu^2 \,,
\end{equation}
where $\delta_v$ is the velocity divergence and $\mu = \hat{{\bf r}} \cdot \hat{\bf k}$. At linear scales, one can simplify the relation by noting that $\delta_g(\veck) = b_g \delta(\veck)$ and
$\delta_v = f(\Omega_m) \delta(\veck)$ to obtain
\begin{equation}
\delta_g^z(\veck) = \delta_g(\veck)[1+\beta\mu^2]\ , ,
\end{equation}
where $\beta = f(\Omega_m)/b_g$; this parameter is of
traditional interest in cosmology as it allows constraints to be
placed on the density parameter $\Omega_m$ through clustering
in galaxy redshift surveys. We refer the reader to a 
review by Strauss \& Willick \cite{StrWil95}), with recent
applications related to the 2dFGRS survey in
\cite{Outetal01,Peaetal01}.
At linear scales, the distortions increase the power by a factor 
$(1+2/3\beta+1/5\beta^2)$, which when $b_g=1$ is 1.41 
for $\Omega_m=0.35$. At non-linear scales, virial velocities within halos modify clustering properties. With the description of the one dimensional virial motions, $\sigma$, which can be described by a Gaussian, we write
\begin{equation}
\delta_g^z(\veck) = \delta_g e^{-(k\sigma \mu)^2/2} \, .
\end{equation}

\begin{figure}[t]
\centerline{\psfig{file=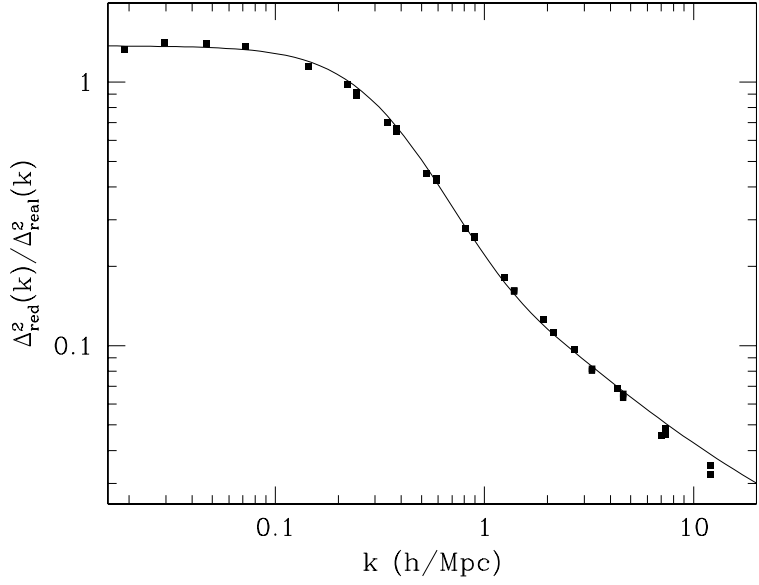,width=4.2in}}
\caption{The ratio of power in redshift space compared to real space from the
halo model (solid line) and from N-body simulations (data point). This figure is reproduced from \cite{Whi01}.}
\label{fig:whitez}
\end{figure}

This allows us to write the power spectrum in redshift space as \cite{Sel01}
\begin{eqnarray}
 P_\gal^z(k) &=& P^{1h}_{\gal}(k) + P^{2h}_{\gal}(k)
               \qquad\qquad{\rm where}\nonumber\\
 P^{1h}_{\gal}(k) &=& \int dm \, n(m) \,
                    \frac{\left< N_\gal(N_\gal-1)|m\right>}{\bar{n}_\gal^2}\,
                   R_p(k\sigma) |u_\gal(k|m)|^p \,,\nonumber\\
P^{2h}_{\gal}(k) &\approx& \left(F_g^2+\frac{2}{3}F_vF_g+\frac{1}{5}F_v^2\right)P^\lin(k)
\label{eqn:galz}
\end{eqnarray}
with 
\begin{eqnarray}
F_g &=& \int dm\, n(m)\, b_1(m)\,
\frac{\left< N_\gal|m\right>}{\bar{n}_\gal}\, R_1(k\sigma) u_\gal(k|m)  \nonumber \\
F_v &=& f(\Omega_m) \int dm\, n(m)\, b_1(m)\,
R_1(k\sigma) u(k|m)  \, ,
\end{eqnarray}
and
\begin{equation}
R_p(\alpha=k\sigma \sqrt{p/2}) = \frac{\sqrt{\pi}}{2} \frac{{\rm erf}(\alpha)}{\alpha} \, ,
\end{equation}
for $p=1,2$. In equation~(\ref{eqn:galz}), $\bar{n}_\gal$ denotes the mean number density of galaxies (equation~\ref{eqn:barngal}).

Even though peculiar velocities increase power at large scales, virial motions within halos lead to a suppression of power. In figure~\ref{fig:whitez}, we show the ratio of power in redshift space to that of real space for dark matter alone. Note the sharp reduction in power at scale corresponding to the 1-halo term of the power spectrum. When compared to the real space 1-halo contribution to the power spectrum, the redshift space 1-halo term is generally reduced. This partly explains the reason why perturbation theory works better in redshift space than in real space \cite{Whi01}.

\section{Weak Gravitational Lensing}
\label{sec:lensing}

\subsection{Introduction}

Weak gravitational lensing of faint galaxies probes the
distribution of matter along the line of sight.  Lensing by
large-scale structure (LSS) induces correlation in the galaxy 
ellipticities at the percent level
(e.g., \cite{Blaetal91,Mir91,Kai92} and recent reviews 
by \cite{BarSch01,Mel99}). 
Though challenging to measure, these
correlations provide important cosmological information that is
complementary to that supplied by the cosmic microwave background and 
potentially as precise
(e.g., 
\cite{JaiSel97,Beretal97,Kametal97b,Kai98,Schetal98,HuTeg99,Coo99,Vanetal99}).
Indeed several recent studies have provided the first clear evidence
for weak lensing in so-called blank fields (e.g., 
\cite{Vanetal00,Bacetal00,Witetal00,Wiletal01}), though more work is 
clearly needed to understand even the statistical errors
(e.g. \cite{Cooetal00b}).

Given that weak gravitational lensing results from the projected mass
distribution, the statistical properties of weak lensing convergence
reflect those of the dark matter.  Current measurements of weak lensing 
involve the shear, which is directly measurable through galaxy 
ellipticities, and constructed through a correction for the
anisotropic point-spread function \cite{Kaietal95}, 
or via a series of  basis functions, called
"shapelets", that make use of information from higher order multipoles, beyond the quadrupole, to represent the galaxy shape \cite{Ref01,RefBac01}.
In such galaxy shear data, statistical measurements include variance and 
shear-shear correlation functions; as we will soon discuss, these measurements 
are related to the power spectrum of convergence. Additionally, in shear 
data,  the convergence can be constructed through approaches such as the 
aperture mass \cite{Kaietal94,Sch96}. Such a construction allows direct 
measurements of statistics related to convergence  such as its power 
spectrum and higher order correlations.

The halo approach to non-linear clustering considered in this review 
allows one to study various statistical 
measurements related to weak lensing. Additionally, one can use the halo model to investigate various statistical and systematic effects in current and upcoming data.
For example, weak lensing
surveys are currently limited to small fields which may not be 
representative of the universe as a whole owing to sample variance. In 
particular, rare massive objects can contribute strongly to the mean power 
in the shear or convergence but not be present in the observed fields. The 
problem is compounded if one chooses blank fields subject to the condition 
that they do not contain known clusters of galaxies. Through the halo mass 
function, we can quantify the extent to which massive halos dominate the 
cosmological weak lensing effect and, thus,  the required survey volume, or projected area 
on the sky, needed to obtain a fair sample of the large scale structure 
\cite{Cooetal00b}.

Non-linearities in the mass distribution also induce non-Gaussianity
in the convergence distribution. These non-Gaussianities contribute to 
higher order correlations in convergence, such as a measurable skewness, 
and also contribute to the covariance of the power spectrum measurements. 
With growing observational and theoretical interest in weak lensing, 
statistics such as skewness have been suggested as probes of cosmological 
parameters and the non-linear evolution of large scale structure 
\cite{Beretal97,Jaietal00,Hui99,MunJai99,Vanetal99}. Similarly, we can 
also consider the bispectrum of convergence, the Fourier analog of the 
three-point correlation function. Since lensing probes non-linear scales, 
the bispectrum or the skewness cannot be considered in perturbation theory 
alone as it is only applicable in the large linear scales. In fact, it has 
been well known that predictions based on perturbation theory 
underestimates the measured skewness in numerical simulations of lensing convergence
\cite{WhiHu99}. The halo model provides a simple analytic technique to 
extend the calculations to the non-linear regime and predictions based on 
the halo model are consistent with the numerical 
simulations \cite{CooHu00}.

In terms of the power spectrum covariance, the non-Gaussian contribution 
arise from the four-point correlation function or the trispectrum in 
Fourier space. These non-Gaussian contributions are especially significant 
if observations are limited to small fields of view such that power 
spectrum measurements are made with wide bins in multipole, or Fourier, 
space. Similar to the PThalo approach that allowed measurements of the 
covariance related to galaxy two-point correlation function 
\cite{Scretal01}, the halo approach provides an analytical scheme to 
estimate the covariance of binned power spectrum of shear or convergence, 
based on the non-Gaussian contribution.
The calculation related to the convergence covariance requires detailed
knowledge on the dark matter density trispectrum, which can be
obtained analytically through perturbation theory (e.g.,
\cite{Beretal97}) or numerically through simulations (e.g.,
\cite{Jaietal00,WhiHu99}). Since numerical simulations are limited by 
computational
expense to a handful of realizations of cosmological models with modest
dynamical range, approaches such as the halo model is useful for speedy 
calculations with accuracies at the level of few tens of percent or 
better.  

\begin{figure}[t]
\centerline{\psfig{file=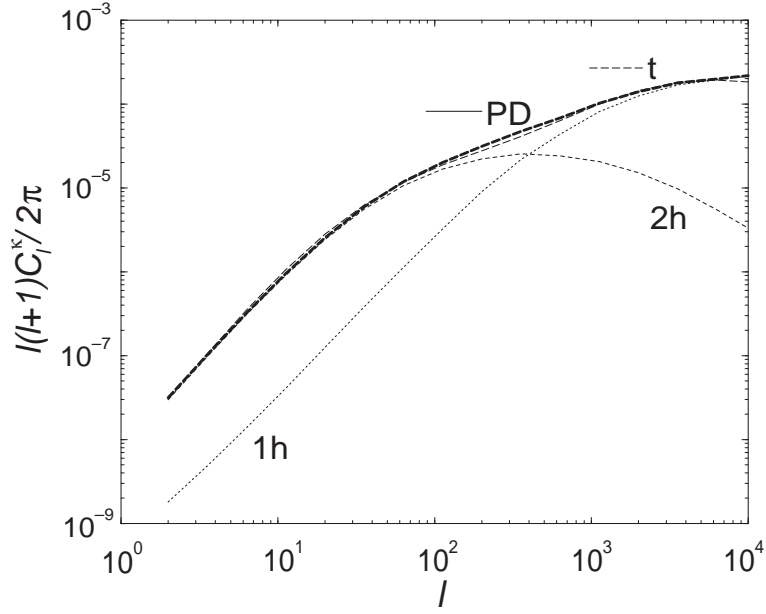,width=4.0in}}
\caption{Weak lensing convergence power spectrum  under the halo 
description.
Also shown is the prediction from the PD non-linear power
spectrum fitting function. We have separated individual contributions
under the halo approach. We have assumed that all sources
are at $z_s=1$.}
\label{fig:weakpower}
\end{figure}

\begin{figure}[t]
\centerline{\psfig{file=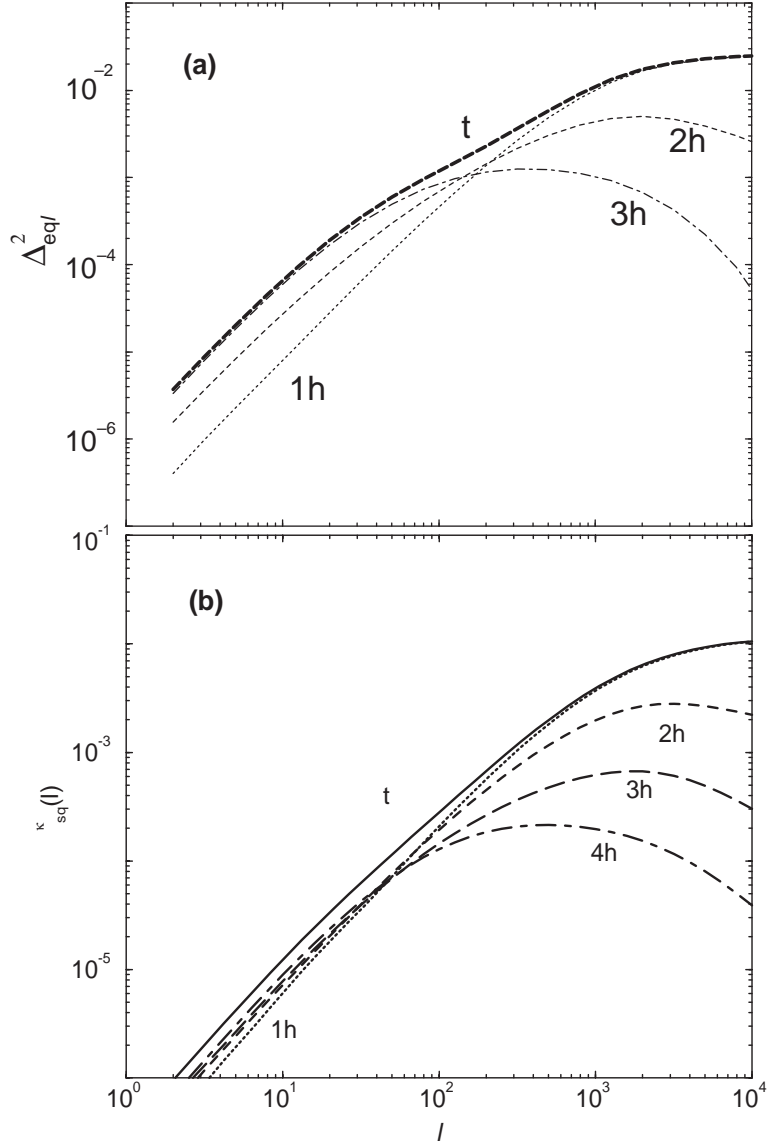,width=4.0in}}
\caption{Weak lensing convergence (a) bispectrum  and (b) trispectrum 
under the halo description.
 We have separated individual contributions
under the halo approach to 3 halos in the case of bispectrum and 4 halos 
in the case of trispectrum. We have assumed that all sources are at 
$z_s=1$.}
\label{fig:weakbitri}
\end{figure}

\subsection{Convergence}
\label{sec:convergence}

Weak lensing probes the statistical properties of shear field and
we can write the deformation matrix that maps  $\Delta {\bf x}$  
separation vector between source, $s$, and image, $i$,  planes, 
$\Delta {\bf x_i}^s = {\bf A}_{ij} \Delta {\bf x_j}^i$, as
\begin{eqnarray}
{\bf A}_{ij} &=& \delta_{ij} - \psi_{ij} \nonumber \\
&=&\left(
\begin{array}{cc}
1-\kappa-\gamma_1 & -\gamma_2-\omega\\
-\gamma_2+\omega & 1-\kappa+\gamma_1
\end{array}
\right).
\label{A}
\end{eqnarray}
Here $\kappa$ is the 
convergence, responsible for magnification and demagnification, 
$w$ is the net rotation of the image, and we 
have separated the components of the shear,
 $\gamma \equiv \gamma_1 +i 
\gamma_2$, which translates as a spin-2 field $\gamma = |\gamma| \e^{2i 
\phi}$ and is a pseudo vector field. 
The shear components are
\begin{eqnarray}
\gamma_1 &=& \frac{1}{2}\left(\psi_{11} - \psi_{22}\right)\nonumber \\
\gamma_2 &=& \frac{1}{2}\left(\psi_{12} +\psi_{21}\right)\,
\end{eqnarray}
where,
\begin{equation}
\psi_{ij} = 2 \int_0^{\rad_s} d\rad  
\frac{\da(\rad_d)\da(\rad_d-\rad_s)}{\da(\rad_s)} \partial_i\partial_j\Phi 
\, .
\end{equation}
To the smallest order in potential fluctuations, $w \approx 0$ and we can ignore the asymmetry associated with the deformation matrix; thus, $\gamma_2 =
\psi_{12}$ as commonly known.

We can write the convergence using the trace of the deformation matrix with 
$\kappa = \frac{1}{2}  (\psi_{11}+\psi_{22})$:
\begin{equation}
\kappa(\bn) = \int_0^{\rad_s} d\rad\, W^\lens(\rad) \nabla_\perp^2 
\Phi(\rad,\rad\bn) \, ,
\label{eqn:kappa}
\end{equation}
where the lensing visibility function for a radial distribution of 
background sources, $n_s(\rad)$, is
\begin{equation}
W^\lens(\rad) =  \int_\rad^{\rad_s} d\rad' \frac{\da(\rad) 
\da(\rad'-\rad)}{\da(\rad')} n_s(\rad') \, .
\label{eqn:lensradial}
\end{equation}
Here, $\rad$ is the comoving radial distance (equation~\ref{eqn:rad}) and  
$d_A$ is the angular
diameter distance (equation~\ref{eqn:da}).

\subsection{Power spectrum}

We can write the 
angular power spectrum of convergence by taking the spherical harmonics
\begin{equation}
\kappa(\bn) = \sum \kappa_{l m} \Ylmn(\bn) \, ,
\label{eqn:klmn}
\end{equation}
with spherical moments of the convergence field defined such that
\begin{eqnarray}
\kappa_{lm} &=& i^l \int \frac{d^3\veck}{2 \pi^2} k_\perp^2 \Phi(\veck)   
\int d\rad  W(\rad)j_{l}(k\rad)  \Ylmn(\hat{\veck}) \, , \nonumber\\
\label{eqn:klmnint}
\end{eqnarray}
where $W(k,\rad)$ is the visibility function associated with weak lensing 
defined in equation~(\ref{eqn:lensradial}).
Here, we have simplified using the Rayleigh expansion of a plane wave
\begin{equation}
e^{i{\bf k}\cdot \hat{\bf n}\rad}=4\pi\sum_{lm}i^lj_l(k\rad)Y_l^{m 
\ast}(\bk) \Ylmn(\bn)\, .
\label{eqn:Rayleigh}
\end{equation}

In the small scale limit, only the modes perpendicular to the radial 
direction contribute to the integral in equation~(\ref{eqn:klmnint}) while 
others are suppressed through near-perfect cancellation of positive and 
negative oscillations along the line of sight. Thus, we can replace 
$k_\perp \approx k$, which only makes an error of order $\Phi \sim 
10^{-5}$. 
Further, we can use the Poisson equation (equation~\ref{eqn:poisson}) to
relate potential fluctuations to those of density, $k^2\Phi(k) = 4 \pi G 
\bar{\rho} a^2 \delta(k)$. These allow us to construct the angular power 
spectrum of the convergence, defined in
terms of the multipole moments, $\kappa_{l m}$, as
\begin{equation}
\left< \kappa_{l m}^* \kappa_{l' m'}\right> = C_l^\kappa \delta_{l l'}
\delta_{mm'}\,,
\label{eqn:twopoint}
\end{equation}
to obtain
\begin{equation}
C_l = \frac{2}{\pi} \int_0^\infty k^2 dk P(k) \int_0^{\rad_0} d\rad_1 \int_0^{\rad_0} 
d\rad_2  W^\lens(\rad_1) W^\lens(\rad_2) j_l(k\rad_1) j_l(k\rad_2) \,
\end{equation}
where 
\begin{equation}
W^\lens(\rad) = \frac{3}{2} \Omega_m \frac{H_0^2}{c^2} \int_\rad^{\rad_0}
d\rad' \frac{d_A(\rad) d_A(\rad' -\rad)}{ad_A(\rad')} n_s(\rad')\, .
\label{eqn:weight}
\end{equation}
When all background sources are at a distance of 
$\rad_s$, $n_s(\rad')=\delta_D(\rad'-\rad)$, 
the weight function reduces to
\begin{equation}
W^\lens(\rad) = \frac{3}{2} \Omega_m \frac{H_0^2}{c^2 a} \frac{
d_A(\rad) d_A(\rad_s -\rad)}{d_A(\rad_s)} \, .
\end{equation}

In the case of a non-flat geometry, one needs to introduce curvature 
corrections to the Poisson equation (see, equation~\ref{eqn:poisson}), and 
replace the radial Bessel functions, $j_l$, with hyperspherical Bessel 
functions.  In the small scale limit, 
for efficient calculational purposes, we can simplify further by using
the Limber, or small angle, 
approximation \cite{Lim54} where one can neglect the
radial component of the Fourier mode $\veck$ compared to the transverse
component. Here, we employ a version based on
the completeness relation of spherical Bessel functions (see,
\cite{CooHu00,HuWhi96} for details)
\begin{equation}
\int dk k^2 F(k) j_l(kr) j_l(kr')  \approx {\pi \over 2} \da^{-2} 
\deld(r-r')
                                                F(k)\big|_{k={l\over 
d_A}}\,,
\label{eqn:ovlimber}
\end{equation}
where the assumption is that $F(k)$ is a slowly-varying function. 
Under this assumption, the contributions to the power spectrum come only 
from correlations at equal time surfaces. 
Finally, we can write the convergence power spectrum 
as \cite{Kai92,Kai98}:
\begin{equation}
C^\kappa_l = \int d\rad \frac{W^\lens(\rad)^2}{d_A^2} 
P\left(\frac{l}{d_A};\rad\right) \, .
\label{eqn:lenspower}
\end{equation}

\subsection{Relation to Shear Correlations}

We can consider the relation between convergence power spectrum and shear 
correlation functions by considering the Fourier decomposition of the 
shear field \cite{HuWhi01} to a gradient-like (E-modes) and curl-like 
(B-modes) components:
\begin{equation}
\gamma_1(\bn) \pm i\gamma_2(\bn) = \int \frac{d^2\vecl}{(2\pi)^2} 
[\epsilon(\vecl)\pm \beta(\vecl)]
\e^{\pm 2 i \phi_l} \e^{i\vecl \cdot \bn} \, ,
\end{equation}
and consider the correlations between $\langle \gamma_1 \gamma_1 \rangle$, 
$\langle \gamma_1 \gamma_2 \rangle$ and $\langle \gamma_2 \gamma_2 
\rangle$.
We can write these correlation functions as
\begin{eqnarray}
\langle \gamma_1(\bn_i) \gamma_1(\bn_j) \rangle &=& \int 
\frac{d^2\vecl}{(2\pi)^2} \left[ C_l^{\epsilon \epsilon} \cos^2 2\phi_l + 
C_l^{\beta \beta} \sin^2 2\phi_l - C_l^{\epsilon \beta} \sin 4\phi_l 
\right] \e^{i \vecl \cdot (\bn_i - \bn_j)} \nonumber \\
\langle \gamma_1(\bn_i) \gamma_2(\bn_j) \rangle &=& \int 
\frac{d^2\vecl}{(2\pi)^2} \left[ \frac{C_l^{\epsilon \epsilon}}{2} \sin 
4\phi_l  -\frac{C_l^{\beta \beta}}{2} \sin 4\phi_l + C_l^{\epsilon \beta} 
\cos 4\phi_l \right] \e^{i \vecl \cdot (\bn_i - \bn_j)} \nonumber \\
\langle \gamma_2(\bn_i) \gamma_2(\bn_j) \rangle &=& \int 
\frac{d^2\vecl}{(2\pi)^2} \left[ C_l^{\epsilon \epsilon} \sin^2 2\phi_l + 
C_l^{\beta \beta} \cos^2 2\phi_l + C_l^{\epsilon \beta} \sin 4\phi_l 
\right] \e^{i \vecl \cdot (\bn_i - \bn_j)} \nonumber \\
\end{eqnarray}
Using the expansion of $e^{i \vecl \cdot (\bn_i -\bn_j)}= \sum_m i^m J_m 
(l\theta) \e^{i m (\phi-\phi_l)}$, in terms of the magnitude $\theta$ and 
orientation $\phi$ of vector $\bn_i-\bn_j$,
we write
\begin{eqnarray}
\langle \gamma_1 \gamma_1 \rangle_{\theta,\phi} &=& \int \frac{ldl}{4\pi}  
\Big\{ C_l^{\epsilon \epsilon} [J_0(l\theta)+\cos(4\phi)J_4(l\theta)]+ 
C_l^{\beta \beta} [J_0(l\theta)-\cos(4\phi)J_4(l\theta) ]\nonumber \\
&& \quad \quad -2C_l^{\epsilon \beta} \sin (4\phi)  J_4(l\theta)\Big\} 
\nonumber \\
\langle \gamma_1 \gamma_2 \rangle_{\theta,\phi} &=& \int \frac{ldl}{4\pi} 
\Big\{ C_l^{\epsilon \epsilon} \sin(4\phi)J_4(l\theta) - C_l^{\beta 
\beta}  \sin(4\phi)J_4(l\theta) +C_l^{\epsilon \beta} 2\cos(4\phi) 
J_4(l\theta) \Big\} \nonumber \\
\langle \gamma_2 \gamma_2 \rangle_{\theta,\phi} &=&  \int \frac{ldl}{4\pi} 
\Big\{ C_l^{\epsilon \epsilon} [J_0(l\theta)-\cos(4\phi)J_4(l\theta)]+ 
C_l^{\beta \beta} [J_0(l\theta)+\cos(4\phi)J_4(l\theta) ]\nonumber \\
&&\quad \quad +2C_l^{\epsilon \beta} \sin (4\phi)  J_4(l\theta)\Big\} 
\,.\nonumber \\
\end{eqnarray}

One can choose an appropriate coordinate system such that measured 
correlation functions in the coordinate frame are independent of the 
choice of coordinates; in the above derivation, this is equivalent to 
setting $\phi=0$ (e.g, \cite{Ste96}). To do this in practice, in analogy 
with CMB polarization (see, \cite{Kametal97a}), shear can be measured 
parallel and perpendicular to the line joining the two points, such that 
$\bn_i-\bn_j \parallel \hat{\bf x}$. In such a coordinate system 
correlation functions reduce to the well known result of 
\cite{Mir91,Kai92}:
\begin{eqnarray}
\langle \gamma_1 \gamma_1 \rangle_{\theta} &=& \int \frac{ldl}{4\pi}  
\left\{C_l^{\epsilon \epsilon} [J_0(l\theta)+J_4(l\theta)]+ C_l^{\beta 
\beta} [J_0(l\theta)-J_4(l\theta) ]\right\} \nonumber \\
\langle \gamma_1 \gamma_2 \rangle_{\theta} &=& \int \frac{ldl}{4\pi} 
2C_l^{\epsilon \beta} J_4(l\theta)  \nonumber \\
\langle \gamma_2 \gamma_2 \rangle_{\theta} &=&  \int \frac{ldl}{4\pi} 
\left\{ C_l^{\epsilon \epsilon} [J_0(l\theta)-J_4(l\theta)]+ C_l^{\beta 
\beta} [J_0(l\theta)+J_4(l\theta) ]\right\} \,.\nonumber \\
\end{eqnarray}
To the first order, contributions to the shear correlations primarily come 
from perturbations involving scalars, or gradient-like modes, with 
$C_l^{\epsilon \epsilon}=C_l^\kappa$  and $C_l^{\beta \beta} = C_l^{\beta 
\epsilon} =0$; even if  $C_l^{\beta \beta}$ contributions are non-zero, 
the latter $C_l^{\beta \epsilon}$ is zero due to parity invariance.

The curl-like modes in shear can be generated by tensor perturbations such 
as gravity-waves. Since there is no appreciable source of primordial 
gravity-wave perturbations at late times (see, \cite{KamKos99} for a 
review), it is unlikely that there is a significant contribution to 
$C_l^{\beta \beta}$, except in two cases:

(1) The first order calculation  of weak lensing distortion matrix and 
convergence is that we have  implicitly integrated over the unperturbed 
photon paths (the use of so-called Born approximation, see 
\cite{Beretal97,Schetal98}).   Similarly, second-order effects such as 
lens-lens coupling involving lenses at two-different redshifts can 
generate a curl-like contribution. With second-order corrections to Born
approximation and lens-lens coupling, we can write the 
deformation matrix associated with weak lensing as
\begin{eqnarray}
A_{ij} =\delta_{ij} -\psi_{ij}^{(1)} -\psi_{ij}^{(2)}\,
\end{eqnarray}
with
\begin{eqnarray}
\psi_{ij}^{(2)} &=& 4 \int d\chi' \frac{d_A(\chi') d_A(\chi-\chi')}{d_A(\chi)} \nonumber \\
&&\quad \times \int d\chi'' \frac{d_A(\chi'') d_A(\chi'-\chi'')}{d_A(\chi')}
\partial_i\partial_k\Phi(\chi') \partial_k\partial_j\Phi(\chi'') \, ,
\end{eqnarray}
due to lens-lens coupling and
\begin{eqnarray}
\psi_{ij}^{(2)} &=& 4 \int d\chi' \frac{d_A(\chi') d_A(\chi-\chi')}{d_A(\chi)}\nonumber \\
&& \quad \times \int d\chi'' d_A(\chi'-\chi'')
\partial_i\partial_j\partial_k\Phi(\chi') \partial_k\Phi(\chi'') \, ,
\end{eqnarray}
due to a correction to Born approximation, respectively \cite{Schetal98,Beretal97,CooHu01c}. The resulting deformation matrix due to these second-order
corrections is asymmetric and results in a contribution to 
$C_l^{\beta \beta}$, as well as a contribution to the net rotation; the latter is equivalent to the Stokes-V contribution in
a polarization field or, equivalently, circular polarization.

The Born approximation and lens-lens  coupling have been tested in numerical simulations by \cite{Jaietal00}
where they evaluated contribution to the
convergence power spectrum resulting from
second order effects. Here, the
rotational contribution to angular power spectrum, due to lens-lens coupling, is roughly 3 orders of
magnitude smaller. In \cite{CooHu01c}, it was shown that the corrections due to the Born approximation is also
smaller compared to the first order result that $C_l^{\beta \beta}=0$.

\begin{figure}[t]
\centerline{\psfig{file=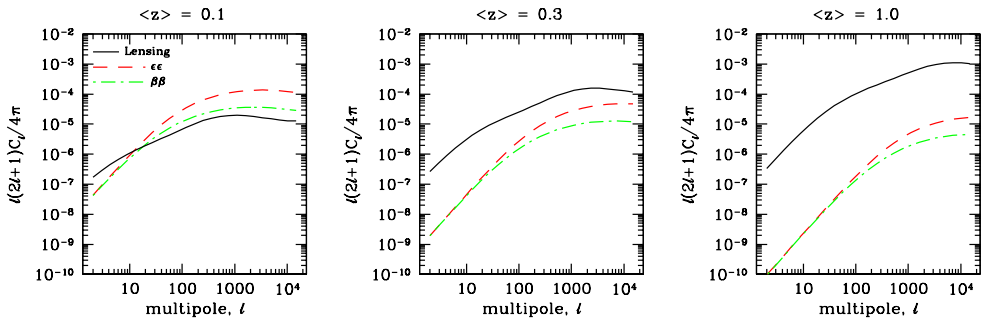,width=5.6in}}
\caption{The power spectra of lensing convergence (solid) and 
$C_l^{\epsilon \epsilon}$ (dashed) and $C_l^{\beta \beta}$ (dot-dashed) 
due to galaxy ellipticity correlations induced by tidal torques, as a 
function of redshift. The tidal torques induce significant correlations at 
low redshifts, while these can be ignored for deep weak lensing surveys 
with background sources at $z \geq 1$. The figure is from 
\cite{Macetal01}.}
\label{fig:macintrinsic}
\end{figure}

(2) The intrinsic correlations between individual background galaxy 
shapes, due to 
long range correlations in the tidal gravitational field in which the 
halos containing galaxies formed, can generate a contribution to 
$C_l^{\beta \beta}$ \cite{CroMet00,Heaetal00,Catetal01,Macetal01}.   The 
intrinsic correlations have a redshift dependence such that they are 
significant at low redshifts.  In figure~\ref{fig:macintrinsic}, we show 
the resulting $C_l^{\epsilon \epsilon}$ and $C_l^{\beta \beta}$
power spectra due to ellipticity alignments in background galaxies arising 
from tidal torques  and a comparison to convergence power spectrum 
associated with ellipticity correlations due to lensing following 
\cite{Macetal01}.
In order to avoid the confusion between lensing generated ellipticity 
correlations vs. tidal torques induced correlations, the results from 
intrinsic alignment calculations generally indicate that deep surveys are 
preferred over shallow ones for cosmological purposes. We will return to 
this issue again based on the  non-Gaussian contribution to convergence 
power spectrum
covariance.

\begin{figure}[t]
\centerline{\psfig{file=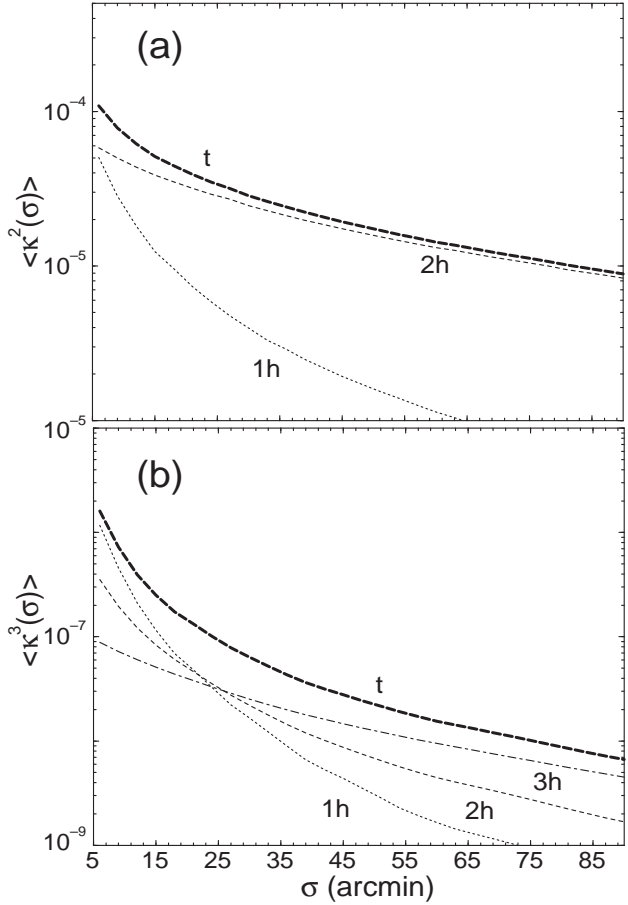,width=4.2in}}
\caption{Moments of
the convergence field as a function of
top-hat smoothing scale $\sigma$ with (a) Second moment broken into 
individual contributions and (b) Third moment broken into individual 
contributions.}
\label{fig:moments}
\end{figure}

In addition to shear correlations, one can also measure the shear 
variance, which can be 
related to the convergence power spectrum by
\begin{equation}
\left< \gamma^2(\sigma) \right>  \equiv \left< \kappa^2(\sigma) \right> =
{1 \over 4\pi} \sum_l (2l+1) C_l^\kappa W_l^2(\sigma)\,,
\label{eqn:secondmom}
\end{equation}
where $W_l$ are the multipole moments, or Fourier transform in a
flat-sky approximation, of the window.   In figure~\ref{fig:moments}(a), we
choose a window which is a two-dimensional top hat in real space with a 
window function in multipole space of $W_l(\sigma) = 2J_1(x)/x$ with $x = 
l\sigma$.
As shown, at  5$'$ to 90$'$ angular scales, most of the contribution to 
the second
moment comes from the double halo correlation term and is dominated by the
linear power spectrum instead of the non-linear evolution.

In figure~\ref{fig:weakpower}(a), we show the convergence power
spectrum of the dark matter halos compared with that predicted by
the  \cite{PeaDod96}  fitting function for the non-linear dark matter
power spectrum.
The lensing power spectrum due to halos has the same behavior as the
dark matter power spectrum. At large angles, $l \lesssim 100$,
the correlations between halos dominate. The transition from linear to 
non-linear is at $l \sim
500$ where halos of mass similar to $M_{\star}(z)$ contribute.
The single halo contributions start dominating at $l > 1000$.
When $l \gtrsim$ few thousand, at small scales corresponding to deeply
non-linear regime, the shot-noise behavior of the background sources 
contribute 
to the convergence power spectrum via a noise term
\begin{equation}
C^{\rm SN}_l = \frac{\langle \gamma_{\rm int}^2\rangle}{\bar{n}} \, .
\end{equation}
Here, $\langle \gamma_{\rm int} \rangle^{1/2} $ is the
rms noise per component introduced by intrinsic ellipticities, typically 
$\sim 0.6$ for
best ground based surveys, and $\bar{n}$ is the surface number density of 
background
source galaxies. 

Note that the shot-noise term is effectively reduced 
by the number of independent modes one measures at each multipole. 
Including the sample variance, the total error expected for a measurement 
of the power spectrum, as a function of multipole, is
\begin{equation}
\Delta C^\kappa_l = \sqrt{\frac{2}{f_\sky(2l+1)}}\left[C^\kappa_l + C^{\rm SN}_l\right] \, .
\label{eqn:kappapoerror}
\end{equation}
Here, the first term represents the sample variance under the Gaussian 
approximation for 
 the convergence field, $\kappa_{l m}$ and is the dominant source of noise 
at  large angular scales. The factor $f_\sky$, fraction of the sky 
observed,
accounts for the reduction in the number of independent modes under 
the partial sky coverage.
In the absence of noise for an all-sky experiment, at a multipole of 
$\sim$ 100, the error on the power spectrum due to sample variance is 
$\sim$ 10\% and is usually reduced with binned measurements of the power 
spectrum  in multipole space. 

For surveys that reach a  limiting magnitude in $R\sim 25$,
the surface density is consistent with
 $\bar{n} \sim 6.9 \times 10^{8}$ sr$^{-1}$ ($\approx 56$ gal 
arcmin$^{-2}$) 
\cite{Smaetal95}, such that $C^{\rm SN}_l \sim  2.3 \times 10^{-10}$. This 
shot-noise
contribution reaches the power due to convergence at multipoles of $\sim$ 
2000 and dominates the cosmological weak lensing signal at  multipoles 
thereafter.
It is clear that the convergence power spectrum at multipoles of few 
thousand probe the small scale behavior of the dark matter power spectrum.
The presence of significant shot-noise, however, complicate studies that 
can 
potentially test assumptions related to large scale structure, such as the
stable clustering hypothesis, or the halo model, such as the use of smooth 
profiles in the presence of substructure seen in numerical simulations.

\begin{figure}[t]
\centerline{\psfig{file=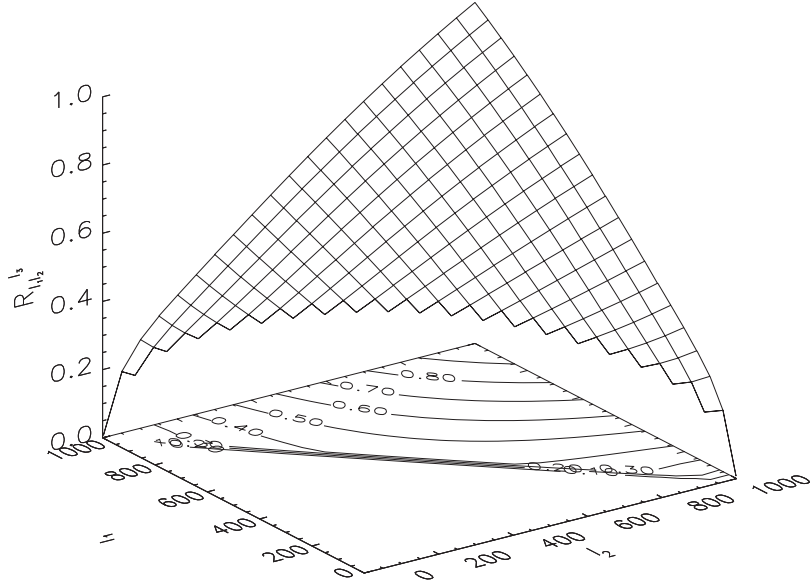,width=4.2in}}
\caption{The bispectrum
configuration dependence $R_{l_1l_2}^{l_3}$
 as a function of $l_1$ and $l_2$ with
$l_3=1000$.  Due to triangular conditions associated with $l$'s, only
the
upper triangular region in $l_1$-$l_2$ space contribute to the
bispectrum.}
\label{fig:bispecsurface}
\end{figure}

\subsection{Bispectrum}

Using the spherical harmonic moments of convergence defined in
equation~(\ref{eqn:klmn}),
the angular bispectrum of the convergence is defined following 
\cite{CooHu00,SpeGol99}
as
\begin{equation}
\left< \kappa_{l_1 m_1} \kappa_{l_2 m_2} \kappa_{l_3 m_3} \right> 
= \wjm B_{l_1 l_2 l_3}^\kappa \,.
\end{equation}
Here, the quantity in parentheses is the Wigner 3$j$ symbol. Its 
orthonormality relation implies
\begin{equation}
 B_{l_1 l_2 l_3}^\kappa = \sum_{m_1 m_2 m_3} \wjm \left< \kappa_{l_1 m_1} 
\kappa_{l_2 m_2} \kappa_{l_3 m_3} \right>  \, .
\label{eqn:bispectrum}
\end{equation}
The angular bispectrum, $B_{l_1 l_2 l_3}^\kappa$, contains all the 
information available in the three-point correlation function, For 
example, 
the third moment or the skewness, the collapsed three-point expression of 
\cite{Hinetal95} and the equilateral configuration statistic of 
\cite{Feretal98} can all be expressed as linear combinations of the 
bispectrum terms (see, \cite{Ganetal94} for explicit expressions).

Similar to our discussion related to the convergence power spectrum, we 
can write spherical moments of the convergence field defined with respect 
to the density field as
\begin{eqnarray}
\kappa_{lm} = i^l \int \frac{d^3\veck}{2 \pi^2}
\delta(\veck)  I_l^\lens(k) \Ylmn(\hat{\veck}) \quad {\rm and} \quad
I_l^\lens(k) = \int d\rad  W^\lens(k,\rad)j_{l}(k\rad) \, , \nonumber \\
\label{eqn:secondaryform}
\end{eqnarray}
where $W(k,\rad)$ is the source function associated with weak lensing
(see equation~\ref{eqn:weight}). 

The bispectrum can be constructed through
\begin{eqnarray}
&&\left< \kappa_{l_1 m_1}\kappa_{l_2 m_2}\kappa_{l_3 m_3}  \right>
=  i^{l_1+l_2+l_3}  \int \frac{d^3\veck_1}{2 \pi^2}
\int \frac{d^3\veck_2}{2 \pi^2} \int \frac{d^3\veck_3}{2 \pi^2}
\left<\delta(\veck_1)  \delta(\veck_2)
\delta(\veck_3)  \right> \nonumber \\ 
&& \quad \quad \times I_l^\lens(k_1) I_l^\lens(k_2) I_l^\lens(k_3)
\Ylm{1}(\hat{\veck_1})  \Ylm{2}(\hat{\veck_2})  \Ylm{3}(\hat{\veck_3})
\, ,
\end{eqnarray}
and can be simplified further by using the bispectrum of density
fluctuations to write the convergence bispectrum  as
\begin{eqnarray}
B^\kappa_{l_1 l_2 l_3} &=& \sum_{m_1 m_2 m_3} \wjm
\left< \kappa_{l_1 m_1}\kappa_{l_2 m_2}\kappa_{l_3 m_3}  \right>
\nonumber \\
&=&  \sqrt{\frac{\prod_{i=1}^3(2l_i +1)}{4 \pi}}
\left(
\begin{array}{ccc}
l_1 & l_2 & l_3 \\
0 & 0  &  0
\end{array}
\right) b_{l_1,l_2,l_3} \, ,
\label{eqn:bigeneral}
\end{eqnarray}
with
\begin{eqnarray}
&& b_{l_1,l_2,l_3} = \frac{2^3}{\pi^3}\int k_1^2 dk_1 \int k_2^2 dk_2
\int k_3^2 dk_3 B(k_1,k_2,k_3) \nonumber  \\
&\times& I^\lens_{l_1}(k_1) I^\lens_{l_2}(k_2) I^\lens_{l_3}(k_3)
\int x^2 dx  j_{l_1}(k_1x) j_{l_2}(k_2x) j_{l_3}(k_3x) \, . \nonumber
\\
\end{eqnarray}

In general, the calculation of $b_{l_1,l_2,l_3}$ involves seven
integrals involving the mode coupling integral and three integrals
involving distances and Fourier modes, respectively. We can simplify 
further by employing the Limber approximation similar to our derivation of 
the power spectrum. Applying
 equation~(\ref{eqn:ovlimber}) 
to the integrals involving $k_1$, $k_2$ and 
$k_3$
allows us to write the angular bispectrum of lensing convergence as
\begin{eqnarray}
\bi^\kappa &=& \sqrt{\prod_{i=1}^3(2l_i+1) \over 4\pi} \wj      \int dr 
{[W^\lens(r)]^3 \over \da^4}  B\left({l_1
\over
        \da},{l_2 \over \da},{l_3\over \da};r\right)\, . \nonumber \\
\label{eqn:kappabispectrum}
\end{eqnarray}
Through angular momentum selection rules, the Wigner-3$j$ symbol restricts 
$l_i$ to form a triangle such that $l_i \leq |l_j -l_k|$. Additional 
properties of the Wigner 3$j$ symbol
can be found in the Appendix of \cite{CooHu00}.

The more familiar flat-sky bispectrum is  \cite{CooHu01a,Hu00b}:
\begin{equation}
B^\kappa(\vecl_1,\vecl_2,\vecl_3) =   \int dr {[W^\lens(r)]^3 \over 
\da^4}  B\left({\vecl_1
\over
        \da},{\vecl_2 \over \da},{\vecl_3\over \da};r\right) \, ,
\label{eqn:kappaflat}
\end{equation}
where $\vecl_i$ are now two-dimensional vectors. In the case of the 
flat-sky bispectrum, the Wigner 3$j$ symbol in the all sky expression 
becomes a triangle equality relating the two-dimensional vectors. The 
implication is that the triplet $(l_1,l_2,l_3)$ can be considered to 
contribute to the triangle configuration of 
$\vecl_1,\vecl_3,\vecl_3=-(\vecl_1+\vecl_2)$ where the multipole number is 
taken as the length of the vector. The correspondence between the
all-sky derivation, equation~(\ref{eqn:kappabispectrum}),
 and the flat-sky approximation, equation~(\ref{eqn:kappaflat}),
 can be noted by expanding the delta function
involved with $\vecl_1+\vecl_2+\vecl_3=0$ \cite{Hu00b}.

In the flat-sky case, we can generalize our result for a any n-point 
Fourier space correlation as
\begin{equation}
P_N^\kappa(\vecl_1,...,\vecl_N) =   \int dr {[W^\lens(r)]^N \over \da^{(2 
N - 2)}}  P_N\left({\vecl_1
\over
        \da}, ... ,{\vecl_N\over \da};r\right) \, ,
\label{eqn:kappanpoint}
\end{equation}
where vectors $\vecl_1+...+\vecl_N=0$.

Similar to the density field bispectrum,
we define
\begin{equation}
\Delta^2_{{\rm eq}l} = \frac{l^2}{2 \pi}
\sqrt{B^\kappa_{l l l}} \, ,
\end{equation}
involving equilateral triangles in $l$-space.
In figure~\ref{fig:weakbitri}(a), we show $\Delta^2_{{\rm eq}l}$.
The general behavior of the lensing bispectrum can be
understood through the individual contributions to the
density field bispectrum: at small multipoles, the triple halo
correlation term  dominates, while at high multipoles,
the single halo term dominates. The double halo term
contributes at intermediate $l$'s corresponding to angular scales of a
few tens of arcminutes.  In figure~\ref{fig:bispecsurface}, we plot the 
configuration dependence 
\begin{equation}
R_{l_1l_2}^{l_3} = 
\frac{l_1l_2}{2\pi}\frac{\sqrt{B^\kappa_{l_1l_2l_3}}}{\Delta^2_{{\rm eq}l} 
}
\end{equation}
as a function of $l_1$ and $l_2$ when $l_3=1000$. The surface, and 
associated contour plot, shows the contribution to the bispectrum from 
triangular configurations in $l$-space relative to that from the 
equilateral configuration. Because of the triangular conditions associated 
with $l$'s, only upper triangular region of $l_1$-$l_2$ space contribute 
to the bispectrum. The symmetry about the $l_1=l_2$ line is due to the 
intrinsic symmetry associated with the bispectrum. Although the weak 
lensing bispectrum peaks for equilateral configurations, the configuration 
dependence is weak. In the case of dark matter bispectrum, it is  now 
known that the halo model somewhat overestimates the configuration 
dependence due to the spherical assumption for halos \cite{Scoetal01}. 
This overestimate should also be present in the projected dark matter 
statistics such as lensing convergence bispectrum.

As discussed in the case of the second moment, it is likely that the
first measurements of higher order correlations in lensing would be
through real space statistics. Thus, in addition to the bispectrum, we
also consider skewness, which is
associated with the third moment  of the smoothed map (c.f. 
equation.~[\ref{eqn:secondmom}]) 
\begin{eqnarray}
\left< \kappa^3(\sigma) \right> &=&
                {1 \over 4\pi} \sum_{l_1 l_2 l_3}
                \sqrt{\prod_{i=1}^3(2l_i+1) \over 4\pi} \wj  \bi^\kappa
                W_{l_1}(\sigma)W_{l_2}(\sigma)W_{l_3}(\sigma)
                \,. \nonumber \\
\end{eqnarray}
One can then construct the skewness as
\begin{equation}
S_3(\sigma) =
\frac{\left<\kappa^3(\sigma)\right>}{\left<\kappa^2(\sigma)\right>^2}
\, ,
\end{equation}
where $\left<\kappa^2(\sigma)\right>^2$ is the second moment of the 
convergence field defined
in equation~(\ref{eqn:secondmom}).

\begin{figure}[t]
\centerline{\psfig{file=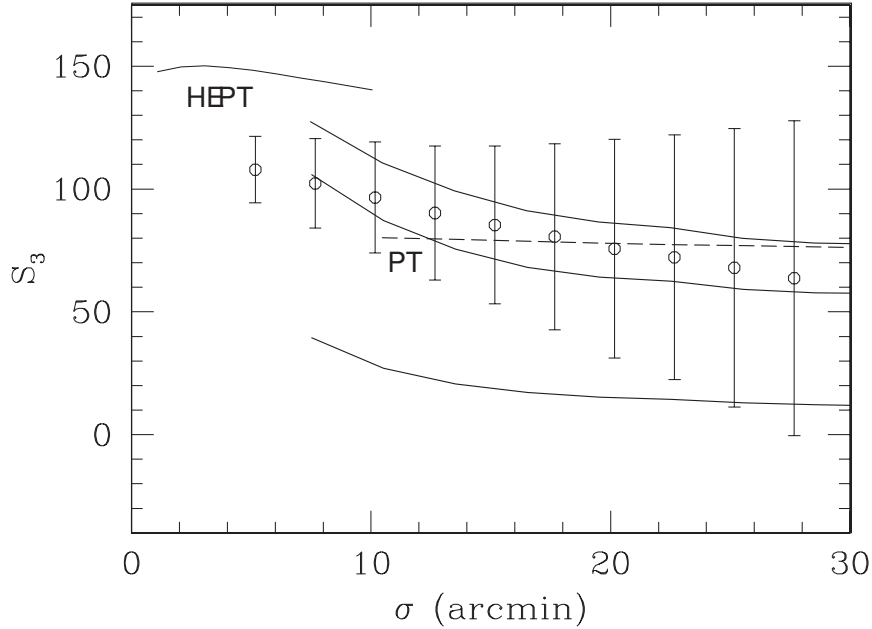,width=4.5in}}
\caption{The skewness,
$S_3(\sigma)$, as a function of angular scale. The filled symbols indicate 
the mean and variance computed from a set of $\kappa$ planes generated in 
particle-mesh (PM) simulations by \cite{WhiHu99}.
Under the halo model, shown here is the skewness with varying maximum mass 
used in the calculation (solid lines ranging from 10$^{14}$  to 10$^{16}$ 
M$_{\sun}$). For comparison, we also show
skewness values  as predicted by hyper-extended perturbation
theory (HEPT) and second-order perturbation theory (PT). Figure is 
reproduced based on \cite{WhiHu99} and \cite{CooHu01a}.}
\label{fig:skewness}
\end{figure}

In figure~\ref{fig:skewness}, we plot the
skewness based on the halo model. Here, we show skewness
as a function of maximum mass, ranging from
$10^{14}$ to $10^{16}$ M$_{\sun}$ (from increasing values of skewness). 
The assumption is that certain surveys, either by design in the case of 
so-called {\it blank-fields} or by chance, will not contain massive halos 
in the universe.
Thus, by arbitrarily cutting off the maximum mass when integrating over 
the mass function, one can estimate how the statistics are sensitive to 
the presence of the massive and rare objects in the universe. Our total 
maximum skewness agrees with what is predicted by numerical
particle mesh simulations \cite{WhiHu99} and yields a value of $\sim$ 116 
at 10$'$. However, it is lower than predicted by HEPT arguments and
simulations of \cite{Jaietal00}, which suggest a skewness of $\sim$ 140 at 
angular
scales of 10$'$ \cite{Hui99}. The HEPT prediction generally 
overpredicts skewness as it is extended to the mildly non-linear regime of clustering, where contributions to the skewness
come from at arcminute scales, from the deeply non-linear regime, corresponding to angular scales of few arcseconds, where it is expected to be valid.
The skewness based on second-order PT 
\cite{Beretal97} is lower than the maximum skewness predicted by halo
calculation, and by construction, agrees with the skewness in the linear 
regime.

The effect of maximum mass on the skewness is interesting. 
When the maximum mass is decreased to
$10^{15}$ M$_{\sun}$ from the maximum mass value where skewness
saturates ($\sim 10^{16}$ M$_{\sun}$), 
the skewness decreases from $\sim$ 116 to 98 at an angular scale of
10$'$, though the convergence power spectrum only changes by less than
few percent when the same change on the maximum mass used 
is made.  When the maximum mass used in the calculation is
$10^{14}$ M$_{\sun}$, the skewness at 10$'$ is $\sim 40$, which is
roughly a factor of 4 decrease in the skewness from the total. 

Thus, the absence of rare and massive halos in  observed fields will 
certainly bias the skewness measurement from the cosmological mean, which 
has been suggested as a probe of the cosmological matter density given 
that $S_3 \propto \Omega_m^{-0.8}$ \cite{Beretal97}. 
One, therefore, needs to exercise caution 
in using the skewness to constrain cosmological models \cite{Hui99}. 
Still, this does not mean that non-Gaussianity measured in small fields, 
where there is likely to be a significant bias due to the lack of massive 
halos, will be useless. With the halo approach, one can calculate the 
expected skewness given some information related to the mass distribution 
of halos within the observed fields. This knowledge may come externally, 
such as through X-ray and Sunyaev-Zel'dovich measurements or internally 
from lensing data themselves, independent of cosmology. Alternatively, if 
cosmology is assumed, one can  also used non-Gaussian information from 
weak lensing to constrain some aspect of the large scale structure halo 
mass distribution, such as the high mass end of the mass function.

\subsection{Weak Gravitational lensing Covariance}
\label{sec:covariance}

For the purpose of this calculation, we assume that upcoming weak
lensing convergence power spectrum will measure binned logarithmic
band  powers at several $l_i$'s in multipole space with bins of
thickness $\delta l_i$.
\begin{equation}
\bp_i = 
\int_{\shell i} 
{d^2 l \over{A_{\shell i}}} 
\frac{l^2}{2\pi} \kappa(\bf l) \kappa(-\bf l) \, ,
\end{equation}
where $A_\shell(l_i) = \int d^2 l$ is the area of the two-dimensional
shell in 
multipole and can be written as $A_\shell(l_i) = 2 \pi l_i \delta l_i 
+ \pi (\delta l_i)^2$.

We can now write the signal covariance matrix
as
\begin{eqnarray}
C_{ij} &=& {1 \over A} \left[ {(2\pi)^2 \over A_{\shell i}} 2 \bp_i^2
+ T^\kappa_{ij}\right]\,,\\
\label{eqn:variance}
T^\kappa_{ij}&=&
\int {d^2 l_i \over A_{\shell i}} 
\int {d^2 l_j \over A_{\shell j}} {l_i^2 l_j^2 \over (2\pi)^2}
T^\kappa(\bfl_i,-\bfl_i,\bfl_j,-\bfl_j)\,,
\end{eqnarray}
where 
$A=4\pi f_\sky$ is the area of the survey in steradian, when the fraction 
of sky covered is $f_\sky$.
Again the first
term is the Gaussian contribution to the sample variance and the
second term is the non-Gaussian contribution.
A realistic survey will also have shot noise variance due to
the finite number of source galaxies in the survey.  Note that in the 
Gaussian limit  with 
$T^\kappa_{ij}=0$, when $\delta l_i=1$, equation~(\ref{eqn:variance}) 
reduces to $(\Delta C^\kappa_l)^2$  given in 
equation~(\ref{eqn:kappapoerror}).

Following equation~(\ref{eqn:kappanpoint}), the convergence trispectrum is related to 
the density  trispectrum by the projection \cite{Scoetal99,CooHu01b}
\begin{eqnarray}
T^\kappa   &=& \int d\rad \frac{W(\rad)^4}{d_A^6} T\left( 
\frac{\bfl_1}{d_A},
\frac{\bfl_2}{d_A},
\frac{\bfl_3}{d_A},
\frac{\bfl_4}{d_A};\rad\right) \, ,
\label{eqn:lenstripower}
\end{eqnarray}                
with the weight function defined in  
equation~(\ref{eqn:weight}) and $\vecl_4=-(\vecl_1+\vecl_2+\vecl_3)$. 

Note that the configurations, in Fourier space, that contribute to the 
power spectrum covariance has the form of parallelograms with 
$\vecl_2=-\vecl_1$ and $\vecl_4=-\vecl_3$. Thus, it is useful to consider 
the behavior of the trispectrum for such configurations.
In figure~\ref{fig:weakbitri}(b), we show the scaled trispectrum 
\begin{equation}
\Delta^\kappa_{\rm sq}(l) = \frac{l^2}{2\pi}
T^\kappa(\vecl,-\vecl,\vecl_\perp,-\vecl_\perp)^{1/3} \, .
\end{equation}
where $l_\perp=l$ and $\vecl \cdot \vecl_\perp=0$.
The projected lensing trispectrum again shows the same behavior as the
density field trispectrum with similar conditions on $\veck_i$'s. 

\begin{figure}[t]
\centerline{\psfig{file=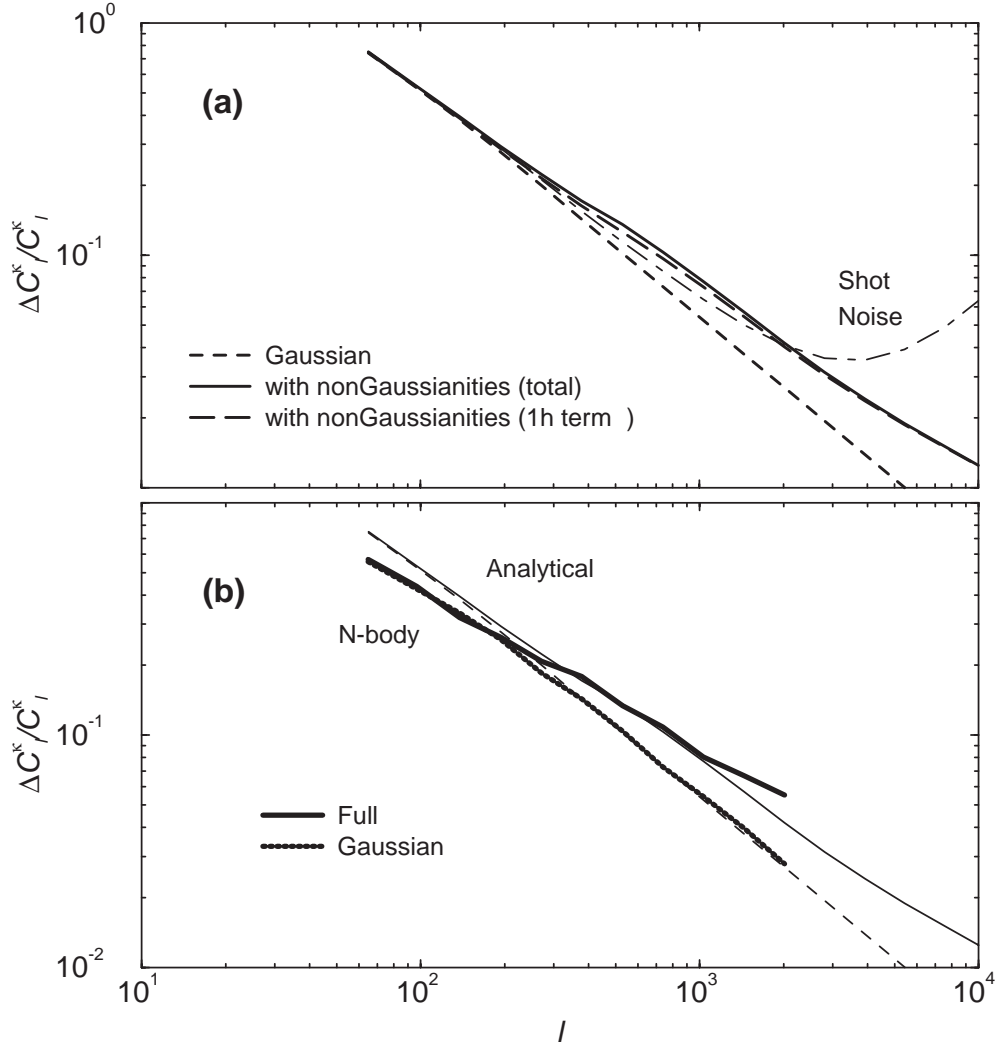,width=5.2in}}
\caption{The 
fractional errors in the measurements of the
convergence band powers.
In (a), we show the fractional errors under the Gaussian
approximation,
the full halo description, the Gaussian plus single halo term, and the
Gaussian plus shot noise term (see equation~\ref{eqn:gaussianerror}). 
As shown, the additional variance can be modeled with the single halo
piece while shot noise generally becomes dominant before non-Gaussian
effects become large. In (b), we compare the halo model with 
simulations from \cite{WhiHu99} (1999). The decrease in the variance
at small $l$ in the simulations is due to the conversion of variance
to covariance by the finite box size of the simulations.}
\label{fig:variance}
\end{figure}

We can now use this trispectrum to study the 
contributions to the covariance, which is what we are primarily
concerned here. In figure~\ref{fig:variance}a, we show the
fractional error, 
\begin{equation}
{\Delta \bp_i  \over \bp_i} \equiv {\sqrt{C_{ii}}  \over \bp_i} \, ,
\end{equation}
for bands $l_i$ given in Table~\ref{tab:cov} following the
binning scheme used by \cite{WhiHu99} on $6^\circ \times
6^\circ$ fields.

The dashed line compares that with the Gaussian errors, 
involving the first term in the covariance (equation~\ref{eqn:variance}).
At multipoles of a few hundred and
greater, the non-Gaussian term begins to dominate the
contributions.  For this reason, the errors are well approximated by
simply taking the Gaussian and single halo contributions.
In figure~\ref{fig:variance}(b), we compare these results
with those of the \cite{WhiHu99} simulations.  The
decrease in errors from the simulations at small $l$ reflects
finite box effects that convert variance to covariance
as the fundamental mode in the box becomes comparable to 
the bandwidth. 

\begin{figure*}[thb]
\centerline{\psfig{file=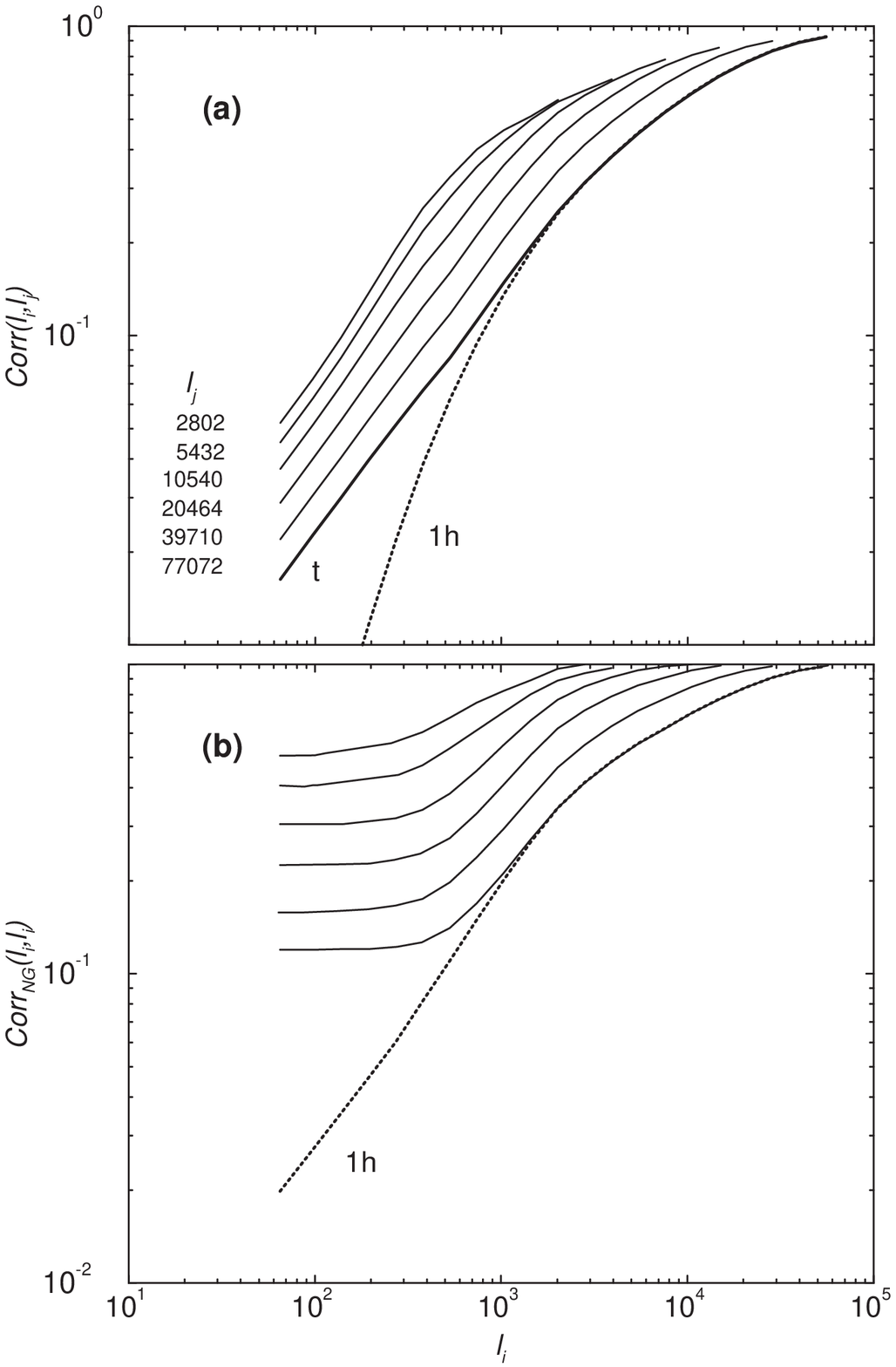,width=3.6in}}
\caption{(a) The correlation coefficient, $\hat C_{ij}$ as a function
of the multipole $l_i$ with $l_j$ as shown in the figure.  We show the
correlations calculated with the full halo model and also with only the 
single halo
term for $l_j=77072$. In (b), we show
the non-Gaussian correlation coefficient $\hat C_{ij}^{\rm NG}$,
which only involves the trispectrum (see,
equation~\ref{eqn:ng}). The transition to full correlation is due to the
domination of the single halo contribution. }
\label{fig:corr}
\end{figure*}

The correlation between the bands is given by
\begin{equation}
\hat C_{ij} \equiv \frac{C_{ij}}{\sqrt{C_{ii} C_{jj}}} \, .
\end{equation}
In table~\ref{tab:cov}, we compare the halo predictions to 
the simulations by \cite{WhiHu99}. 
The upper triangle here is the
correlations under the halo approach, while the lower triangle shows
the correlations found in numerical simulations.
The correlations along individual columns increase, as one goes to
large $l$'s or small angular scales, consistent with simulations.
In figure~\ref{fig:corr}, we show the correlation coefficients with (a)
and without (b) the Gaussian contribution to the diagonal.

\begin{table}
\begin{center}
\caption{\label{tab:cov}}
{\sc Weak Lensing Convergence Power Spectrum Correlations\\}
\begin{tabular}{ccccccccccc}
\hline
$\ell_{\rm bin}$
       & 97      & 138     & 194     & 271     & 378     & 529     &
739 & 1031    & 1440   & 2012 \\
\hline
    97 & 1.00    & 0.04  & 0.05    & 0.07    & 0.08   & 0.09    & 0.09
& 0.09    & 0.08   & 0.08\\
   138 & (0.26) & 1.00   & 0.08    & 0.10   & 0.11    & 0.12    &0.12
& 0.12    & 0.11 & 0.11\\
   194 & (0.12) & (0.31) & 1.00   & 0.14    & 0.17    & 0.18    &0.18
& 0.17    & 0.16 & 0.15\\
   271 & (0.10) & (0.21)  & (0.26) & 1.00  & 0.24   & 0.25     &0.25
& 0.24   & 0.22   & 0.21\\
   378 & (0.02) & (0.09)  & (0.24) & (0.38) & 1.00    & 0.33   &0.33
& 0.32    & 0.30   & 0.28\\
   529 & (0.10) & (0.14)  & (0.28) & (0.33) & (0.45) & 1.00    &0.42
& 0.40    & 0.37  & 0.35\\
   739 & (0.12) & (0.16)  & (0.17)  & (0.34) & (0.38) & (0.50) & 1.00
& 0.48    & 0.45   & 0.42\\
  1031 & (0.15) & (0.18)  & (0.15) & (0.27) & (0.33) & (0.48) & (0.54)
& 1.00    & 0.52  & 0.48\\
  1440 & (0.18) & (0.15) & (0.19) & (0.19) &(0.32) & (0.36) & (0.53) &
(0.57) & 1.00  & 0.54\\
  2012 & (0.19) & (0.22) & (0.16) & (0.32) & (0.27) & (0.46) & (0.50)
& (0.61) & (0.65) & 1.00\\
\hline
\end{tabular}
\end{center}
\footnotesize
NOTES.---%
Covariance of the binned power spectrum when sources are at a redshift
of 1.
Upper triangle displays the covariance found under the halo model.
Lower triangle (parenthetical numbers) displays the covariance found
in numerical simulations by \cite{WhiHu99}. To be consistent
with these simulations, we use the same binning scheme as the one used
there.
\end{table}

We show in figure~\ref{fig:corr}(a) the behavior of the correlation
coefficient between a fixed $l_j$ as a function of $l_i$.  When
$l_i=l_j$
the coefficient is 1 by definition.  Due to the presence of
the dominant Gaussian contribution at $l_i=l_j$, the coefficient has
an apparent
 discontinuity between $l_i=l_j$ and $l_i = l_{j-1}$ that decreases
as $l_j$ increases and non-Gaussian effects dominate.

To better understand this behavior it is useful to isolate
the  purely non-Gaussian correlation
coefficient 
\begin{equation}
\hat C^{\rm NG}_{ij} =
\frac{T_{ij}}{\sqrt{T_{ii} T_{ij}}} \,.
\label{eqn:ng}
\end{equation}
As shown in figure~\ref{fig:corr}(b), 
the coefficient remains constant for $l_i \ll l_j$ and smoothly
increases to unity across a transition scale that is related to where the
single halo terms starts to contribute. 
A comparison of figure~\ref{fig:corr}(b) and \ref{fig:weakbitri}(b), shows
that this transition happens around $l$ of few hundred to 1000.
Once the power spectrum is dominated by correlations in single halos,
the fixed profile of the halos will correlate the power in all the
modes. The multiple halo terms on the other hand correlate linear and
non-linear scales but at a level that is generally negligible compared 
with the
Gaussian variance. 

Note that the behavior seen in the halo based covariance, however, is not
present when the covariance is
calculated with hierarchical arguments for the trispectrum (see,
\cite{Scoetal99}). With hierarchical arguments, which are by
construction only valid in the deeply non-linear regime, one predicts
correlations which are, in general, constant across all scales and
shows no decrease in correlations between very small and very large
scales. Such hierarchical models also violate the Schwarz inequality with
correlations greater than 1 between large and small scales (e.g.,
\cite{Scoetal99,Ham00}).
The halo model, however, shows a decrease in correlations similar
to numerical simulations suggesting that the
halo model, at least qualitatively, provides a better
approach to studying non-Gaussian correlations in the translinear
regime.

\subsubsection{Scaling Relations}
\label{sec:scalings}

To better understand how the non-Gaussian contribution scale with our
assumptions,  we can consider the ratio of
non-Gaussian variance to the Gaussian variance 
\begin{equation}
\frac{C_{ii}}{C_{ii}^{\rm G}} = 1 + R \, ,
\end{equation}
with
\begin{equation}
R \equiv \frac{A_{si} T_{ii}^\kappa}{(2
\pi)^2 2 C_i^2} \, .
\label{eqn:rexact}
\end{equation}
Under the assumption that contributions to lensing convergence can be
written through an effective distance $r_\star$, at half the angular
diameter distance to background sources, and a width $\Delta r$
for the lensing window function,
the ratio of lensing convergence trispectrum and power
spectrum contribution to the variance  can be further simplified to
\begin{equation}
R \sim \frac{A_{si}}{(2 \pi)^2V_{\rm eff}}\frac{
\bar{T}(r_\star)}{2\bar{P}^2(r_\star)} \, .
\label{eqn:rapprox}
\end{equation}
Since the lensing window function peaks at $r_\star$, we have
replaced the integral over the window function of the
density field trispectrum and power spectrum by its value at the peak.
This ratio shows how the relative contribution from non-Gaussianities
scale with survey parameters: (a) increasing the bin size, through
$A_{si}$ ($\propto \delta l$), leads to an increase in the 
non-Gaussian contribution linearly,
(b) increasing the source redshift, through the effective volume of
lenses in the survey 
($V_{\rm eff} \sim r_\star^2 \Delta r$), decreases the non-Gaussian
contribution, while (c)
the growth of the density field trispectrum and power spectrum,
through the ratio $\bar{T}/\bar{P}^2$,
decreases the contribution as one moves to a higher redshift. The
volume factor quantifies the number of foreground halos in the survey
that effectively act as gravitational lenses 
for background sources; as the number of such halos is increased, 
the non-Gaussianities are reduced by the central limit theorem.

\begin{figure}[t]
\centerline{\psfig{file=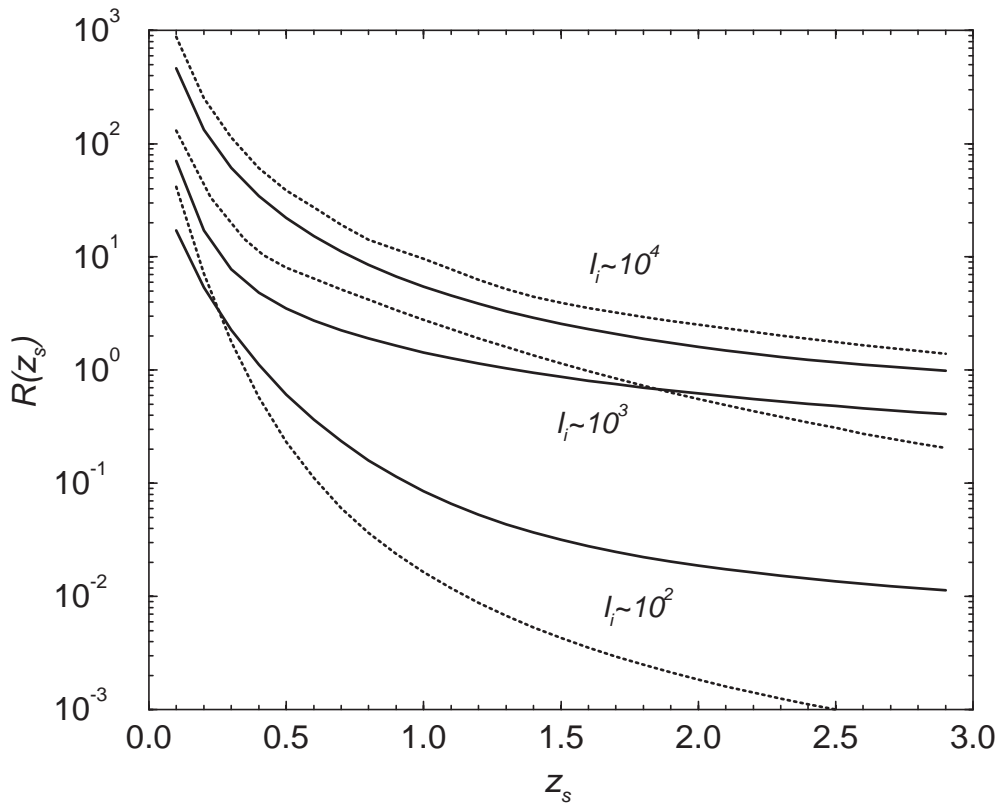,width=5.2in}}
\caption{The ratio of
non-Gaussian to Gaussian contributions,
$R$, as a  function of source redshift ($z_s$). The solid lines
are through the exact calculation (equation~\ref{eqn:rexact}) while the
dotted lines are using the approximation given in
equation~(\ref{eqn:rapprox}). Here, we show the ratio $R$ for three
multipoles corresponding to large, medium and small angular
scales. The multipole binning is kept constant such that $\delta l
\sim
l$. Decreasing this bin size will linearly decrease the value of $R$.}
\label{fig:r}
\end{figure}

In figure~\ref{fig:r}, we summarize our results as a function of
source redshift with $l_i \sim 10^2,10^3$ and 10$^4$ and setting
the bin width such that $A_s(l_i) \sim l_i^2$, or $\delta l \sim l$.
As shown, increasing the source redshift leads to a decrease in the
non-Gaussian contribution to the variance.
The prediction based on the simplifications in equation~(\ref{eqn:rapprox})
tend to overestimate the non-Gaussianity at lower
redshifts while underestimates it at higher redshifts, though the
exact transition depends on the angular scale of interest; this
behavior can be understood due to the fact that we do not
consider the full lensing window function but  only the
contributions at an effective redshift, midway between the 
observer and sources.

In order to determine whether its the increase in volume or the
decrease in the growth of
structures that lead to a decrease in the 
relative importance of non-Gaussianities
as one moves to a higher source redshift, we calculated
the non-Gaussian to Gaussian variance ratio under the halo model 
for several source redshifts and
survey volumes. Up to source redshifts $\sim$ 1.5, the increase in
volume decreases the non-Gaussian contribution significantly. When
surveys are sensitive to sources at redshifts beyond 1.5, 
the increase in volume becomes less significant
and the decrease in the growth of structures
begin to be important in
decreasing the non-Gaussian contribution. Since, in the deeply
non-linear regime, $\bar{T}/\bar{P}^2$ scales with 
redshift as the cube of the growth factor,
this behavior is consistent with the overall redshift scaling of the
volume and growth. 

The importance of the non-Gaussianity to the variance also scales
linearly with bin width.  As one increases the bin width the
covariance induced by the non-Gaussianity manifests itself as increased 
variance
relative to the Gaussian case.  The normalization of $R$ is therefore 
somewhat
arbitrary in that it depends on the binning scheme, i.e. $R \ll 1$ 
does not necessarily mean non-Gaussianity can be entirely neglected 
when summing over all the bins.  The scaling with redshift and the
overall scaling of the variance with the survey area $A$ is not.
One way to get around the increased non-Gaussianity associated with
shallow 
surveys, is to have it sample a wide patch of sky since 
$C_{ii} \propto (1+R)/A$.
This relation tells us the trade off between designing an survey to
go wide instead of deep.   One should bear in mind though that not
only will shallow surveys have decreasing number densities of source 
galaxies
and hence increasing shot noise, they will also suffer more from the 
decreasing
amplitude of the signal itself and the increasing importance 
of systematic effects,  including the intrinsic correlations of 
galaxy shapes (e.g.,
\cite{Catetal01,CroMet00,Heaetal00}). These problems 
tilt the balance more towards deep but narrow surveys than the naive
statistical scaling would suggest.

\subsubsection{The effect of non-Gaussianities}

With steady improvements in the observational front, it is likely that weak
lensing will eventually reach its full ability as a complimentary probe of 
cosmological parameters when compared to angular power spectrum of CMB 
anisotropies (see, e.g., \cite{HuTeg99}).
Thus, for a proper interpretation of observational measurements of lensing 
convergence power spectrum or shear correlation functions, it will be 
essential to include the 
associated full covariance or error matrix in upcoming analyses.
In the absence of many fields where the covariance can be
estimated directly from the data, the halo model provides
a useful, albeit model dependent, quantification of the
covariance.  As a practical approach one could imagine
taking the variances estimated from the survey under
a Gaussian approximation,  but which accounts for uneven 
sampling and edge effects \cite{HuWhi01}, 
and scaling it up by the non-Gaussian
to Gaussian variance ratio of the halo model along with
inclusion of the band power correlations. Additionally, it is in
principle possible to use the expected correlations 
from the halo model to decorrelate individual band power measurements,
similar to studies involving CMB temperature anisotropy and galaxy
power spectra (e.g., \cite{Ham97,HamTeg00}).

The resulting non-Gaussian effects on cosmological parameter
estimation was discussed in \cite{CooHu01b}.
In \cite{HuTeg99}, the potential of wide-field lensing
surveys to measure cosmological parameters was investigated
using the Gaussian approximation of a diagonal covariance
and Fisher matrix techniques.
The Fisher matrix is simply a projection of the covariance
matrix, $\C$, onto the basis of cosmological parameters $p_i$
\begin{equation}
{\bf F}_{\alpha\beta} = \sum_{ij} 
      {\partial \bp_i \over \partial p_\alpha} (\C_{\rm tot}^{-1})_{ij}
{\partial \bp_j \over \partial p_\beta} \, ,
\label{eqn:fisher}
\end{equation}
where the total covariance includes both the signal
and noise covariance.  Under the approximation of Gaussian shot
noise, this reduces to replacing $C^\kappa_l \rightarrow
C^\kappa_l + C^{\rm SN}_l$ in the expressions leading up
to the covariance equation~(\ref{eqn:variance}).  In the case where 
non-Gaussian contribution to the covariance is ignored, 
equation~(\ref{eqn:fisher}) reduces to \cite{HuTeg99,Coo99}
\begin{equation}
{\bf F}_{\alpha\beta} = \sum_{l=l_{\rm min}}^{l_{\rm max}} \frac{f_\sky 
(l+1/2)}{(C^\kappa_l+C^{\rm SN}_l)^2}\frac{\partial C^\kappa_l}{\partial 
p_\alpha} \frac{\partial C^\kappa_l}{\partial p_\beta}\, .
\label{eqn:gaussianerror}
\end{equation}
Under the approximation that there are a sufficient number
of modes in the band powers that the distribution of power
spectrum estimates is approximately Gaussian, the Fisher matrix
quantifies the best possible errors on cosmological parameters that can
be achieved by a given survey.  In particular $F^{-1}$ is
the optimal covariance matrix of the parameters and
$(F^{-1})_{ii}^{1/2}$ is the optimal error on the $i$th parameter.

For a cosmological model involving a set of 5 parameters, 
$\Omega_\Lambda$, normalization of the power spectrum, $\Omega_K = 
1-\Omega_m-\Omega_\Lambda$, $n_s$ and $\Omega_mh^2$, Cooray \& Hu 
\cite{CooHu01a} found that non-Gaussianities increase the uncertainties of 
each of the 5 parameters determined from an all-sky experiment down to the 
25th magnitude, and assuming all sources at a redshift of $\sim$ 1,  by 
about $\sim$ 10 to 15\%. In the case of weak lensing, the shot-noise due 
to finite number of background sources and their intrinsic ellipticity 
becomes the dominant error before the non-Gaussian effects dominate over 
the Gaussian noise. Thus, for the above assumed depth and redshift, the 
non-Gaussian effect on cosmological parameters is some what insignificant. 
For certain planned deeper surveys with better imaging, such as planned 
surveys with Large-Aperture Synoptic Telescope (LSST; \cite{TysAng00}), 
the shot-noise term will be subdominant and the non-Gaussian contributions 
may be more important for a precise determination of the cosmological 
parameters. As discussed with scaling relations, \S~\ref{sec:scalings}, 
the intrinsic non-Gaussian contribution to the onset of non-linearity 
decreases with increasing survey depth, and thus, deeper surveys are in 
fact preferred over shallow ones for the purposes of cosmological lensing 
work.

\subsection{The Galaxy-Mass Cross-Correlation}
\label{sec:galaxy-mass}

\begin{figure}[t]
\centerline{\psfig{file=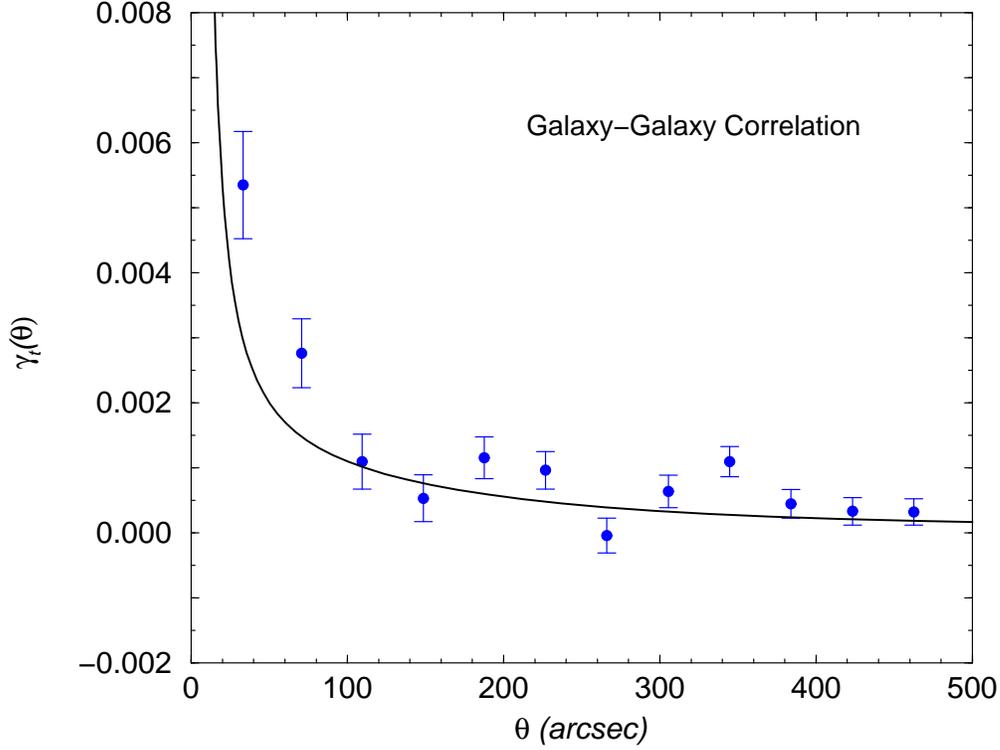,width=5.2in,angle=-90}}
\caption{The SDSS galaxy-mass
cross-correlation using galaxy-shear correlation function. We show the halo
model prediction with a solid line. The data
are from \cite{Fisetal00}.}
\label{fig:sloanshear}
\end{figure}

Our description for the galaxy power spectrum, see
\S~\ref{sec:galaxy}, allows us to extend the discussion to also consider 
the cross-correlation between galaxies and mass. 
Such a cross-power spectrum can be probed through two independent methods:
(1) the weak lensing tangential shear-galaxy correlation function
and (2) the foreground-background source correlation function.
As we find later, these two correlations probe different scales in the
galaxy-mass power spectrum. 

\subsection{Shear-Galaxy correlation}

The shear-galaxy correlation function can be
constructed by correlating tangential shear of background galaxies 
surrounding foreground galaxies. The assumption is that these foreground 
galaxies trace the mass distribution  along the line of sight to 
background sources.  Here, observations
involve the mean tangential shear due to gravitational lensing 
which is related to convergence through
\begin{equation}
\left< \gamma_t(\theta) \right> = -\frac{1}{2} \frac{d
\bar{\kappa}(\theta)}{d {\rm ln} \theta}  \, ,
\label{eqn:tangential}
\end{equation}
where $\bar{\kappa}(\theta)$ is the mean convergence within a circular
radius of $\theta$ \cite{KaiSqu93,SquKai96,GuzSel00}.

Since the shear, averaged over a circular aperture, is
correlated with foreground galaxy positions, one essentially probes the
galaxy-mass correlation discussed in \S~\ref{sec:galaxy-mass} such that
\begin{equation}
\bar{\kappa}(\theta) = \int d\rad W^\lens(\rad)W^\gal(\rad)
\int dk k P_{\gal-\rm DM}(k) \frac{2 J_1(k d_A \theta) }{k d_A \theta} \,  
.
\end{equation}
Following equation~(\ref{eqn:tangential}), we can write the mean 
tangential shear involved with galaxy-galaxy lensing as
\begin{equation}
\left<\gamma_t(\theta)\right> = \int d\rad W^\lens(\rad)W^\gal(\rad)
\int dk k P_{\gal-\rm DM}(k) J_2(k d_A \theta) \, .
\label{eqn:tanshear}
\end{equation}
Here, $W^\lens$ is the lensing window function introduced in 
equation~(\ref{eqn:weight}), while $W^\gal$ is the normalized redshift
distribution of foreground galaxies. Note that, in general, $W^\lens$
involves the redshift distribution of background sources beyond the
simple single source redshift assumption we have considered in prior 
calculations. 

The highest  
signal-to-noise measurement yet of tangential lensing correlation around 
foreground galaxies  comes from the Sloan Digital Sky Survey \cite{Fisetal00}.
We compare these measurements with
a prediction based on the halo model in figure~\ref{fig:sloanshear}. 
Here, for simplicity, we take the same description for galaxy number counts as introduced in 
\S~\ref{sec:galaxy}, and calculate the galaxy-dark matter correlation function
following equations~(\ref{eqn:galdm1h}). In calculating the expected
correlation function, we have used the expected redshift distributions for foreground and background galaxies in the Sloan samples.
The observed measurements shown in  figure~\ref{fig:sloanshear}
comes from the Sloan survey for field galaxies;  
tangential shear around a  selected sample of 42 foreground galaxy clusters in Sloan
data were recently presented by \cite{Sheletal01}. Traditionally, the 
galaxy-galaxy lensing correlation function, similar to the above, was interpreted by a mass and a size distribution for foreground galaxies with foreground galaxies  generally assumed to be
distributed randomly. This, or similar approaches, allow constraints on certain galaxy properties such as mass  and size (see, \cite{Fisetal00} for details).  

The halo model provides an alternative, and perhaps an improved,
 description consistent with our basic ideas of large scale 
structure: since galaxies effectively trace the dark matter halos and it 
is the dark matter that is mostly responsible for the tangential lensing 
of background sources, the constraints on mass and size effectively 
applies to halos that galaxies reside in. If field galaxies are simply selected as foreground sources, then, the constraint on mass and size applies to the dark matter halo of the sample, each of which contains a single galaxy. If the foreground sample contains contributions from a wide variety of dark matter halo mass scales, then more than one galaxy can reside in dark matter halos at the high mass end and a simple interpretation may not be possible. Additionally, since halos distribute the large scale structure, one should account for the clustering component, i.e.,. the 2-halo term of the dark matter-galaxy correlation function,
 when extracting statistical properties related to individual halos that contribute along the line of sight. As shown in figure~\ref{fig:sloanshear}, the total halo model prediction, 
both due to individual halos and their clustering, is consistent with 
observed measurements; the correlation at largest angular scales is due to the intrinsic clustering of halos and cannot be simply interpreted as a large extent for the dark matter halos.
 A more thorough study of the weak lensing
shear-galaxy cross-correlation, under the halo model, is available in
\cite{GuzSel00}  and we refer the reader to this paper for
further details.

\begin{figure*}[t]
\centerline{\psfig{file=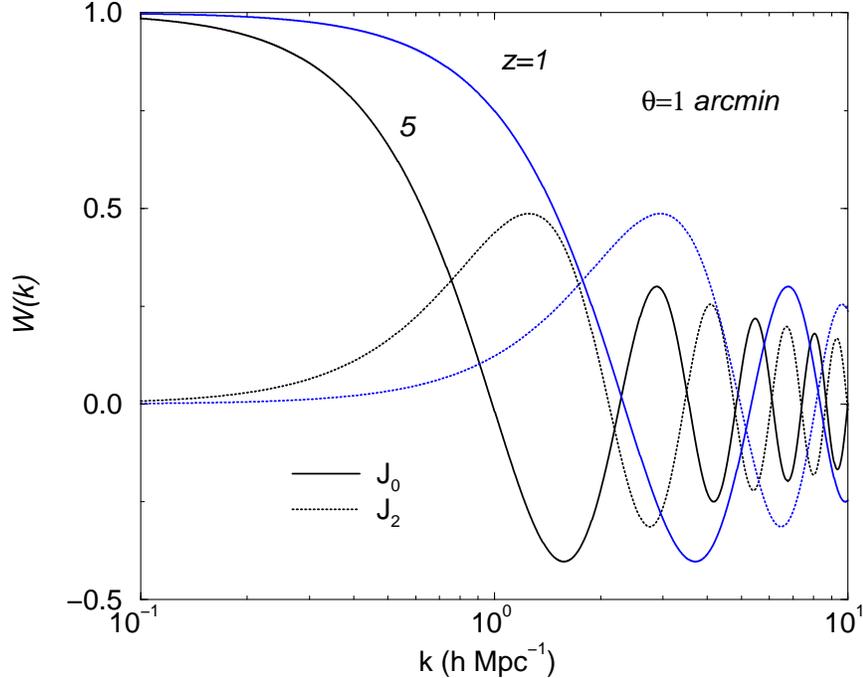,width=4.5in,angle=-90}}
\caption{The window functions involved with the projection of
galaxy-mass power spectrum in producing the tangential shear-galaxy
correlation ($J_2$) and the foreground-background galaxy
correlation ($J_0$). Note that the tangential shear-galaxy correlation
probes smaller physical scales in the galaxy-mass power spectrum and are, thus,
more sensitive to the non-linear aspect of this correlation function,
such as the single-halo contribution.}
\label{fig:j0j2}
\end{figure*}

\subsection{Foreground-background source correlation}

The second observational probe of the galaxy-mass correlation function 
comes from the clustering of background sources around foreground objects.
One can construct a power spectrum by simply counting  the number of objects, 
such as quasars or X-ray sources,
surrounding a sample of foreground sources, such as galaxies.
The dependence on the correlation comes from the fact that
foreground sources trace the mass density field which can potentially
affect the number counts of background sources
by the weak lensing effect.

To understand this correlation, we can consider a sample of background
sources whose number counts can be written as
\begin{equation}
N(s) = N_0 s^{-\alpha} \,
\end{equation}
where $s$ is the flux and $\alpha$ is the slope of number counts\footnote{Similarly, we can describe this calculation with counts based on magnitudes instead of flux. In that case, one should replace $\alpha$ with $2.5\alpha_m$ where 
$\alpha_m = d log N(m)/dm$; the logarithmic slope of the magnitude counts}.
Due to lensing, when the amplification involved is $\mu$,
one probes to a lower flux limit $s/\mu$ while the total 
number of sources are reduced by another factor $\mu$; the latter
results from the decrease in volume such that the total surface brightness
is conserved in lensing. Thus, in the presence of lensing, 
number counts are changed to
\begin{eqnarray}
N(s) &=& \frac{N_0}{\mu} \left(\frac{s}{\mu}\right)^{-\alpha} \nonumber \\
&=& N_0 s^{-\alpha} \mu^{\alpha-1} \, .
\end{eqnarray}
In the limit of weak lensing, as more appropriate for the large scale
structure, $\mu \approx (1+2\kappa)$ where convergence $\kappa$ was defined in
equation~(\ref{eqn:kappa}).
This allows us to write the fluctuations in background number counts, 
$N_b(\bn) = \bar{N}_b[1+\delta N_b(\bn)]$, in the
presence of foreground lensing as \cite{MoeJai98,Moeetal97}
\begin{eqnarray}
\delta N_b(\bn) &=& 2(\alpha-1)\kappa(\bn) \nonumber \\
&=& 2(\alpha-1) 
\int_0^{\rad_0} d\rad W^\lens(r)\delta(\bn,\rad) \, ,
\label{eqn:background}
\end{eqnarray}
where the lensing weight function integrates over the
background source population following equation~(\ref{eqn:weight}).

The foreground sources are assumed to trace the density field and based
on the source clustering, one can write the fluctuations in the
foreground source population, $N_f(\bn) = \bar{N}_f[1+\delta N_f(\bn)]$,
as
\begin{equation}
N_f(\bn) = \int_0^{\rad_0} d\rad n_f(r) \delta_g(\bn,\rad) \,
\label{eqn:foreground}
\end{equation}
where $n_f(\rad)$ is the radial distribution of foreground sources.

We can write the correlation between the foreground and
background sources as 
\begin{eqnarray}
w_{fb}(\theta) &=& \langle N_f(\alpha) N_b(\alpha+\theta) \rangle \nonumber \\
&=& 2(\alpha-1) \int_0^{\rad_0} n_f(\rad) W^\lens(\rad)
	\int_0^\infty \frac{k dk}{2 \pi} P_{\gal-DM}(k) J_0(kd_A\theta) \, ,
\label{eqn:b-f}
\end{eqnarray}				
where we have simplified using the Fourier expansion of 
equations~(\ref{eqn:background}) and (\ref{eqn:foreground}), and have
introduced the galaxy-mass power spectrum.

In the case where foreground and background sources are
not distinctively separated in radial distance, note that there may be
an additional correlation resulting from the fact that background
sources trace the same overlapping density field traced by the foreground 
sources. This leads to a clustering term where
\begin{eqnarray}
w_{fb}^{\rm overlap}(\theta) &=& \langle N_f(\alpha) N_b(\alpha+\theta) \rangle \nonumber \\
&=& \int_0^{\rad_0} n_f(\rad) n_b(\rad)
	\int_0^\infty \frac{k dk}{2 \pi} P_{\gal-source}(k) J_0(kd_A\theta) \, ,
\end{eqnarray}				
where $P_{\gal-source}(k)$ is now the cross power spectrum between foreground
source sample, galaxies in this case, and the population of background sources,
such as quasars. This cross power spectrum can be modeled under the halo
approach by introducing a relationship between how background sources
populate dark matter halos similar to the description for galaxies.
This clustering component usually becomes a source of contamination for the
detection of background source-foreground galaxy correlation due to weak 
lensing alone.

\begin{figure}[t]
\centerline{\psfig{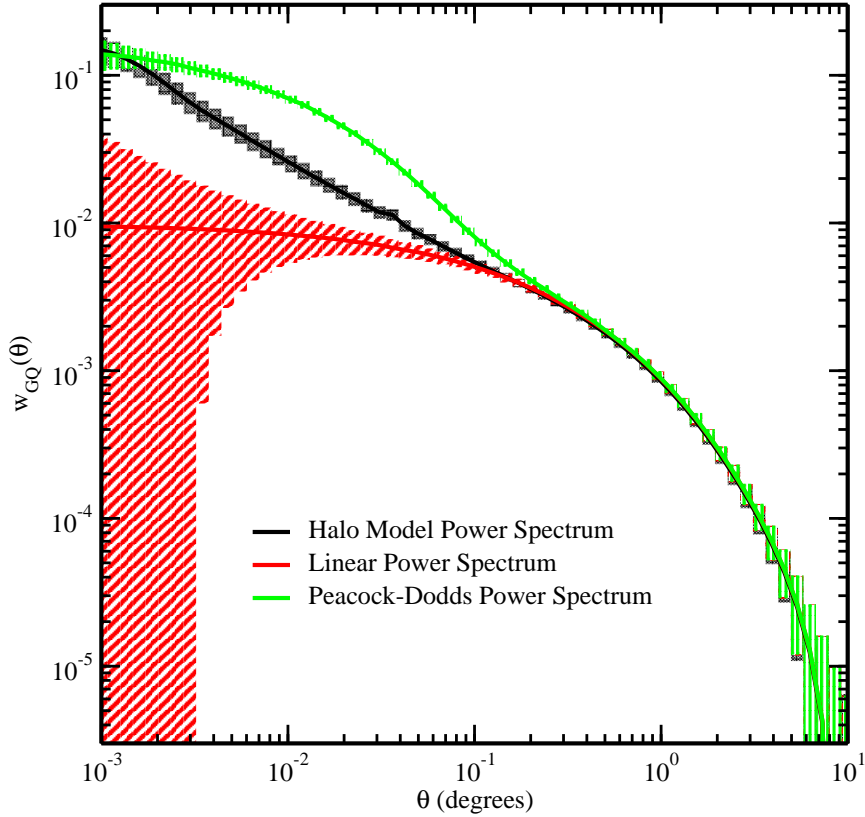}}
\caption{The expected foreground galaxy-background quasar correlation due to
lensing magnification under several descriptions of the galaxy-mass power
spectrum. The expected error bars are for the whole Sloan catalog of galaxies
with $21 < r' < 22$, as foreground sources, and Sloan quasars at redshifts
greater than 1, as background sources. The figure is from R. Scranton (in preparation).}
\label{fig:sdssgalqso}
\end{figure}

Note that background-foreground source correlation, equation~(\ref{eqn:b-f}),
and the tangential shear-foreground galaxy correlation, 
equation~(\ref{eqn:tanshear}), weigh the galaxy-mass cross-power spectrum with
two different window functions involving a $J_0$ and a $J_2$, respectively.
For a given projected distance $d_A\theta$,  the two observational
methods probe the galaxy-mass power spectrum at different scales.
As shown in figure~\ref{fig:j0j2}, the tangential shear-foreground
galaxy correlation function probes the non-linear scales of the
galaxy-mass correlation and, thus, more sensitive to the behavior of the single-halo contribution than the foreground-background correlation function of
sources. The dependence of the non-linear scales in the
shear-galaxy correlation suggests that it is more suitable to
probe the physical aspects of how foreground galaxies trace their dark
matter halos. On the other hand, the foreground-background source
correlation function probes the clustering aspects of
foreground sources that trace the linear density field. 

In figure~\ref{fig:sdssgalqso}, we show the expected correlation between
foreground galaxies in the Sloan Digital Sky Survey  and background quasars
at redshifts greater than 1.
The expected errors suggest that the correlation
will be measured out to angular scales of several degrees. 
Since sufficient statistics will
soon be available, the catalog can be divided in to redshift bins and be
combined with associated data on the shear-galaxy correlation for
detailed studies on galaxy-mass cross clustering.

\section{Halo applications to CMB: Secondary effects}

The angular power spectrum of 
cosmic microwave background (CMB) temperature fluctuations is now a well 
known probe of cosmology. The anisotropies can
be well described through linear physics involving Compton scattering and linearized
general relativity. The well known features in the power spectrum,
the acoustic oscillations at large angular scales 
and the damping tail  at medium angular scales
\cite{PeeYu70,SunZel70,Sil68,Huetal97}, allow the ability to constrain
most, or certain combinations of, parameters that define the currently favored
CDM models with a cosmological constant 
\cite{Kno95,Junetal95,Bonetal97,Zaletal97,Eisetal99}. This has led to a wide number of 
experimental attempts
with results so far suggesting the evidence for
acoustic peaks as expected in models with adiabatic initial
conditions and a scale invariant power spectrum of fluctuations
 \cite{Miletal99,deBetal00,Hanetal00,Haletal01}.

The small angular  scale temperature anisotropies
contain a wide-variety of information related to the growth 
and evolution of large scale structure including non-linear aspects of clustering. Such a
contribution from the low redshifts is partly contrary to
the general assumption that CMB fluctuations are solely described by
linear physics at the last scattering at a redshift $\sim$ 1100. 
There are two methods by which large scale structure of the local universe 
can modify CMB temperature:
gravity and scattering. The gravitational contributions arise from
variations in the frequency of CMB photons 
via gravitational redshifts and blueshifts 
\cite{SacWol67,ReeSci68,Sel96a,Coo01c,Dabetal00,Lasetal99} and
via deflection 
\cite{BlaSch87,Kas88,Lin88,ColEfs89,Sas89,WatTom91,Fuketal92,Cayetal93,Sel96b,Hu00b}
and time-delay \cite{HuCoo00}
effects on CMB photons due to gravitational lensing. 
In the reionized epoch, with a population of free electrons, the CMB photons can also be 
Compton-scattered \cite{OstVis86,Vis87,Kai84b,DodJub95,Efs88,JafKam98,Huetal95}.

The large scale structure contributions to CMB, either due to gravity or 
scattering, can be modeled using the halo 
approach and their statistical properties can be calculated in detail, similar to the application of the halo models to galaxy and weak lensing statistics. Here, we will consider several such secondary 
contributions including the thermal and kinetic
Sunyaev-Zel'dovich (SZ; \cite{SunZel80}) effects, the gravitational 
lensing modification to CMB, and the non-linear contribution to the
integrated Sachs-Wolfe effect (ISW; 
\cite{SacWol67}) at small angular scales. 

The anisotropy power spectrum at
 small angular scales has recently  become the focus of several 
theoretical and experimental  studies. 
Though  there are several upper-limits and an initial detection of 
 anisotropy power at small scales \cite{Dawetal00,Holetal00,Subetal00,Chuetal97}, a 
wide-field CMB image is yet to be produced with resolution necessary for 
studies related to secondary effects. To this end,
several experiments are now working towards obtaining such information either from
direct imaging or interferometric techniques. These experimental attempts include the
proposed 12 deg.$^2$ survey by \cite{Caretal96} at the combined and expanded BIMA and OVRO arrays (CARMA), the  Atacama Telescope (ACT; Lyman Page, private
communication), and the BOLOCAM array on the Caltech Submillimeter
Observatory (Andrew Lange, private communication). In the 
longer term, a few thousand sqr. degrees is proposed to be imaged in a few years with a wide-field
bolometer array at the South Pole Telescope (John Carlstrom, private communication) and
the Planck surveyor will allow detailed studies of certain secondary effects and foreground
via multi-frequency all-sky maps.
 
\subsection{The Thermal SZ effect}

The SZ thermal effect arises from the  inverse-Compton scattering of CMB
photons by hot electrons along the line of sight. This effect has now been 
directly imaged
towards massive galaxy clusters (e.g., \cite{Caretal96,Jonetal93}), 
where temperature of the scattering medium can reach as high as
10 keV producing temperature changes in the CMB of order 1 mK at
Rayleigh-Jeans wavelengths. The SZ effect is now well known for its main 
cosmological application involving measurements of the Hubble constant. 
The basic idea follows from the initial suggestions by Gunn \cite{Gun78} 
and Silk \& White \cite{SilWhi78}: the SZ temperature decrement, $\Delta T 
\propto T_e n_e dl$, towards a given cluster can be combined with thermal 
Bremsstrahlung X-ray emission, $S_x \propto T_e^{1/2} n_e^2 dl$, towards 
the same cluster to obtain an estimate of the line of sight distance 
through the cluster: $L \propto S_x/\Delta T^2$. This requires a 
measurement of $T_e(r)$ across the cluster; the isothermal assumption 
$T_e(r)=T_0$ is generally employed due to limitations on the observational 
front. A comparison of this distance to the projected separation of the 
cluster across the sky determines the angular diameter distance to the 
cluster, independent of cosmological distance ladder (see,
\cite{Masetal01,Mauetal00,Patetal00,Greetal00,Reeetal00,HugBir98} for recent
$H_0$ measurements). Through a 
cosmological model for the distance, one can extract parameters such as 
the Hubble constant and with measurements over a wide range in redshift, 
values for the matter density and the cosmological constant.

There are several limitations that prohibit a reliable measurement of the Hubble constant from the combined SZ and X-ray data, at least in the case of a single cluster. The usual spherical assumption for clusters are inconsistent with observations and can bias the
distance measurement at the level of 10\% to 20\% \cite{Coo00a,Sul99,Puy00}. 
The isothermal assumption for electron temperature has been shown to be inconsistent with numerically simulated galaxy cluster gas distributions, though, this assumption is yet to be tested with observations of galaxy clusters.
In the case of clusters with significant cooling flows, it is clear that a single temperature cannot be used to describe the electron temperature; this again leads to biases at the few tens of percent level \cite{Rotetal97,MajNat00}.
Additional contributions at the 10\% level and less include, the presence of contaminating radio point sources, either in the cluster \cite{Cooetal98} or background sources gravitationally lensed by the cluster potential \cite{LoeRef97}, fluctuations in the background anisotropies lensed through the cluster
\cite{Cen98} and the peculiar velocity contribution to the kinetic SZ effect. 
Though, in general, these effects limit the
reliability of the Hubble constant measured towards a single cluster, a significant sample of clusters is expected to produce a measurement that is within the few percent level.
In the case of projection effects involving ellipsoidal clusters, distributed following ellipticities observed for present-day cluster samples, it can be shown that for a sample of at least 25 or more clusters, the mean Hubble constant is consistent with the true value \cite{Coo00a}.

\begin{figure}[t]
\centerline{\psfig{file=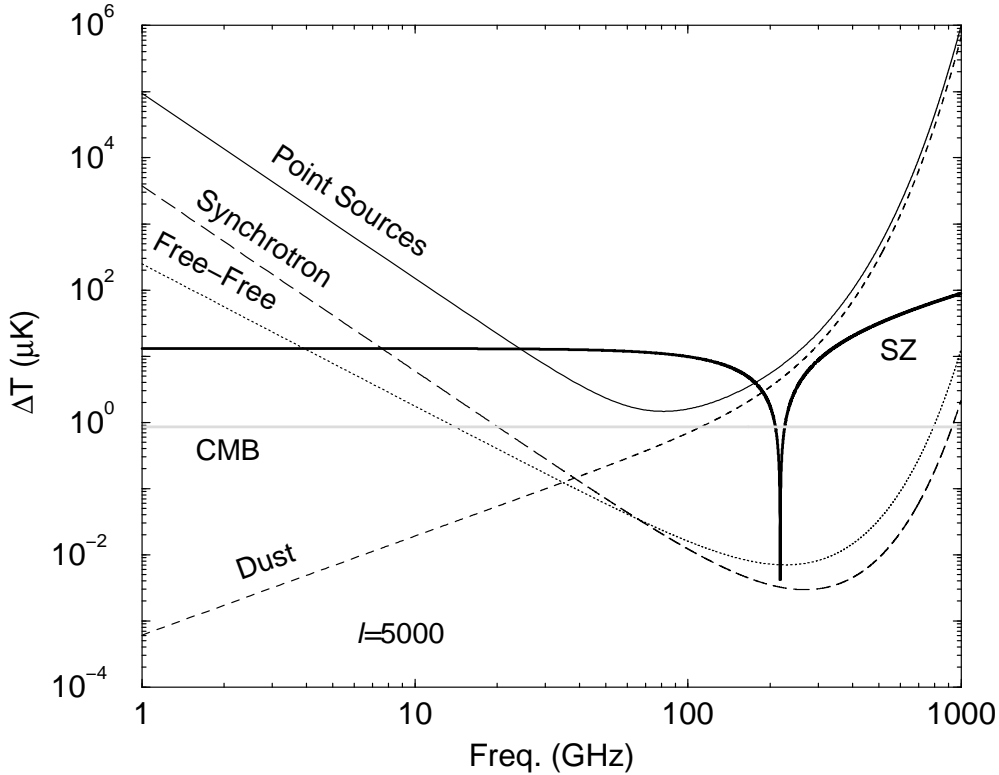,width=5.2in,angle=-90}}
\caption{Frequency dependence
of the SZ effect at a multipole of $l \sim 5000$. Here, we show the
absolute value of temperature relative to the thermal CMB spectrum. 
For comparison, we also
show the temperature fluctuations due to point sources (both radio at
low frequencies and far-infrared sources at high frequencies; solid
line), galactic synchrotron (long dashed line), galactic free-free
(dotted line) and galactic dust (short dashed line). At small angular
scales, frequencies around 50 to 100 GHz is ideal for a SZ
experiment.}
\label{fig:szfreq}
\end{figure}

\begin{figure}[t]
\centerline{
\psfig{file=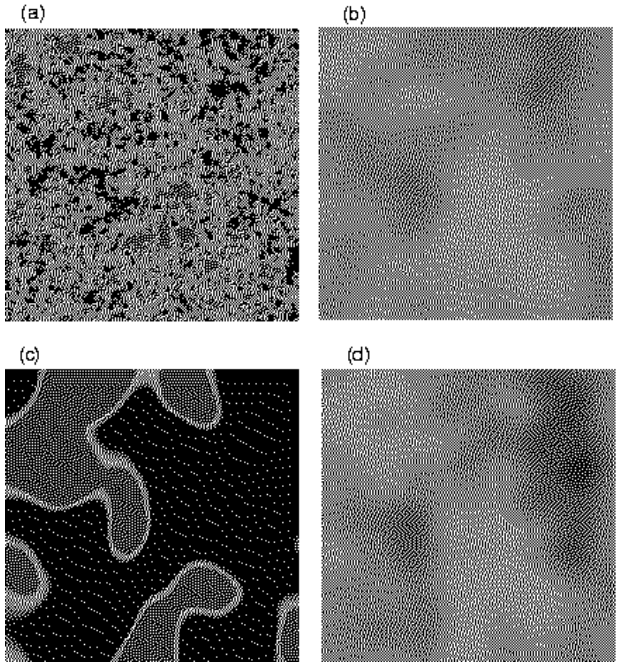,width=4.5in}}
\caption{Recovery of the SZ signal with Planck multifrequency data: (a) A 
line of sight integrated model SZ map with the assumption that pressure 
traces dark matter with a scale independent bias at all scales, (b) The 
map smoothed at 20$'$, (c) this SZ signal+noise from primary anisotropies 
and foregrounds, and (d) final recovered map with a SZ frequency 
spectrum. For Planck, the recovered spectrum is consistent with the input 
spectrum and allows a determination of the SZ power spectrum with a 
cumulative signal-to-noise greater than 100 \cite{Cooetal00a}.}
\label{fig:szmaps}
\end{figure}

In the future when wide-field SZ surveys are available, we are more interested in the statistics of SZ effect, such as the SZ correlation function or power spectrum in real space. Since on top of the SZ effect, one also gets a contribution from the CMB anisotropy fluctuations, it is clear that one requires reliable ways to separate them and also contaminant foregrounds such as radio point sources and galactic dust.
Due to the nature of inverse-Compton scattering, where photons are 
upscattered from low to high frequencies, the SZ effect, fortunately, bears a 
spectral  signature that differs from other  temperature fluctuations 
including the dominant CMB primary component (see, 
figure~\ref{fig:szfreq}). 
 In upcoming multifrequency CMB data, thus, the SZ contribution can be  
separated  using its frequency dependence. This allows statistics related 
to the SZ effect be studied independently of, say, dominant CMB 
temperature fluctuations. As 
discussed in detail in \cite{Cooetal00a}, a multi-frequency approach can 
easily be applied to Planck 
surveyor\footnote{http://astro.estec.esa.nl/Planck/; also, ESA
D/SCI(6)3.} missions (see, figure~\ref{fig:szmaps}).

\subsubsection{SZ Power Spectrum}
\label{sec:szpower}

The temperature decrement along the line of sight  due to SZ effect
can be written as the integral of pressure along the same line of
sight
\begin{equation}
y\equiv\frac{\Delta T}{T_{\rm CMB}} = g(x) \int  d\rad  a(\rad)
\frac{k_B \sigma_T}{m_e c^2} n_e(\rad) T_e(\rad) \,
\end{equation}
where $\sigma_T$ is the Thomson cross-section, $k_B$ is the Boltzmann's 
constant,  $n_e$ is the electron 
number density, $\rad$ is the comoving distance, and $g(x)=x{\rm
coth}(x/2) -4$ with $x=h \nu/k_B
T_{\rm CMB}$ is the spectral shape of the SZ effect. At Rayleigh-Jeans
(RJ) part of the CMB, $g(x)=-2$.
For the rest of this paper, we assume observations in the
Rayleigh-Jeans
regime of the spectrum, though, an experiment such as Planck with 
sensitivity
beyond the peak of the spectrum can separate out these contributions
based on the spectral signature, $g(x)$ \cite{Cooetal00a} (see also,
\cite{Tegetal99,Hobetal98} for frequency separation of CMB from 
foregrounds).

The SZ power spectrum, bispectrum and trispectrum are
defined in the flat sky approximation in the usual way
\begin{eqnarray}
\left< y(\bfl_1)y(\bfl_2)\right> &=&
        (2\pi)^2 \delta_\dirac(\bfl_{12}) C_l^\sz\,,\nonumber\\
\left< y(\bfl_1) y(\bfl_2) 
       y(\bfl_3)\right>_c &=& (2\pi)^2 \delta_\dirac(\bfl_{123})
        B^\sz(\bfl_1,\bfl_2,\bfl_3)\,, \nonumber \\
\left< y(\bfl_1) \ldots
       y(\bfl_4)\right>_c &=& (2\pi)^2 \delta_\dirac(\bfl_{1234})
        T^\sz(\bfl_1,\bfl_2,\bfl_3,\bfl_4)\,.
\end{eqnarray}
These can be written as a redshift projection of the
pressure power spectrum, bispectrum and trispectrum, respectively:
\begin{eqnarray}
C_l^\sz &=& \int d\rad \frac{W^\sz(\rad)^2}{d_A^2}
P_{\Pi}\left(\frac{l}{d_A},\rad\right) \, , \\
B^\sz  &=& \int d\rad \frac{W^\sz(\rad)^3}{d_A^4} B_\Pi\left(
\frac{\bfl_1}{d_A},
\frac{\bfl_2}{d_A},
\frac{\bfl_3}{d_A},
;\rad\right) \, , \nonumber \\
T^\sz  &=& \int d\rad \frac{W^\sz(\rad)^4}{d_A^6} T_\Pi\left(
\frac{\bfl_1}{d_A},
\frac{\bfl_2}{d_A},
\frac{\bfl_3}{d_A},
\frac{\bfl_4}{d_A},
;\rad\right) \, .
\label{eqn:szpower}
\end{eqnarray}
Here, $d_A$ is the angular diameter distance. At RJ part of the
frequency spectrum,  the SZ weight function is
\begin{equation}
W^\sz(\rad) = -2 \frac{k_B \sigma_T \bar{n}_e}{a(\rad)^2 m_e c^2}
\end{equation}
where $\bar{n}_e$ is the mean electron density today. In deriving
equation~(\ref{eqn:szpower}),
we have used the Limber approximation \cite{Lim54} by setting
$k = l/d_A$ and flat-sky approximation. Here, we have written the 
correlations in terms of the large scale structure pressure, denoted by 
$\Pi$, power spectrum, bispectrum and trispectrum.

The halo approach has been widely utilized to make analytical predictions 
on the statistics related to SZ thermal effect from the large scale 
structure such as the power spectrum (e.g., \cite{ColKai88,KomKit99}). 
Other approaches include a biased description of the pressure
power spectrum with respect to the dark matter density field (e.g., 
\cite{Peretal95,Cooetal00a}). 
These analytical calculations are now fully complemented by numerical 
simulations
(e.g., \cite{daS99,Refetal99,Seletal01,Spretal00}) which are now beginning 
to test the assumptions related to the halo based calculations.
So far, comparisons between numerical simulations 
and the halo approach suggest significant agreement better than comparisons involving
dark matter alone \cite{RefTey01}. We will discuss reasons for this below.

First, we will describe the halo based approach to SZ statistics 
by introducing the clustering of large scale structure pressure. This is 
similar to the dark matter power spectrum and its projection along the 
line of sight that leads to weak lensing convergence power spectrum: the 
line of sight projections of the large scale structure pressure leads to 
the SZ effect.

\subsubsection{Clustering Properties of Large Scale Structure Pressure}
\label{sec:pressure}

In order to describe the large scale structure pressure, we make use of the 
hydrostatic equilibrium between the gas and the dark matter  
distributions within halos. 
 The hydrostatic assumption is
supported by various observations of galaxy clusters,
where the existence of regularity relations, such as the size-temperature 
relation \cite{MohEvr97}, between physical properties of dark matter and
baryon distributions suggest simple physical relations between the two
properties.

\begin{figure*}[t]
\centerline{\psfig{file=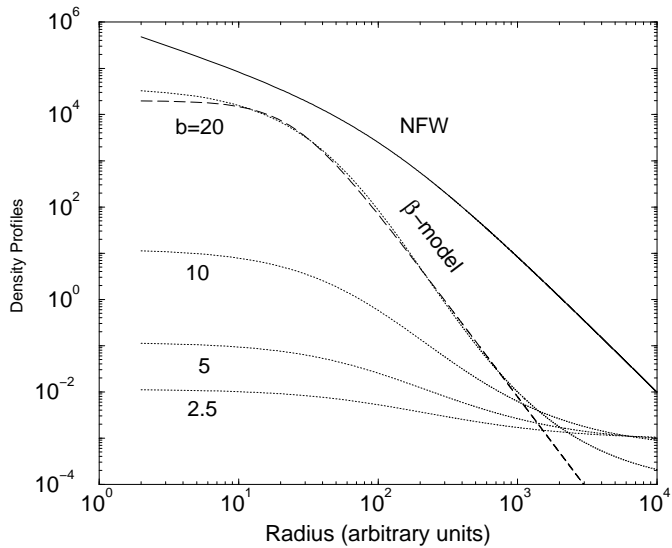,width=3.5in,angle=-90}}
\caption{The dark matter (NFW)
profile and the ones predicted
by the hydrostatic equilibrium for gas, as a function of the $b$
parameter (see equation~\ref{eqn:b}) with $r_s=100$. The relative
normalization between individual parameters is set using a  gas
fraction value of 0.1, though the NFW profile is arbitrarily
normalized with $\rho_s=1$; the gas profiles scale with the same
factor. For comparison, we
also show a typical example of the so-called $\beta$ model
$(1+r^2/r_c^2)^{-3\beta/2}$
which is generally used as a fitting function for X-ray and SZ
observations of clusters. We refer the reader to \cite{Maketal98}
and \cite{Sutetal98} for a detailed comparison of
$\beta$ models and the NFW-gas profiles.}
\label{fig:profiles}
\end{figure*}

The hydrostatic equilibrium for gas with pressure $P$ and density $\rho_g$ 
\begin{equation}
\rho_g^{-1} \frac{dP}{dr} = -\frac{GM_\delta(r)}{r^2} \, 
\end{equation}
can be simplified in the limit gas is ideal, $P=\frac{k_BT_e}{\mu 
m_p}\rho_g$,
and isothermal to obtain
\begin{equation}
\frac{k_BT_e}{\mu m_p} \frac{d\log \rho_g}{dr} = -
\frac{GM_\delta(r)}{r^2} \, ,
\end{equation}
where $\mu=0.59$, corresponding to a hydrogen mass
fraction of 76\%. 
Here, now the $M_\delta(r)$ is the dark matter mass only out to a radius of 
$r$.
Using a NFW profile for dark matter distribution, we can
 analytically calculate the baryon density profile $\rho_g(r)$
\begin{equation}
\rho_g(r) = \rho_{g0} e^{-b} \left(1+\frac{r}{r_s}\right)^{br_s/r} \,
,
\label{eqn:gasprofile}
\end{equation}
where $b$ is a constant, for a given mass
\cite{Maketal98,Sutetal98}:
\begin{equation}
b = \frac{4 \pi G \mu m_p \rho_s r_s^2}{k_B T_e} \, .
\label{eqn:b}
\end{equation}
The normalization, $\rho_{go}$, can be set 
to obtain a constant gas mass fraction for halos comparable with the 
universal
baryon to dark matter ratio: $f_g \equiv M_g/M_\delta
=\Omega_b/\Omega_m$. The total gas mass present in a dark matter halo 
within the virial radius, $r_v$, is
\begin{equation}
M_g(r_v) = 4 \pi \rho_{g0} e^{-b} r_s^3 \int_0^{c} dx \, x^2
(1+x)^{b/x} \, .
\label{eqn:gasmass}
\end{equation}

The electron temperature can be calculated based on the virial
theorem or similar arguments as discussed in \cite{Coo00b}.
Using the virial theorem, we can write
\begin{equation}
k_B T_e = \frac{\gamma G \mu m_p M_\delta}{3 r_v} \, ,
\end{equation}
with $\gamma=3/2$. Since $r_v \propto M_\delta^{1/3}(1+z)^{-1}$ in
physical coordinates, $T_e \propto
M^{2/3}(1+z)$. The average density weighted temperature is
\begin{equation}
\left<T_e\right>_\delta  = \int dM\, \frac{M}{\rho_b}
\frac{dn}{dM}(M,z) T_e(M,z) \, 
.
\label{eqn:etemp}
\end{equation}
In figure, we show the evolution of density weighted temperature from 
\cite{RefTey01}. The results from numerical simulations are well 
reproduced with a Press-Schechter mass distribution for halos. For the 
$\Lambda$CDM cosmology, the halo model predicts a density weighted 
temperature for large scale structure electrons of $\sim$ 0.5 keV today; 
if halos out to a mass of $8 \times 10^{14}$ M$_{\sun}$ only included, this 
mean density weighted temperature decreases to 0.41 keV. 

\begin{figure}[t]
\centerline{\psfig{file=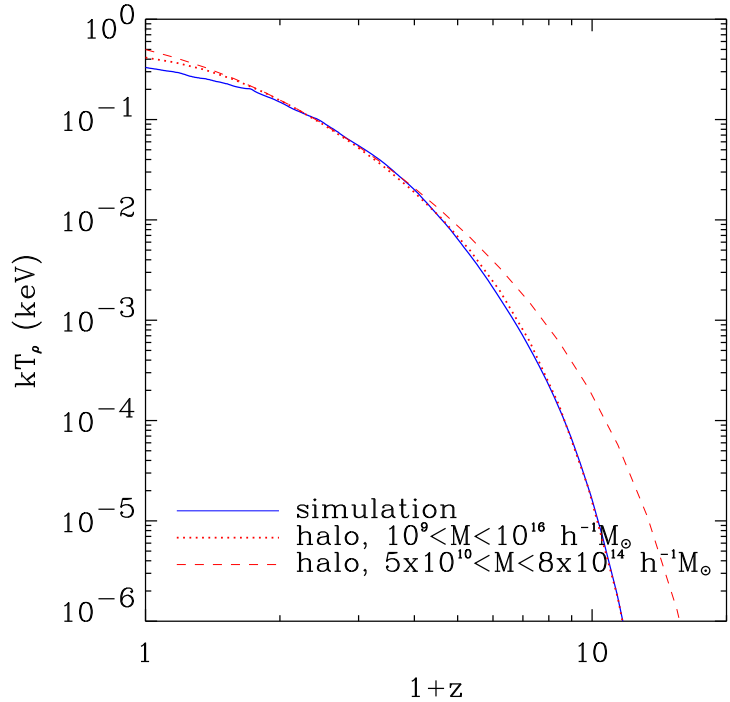,width=4.2in}}
\caption{The
variation in the density weighted temperature  of
electron as a function of redshift. The solid line shows the
redshift evolution of the temperature in hydrodynamical simulations while 
the halo models, with varying halo masses, are shown in dotted and dashed 
lines. The figure is from \cite{RefTey01}.}
\label{fig:etemp}
\end{figure}

In figure~\ref{fig:profiles}, we show the NFW profile for the dark
matter and arbitrarily normalized gas profiles predicted by the
hydrostatic equilibrium and virial theorem for several values of $b$.
As $b$ is decreased, such that the temperature is increased, the turn
over radius of the gas distribution shifts to higher radii. As an
example, we also show the so-called $\beta$ model that is commonly
used to describe X-ray and SZ observations of galaxy clusters and for
the derivation purpose of the Hubble constant by combined SZ/X-ray
data. The $\beta$ model describes the underlying gas distribution
predicted by the gas profile used here in equilibrium with the NFW
profile, though, we find differences especially at the outer most
radii of halos. This difference can be used as a way to establish the
hydrostatic equilibrium of clusters, though, any difference of gas
distribution at the outer radii should be accounted in the context of
possible substructure and mergers.

A discussion on the comparison between the gas profile used here and the 
$\beta$ model is available
in \cite{Maketal98} and \cite{Sutetal98}.
In addition, we refer the reader to \cite{Coo00b} 
for full detailed discussion on issues
related to modeling of pressure power spectrum using halo and
associated systematic errors. Comparisons of the halo model
predictions with numerical simulations are available in
\cite{Seletal01}  and \cite{RefTey01}.

As discussed in \cite{KomSel01}, one can make several improvements to the 
above gas profile. One can constrain the gas distribution such that at 
outer most radii of halos, gas distributions follows that of the dark 
matter. This can be done by setting the slopes of dark matter and gas 
profiles to be the same beyond some radius. If gas is assumed to be in 
hydrostatic equilibrium, a gas profile that traces dark matter produces a 
temperature profile that varies with redshift.
In general, one can obtain consistent solutions by assuming a polytropic 
form for pressure,
$P \propto \rho_g T_e \propto \rho_g^\gamma$. As discussed in 
\cite{KomSel01},
predictions based on this prescription for cluster gas are more 
consistent with observations than the simple description involving an 
isothermal electron distribution

\begin{figure}[t]
\centerline{\psfig{file=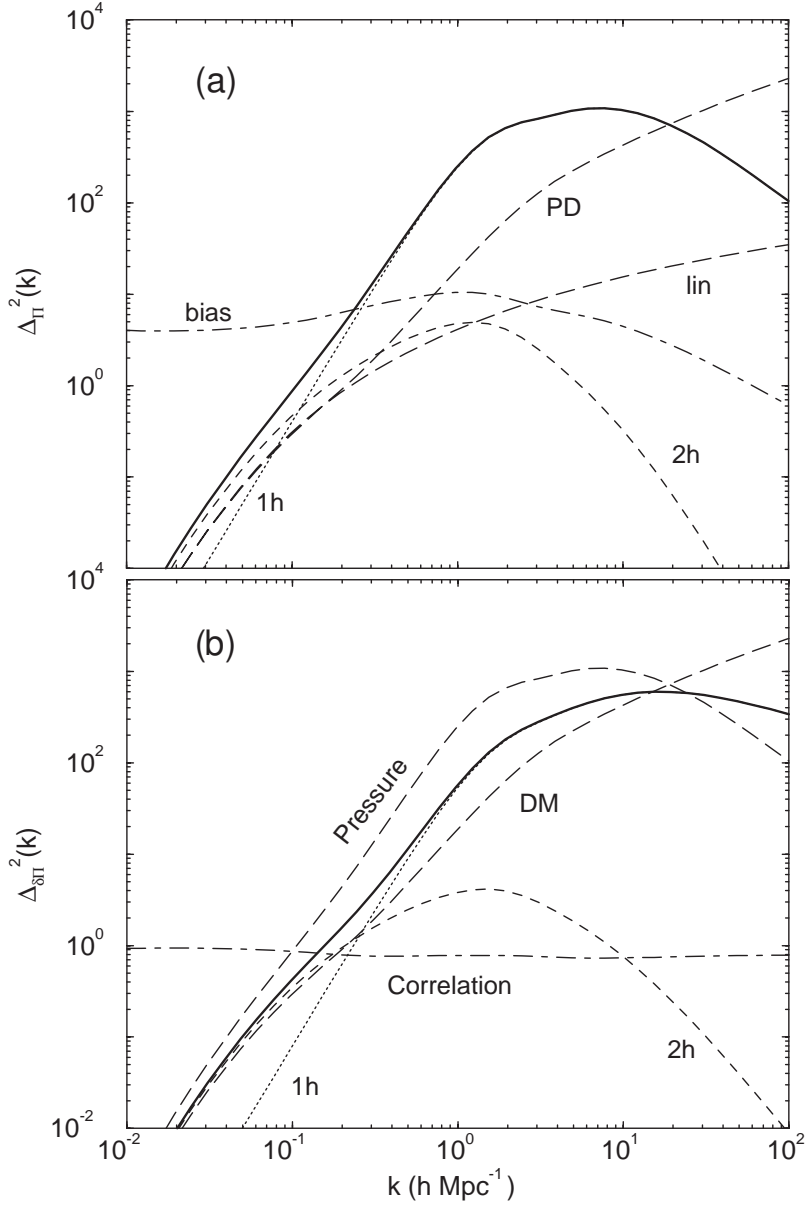,width=4.2in}}
\caption{The (a) pressure and (b) pressure-dark matter cross power 
spectrum  today
broken into individual contributions under the halo description. 
For comparison, we also show the dark matter power spectrum
under the halo model and in (a) pressure bias and in (b) pressure-dark 
matter correlation.}
\label{fig:pressurepower}
\end{figure}

Given a description of the halo electron (or gas) profile and their
temperature distribution, we can write the power spectrum of
large scale structure pressure as
\begin{eqnarray}
P_\Pi(k) &=& P^{1h}(k) +  P^{2h}(k) \,, \\
P^{1h}(k) & = & M^\Pi_{02}(k,k) \,, \\
P^{2h}(k) & = &\left[  M^\Pi_{11}(k) \right]^2 P^\lin(k)\,,
\end{eqnarray}
where the two terms represent contributions from two points in
a single halo (1h) and points in two different halos (2h)
respectively.

Here, we redefine the integral in equation~(\ref{Iij}) for
dark matter to account for pressure as
\begin{eqnarray}
 M^\Pi_{ij}(k_1,\ldots,k_j;z) &\equiv&
 \int dm \left(\frac{M}{\bar{\rho}}\right)^j \tilde \frac{dn}{dm}(m;z) 
b_i(m;z) T_e^\j(m;z) \nonumber \\
&& \quad \quad \times [u_\Pi(k_1|m;z)\ldots u_\Pi(k_j|m;z)] \, ,
 \label{Pij} 
\end{eqnarray}
with the three-dimensional Fourier transform of the gas profile substituted
in equation~(\ref{eqn:yint}) to obtain $u_\Pi(k|m;z)$.
We define the bias and correlation of pressure, relative to dark matter,
as \begin{equation}
{\rm bias}_\Pi(k) = \sqrt{\frac{P_\Pi(k)}{P_\delta(k)}} \, ,
\end{equation}
and
\begin{equation}
r_{ij}(k) = \frac{P_{\Pi-\delta}(k)}{\sqrt{P_\Pi(k)P_\delta(k)}} \, ,
\end{equation}
respectively. Here, $P_\delta$ is the dark matter power spectrum and
$P_{\Pi-\delta}$ is the pressure-dark matter cross power spectrum.
As presented for dark matter, 
we can similarly extend the derivation to calculate pressure bispectrum 
and trispectrum.

In figure~\ref{fig:pressurepower}(a), we show the logarithmic
power spectrum of pressure and dark matter such that
$\Delta^2(k)=k^3 P(k)/2\pi^2$ with
contributions broken down to the $1h$ and $2h$ terms today. 
As shown, the pressure power spectrum depicts an increase in power
relative to the dark matter at scales out to few h
Mpc$^{-1}$, and a decrease thereafter.

\begin{figure}[t]
\centerline{\psfig{file=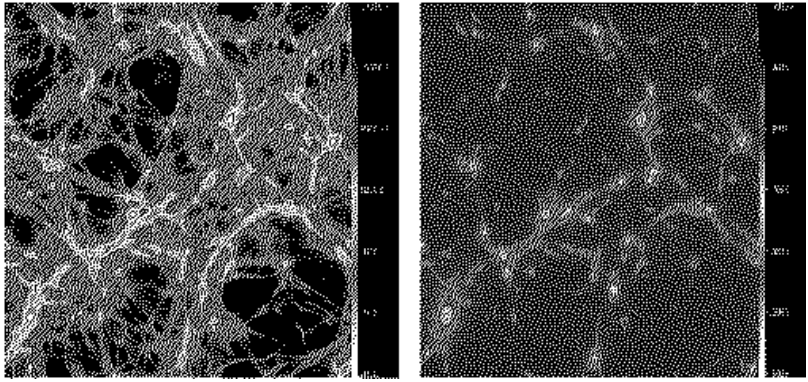,width=4.2in}}
\caption{The baryon density (left) and temperature weighted density, or 
pressure (right), in a time-slice of a hydrodynamical simulation by 
\cite{Seletal01}. As shown, most of the contribution to large scale 
structure pressure comes from massive halos while the baryon density is 
distributed over a wide range of mass scales and trace the filamentarity 
structures defined by the dark matter distribution. The figure is from U. 
Seljak based on simulations by \cite{Seletal01}.}
\label{fig:gaspressure}
\end{figure}

\begin{figure}[t]
\centerline{\psfig{file=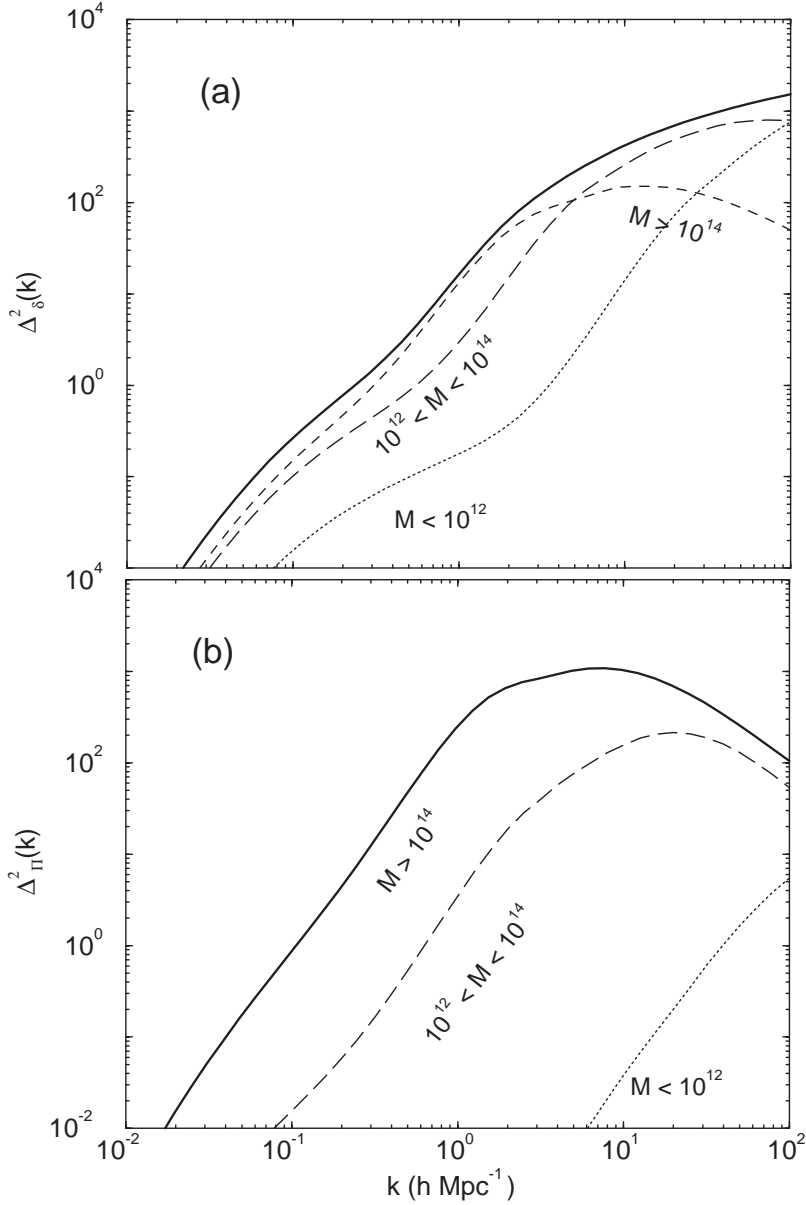,width=4.2in}}
\caption{
The mass dependence on the dark matter power spectrum (a) and
pressure power spectrum (b). Here, we show the total contribution
broken in mass limits as written on the figure.
As shown in (a), the large scale contribution to the dark matter power
comes from massive halos while small mass halos contribute at small
scales. For the pressure, in (b), only massive halos above a mass of
$10^{14}$ M$_{\rm sun}$ contribute to the power.}
\label{fig:powermass}
\end{figure}

 The decrease in power at small scales can be understood through the 
relative contribution to pressure as a function of the halo mass. In 
figure~\ref{fig:powermass}, we break the total dark matter power
spectrum (a) and the total pressure power spectrum (b), to a function of
mass. As shown, contributions to both dark matter and pressure comes from
massive halos at large scales and by small mass halos at small scales.
The pressure power spectrum is such that through temperature weighing,
with $T_e \propto M^{2/3}$ dependence, the contribution from
low mass halos to pressure is suppressed relative to that from the
high mass end. 

In figure~\ref{fig:gaspressure}, we show two images of a time slice through 
numerical simulations by \cite{Seletal01}. The gas, or baryon, density 
distributions is such that it is highly filamentary and traces the large 
scale dark matter distribution. The pressure, however, is confined to 
virialized halos in the intersections between filaments. These are the 
massive clusters in the simulation box: the density weighted temperature, 
or pressure, of large scale structure is clearly dominant in massive 
clusters.  Thus, the pressure power spectrum, at all scales of
interest, can be easily described with halos of mass greater than
$10^{14}$ $M_{\sun}$.
A comparison of the dark matter and pressure
power spectra, as a function of mass, in figure~\ref{fig:powermass} 
reveals that the turn
over in the pressure power spectrum results in an effective scale
radius for halos with mass greater than $10^{14}$ M$_{\sun}$.
We refer the reader to \cite{Coo00b} for  further
details on the pressure power spectrum and its properties.

\begin{figure}[t]
\centerline{\psfig{file=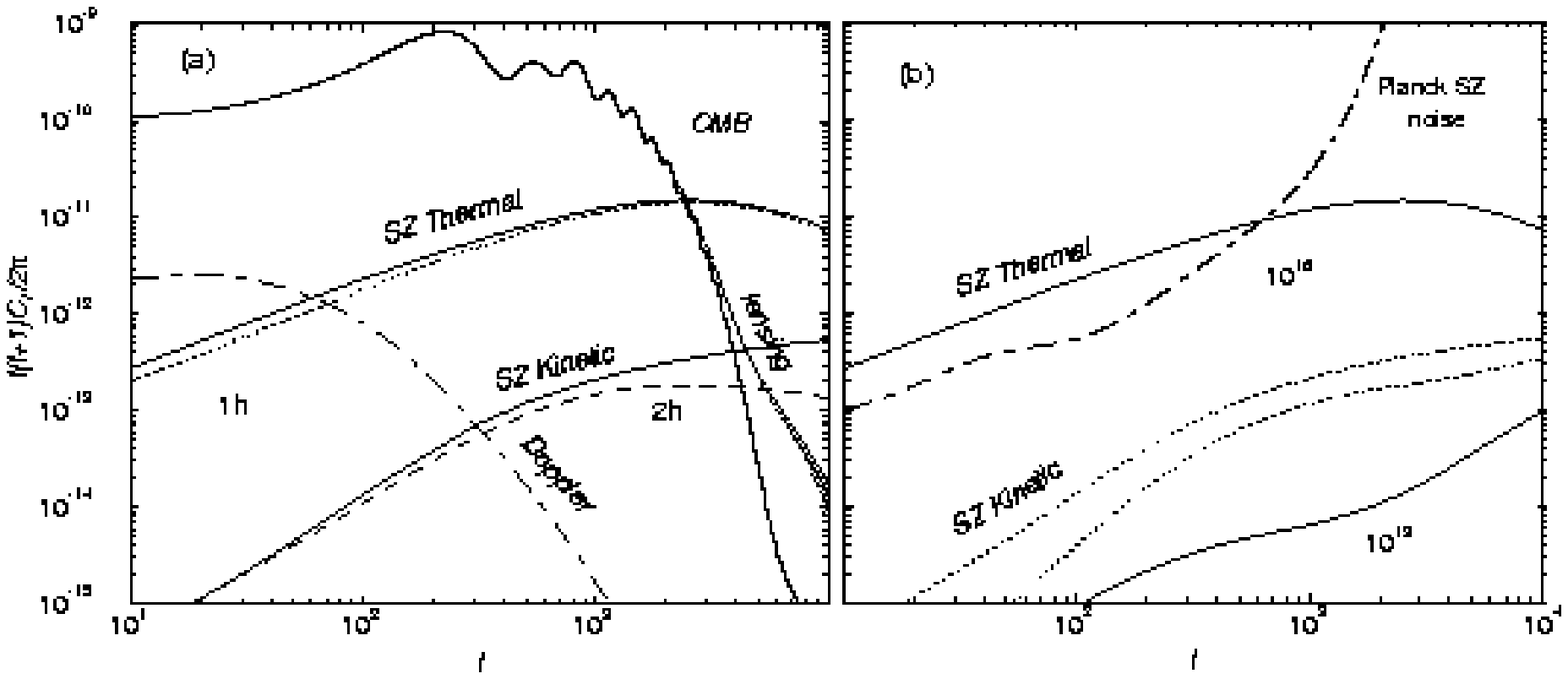,width=6.0in}}
\caption{
The angular power spectra of SZ thermal and kinetic
effects. As shown in (a), the thermal
SZ effect is dominated by individual halos, and thus, by the single
halo term, while the kinetic effect is dominated by the large scale
structure correlations depicted by the 2-halo term. In (b), we show
the mass dependence of the SZ thermal and kinetic effects with a
maximum mass of $10^{16}$ and $10^{13}$ M$_{\sun}$. The SZ thermal
effect is strongly dependent on the maximum mass, while due to large
scale correlations, kinetic effect is not.}
\label{fig:szpower}
\end{figure}

We can now use the pressure power spectrum to calculate the SZ angular 
power spectrum by projecting it along the line of sight following
equation~(\ref{eqn:szpower}).
In figure~\ref{fig:szpower}(a), we show the  SZ power spectrum due to
baryons present in virialized halos.
As shown, most of the contributions to SZ power
spectrum comes from individual massive halos, while the halo-halo
correlations only contribute at a level of 10\% at large angular
scales. This is contrary to, say, the lensing convergence power
spectrum, where most of the
power at large angular scales is due to halo-halo
correlations. The difference is effectively due to the dependence of 
pressure on most massive halos in the large scale structure and to a 
lesser, but somewhat related,
reason that SZ  weight function increases towards
low redshifts. Note that the lensing weight function selectively probes
the large scale dark matter density power spectrum at comoving
distances half to that of background sources ($z \sim 0.2$ to 0.5 when
sources are at a redshift of 1), but has no extra dependence on mass when 
compared to the SZ weight function. 

\begin{figure}[t]
\centerline{\psfig{file=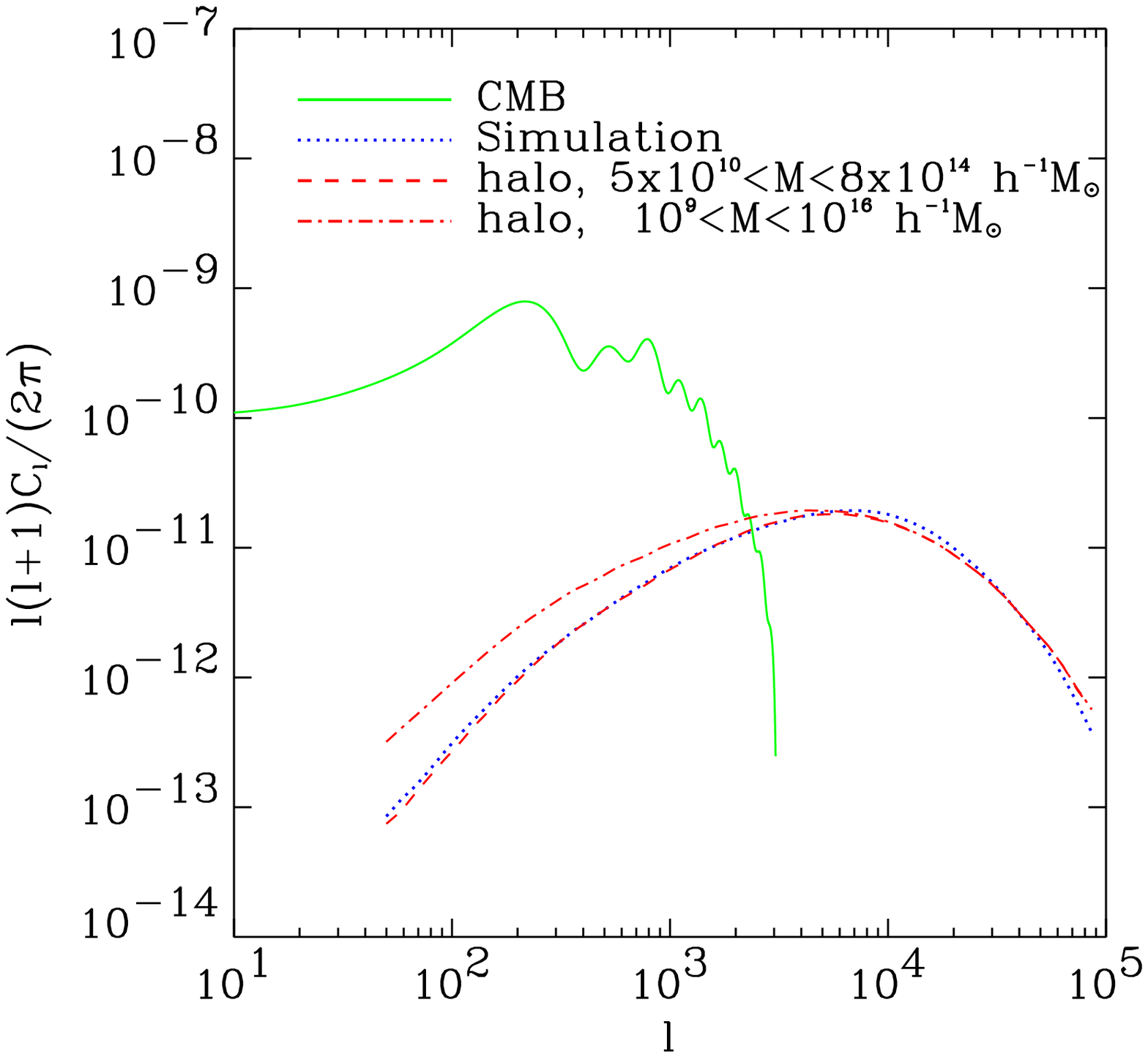,width=4.2in}}
\caption{The SZ power spectrum based on numerical simulations and the
analytical calculations based on the halo model. The simulations
are consistent with the mass distribution of halos in the simulated box.
The decrease in power at largest scales is due to the lack of most massive 
halos, which are rare. The simulations are in good agreement with the
halo based calculations. The figure is from \cite{RefTey01}.}
\label{fig:reftey}
\end{figure}

The predictions based on halo model are consistent with numerical simulations.
In figure~\ref{fig:reftey}, we show the angular power spectrum of 
SZ effect as measured in numerical simulations by \cite{RefTey01} and
a comparison to the halo calculation following \cite{Coo00b}. 
Note that simulations show a slight decrease in signal 
when the total mass included in the calculation is $10^{16}$ h$^{-1}$ 
M$_{\sun}$. The measurements are best described with a halo mass distribution
out to a maximum mass of $8 \times 10^{14}$ h$^{-1}$ M$_{\sun}$,
consistent with the expectation that highest mass halos are rare and 
are not present in the simulated box. 

\begin{figure}[t]
\centerline{\psfig{file=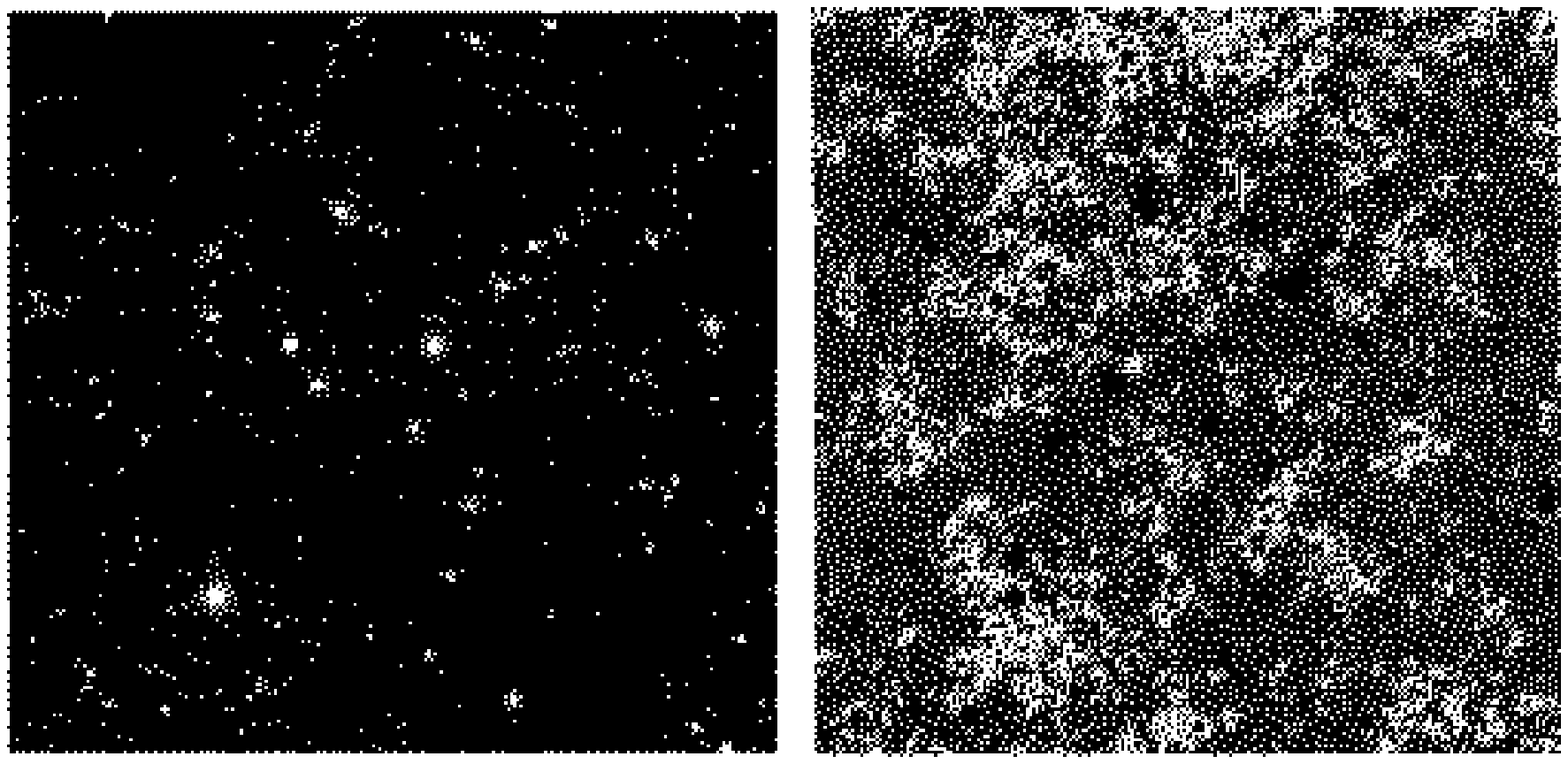,width=4.2in}}
\caption{Line of sight projected maps of the thermal (left) and kinetic 
(right) SZ effects. The maps are $1^\circ$ on a side and cover the same 
field of view. Note that the thermal SZ map picks out massive halos while 
contributions to kinetic SZ effect comes from wide range of masses. Unlike 
thermal SZ, which produces a negative decrement at Rayleigh-Jeans 
wavelengths, the kinetic SZ effect oscillates from negative and positive 
values depends on the direction of the velocity field.
Here, structures in red are moving towards the observer while those in 
blue are moving away.
This figure is from \cite{Spretal00}.}
\label{fig:tszksz}
\end{figure}

As we discuss later, the kinetic SZ effect has no such dependence on the 
massive halos and contributions to kinetic SZ effect comes from masses 
over a wide range. In figure~\ref{fig:tszksz}, we show projected maps of 
the SZ thermal and SZ kinetic effect produced in simulations by 
\cite{Spretal00}. The maps clearly show that the SZ thermal effect may be 
a useful way to map the massive structures in the universe. 

The fact that the SZ power spectrum results mainly from the single
halo term also results in a sharp reduction of power when the maximum
mass used in the calculation is varied. For example, as discussed in
\cite{Coo00b} and illustrated in figure~\ref{fig:szpower}(b), with
the maximum mass decreased from $10^{16}$ to $10^{13}$ M$_{\sun}$, the
SZ power spectrum reduced by a factor nearly two orders of magnitude
in large scales and an order of magnitude at $l \sim 10^{4}$. The same 
dependence also suggests a significant sample variance for the SZ effect 
as massive halos are rare; as discussed in \cite{Coo01a}, the SZ  
statistics from small fields are likely to be heavily biased based on the 
mass distribution of halos.  The same effect was found in numerical
simulations where the power spectrum was observed to vary over
a factor of $\sim$ 2 from 4 deg.$^2$ field to field over all scales probed
\cite{Seletal01}. 
For similar reasons, there is also a significant non-Gaussian 
contribution to the covariance of the
SZ effect that may complicate the use of SZ power spectrum as a probe of 
cosmology or galaxy cluster physics \cite{Coo01a}.

Following \cite{Yosetal01},  one can calculate the number counts of 
SZ halos under the approximation that gas traces dark matter 
and that the temperature of electrons can be related to velocity 
dispersion of the halo through virial arguments.  
This allows one to simplify the expected temperature decrement due to
the SZ effect at RJ wavelengths
\begin{equation}
 \frac{\Delta T}{T_{\rm CMB}} = -2 \int  d\rad  a(\rad)
                           \frac{k_B \sigma_T}{m_e c^2} n_e(\rad) T_e(\rad) 
\approx -2\frac{\sigma_T}{m_e} \frac{\Omega_b}{\Omega_m} 
         \int d\rad \frac{\sigma^2_{\rm dm}(\rad)}{2c^2}\rho_{\rm dm}(\rad) \,
\end{equation}
where the temperature of electrons has been approximated via line of sight
velocity dispersion of dark matter particles, $\sigma^2_{\rm dm}$.

\begin{figure}[t]
\centerline{\psfig{file=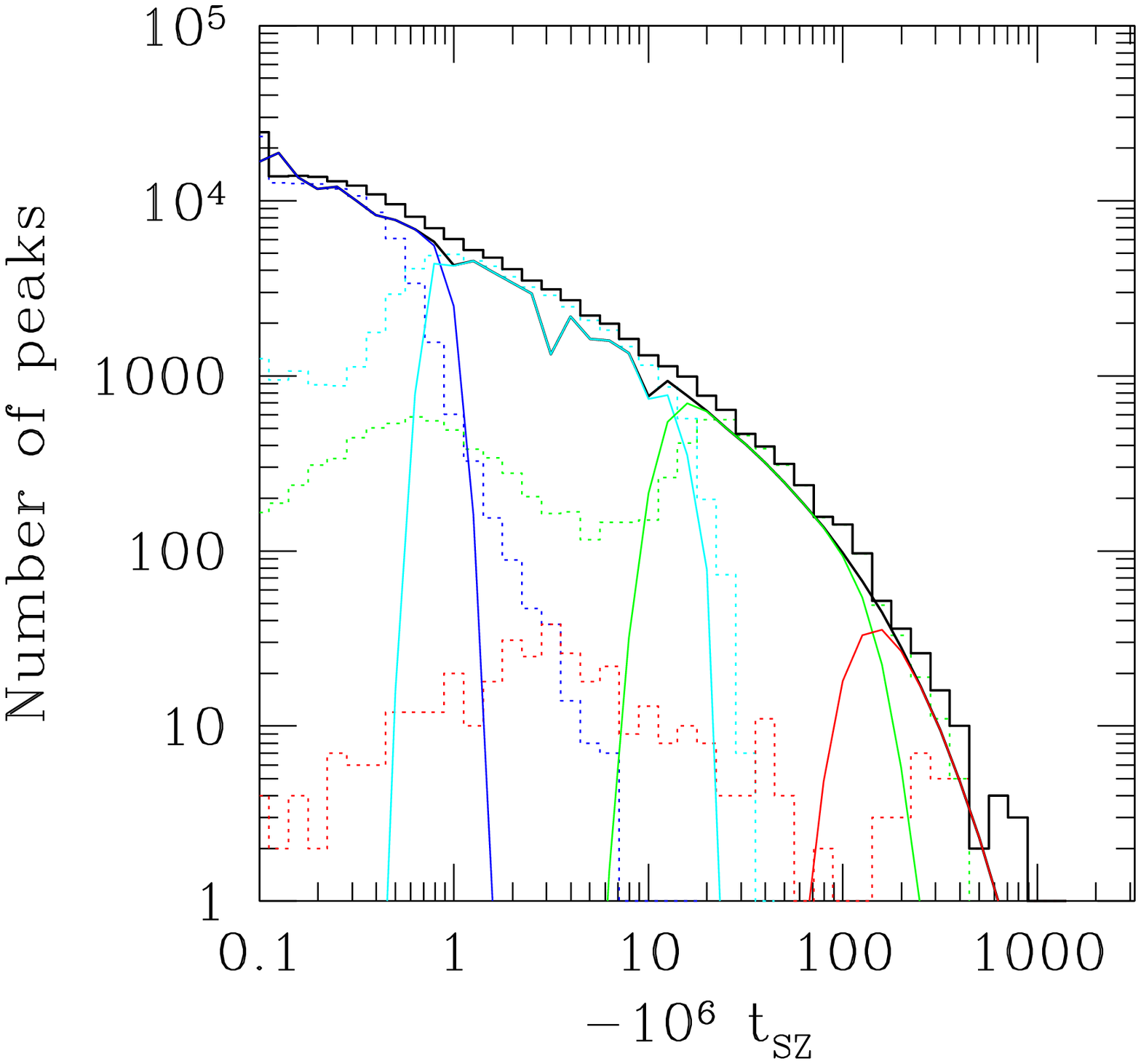,width=3.8in}}
\caption{Distribution of peak heights or number counts of thermal SZ
temperature decrements in simulations (solid histograms) and in
analytical calculations (solid curves). Dashed lines show the contributions
to the total from halos with mass in the range $10^{13} -10^{14}$,
$10^{14}-10^{15}$ and above $10^{15}$ from increasing temperature decrement
values. Here $t_\sz =\Delta T^\sz/T_\cmb$. The figure is from 
\cite{Yosetal01}.}
\label{fig:szcounts}
\end{figure}

The expected number of peaks due to the thermal SZ effect can be evaluated 
by  determining the expected SZ flux, integrated over the cluster, as a 
function of mass and then integrating over the mass function:
\begin{equation}
N(\sz) = \int dm n(m) p(\Delta T^\sz|m)d\Delta T^\sz \, ,
\end{equation}
where the probability distribution of temperature fluctuations arises from the
lognormal scatter in the concentration-mass relation \cite{Yosetal01}.
Figure~\ref{fig:szcounts} shows that the counts predicted by this 
model are in good agreement with numerical simulations.  Note, however, 
that the simulations were of dark matter only, so they also assumed 
that gas traces density.  Hydrodynamical simulations have been used 
to test the extent to which gas traces dark matter; they show that 
gas pressure effects can be important at the low mass end.  
Therefore, one expects modifications to Figure~\ref{fig:szcounts} 
at the low mass end; counts based on hydrodynamical simulations can 
be found in e.g., \cite{Spretal00,daS99}.

\subsection{The kinetic SZ effect}
\label{sec:ksz}

Extending our calculation on the contribution of large scale
structure gas distribution to CMB anisotropies through SZ effect, 
we can also study an  
associated effect involving baryons associated with halos in the large
scale structure.

The bulk flow of electrons, that scatter CMB photons, lead to temperature
fluctuations through the well known Doppler effect
\begin{equation}
T(\bn) = \int d\rad g(\rad) \bn \cdot {\bf v}(r\bn,r) \, ,
\end{equation}
where ${\bf v}$ is the baryon velocity.
In figure~\ref{fig:szpower}, we show the general Doppler effect due to the
velocity field. The power spectrum is such that it peaks around
the horizon at the scattering event projected on the sky today.
On scales smaller than the horizon at scattering, the contributions are
significantly canceled as photons scatter against the crests and troughs of
the perturbation. As a result, the Doppler effect is moderately sensitive to
how rapidly the universe reionizes since contributions from
a sharp surface of reionization do not cancel \cite{CooHu00}.
Also important are the double scattering events, which first scatter out of
the line of sight and the scatter back in, that do not necessarily cancel
\cite{Kai84b,CooHu00}.

The cancellations can be avoided by modulating the velocity field with
electron number density fluctuations. This is the so-called Ostriker-Vishniac
\cite{OstVis86,Vis87} effect.
The OV effect has been described as the
contribution to temperature anisotropies due to baryon modulated
Doppler effect in the linear regime of fluctuations.
At non-linear scales,
it is well known that the peculiar velocity of galaxy clusters, along
the line of sight, also lead to a contribution to temperature
anisotropies. This effect is commonly known as the kinetic
Sunyaev-Zel'dovich effect and arises from the halo
modulation of the Doppler effect 
associated with the velocity field \cite{SunZel80}.
The kinetic SZ effect can be considered as the OV effect extended to
the non-linear regime of baryon fluctuations \cite{Hu00a}, however,
it should be understood that the basic physical
mechanism responsible for the two effects is the same and
that there is no reason to describe them as separate contributions.

\subsubsection{Kinetic SZ power spectrum}

The kinetic SZ temperature fluctuations, denoted as $\dsz$,
 can be written as a product of the
line of sight velocity, under linear theory, and density fluctuations
\begin{eqnarray}
&&T^\dsz(\hat{\bf n})=  \int d\rad
        g(r) \hat{\bf n} \cdot {\bf v}_g(r,\bn r) \delta_g(r, \bn r)
\nonumber\\
&& \quad = -i \int d\rad g \dot{G} G
\int \frac{d^3{\bf k}}{(2\pi)^3} \int \frac{d^3{\bf k}'}{(2\pi)^{3}}
 \delta_\delta^\lin({\bf k}-{\bf k}')\delta_g({\bf k'})
e^{i{\bf k}\cdot \hat{\bf n}\rad}  \left[ \hat{\bf n} \cdot 
\frac{\veck - \veck'}{|\veck - \veck'|^2}\right] \, , \nonumber \\
\end{eqnarray}
Here, we have used linear theory to write the large scale velocity 
field in terms of the linear dark matter density field. The
multiplication between the velocity and density fields in real space
has been converted to a convolution between the two fields in Fourier
space.
We can now expand the temperature perturbation due to the kinetic SZ
effect, $T^\dsz$, using spherical harmonics:
\begin{eqnarray}
a_{lm}^\dsz &=& -i \int d\hat{\bf n}
\int d\rad\; (g\dot{G} G)
\int \frac{d^3{\bf k_1}}{(2\pi)^3}\int \frac{d^3{\bf
k_2}}{(2\pi)^3}
\delta_\delta^\lin({\bf k_1})\delta_g({\bf k_2}) \nonumber \\
&&\times e^{i({\bf k_1+k_2})
\cdot \hat{\bf n}\rad} \left[ \frac{\hat{\bf n} \cdot \veck_1}{k_1^2}
\right]
Y_l^{m\ast}(\hat{\bf n}) \, ,
\end{eqnarray}
where we have symmetrized by using $\veck_1$ and $\veck_2$
to represent $\veck-\veck'$ and $\veck'$ respectively.
Using
\begin{equation}
\hat{\bf n} \cdot \veck = \sum_{m'} \frac{4\pi}{3} k
Y_1^{m'}(\hat{\bf n}) Y_1^{m'\ast}(\hat{\veck}) \, ,
\end{equation}
and the Rayleigh expansion (equation~\ref{eqn:Rayleigh}),
 we can further simplify and rewrite the multipole moments as
\begin{eqnarray}
&&a_{lm}^\dsz = -i \frac{(4 \pi)^3}{3}
\int d\rad
\int \frac{d^3{\bf k}_1}{(2\pi)^3} \int \frac{d^3{\bf
k}_2}{(2\pi)^3}
\sum_{l_1 m_1}\sum_{l_2 m_2}\sum_{m'} \nonumber\\
&& \times
(i)^{l_1+l_2}
(g\dot{G}G)
\frac{j_{l_1}(k_1\rad)}{k_1}
j_{l_2}(k_2\rad)
\delta_\delta^\lin({\bf k_1})\delta_g({\bf k_2})
Y_{l_1}^{m_1}(\hat{\veck}_1) Y_1^{m'}(\hat{\veck}_1)
Y_{l_2}^{m_2}(\hat{\veck}_2) \nonumber \\
&\times& \int d\hat{\bf n}
Y_l^{m\ast}(\hat{\bf n}) Y_{l_1}^{m_1\ast}(\hat{\bf n})
Y_{l_2}^{m_2\ast}(\hat{\bf n})
Y_1^{m'\ast}(\hat{\bf n})\,.
\label{eqn:almdsz}
\end{eqnarray}

We can construct the angular power spectrum by considering
$\langle a_{l_1m_1} a^*_{l_2m_2} \rangle$.
Under the  assumption that the temperature field is statistically
isotropic, the correlation is independent of $m$, and we can write the
angular power spectrum as
\begin{eqnarray}
\langle \alm{1}^{*, \dsz} \alm{2}^\dsz\rangle = \deld_{l_1 l_2}
\deld_{m_1 m_2}
        C_{l_1}^\dsz\, .
\end{eqnarray}

The correlation can be written using
\begin{eqnarray}
&& \langle a^{\ast, \dsz}_{l_1m_1} a^\dsz_{l_2m_2} \rangle= \frac{(4
\pi)^6}{9}
\int d\rad_1 g \dot{G} G \int d\rad_2 g \dot{G} G  \nonumber \\
&\times& \int \frac{d^3{\bf k_1}}{(2\pi)^3}\frac{d^3{\bf
k_2}}{(2\pi)^3}
\frac{d^3{\bf k_1'}}{(2\pi)^3}\frac{d^3{\bf k_2'}}{(2\pi)^3}
\langle \delta_\delta^\lin({\bf k_1'})\delta_g({\bf k_2'})
\delta_\delta^{\ast \lin}({\bf k_1})\delta_g^\ast({\bf k_2})  \rangle
\nonumber \\
&\times& \sum_{l_1'm_1' l_1'' m_1'' m_1''' l_2'm_2' l_2'' m_2''
m_2'''}  (-i)^{l_1'+l_1''} (i)^{l_2'+l_2''}
j_{l_2'}(k_1'\rad_2) \frac{j_{l_2''}(k_2'\rad_2)}{k_2'}
\frac{j_{l_1'}(k_1\rad_1)}{k_1}
j_{l_1''}(k_2\rad_1) \nonumber \\
&\times& \int d\hat{\bf m}
Y_{l_2m_2}(\hat{\bf m}) Y_{l_2'm_2'}^\ast(\hat{\bf m})
Y_{l_2''m_2''}^\ast(\hat{\bf m}) Y_{1m_2'''}^\ast(\hat{\bf m})
\nonumber \\
&\times& \int d\hat{\bf n}
Y_{l_1m_1}^\ast(\hat{\bf n}) Y_{l_1'm_1'}(\hat{\bf n})
Y_{l_1''m_1''}(\hat{\bf n})
Y_{1m_1'''}(\hat{\bf n})  \nonumber \\
&\times& \int d\hat{\veck_1'} \int d\hat{\veck_2'}
 Y_{l_2'm_2'}(\hat{\veck_1'}) Y_{1m_2'''}(\hat{\veck_2'})
Y_{l_2''m_2''}(\hat{\veck_1'}) \nonumber \\
&\times&\int d\hat{\veck_1} \int d\hat{\veck_2}
Y_{l_1'm_1'}^\ast(\hat{\veck_1})
Y_{1m_1'''}^\ast(\hat{\veck_1}) Y_{l_1''m_1''}^\ast(\hat{\veck_2}) \,
.
\end{eqnarray}
We can separate out the contributions such that the total is made of
correlations
following $\langle v_g v_g\rangle \langle \delta_g \delta_g \rangle$
and $\langle v_g \delta_g \rangle \langle v_g \delta_g \rangle$
depending on
whether we consider cumulants by combining $\veck_1$ with $\veck_1'$
or $\veck_2'$ respectively. After some straightforward but tedious
algebra, and noting that 
\begin{equation} 
\sum_{m_1' m_2'} 
\left(
\begin{array}{ccc}
l_1' & l_2' & l_1 \\
m_1' & m_2'  &  m_1
\end{array}
\right)
\left(
\begin{array}{ccc}
l_1' & l_2' & l_2 \\
m_1' & m_2'  &  m_2
\end{array}
\right) 
= \frac{\delta_{m_1 m_2} \delta_{l_1 l_2}}{2l_1+1} \, ,
\label{eqn:w3jsimplify}
\end{equation} 
we can write
\begin{eqnarray}
&&C_l^\dsz = \frac{2^2}{\pi^2} \sum_{l_1 l_2}
\left[\frac{(2l_1+1)(2l_2+1)}{4\pi}\right]
\left(
\begin{array}{ccc}
l & l_1 & l_2 \\
0 & 0  &  0
\end{array}
\right)^2 \nonumber \\
&\times& \int d\rad_1 g \dot{G} G
\int d\rad_2 g \dot{G} G
\int k_1^2 dk_1 \int k_2^2 dk_2 \nonumber \\
&\times& \Big[ P_{\delta\delta}^\lin(k_1)
P_{gg}(k_2)j_{l_1}(k_2\rad_2) j_{l_1}(k_2\rad_1) 
\frac{j_{l_2}'(k_1\rad_1)}{k_1} \frac{j_{l_2}'(k_1\rad_2)}{k_1}
\nonumber \\
&+& P_{\delta b}(k_1) P_{\delta b}(k_2) j_{l_2}(k_2\rad_1)
\frac{j_{l_1}'(k_1\rad_1)}{k_1}j_{l_1}(k_1\rad_2)
\frac{j_{l_2}'(k_2\rad_2)}{k_2}
  \Big] \, . \nonumber \\
\end{eqnarray}

Here, the first term represents the contribution from $\langle v_g v_g
\rangle
\langle \delta_g  \delta_g \rangle$  while the second term is
the $\langle v_g \delta_g \rangle \langle v_g \delta_g \rangle$
contribution, respectively.
In simplifying the integrals involving spherical harmonics, we have
made use of the properties of Clebsh-Gordon coefficients, in
particular, those involving $l=1$.
The integral involves two distances
and two Fourier modes and is summed over the Wigner-3$j$ symbol to
obtain the power spectrum. 
Since we are primary interested in the contribution at small
angular scales here, we can ignore the contribution to the kSZ effect
involving the correlation between linear density field and baryons and
only  consider the contribution that results from baryon-baryon and
density-density
correlations. In fact, under the halo description provided here, there
is no correlation of the baryon field within halos and the velocity
field traced by individual halos (see \S~\ref{sec:velocities}).
Thus, contribution to the
baryon-velocity correlation only comes from the 2-halo term of the
density field-baryon correlation. This correlation is suppressed at
small scales and is not a significant contributor to the kinetic SZ
power spectrum \cite{Hu00a}. 

Similar to the Limber approximation \cite{Lim54},
in order to simplify the calculation associated with $\langle v_g v_g
\rangle \langle \delta_g \delta_g \rangle$, we use an equation
involving completeness of spherical Bessel functions
(equation~\ref{eqn:ovlimber}) and apply 
it to the integral over $k_2$  to obtain
\begin{eqnarray}
&&C_l^\dsz = \frac{2}{\pi} \sum_{l_1 l_2}
\left[\frac{(2l_1+1)(2l_2+1)}{4\pi}\right]
\left(
\begin{array}{ccc}
l & l_1 & l_2 \\
0 & 0  &  0
\end{array}
\right)^2 \nonumber \\
&\times& \int d\rad_1 \frac{(g \dot{G})^2}{d_A^2}
\int k_1^2 dk_1 
P_{\delta\delta}^\lin(k_1) P_{gg}\left[ \frac{l_1}{d_A}; \rad_1
\right]
\left(\frac{j_{l_2}'(k_1\rad_1)}{k_1}\right)^2\, . \nonumber \\
\label{eqn:redallsky}
\end{eqnarray}

The alternative approach, which has been the calculational method in
many of the previous papers \cite{Vis87,Efs88,JafKam98,DodJub95,Hu00a}, 
is to
use the flat-sky approximation
with  the kinetic SZ power spectrum written as
\begin{eqnarray}
&&C_l^\dsz = \frac{1}{8\pi^2} \int d\rad \frac{(g \dot{G}G)^2}{d_A^2}
P_{\delta\delta}(k)^2
I_v\left(k=\frac{l}{d_A}\right) \, ,
\end{eqnarray}
with the mode-coupling integral given by
\begin{eqnarray}
I_v(k) = \int dk_1 \int_{-1}^{+1} d\mu \frac{(1-\mu^2)(1-2\mu
y_1)}{y_2^2} \frac{P_{\delta\delta}(k y_1)}{P_{\delta\delta}(k)}
\frac{P_{\delta\delta}(k y_2)}{P_{\delta\delta}(k)} \, . \nonumber \\
\end{eqnarray}
We refer the reader to \cite{Vis87} and \cite{DodJub95} 
for details on this derivation. In above, 
$\mu = \hat{\bf k} \cdot \hat{\bf k_1}$, $y_1 = k_1/k$ and
$y_2 = k_2/k = \sqrt{1-2\mu y_1+y_1^2}$. This flat-sky approximation
makes use of the Limber approximation \cite{Lim54} to further
simplify
the calculation with the replacement of $k = l/d_A$. The power spectra
here
represent the baryon field power spectrum 
and the velocity field power spectrum; the former assumed to trace the
dark matter density field while the latter  is generally 
related to the linear dark matter density field through the use of
linear
theory arguments. 

The correspondence between the flat-sky and all-sky
formulation can be obtained by noting that in the small scale limit
contributions to the flat-sky effect comes when $k_2 = |\veck -
\veck_1| \sim k$ such that $y_1 \ll 1$. In this limit, the flat sky
Ostriker-Vishniac effect reduces to a simple form given by \cite{Hu00a} 
\begin{equation}
C_l^\dsz = \frac{1}{3}
\int d\rad \frac{(g \dot{G} G)^2}{d_A^2} P_{gg}(k) v_\rms^2 \, .
\label{eqn:redflatsky}
\end{equation}
Here, $v_\rms^2$ is the rms of the uniform bulk velocity
form
large scales
\begin{equation}
v_\rms^2 = \int dk \frac{P_{\delta\delta}(k)}{2\pi^2} \, .
\end{equation}
The $1/3$ arises from the fact that rms in each component is $1/3$rd
of the total velocity. Similarly, one can reduce the all sky expression, 
equation~(\ref{eqn:redallsky}),
 to that of the flat--sky,
equation~(\ref{eqn:redflatsky}), in the small scale limit of $l \sim l_1 
\gg l_2$,
with $l_1$ probing the density field \cite{Coo01a}.
 
\begin{figure}[t]
\centerline{\psfig{file=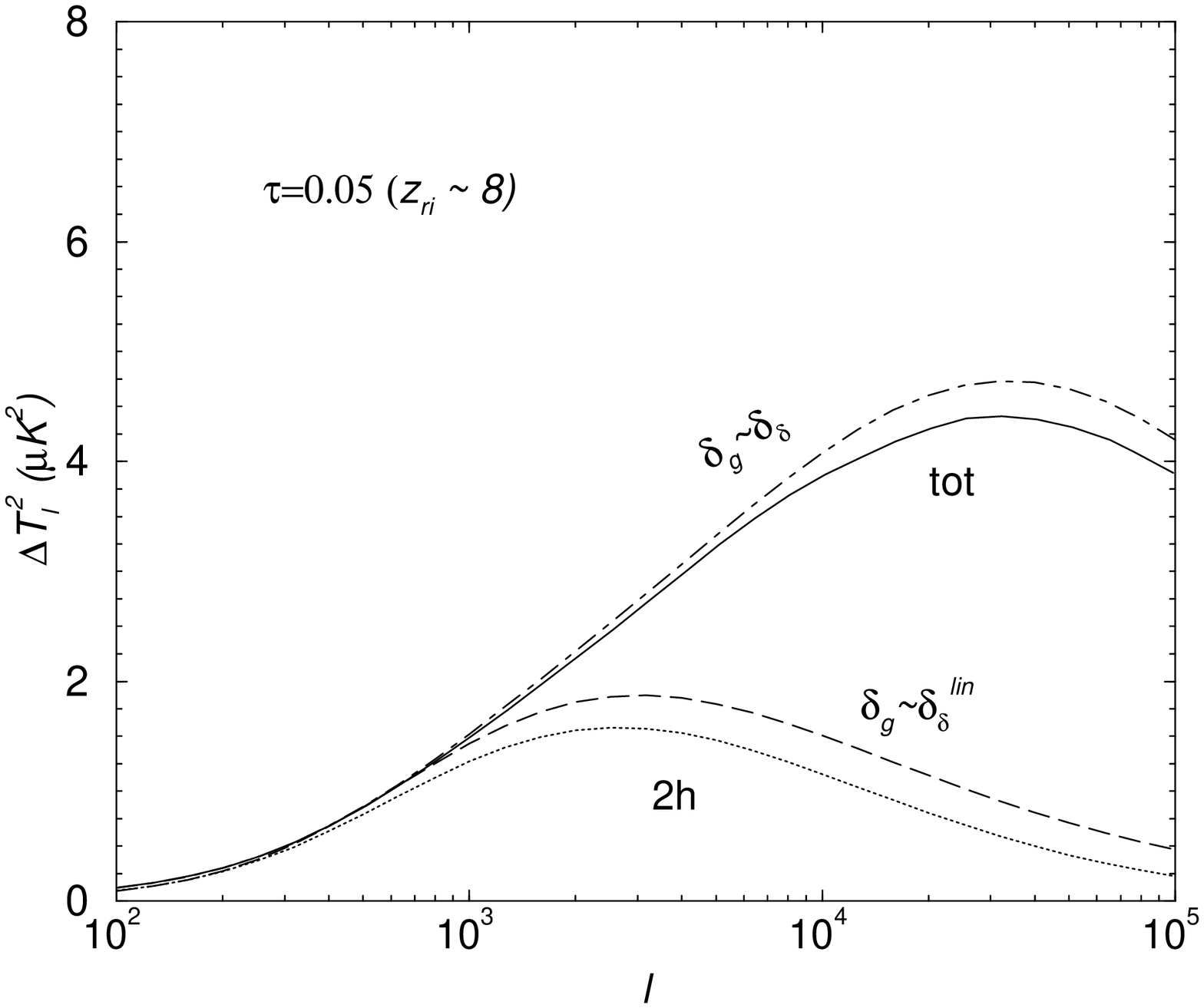,width=4.2in,angle=-90}}
\caption{The temperature
fluctuation power $(\Delta T^2_l =
l(l+1)/(2\pi) C_l T_{\rm CMB}^2)$ for a variety of methods to
calculate the kinetic SZ effect. Here, we show the contribution for a
reionization redshift of $\sim$ 8 and an optical depth to reionization
of 0.05. The contributions are calculated under the assumption that
the baryon field traces the non-linear dark matter ($P_g(k) =
P_\delta(k)$ with $P_\delta(k)$ predicted by the halo model), 
the linear density field ($P_g(k) = P^\lin(k)$), and the halo model
for gas, with total and the 2-halo contributions  shown separately.
For the most part, the kinetic SZ effect can be described using linear
theory, and the non-linearities only increase the temperature
fluctuation power by a factor of a few at $l \sim 10^5$.}
\label{fig:ovtemp}
\end{figure}

In figure~\ref{fig:szpower}, we show our prediction for the SZ kinetic
effect and a comparison with the SZ thermal contribution.
As shown, the SZ kinetic contribution is roughly an order of magnitude
smaller than the thermal SZ contribution. 
There is also a more fundamental difference between the two:
the SZ thermal effect, due to its dependence on highest temperature
electrons is more dependent on the most massive halos in the universe,
while the SZ kinetic effect arises more clearly due to 
large scale correlations of the halos that make the
large scale structure. 

 The difference between the two effects arises from that fact that
kinetic SZ effect is mainly due to the baryons and not the temperature
weighted baryons that trace the pressure responsible for the thermal
effect. Contributions to the SZ kinetic effect comes from baryons
tracing all scales and down to small mass halos.
The difference associated with mass dependence 
between the two effects suggests that a wide-field 
SZ thermal effect map and a wide-field 
SZ kinetic effect map will be different from each other in that
massive halos, or clusters, will be clearly visible in a SZ thermal
map while the large scale structure will be more evident in
a SZ kinetic effect map. As shown with the thermal and kinetic SZ maps
in figure~\ref{fig:tszksz} from \cite{Spretal00},
numerical simulations are in fact consistent
with this picture (see, also \cite{daS01}).

As shown in figure~\ref{fig:szpower}(b), the
variations in maximum mass used in the calculation does not lead to
orders of magnitude changes in the total kinetic SZ contribution,
which is considerably less than the changes in the total thermal SZ
contribution as a function of maximum mass. This again is consistent
with our basic result that most contributions come from the large scale 
linear
velocity modulated by baryons in halos. 
Consequently, while the thermal SZ effect is dominated by shot-noise
contributions, and is heavily affected by the sample variance, the same is 
not 
true for the kinetic SZ effect. 

In figure~\ref{fig:ovtemp}, we show several additional predictions for
the kinetic
SZ effect, following the discussion in \cite{Hu00a}. 
Due to the density weighting,
the kinetic SZ effect peaks at small scales: arcminutes for $\Lambda$CDM. 
For a fully ionized universe, contributions 
are moderately dependent on the optical depth
$\tau$. Here, we assume an optical depth to ionization of 0.05,
consistent with current upper limits on the reionization redshift from
CMB \cite{Grietal99} and other observational data (see, e.g., \cite{HaiKno99} and
references therein). In figure~\ref{fig:ovtemp}, 
we have calculated the kinetic SZ power spectrum under several
assumptions, including the case when gas is assumed to trace 
the non-linear density field and the linear density field. 
We compare predictions based on such assumptions
to those calculated using the halo model. As shown, the halo model
calculation shows slightly less power than when using the non-linear
dark matter density field to describe clustering of
baryons. This difference arises from the fact that baryons do not
fully trace the dark matter in halos. Due to small differences,
one can safely use the non-linear dark matter power spectrum to
describe baryons. Using the linear theory only, however, leads to an
underestimate of power by a factor of 3 to 4 at scales corresponding to 
multipoles of
$l \sim 10^4$ to $10^5$ and may not provide an accurate description of
the total kinetic SZ effect.

In addition to the contribution due to the line of sight motion of halos,
there is an additional effect resulting from halo rotations as discussed by
\cite{CooChe01}. Here, the resulting rotational contribution to kinetic
SZ effect was evaluated under the assumption that baryons in halos are
corotating with dark matter; this assumption is primarily due to the lack 
of
knowledge on angular momentum of gas in virialized halos from numerical
simulations. In terms of the dark matter, recent high resolution numerical 
simulations show that the spatial distribution
of angular momentum in dark matter halos has a universal profile (see e.g.
\cite{Buletal01,Vitetal01}). This profile is consistent with
that of solid body rotation, but saturates at large values for angular 
momentum.
The spatial distribution of angular momentum in most halos (80\%) tend to 
be
cylindrical and well-aligned with
the spin of a halo. Also, angular momentum is 
almost independent with the mass of
the halo and does not evolve with redshift except after major
mergers. 

For an individual cluster at a redshift $z$ with an angular
diameter distance $d_A$, one can write the temperature fluctuation
as an integral of the electron density, $n_e(r)$, weighted by
the rotational velocity component, $\omega r \cos \alpha$,
along the line of sight. Introducing the fact the line of sight
velocity due to rotation is proportional to sine of the inclination
angle of the rotational axis with respect to the observer, $i$, we write
\begin{equation}
\frac{\Delta T}{T}(\theta,\phi) = \sigma_T e^{-\tau} \eta(\theta)
\cos \phi \sin i \,
\label{eqn:rotation}
\end{equation}
where
\begin{equation}
\eta(\theta) =  \int_{d_c\theta}^{R_{\vir}} \frac{2 r
dr}{\sqrt{r^2 - d_c^2\theta^2}}
n_e(r) \omega d_c \theta .
\label{eqn:cluster}
\end{equation}
Here, $\theta$ is the line of sight angle relative to the cluster
center and $\phi$ is an azimuthal angle measured relative to an axis
perpendicular to the spin axis in the plane of the sky.
In simplifying, we have introduced the fact that
the angle between the rotational velocity and line of sight, $\alpha$, is 
such
that  $\alpha = \cos^{-1} d_A \theta/r$.
In equation~(\ref{eqn:cluster}), $R_\vir$ is the cluster virial radius.

\begin{figure}[t]
\centerline{\psfig{file=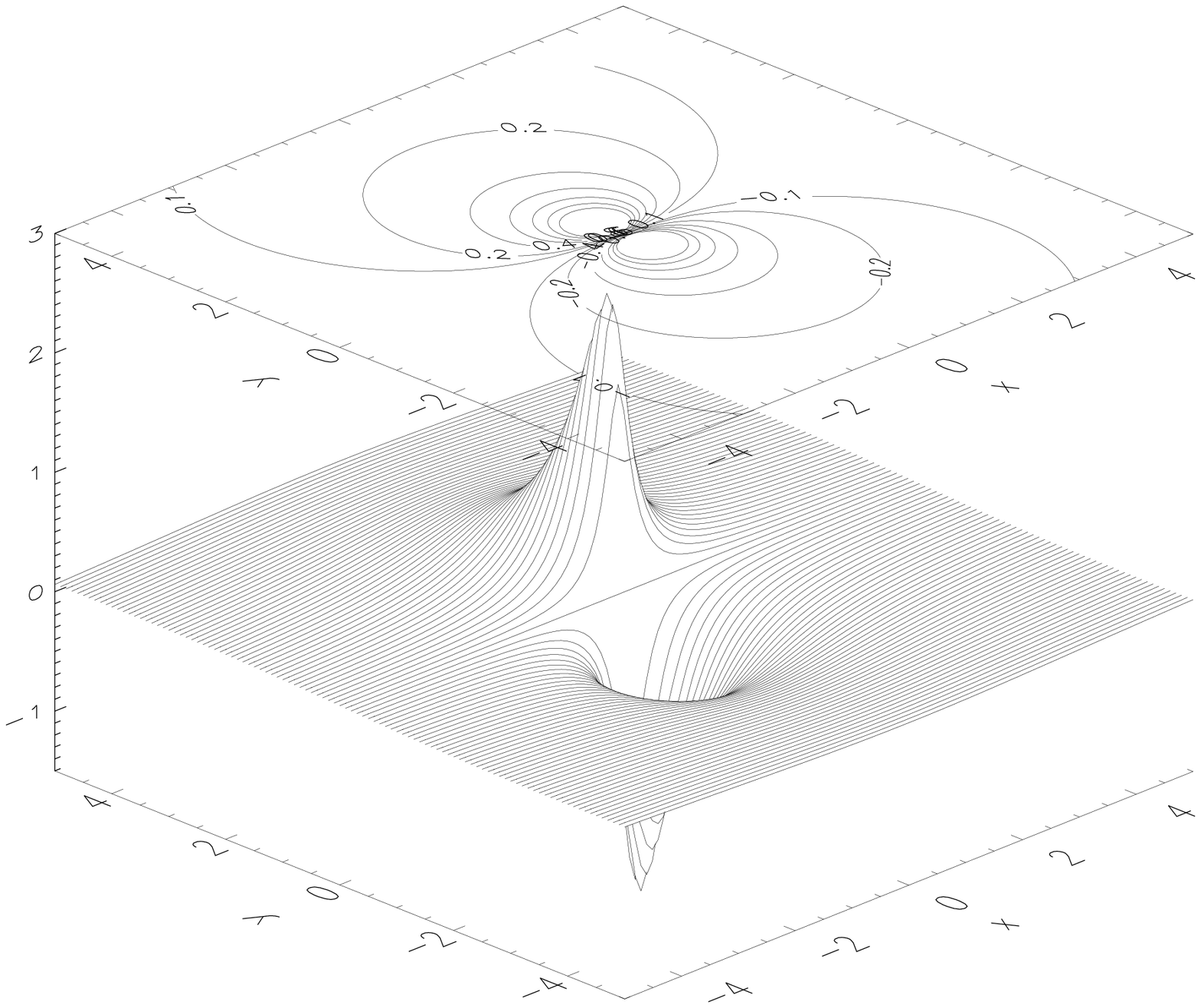,width=4.5in,angle=90}}
\caption{
Contribution to temperature fluctuations through halo
rotation for a cluster of mass $5 \times 10^{14}$ M$_{\sun}$ at a
redshift of 0.5. The temperature fluctuations produce a distinct
bipolar-like pattern on the  sky with a maximum of $\sim$ 2.5 $\mu$K.
Here, rotational axis is perpendicular to the line of sight and
$x$ and $y$ coordinates are in terms of the
scale radius of the cluster, based on the NFW profile.}
\label{fig:rotational}
\end{figure}

To describe the halo rotations, we write the dimensionless spin
parameter $\lambda (= J \sqrt{E}/GM^{5/2})$ following \cite{Buletal01}
as
\begin{equation}
\lambda = \frac{J}{2 V_c M_\vir R_\vir} \frac{\sqrt{cg(c)} }{f(c)} \,
\label{eqn:lambda}
\end{equation}
where the virial concentration for the NFW profile is
$c=R_\vir/r_s$, $J$ is the total angular momentum, and $V_c^2= G 
M_\vir/R_\vir$.
In Ref. \cite{Buletal01}, the
probability distribution function for $\lambda$ was measured through
numerical simulations and was found to be well described by a log
normal distribution with a mean, $\bar{\lambda}$, of $0.042 \pm 0.006$
and a width, $\sigma_\lambda$ of $0.50 \pm 0.04$. 

To relate angular velocity, $\omega$, to spin, we first
integrate the NFW profile over a cluster to calculate $J$, and substitute 
in above to find
\begin{equation}
\omega = \frac{3 \lambda V_c c^2 f^2(c)}{R_\vir h(c) \sqrt{c g(c)}} \, .
\label{eqn:omega}
\end{equation}
The functions $f(c)$, $g(c)$ and
$h(c)$, in terms of the concentration, follows as
\begin{eqnarray}
f(c) &=& \ln(1+c) -\frac{c}{1+c}\nonumber \\
g(c) &=& 1-\frac{2\ln(1+c)}{1+c} -\frac{1}{(1+c)^2} \nonumber \\
h(c) &=& 3\ln(1+c) +\frac{c(c^2-3c-6)}{2(1+c)} \, .
\end{eqnarray}

In figure~\ref{fig:rotational}, we show the temperature fluctuation
produced by the rotational component for a typical cluster with mass
$5 \times 10^{14}$ M$_{\sun}$ at a redshift of 0.5. The maximal effect, 
with the mean spin parameter measured by \cite{Buletal01},
is on the order of $\sim$ 2.5 $\mu$K. The sharp drop towards the center
of the cluster is due to the decrease in the rotational velocity. As
shown, the effect leads to a distinct temperature distribution with a
dipole like pattern across clusters. Here, we have taken the cluster
rotational axis to be aligned perpendicular to the line of sight; as
it is clear, when the axis is aligned along the line of sight, there
is no resulting contribution to the SZ kinetic effect through
scattering. 

The order of magnitude  of this rotational contribution can be
understood by estimating the rotational velocity where the effect peaks.
In equation~(\ref{eqn:omega}),
rotational velocity is $\omega \sim 3 \lambda V_c/R_\vir$ with functions
depending on the concentration in the order of a few ($\approx 2.4$
when $c=5$). Since the circular velocity for typical cluster is of
order $\sim$ 1500 km s$^{-1}$, with $R_\vir \sim$  Mpc and
$\bar{\lambda} \sim 0.04$, at typical inner radii of order $\sim 1/5
R_\vir$, we find velocities of order $\sim$ 30 km s$^{-1}$. Since, on 
average,
peculiar velocities for clusters are of order $\sim$ 250 km s$^{-1}$, the
rotational velocity is lower by a factor of $\sim$ 8,  when compared
with the peculiar velocity of the typical cluster. Furthermore, since the
kinetic SZ due to peculiar motion peaks in the center of the halo where
the density is highest, while the
rotational effect peaks away from the center, the difference between 
maximal peculiar
kinetic SZ and rotational kinetic SZ temperature fluctuations is
even greater. Note, however, each individual cluster has a different 
orientation
and magnitude of peculiar velocity and rotation, thus the 
velocity-to-rotation
ratio could vary a lot. In favorable cases where the peculiar velocity is
aligned mostly across the line of sight, the rotational contribution
may be important.

\begin{figure}[!h]
\centerline{\psfig{file=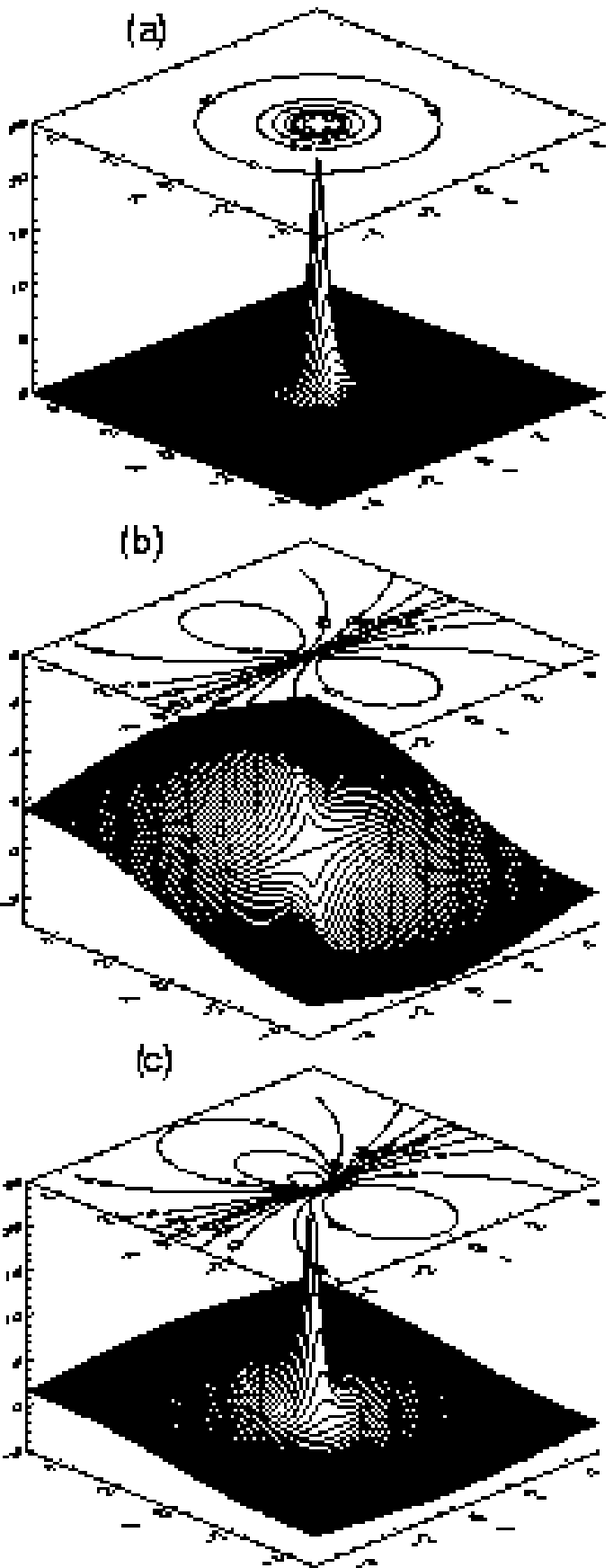,width=2.5in,angle=0}}
\caption{Temperature fluctuations due to galaxy clusters: (a) kinetic SZ
effect involving peculiar motion,
(b) lensing of CMB primary temperature fluctuations, and (c)
the total contribution from kinetic SZ, lensing and rotational
velocity. The total contribution leads asymmetric bipolar pattern with
a sharp rise towards the center. We have not included the thermal SZ
effect as its contribution can be separated from these effects, and
primary temperature fluctuations, based on its frequency dependence.
We use the same cluster as shown in figure~\ref{fig:rotational}.}
\label{fig:secondary}
\end{figure}

In figure~\ref{fig:secondary}, we show the
kinetic SZ effect towards the same cluster due to the peculiar motion
and the contribution resulting from the lensed CMB towards the same
cluster. The latter contribution is sensitive to the gradient of the
dark matter potential of the cluster along the large scale CMB
gradient. In this illustration, we  haven taken the CMB gradient
to be the rms value with 13 $\mu$K arcmin$^{-1}$ following
\cite{SelZal01}. Previously, it was suggested that the lensed
CMB contribution can be extracted based on its dipole like
signature. Given the fact that the rotational contribution also leads
to a similar pattern, any temperature distribution with a dipole
pattern across a cluster cannot easily be prescribed to the lensing 
effect.
However, as evident from figures~\ref{fig:rotational} and
\ref{fig:secondary}, the dipole signature associated with the
rotational scattering is limited to the inner region of the cluster
while the lensing effect, due to its dependence on the gradient of the
dark halo potential, covers a much larger extent. Also, the two
dipoles need not lie in the same direction as the background gradient
of the primary CMB fluctuations and the rotational axis of halos may
be aligned differently. Thus, to separate the lensed effect and the
rotational contribution from each other and from dominant kinetic SZ
one can consider various filtering schemes (see,
discussion in \cite{SelZal01}). 
In figure~\ref{fig:secondary}, we have not included the
dominant thermal SZ contribution since it can be separated  from
other contributions reliably if
multifrequency data are available.

The interesting experimental possibility here is whether one can obtain
a wide-field map of the SZ kinetic effect. Since it is now well
known that the unique spectral dependence of the thermal SZ effect can
be used to separate its contribution \cite{Cooetal00a}, 
it is likely that after such a separation, the SZ kinetic effect 
will be the dominant signal at small angular scales. 
To separate the SZ thermal effect,
observations, at multifrequencies, are needed to arcminute scales.
Upcoming interferometers and similar experiments
will allow such studies to be eventually carried out. A wide-field
kinetic SZ map of the large scale structure will allow an 
understating of the large scale velocity field of baryons, as the  density fluctuations
can be identified through cross-correlation of such a map with the thermal SZ map \cite{Coo01a}.

\subsection{Non-Linear Integrated Sachs-Wolfe Effect}
\label{sec:isw}

The integrated Sachs-Wolfe effect \cite{SacWol67} results from
the late time decay of gravitational potential fluctuations. The
resulting
temperature fluctuations in the CMB can be written as
\begin{equation}
T^\isw(\bn) = -2 \int_0^{\rad_0} d\rad \dot{\Phi}(\rad,\bn \rad) \, ,
\end{equation}
where the overdot represent the derivative with respect to conformal
distance (or equivalently look-back time).
Writing multipole moments of the temperature fluctuation field
$T(\hat{\bf n})$,
\begin{equation}
a_{lm} = \int d\bn T(\bn) \Ylmn {}^*(\bn)\,,
\end{equation}
we can formulate the angular power spectrum as
\begin{eqnarray}
\langle \alm{1}^* \alm{2}\rangle = \deld_{l_1 l_2} \deld_{m_1 m_2}
        C_{l_1}\,.
\end{eqnarray}

For the ISW effect, multipole moments are
\begin{eqnarray}
a^{\rm ISW}_{lm} &=&i^l \int \frac{d^3\veck}{2 \pi^2}
\int d\rad   \dot{\Phi}(\veck) I_l(k)  \Ylmn(\hat{\veck}) \, ,
\nonumber\\
\label{eqn:moments}
\end{eqnarray}
with $I_l(k) = \int d\rad W^\isw(k,\rad) j_l(k\rad)$, and the window
function for the ISW effect, $W^\isw=-2$.
The angular power spectrum is then given by
\begin{equation}
C_l^\isw = {2 \over \pi} \int k^2 dk P_{\dot{\Phi}\dot{\Phi}}(k)
                \left[I_l(k)\right]^2 \,,
\label{eqn:clexact}
\end{equation}
where the three-dimensional power spectrum of the time-evolving
potential fluctuations are defined as
\begin{equation}
\langle \dot{\Phi}({\bf k_1}) \dot{\Phi}({\bf k_2}) \rangle = (2\pi)^3
\delta_D(\veck_1+\veck_2) P_{\dot{\Phi} \dot{\Phi}}(k_1) \, .
\label{eqn:phidotpower}
\end{equation}
 
The above expression for the angular power spectrum can be
evaluated efficiently under the Limber approximation \cite{Lim54} for
sufficiently high $l$ values, usually in the order of few tens, as
\begin{equation}
C_l^\isw = \int d\rad \frac{\left[W^\isw\right]^2}{\da^2}
                P_{\dot{\Phi}\dot{\Phi}}\left[k=\frac{l}{\da},\rad\right]
                \, .
\label{eqn:cllimber}
\end{equation}
 
In order to calculate the power spectrum of
time-derivative of potential fluctuations, we make use of the
cosmological Poisson equation in equation~(\ref{eqn:poisson}) and
write the derivative of the potential through a derivative
of the density field and the scale factor $a$.
Considering a flat universe
with $\Omega_K=0$, we can write the full expression for the power spectrum
of time-evolving potential fluctuations, as necessary for the ISW effect
valid in all
regimes of density fluctuations, as
\begin{eqnarray}
P_{\dot{\Phi}\dot{\Phi}}(k,\rad) &=& 
{9 \over 4} \left(\frac{\Omega_m}{a}\right)^2 \left({H_0 \over
k}\right)^4 \nonumber \\
&& \quad \quad \times \left[\left(\frac{\dot{a}}{a}\right)^2P_{\delta\delta}(k,\rad)
-2\frac{\dot{a}}{a} P_{\delta\dot{\delta}}(k,\rad)
+ P_{\dot{\delta}\dot{\delta}}(k,\rad)\right] \, .
\label{eqn:non-linear}
\end{eqnarray}

To calculate the power spectrum involving the correlations between
time derivatives of density fluctuations, $P_{\dot{\delta}\dot{\delta}}$,
 and the cross-correlation
term involving the density and time-derivative of the density fields,
$P_{\delta \dot{\delta}}$, we make use of
the continuity equation in \ref{eqn:continuity},
 which can be written in the form:
\begin{equation}
\dot{\delta}({\bf x},\rad) = -\nabla \cdot
\left[1+\delta({\bf x},\rad)\right]{\bf v}({\bf x},\rad) \, .
\end{equation}
In the linear regime of fluctuations, when $\delta({\bf x},\rad) =
G(\rad) \delta({\bf x},0) \ll 1$,
the time derivative is simply
$\dot{\delta}^\lin({\bf x},\rad) =-\nabla \cdot {\bf v}({\bf x},\rad)$ 
leading to the
 well-known
result for linear theory velocity field (equation~\ref{eqn:linvel}).
Thus, in linear theory, from equation~(\ref{eqn:velpk}),
$P_{\dot{\delta} \dot{\delta}} \equiv k^2P_{vv}(k,r) = \dot{G}^2 P_{\delta
\delta}^\lin(k,0)$
and
$P_{\delta \dot{\delta}} \equiv kP_{\delta v}(k,r) = G\dot{G} P_{\delta
\delta}^\lin(k,0)$.

These lead to the well-known results
for the linear ISW effect, with a power spectrum for $\dot{\Phi}$ as
\begin{equation}
P_{\dot{\Phi}\dot{\Phi}}^\lin(k,\rad) = {9 \over 4}
\left(\frac{\Omega_m}{a} \right)^2
\left({H_0 \over k}\right)^4
\left[-\frac{\dot{a}}{a} G(\rad) + \dot G \right]^2
P_{\delta\delta}^\lin(k,0)
\, .
\label{eqn:linear}
\end{equation}
The term within the square bracket is $\dot{F}^2$ where $F=G/a$ following
derivation for the linear ISW effect in \cite{CooHu00}.
Even though, we have replaced the divergence of the velocity field
with a time-derivative of the growth function,
it should be understood that the contributions to the ISW effect
comes from the divergence of the velocity field and not directly from the
density field. Thus, to some extent, even the linear ISW effect
reflects statistical properties of the large scale structure
velocities.       

In the mildly non-linear to fully non-linear regime of fluctuations,
the approximation in equation~(\ref{eqn:continuity}),
involving $\delta \ll 1$, is no longer valid and a
full calculation of the time-derivative of density perturbations is
required. This can be achieved in the second order perturbation
theory, though, such an approximation need not be fully applicable as
the second order perturbation theory fails to describe even the weakly
non-linear regime of fluctuations exactly. 
Motivated by applications of the halo 
approach to large scale structure 
and results from numerical simulations 
\cite{Sel96a,MaFry01,Sheetal01b}, we consider a description for the
time-derivative of density fluctuations and rewrite
equation~(\ref{eqn:continuity}) as
\begin{equation}
\dot{\delta}({\bf x},\rad) = -\nabla \cdot {\bf v}({\bf x},\rad)
-\nabla \cdot \delta({\bf x},\rad){\bf v}({\bf x},\rad) \, ,
\end{equation}
where we have separated the momentum term involving $p=(1+\delta)v$
to a velocity contribution and a density velocity product.
In Fourier space, the power spectrum is simply $\dot{\delta}(k) =i\veck \cdot p(\veck)$ and the power spectrum of $\dot{\delta}$ can be calculated following
the halo model description of the momentum-density field (\S~\ref{sec:mom}).

\begin{figure}[t]
\centerline{\psfig{file=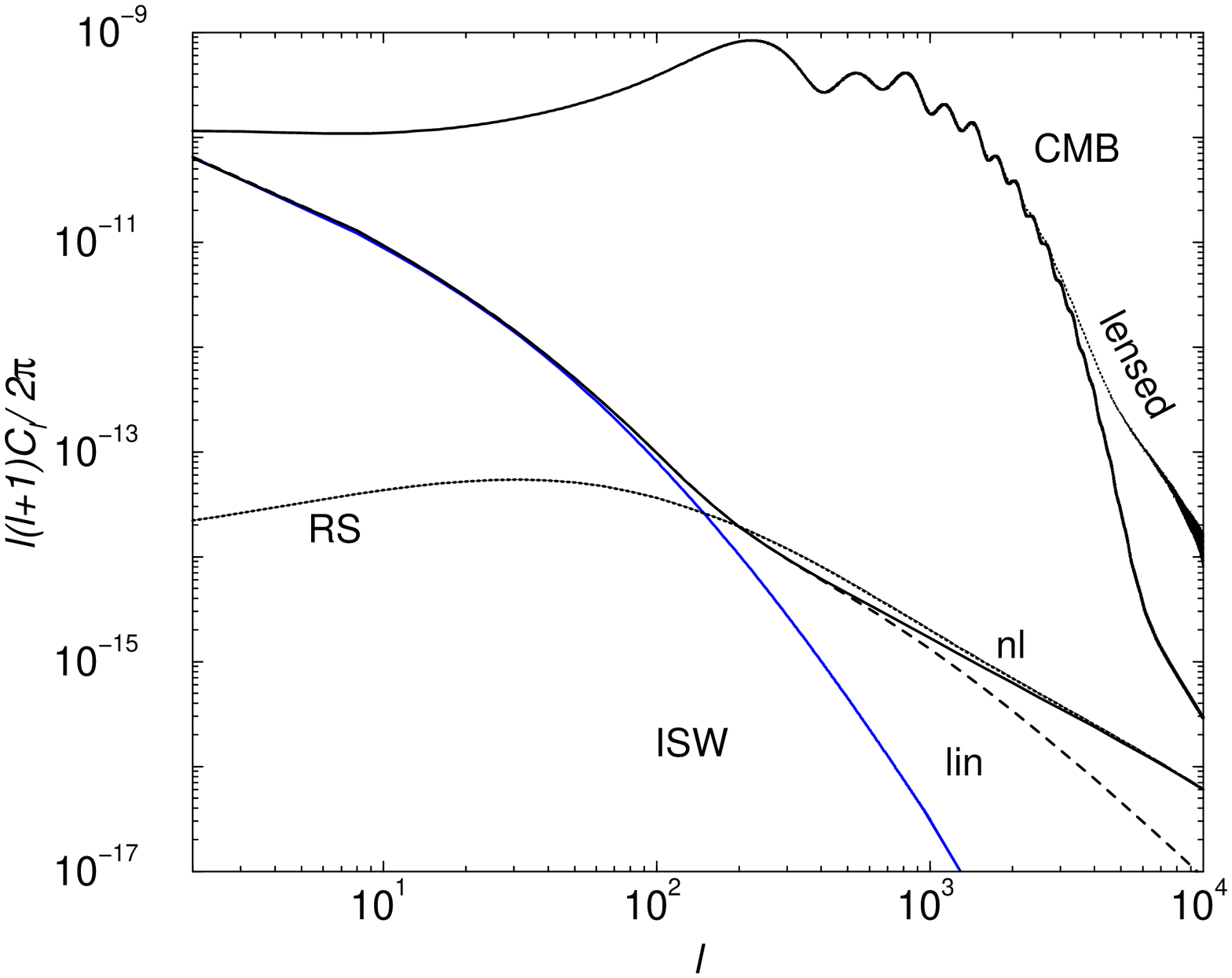,width=3.6in,angle=-90}}
\caption{The angular power spectrum of the full ISW effect, including
non-linear contribution. The contribution called Rees-Sciama (RS)
shows the non-linear extension, though for the total contribution, the
cross term between the momentum field and the density field leads to a
slight suppression between $l$ of 100 and 1000. The curve labeled ``nl''
is the full non-linear contribution while the curve labeled ``lin'' is
the contribution resulting from the momentum field under the second
order perturbation theory.}
\label{fig:clisw}
\end{figure}

In addition to the power spectrum of density derivatives, 
in equation~(\ref{eqn:non-linear}), we also
require the cross power spectrum between density derivatives and
density field itself $P_{\delta \dot{\delta}}$. In \S~\ref{sec:mom}, using
the halo approach as a description of the momentum density field,
we suggested that the cross-correlation between the density field and
the momentum field can be well described as
\begin{equation}
P_{p\delta}(k) = \sqrt{P_{pp}(k)P_{\delta\delta}(k)} \, .
\end{equation}
This is equivalent to the statement that the density and momentum density
fields are perfectly correlated
with a cross-correlation coefficient of 1; this relation is
exact at mildly-linear scales while at deeply non-linear scales
this perfect cross-correlation requires mass independent peculiar
velocity for individual halos \cite{Sheetal01b}.
Using this observation, we make the
assumption that $P_{\delta \dot{\delta}} \sim \sqrt{P_{\delta \delta}
P_{\dot{\delta} \dot{\delta}}}$, which is generally reproduced under
the halo model description of the cross-correlation between density
field and density field derivatives. This cross-term leads to a 10\%
reduction of power at multipoles between 100  and 1000, when compared
to the total when linear and non-linear contributions are
simply added.

In figure~\ref{fig:clisw}, we show the angular power spectrum of the ISW
effect with its
non-linear extension (which we have labeled RS for Rees-Sciama effect 
\cite{ReeSci68}).
The curve labeled ISW effect is the simple linear theory calculation with
a power
spectrum for potential derivatives given in
equation~(\ref{eqn:linear}). The curves labeled ``lin'' and ``nl'' shows 
the full non-linear
calculation following the description given in 
equation~(\ref{eqn:non-linear})
and using the linear theory or full non-linear power spectrum,
in equation~(\ref{eqn:momdiv}), for the density field, respectively.
For the non-linear density field power spectrum, we use the
halo approach for large scale structure clustering 
and calculate the power spectrum through a distribution of dark matter
halos. We use linear theory to
describe the velocity field in both linear and non-linear cases;
since the velocity field only contributes as an overall normalization,
through $v_{\rm rms}$,
its non-linear effects, usually at high $k$ values,
are not important due to the shape of the velocity power spectrum.

As shown in figure~\ref{fig:clisw},
the overall correction due to the non-linear ISW effect leads roughly two
orders of magnitude increase in power at $l \sim 1000$. The difference between
linear and non-linear theory density field power spectrum in
equation~(\ref{eqn:momdiv}), only
leads to at most an order of magnitude change in power.
Note that the curve labeled ``lin'' agrees with previous second order
perturbation theory calculations of the Rees-Sciama effect \cite{Sel96a},
while the
curve labeled ``nl'' is also consistent with previous estimates based on
results from
numerical simulations.

\section{Summary}

We have presented the halo approach to 
large scale structure clustering where we described the dark matter
distribution of the local universe through a 
collection of collapsed and virialized halos.
The statistical properties of the large scale structure 
can now be described through properties associated with these
halos, such as their spatial distribution and the distribution
of dark matter within these halos.
These halo properties are well studied either through analytical 
models or numerical simulations and include such necessary information
as the halo mass function, halo bias relative to linear density field
and the halo dark matter profile.

The halo approach to clustering essentially allows 
one to bridge the linear regime described by
perturbation theories to the non-linear regime described 
by clustering of dark matter
within halos. The perturbation theories fail to describe the
weakly to strongly non-linear regime completely, while, the
halo model predictions are in better agreement with
numerical results based on simulations.
Though statistically averaged measurements are
well produced by the halo based calculations, in detail, 
individual configurations of higher order correlations
are only produced at the 20\% level.
The uncertainties here are mostly due to assumptions in the 
current halo model calculations, such as the
use of spherical halos or ignoring the substructure within halos.

Though such uncertainties limit the accuracy of halo based calculations,
the approach has the advantage that
it can be easily extened to describe a wide variety  of large scale
structure properties. In this review, we have discussed stastical
aspects  involving the galaxy distribution, velocities and pressure.
In order to calculation statistical aspects associated with
these physical properties, we have introduced simple
descriptions involving how they relate
to dark matter within halos; almost all 
of these relations are  based on numerical simulation results.
Using these descriptions, we have discussed a wide number of 
applications of the halo model for non-linear
clustering including observations of the dark matter distribution 
via weak lensing,
galaxy properties via wide-field redshift and imaging 
surveys and applications to upcoming
cosmic microwave background anisotropy experiments.
The halo model has already become useful for several purposes, including
(1) understand why the galaxy clustering essentially produces a power-law
correlation function or a power spectrum, 
(2) estimate statistical biases in current and upcoming large
scale structure weak lensing surveys, and (3) calculate 
the full covariance matrix associated with certain large scale structure
observations, such as the angular correlation function of galaxies
in the Sloan Digital Sky Survey, among others.

\section*{Acknowledgments}   
We would like to acknowledge contributions from many of our 
collaborators, especially, Antonaldo Diaferio, Wayne Hu, 
Roman Scoccimarro and Giuseppe Tormen.  We thank Joerg Colberg, 
Andrew Connolly, Antonaldo Diaferio, Vincent Eke, Adrian Jenkins, 
Chung-Pei Ma, Julio Navarro, Alexandre Refregier,Roman Scoccimarro, 
Ryan Scranton, Uros Seljak and Martin White for use of their figures. 
We thank all the participants of the ``Workshop on Structure Formation 
and Dark Matter Halos'' at Fermilab in May, 2001 for useful discussions 
and initial suggestions with regards to topics covered in this review. 
We thank Marc Kamionkowski for inviting us to submit a review article 
on the halo model and for his help during the writing and editorial process.
At the initial stages of this work RKS was supported by the DOE and 
NASA grant NAG 5-7092 at Fermilab.  
AC is supported at Caltech by the Sherman Fairchild foundation and by
the DOE grant DE-FG03-92-ER40701.   We acknowledge extensive use of
the abstract server at the NASA's Astrophysics Data System
and the astro-ph preprint server and its archive.

\end{document}